\pdfoutput=1

\documentclass[11pt,a4paper]{article}
\usepackage[T1]{fontenc}
\usepackage[utf8]{inputenc}
\usepackage{lmodern,amsmath,amssymb,amsthm,mathtools}
\usepackage[margin=1.08in]{geometry}
\usepackage{booktabs,array,enumitem,microtype,xcolor,needspace,upquote,multicol}
\usepackage{tikz}
\usetikzlibrary{arrows.meta,positioning,calc}
\usepackage[hypcap=false]{caption}
\usepackage[unicode,colorlinks=true,linkcolor=blue!45!black,citecolor=blue!45!black,urlcolor=blue!45!black]{hyperref}
\setlist{nosep}
\makeatletter
\renewcommand*\@pnumwidth{2em}
\renewcommand*\l@section[2]{\addvspace{.15em}\@dottedtocline{1}{0pt}{2.4em}{\bfseries #1}{\bfseries #2}}
\makeatother
\newtheorem{theorem}{Theorem}[section]
\newtheorem{lemma}[theorem]{Lemma}
\newtheorem{proposition}[theorem]{Proposition}
\newtheorem{corollary}[theorem]{Corollary}
\newtheorem{maintheorem}{Theorem}

\theoremstyle{definition}
\newtheorem{definition}[theorem]{Definition}
\theoremstyle{remark}
\newtheorem{remark}[theorem]{Remark}
\numberwithin{equation}{section}
\DeclareMathOperator{\supp}{supp}
\DeclareMathOperator{\Tr}{Tr}

\newcommand{\Z}{\mathbb Z}
\newcommand{\N}{\mathbb N}
\newcommand{\R}{\mathbb R}
\newcommand{\eps}{\varepsilon}
\newcommand{\cE}{\mathcal E}
\newcommand{\cD}{\mathcal D}
\newcommand{\one}{\mathbf 1}
\colorlet{rxaccent}{orange!80!black}
\setlist[enumerate]{label=\textup{(\roman*)},leftmargin=2em,itemsep=3pt,topsep=4pt}
\setlist[itemize]{leftmargin=1.7em,itemsep=2pt,topsep=4pt}
\allowdisplaybreaks[2]
\hypersetup{pdftitle={Undecidable quantum storage: no algorithm can tell from the noise rate whether a fixed circuit remembers},pdfauthor={Seth Douglas and Nidhal Mghirbi},pdfsubject={Quantum storage and undecidability},pdfkeywords={quantum memory, fault tolerance, undecidability, noise threshold, cellular automata}}
\title{Undecidable quantum storage:\\[.3em]\small no algorithm can tell from the noise rate whether a fixed circuit remembers}
\author{\begin{tabular}{@{}c@{\qquad}c@{}}Seth Douglas & Nidhal Mghirbi\\[-1pt]\small\href{https://orcid.org/0009-0007-4708-3252}{0009-0007-4708-3252} & \small\href{https://orcid.org/0009-0005-6534-1118}{0009-0005-6534-1118}\end{tabular}\\[3pt]\normalsize\href{https://finite.works}{finite.works}}
\date{November 2026}
\begin{document}
\maketitle
\begin{abstract}
Below a threshold noise rate, error correction lets larger quantum memories store a qubit for longer, without limit. Can the noise rate alone tell us whether a given noisy circuit does this? We construct fixed local circuits for which no algorithm can, already in one dimension under single-qubit noise. Each circuit is translation-invariant and time-periodic, with depolarizing noise on every wire after every layer, including all classical control. A computable list of disjoint rate intervals is indexed by the inputs $n$ of a universal Turing machine; local noise statistics select the input that the circuit simulates, and halting triggers erasure. If the machine runs forever on $n$, a qubit is stored throughout the $n$-th interval at entanglement fidelity at least $15/16$ for a time that grows without bound with size: linearly on rings, exponentially in six dimensions. The encoding and decoding maps are noiseless and global; they depend on $n$, the system size and, for decoding, the time, but not on the rate. If it halts, the fidelity of every encoding falls below $1/2+\eta$, for any fixed $\eta>0$, after a time independent of size and rate. The labels of the intervals with unbounded storage form a $\Pi^0_1$-complete set, and intervals of both kinds accumulate at zero, so these circuits have no noise threshold. Without rate selection, the ring construction is a translation-invariant, time-periodic memory: from an initialized classical hierarchy it stores a qubit for a time linear in the ring length at every sufficiently small rate, addressing the memory part of a one-dimensional question raised by D\"unnweber, Styliaris and Trivedi. These are existence results: registers and interaction ranges are very large and the rates astronomically small.
\end{abstract}
\begin{center}\small Claude Opus 5.5, GPT-6.1 Sol, GPT-6.2 Sol and GPT-6.2 Astra assisted with this work; the authors take responsibility for every claim. Archive and Lean sources: \href{https://doi.org/10.5281/zenodo.24254845}{doi:10.5281/zenodo.24254845}.\end{center}
\makeatletter
\newcommand{\papercontents}{%
  \section*{\contentsname}%
  \begin{multicols}{2}%
  \footnotesize\renewcommand\@tocrmarg{2.55em plus 1fil}\hyphenpenalty=10000\exhyphenpenalty=10000
  \@starttoc{toc}%
  \end{multicols}%
}
\makeatother
\papercontents
\expandafter\gdef\csname r@cl:lem:ascent\endcsname{{6.3}{}{}{}{}}
\expandafter\gdef\csname r@cl:lem:cap-cut\endcsname{{6.2}{}{}{}{}}
\expandafter\gdef\csname r@cl:lem:clean\endcsname{{3.1}{}{}{}{}}
\expandafter\gdef\csname r@cl:lem:joint\endcsname{{4.6}{}{}{}{}}
\expandafter\gdef\csname r@cl:lem:locking\endcsname{{6.1}{}{}{}{}}
\expandafter\gdef\csname r@cl:lem:past\endcsname{{6.4}{}{}{}{}}
\expandafter\gdef\csname r@cl:lem:receiver-mixing\endcsname{{4.4}{}{}{}{}}
\expandafter\gdef\csname r@cl:lem:sparse-witness\endcsname{{4.2}{}{}{}{}}
\expandafter\gdef\csname r@cl:lem:stripes\endcsname{{4.5}{}{}{}{}}
\expandafter\gdef\csname r@cl:lem:two-period\endcsname{{5.24}{}{}{}{}}
\expandafter\gdef\csname r@cl:prop:organization\endcsname{{5.3}{}{}{}{}}
\expandafter\gdef\csname r@cl:def:sparse\endcsname{{4.1}{}{}{}{}}
\expandafter\gdef\csname r@cl:def:observations\endcsname{{5.1}{}{}{}{}}
\expandafter\gdef\csname r@cl:lem:endpoint\endcsname{{5.18}{}{}{}{}}
\expandafter\gdef\csname r@cl:lem:clock-witness\endcsname{{6.5}{}{}{}{}}
\expandafter\gdef\csname r@cl:sec:construction\endcsname{{3}{}{}{}{}}
\expandafter\gdef\csname r@cl:sec:native\endcsname{{3.1}{}{}{}{}}
\expandafter\gdef\csname r@cl:sec:clock\endcsname{{3.2}{}{}{}{}}
\expandafter\gdef\csname r@cl:sec:packets\endcsname{{3.3}{}{}{}{}}
\expandafter\gdef\csname r@cl:sec:verifier\endcsname{{3.4}{}{}{}{}}
\expandafter\gdef\csname r@cl:sec:parameter\endcsname{{3.5}{}{}{}{}}
\expandafter\gdef\csname r@cl:sec:embedding\endcsname{{3.6}{}{}{}{}}
\expandafter\gdef\csname r@cl:sec:receiver\endcsname{{4.2}{}{}{}{}}
\expandafter\gdef\csname r@cl:sec:stripes\endcsname{{4.3}{}{}{}{}}
\expandafter\gdef\csname r@cl:sec:joint\endcsname{{4.4}{}{}{}{}}
\expandafter\gdef\csname r@cl:sec:constants\endcsname{{A}{}{}{}{}}

\makeatletter
\renewcommand\@cite[2]{[{#1\if@tempswa ,\nobreakspace #2\fi}]}
\makeatother
\hyphenation{Corollary Corollaries}
\section{Introduction}\label{s:intro}
\subsection{The question and the answer}\label{intro:question}\label{intro:results}
Threshold theorems show that below a sufficiently small noise rate, error correction can extend quantum storage indefinitely as resources grow \cite{ABO2008,AGP2006,DKLP2002,Kitaev1997,KLZ1998}. For a fixed noisy circuit, can we tell from the noise rate whether storage time grows without bound as the system grows? We construct fixed circuits for which no algorithm can answer this question. Each circuit has a computable sequence of disjoint rate windows \(I_n\), nonempty open intervals with rational endpoints. Given a rational noise rate in these windows, it is undecidable whether the optimal lifetime, the last time at which some encoding and decoding still reach fidelity \(7/8\), is unbounded in the side length \(L\) (Corollary~\ref{cor:window-undecidable}). Windows of both kinds accumulate at zero, so the circuits have no positive noise threshold (Corollary~\ref{cor:no-threshold}).

The circuit never receives the rate as an input. Local noise statistics select a computation whose halting triggers erasure of the quantum data. The encoder is told the window index \(n\), but this does not let it defeat the rate selection: the forgetting bound holds for every encoding, including encodings that write another index or arbitrary data into the controller (Section~\ref{intro:nontrivial}). Threshold theorems are untouched, since they concern particular fault-tolerant constructions. What fails is the hope that storage is a computable function of a local circuit and its noise rate: the storage diagram of one fixed circuit switches between unbounded storage and forgetting infinitely often as the rate decreases. For prescribed encoding and decoding maps, failure of a storage guarantee at a rational rate has a finite exact witness; the universal guarantee cannot, in general, be certified by finitely many such tests (Section~\ref{a:discussion}). The circuits are existence proofs, not designs: their registers and ranges are very large, and their windows lie at astronomically small rates (Section~\ref{intro:scope}).

\paragraph{Contributions.} Every rate in a window selects the same machine input. On rings under independent single-qubit noise, one fixed circuit stores a qubit on nonhalting windows for a time linear in the ring length and forgets every encoding on halting windows (Theorem~\ref{q1:main}). Without rate selection, the same construction is a translation-invariant, time-periodic memory on rings in which every wire is noisy (Corollary~\ref{q1:memory}). In higher dimensions the storage lower bounds are exponential in six dimensions, stretched-exponential with a small exponent in four dimensions under whole-site noise, and linear in three dimensions under either noise model (Theorems~\ref{thm:exponential}--\ref{thm:main}). Every construction protects quantum data under faults correlated through the controller, and on halting windows every encoding forgets, including encodings into the controller or other work registers. The rate selection builds on our classical noise diagram, using version~1.4.0 \cite{DMNoise2026} in three to six dimensions and version~2.0.0 \cite{DMNoise1D2026} of the same archived preprint on rings, which also supplies the one-dimensional hierarchical clock; Appendix~\ref{app:imports} lists every imported statement. The observable here is a quantum channel's storage lifetime as size grows, distinct from the spectral gaps and ground-state phases studied for parameterized Hamiltonians \cite{BCW2021,CPGW2015}.

\paragraph{Setting.} Fix a universal Turing machine \(U\), and write \(U(n)\uparrow\) for nonhalting and \(U(n)\downarrow\) for halting. The circuit acts on the periodic lattice \((\Z/L\Z)^d\), repeating a fixed schedule of elementary layers in each round. Time counts rounds in Theorems~\ref{thm:exponential} and~\ref{thm:main} and elementary layers in Theorems~\ref{thm:four-dimensional} and~\ref{q1:main}. Fidelity is the overlap with the original Bell pair after encoding, storage and decoding on one half, leaving the reference untouched (Definition~\ref{def:fidelity}). Write \(F^*_{L,\eps}(t)\) for the best fidelity over all encoders and decoders on the full register. Our constructions guarantee \(15/16\) during storage, use \(7/8\) to define the optimal lifetime \(\tau_{L,\eps}\) and test certification, and bound halting-side fidelity eventually by \(1/2+\eta\) for every \(\eta>0\), near the measure-and-prepare benchmark \(1/2\) (Definitions~\ref{def:lifetime} and~\ref{def:windows}). The explicit encoding and final decoding maps are global, noiseless and total computable. They depend on \(n,L\) and, for decoding, on \(t\), independently of the rate and the halting answer. The encoder prepares the classical control configuration as well as the code state. Correction during storage is local.

Whole-site depolarization (WS) replaces the entire site register with a maximally mixed state with probability \(\eps\), independently at each site after each elementary layer. Independent single-qubit depolarization (Q) does the same separately for each wire. Both act on every factor, including idle, controller and work factors (Section~\ref{s:local-circuits}). An \emph{admissible circuit} on \(\Z^d\) is computably specified, translation-invariant, time-periodic and of finite range, with elementary operations in \(\{X,H,\mathrm{CNOT},\mathrm{Toffoli},R_0,D_Z,I\}\) (Section~\ref{s:local-circuits} defines reset \(R_0\) and dephasing \(D_Z\)).

\begin{maintheorem}[Six dimensions, exponential storage]\label{thm:exponential}
There is a rational \(c_{\mathrm e}>0\) such that, for each of \textup{WS} and \textup{Q} applied after every elementary layer, there is a separately compiled admissible circuit on \(\Z^6\) with computable disjoint rational windows \(I_n\) for which the following hold.
\begin{enumerate}
\item If \(U(n)\uparrow\), the explicit noiseless encoder \(\cE_{n,L}\) and time-dependent decoder \(\cD_{n,L,t}\), total computable and independent of the rate, achieve entanglement fidelity at least \(15/16\) at every rate in \(I_n\) and every integer round \(0\le t\le 2^{\lfloor c_{\mathrm e}L\rfloor}\) for all \(L\) at or above a computable size cutoff \(L_{\mathrm e}(n)\).
\item If \(U(n)\downarrow\), then for every \(\eta>0\) there is a finite \(C_{n,\eta}\), independent of \(L\) and the rate in \(I_n\), such that \(F^*_{L,\eps}(t)\le 1/2+\eta\) for every \(L\), every \(\eps\in I_n\) and every integer \(t\ge C_{n,\eta}\).
\end{enumerate}
\end{maintheorem}

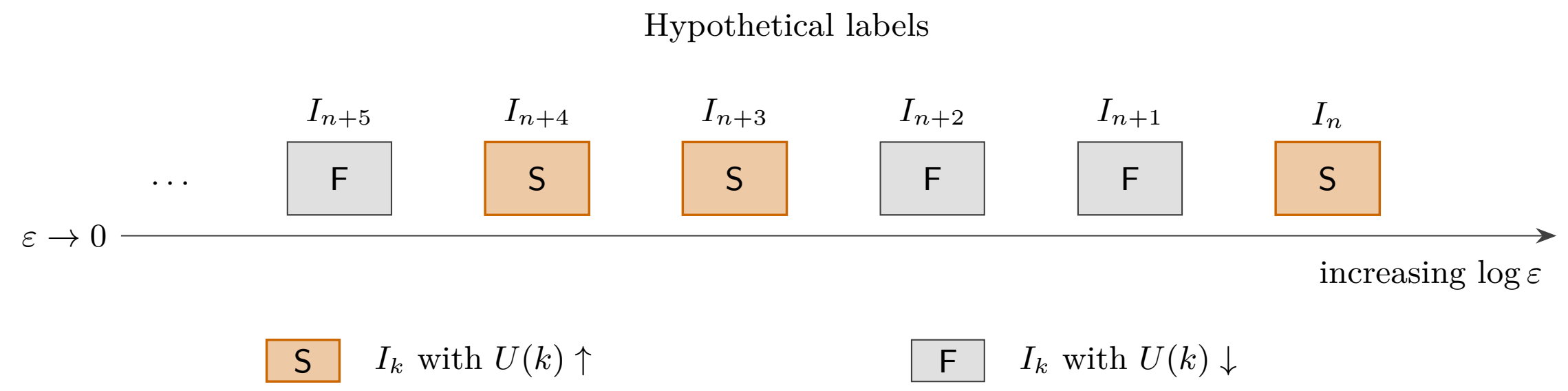
\begin{figure}[!tb]
\centering\begin{tikzpicture}[x=1mm,y=1mm,font=\footnotesize,bar/.style={draw=black!75,fill=black!12,line width=.4pt},store/.style={draw=rxaccent,fill=rxaccent!35,line width=.65pt}]
\draw[-{Stealth[length=2mm]},black!75] (7,0)--(145,0);
\node[anchor=east] at (7,0) {$\eps\to0$};
\node[anchor=north east] at (145,-1) {increasing $\log\eps$};
\foreach \x/\lab/\state/\sym in {23/{n+5}/bar/F,42/{n+4}/store/S,61/{n+3}/store/S,80/{n+2}/bar/F,99/{n+1}/bar/F,118/{n}/store/S}{
 \draw[\state] (\x,2) rectangle +(10,7);
 \node at (\x+5,5.5) {\textsf{\sym}};
 \node[above] at (\x+5,9) {$I_{\lab}$};
}
\node at (71,20) {Hypothetical labels};
\node at (12,5) {$\cdots$};
\draw[store] (21,-14) rectangle +(7,4);
\node at (24.5,-12) {\textsf{S}};
\node[anchor=west] at (30,-12) {$I_k$ with $U(k)\uparrow$};
\draw[bar] (83,-14) rectangle +(7,4);
\node at (86.5,-12) {\textsf{F}};
\node[anchor=west] at (92,-12) {$I_k$ with $U(k)\downarrow$};
\end{tikzpicture}
\caption{Six windows of Theorem~\ref{thm:exponential} on a schematic logarithmic axis, with hypothetical outcomes (S: storing, F: forgetting). Windows of both kinds accumulate at zero; rates between windows are unclassified.}\label{fig:rates}
\end{figure}

The full quantifiers appear in Theorem~\ref{v:detailed}. The estimate in Lemma~\ref{v:bridge} also gives fidelity tending to one with size throughout the stated horizon. Proposition~\ref{v:certification} gives the upper bound and certification for these circuits. Figure~\ref{fig:rates} illustrates the rate windows.

\begin{maintheorem}[Four dimensions, stretched-exponential storage]\label{thm:four-dimensional}
For \textup{WS} applied after every elementary layer, there is an admissible circuit on \(\Z^4\) with computable disjoint rational windows \(I_n\) and rationals \(c_4>0\) and \(0<\alpha_4<2\times10^{-4}\), independent of \(n\), such that the following hold.
\begin{enumerate}
\item If \(U(n)\uparrow\), the explicit noiseless encoder \(\cE_{n,L}\) and time-dependent decoder \(\cD_{n,L,t}\), total computable and independent of the rate, achieve entanglement fidelity at least \(15/16\) at every rate in \(I_n\) and every integer elementary-layer time \(0\le t\le \exp(c_4L^{\alpha_4})\) for all \(L\) at or above a computable size cutoff \(L_4(n)\).
\item If \(U(n)\downarrow\), then for every \(\eta>0\) there is a finite \(C_{n,\eta}\), independent of \(L\) and the rate in \(I_n\), such that \(F^*_{L,\eps}(t)\le 1/2+\eta\) for every \(L\), every \(\eps\in I_n\) and every integer \(t\ge C_{n,\eta}\).
\end{enumerate}
\end{maintheorem}

The detailed statement is Theorem~\ref{f4:detailed}, and the bound on \(\alpha_4\) is computed in Section~\ref{f4:constants}. The lower bound includes intermediate elementary layers as well as round boundaries.

\begin{maintheorem}[Three dimensions, linear scale]\label{thm:main}
For each of \textup{WS} and \textup{Q} applied after every elementary layer, there is a separately constructed admissible circuit on \(\Z^3\) with computable disjoint rational windows \(I_n\). For the rational \(c>0\) and computable \(C_0\ge 2\) of Theorem~\ref{thm:detailed}, both independent of \(n\), the optimal lifetime \(\tau_{L,\eps}\) in rounds (Definition~\ref{def:lifetime}) and the optimal fidelity satisfy the following bounds.
\begin{enumerate}
\item If \(U(n)\uparrow\), then \(\lfloor cL\rfloor\le\tau_{L,\eps}<201L+C_0\) for every \(L\ge L_C(n)\), uniformly in \(\eps\in I_n\), where \(L_C(n)\) is the computable size cutoff defined after Theorem~\ref{thm:detailed}.
\item If \(U(n)\downarrow\), then for every \(\eta>0\) there is a finite \(C_{n,\eta}\), independent of size and rate, such that \(F^*_{L,\eps}(t)\le1/2+\eta\) for every \(L\), every \(\eps\in I_n\) and every integer \(t\ge C_{n,\eta}\).
\end{enumerate}
For the constants \(c\) and \(L_0(n)\) of Theorem~\ref{thm:detailed}, the set of indices \(n\) for which the explicit encoding and recovery achieve fidelity at least \(7/8\) at every rational rate in \(I_n\), every integer \(L\ge L_0(n)\) and every integer \(0\le t\le\lfloor cL\rfloor\) is \(\Pi^0_1\)-complete.
\end{maintheorem}
The certification statement uses the cutoff \(L_0(n)\) of the explicit storage guarantee alone; \(L_C(n)\ge L_0(n)\) also covers the lifetime upper bound.

\Needspace{24\baselineskip}
\begin{maintheorem}[One dimension, single-qubit noise]\label{q1:main}
There is one admissible circuit on rings, with finitely many qubits per site, independent single-qubit depolarization after every elementary layer, computable pairwise disjoint nonempty open rational rate windows \(I_n\), a rational \(c_1>0\), and a computable size cutoff \(L_1(n)\), with these properties.
\begin{enumerate}
\item If \(U(n)\uparrow\), the explicit noiseless encoding and decoding maps of Section~\ref{q1:storage}, total computable from \(n,L\) and, for decoding, \(t\), and independent of the rate, achieve entanglement fidelity at least \(15/16\) for every \(\eps\in I_n\), every \(L\ge L_1(n)\), and every integer elementary-layer time \(0\le t\le\lfloor c_1L\rfloor\).
\item If \(U(n)\downarrow\), then for every \(\eta>0\) there is a finite \(C_{n,\eta}\), uniform in size and rate, such that \(F^*_{L,\eps}(t)\le1/2+\eta\) for every \(L\ge1\), every \(\eps\in I_n\), and every integer elementary-layer time \(t\ge C_{n,\eta}\).
\end{enumerate}
The scalable-memory window labels, in the sense of Definition~\ref{def:windows}, are \(\Pi^0_1\)-complete. The same is true of the labels for which these encoding and decoding maps have entanglement fidelity at least \(7/8\) at every rational rate in the window, every \(L\ge L_1(n)\), and every \(0\le t\le\lfloor c_1L\rfloor\).
\end{maintheorem}
The proof gives the larger fidelity \(123/128\): it loses \(1/128\) in preparing the selector's helper marginal (Lemma~\ref{q1:l15}) and at most \(1/32\) to failures at the top of the hierarchy (Lemma~\ref{q1:l18}).

\begin{corollary}[A translation-invariant periodic memory on rings]\label{q1:memory}
There is an admissible circuit on rings, with a fixed finite number of qubits per site, rationals \(c_{\rm m},\eps_{\rm m}>0\), and noiseless encoders and decoders, total computable from \(L\) and, for decoding, \(t\), and independent of the rate, and a size cutoff \(L_{\rm m}\), such that for every rate \(0<\eps\le\eps_{\rm m}\) of independent single-qubit depolarization after every elementary layer, every \(L\ge L_{\rm m}\) and every integer elementary-layer time \(0\le t\le\lfloor c_{\rm m}L\rfloor\) they preserve entanglement fidelity at least \(15/16\). The encoder initializes the classical clock hierarchy; all storage-time control and correction are local, and every wire, including all control, is subject to the same noise.
\end{corollary}

The proofs occupy Sections~\ref{q1:model}--\ref{q1:effective}. Both classical lanes run the one-dimensional construction of \cite{DMNoise1D2026}, which adapts G\'acs's one-dimensional fault-tolerant automaton \cite{Gacs1986} and is used here from an initialized hierarchy. Its properties are proved in \cite{DMNoise1D2026}. For the ring proof,
Lean checks Lemmas~\ref{q1:l1} and~\ref{q1:l1.1}, selected exponent
calculations from Lemmas~\ref{q1:l11} and~\ref{q1:l12}, and the
recursion and horizon arithmetic and a conditional fidelity calculation
from Lemma~\ref{q1:l18} (Appendix~\ref{app:lean}).
One-dimensional fault tolerance was proved earlier for compiled, non-uniform schedules \cite{ABO2008,Gottesman2000,GB2025}. D\"unnweber, Styliaris and Trivedi give the closest architecture, a fixed, local, translation-invariant two-dimensional automaton with noisy hierarchical classical control and a measurement-free concatenated code \cite{DST2026}. They ask whether a similar scheme exists in one dimension and point to G\'acs's one-dimensional automaton. Corollary~\ref{q1:memory} addresses the memory part of that question from an initialized hierarchy, with a lifetime bound linear in \(L\); their two-dimensional scheme also computes, and suppresses logical errors exponentially in system size. Section~\ref{intro:related} compares Corollary~\ref{q1:memory} with the other one-dimensional constructions.

Table~\ref{tab:results} summarizes the four guarantees.

\begin{table}[!tb]
\caption{Storage guarantees at fidelity \(15/16\) and upper bounds on optimal lifetime at fidelity \(7/8\), on nonhalting windows. Each lower bound holds above a computable size cutoff. The displayed constants are fixed for the indicated circuits, and \(O_n\) permits dependence on the window index \(n\). On a halting window each circuit satisfies \(F^*\le1/2+\eta\) after a finite time uniform in size and rate, for every \(\eta>0\). The scalable-memory labels column concerns scalability of the optimal lifetime (Corollary~\ref{cor:window-undecidable} and Theorem~\ref{q1:main}).}\label{tab:results}
\centering\small\setlength{\tabcolsep}{5pt}
\begin{tabular}{@{}>{\raggedright\arraybackslash}p{.16\linewidth}>{\raggedright\arraybackslash}p{.23\linewidth}>{\raggedright\arraybackslash}p{.29\linewidth}>{\raggedright\arraybackslash}p{.23\linewidth}@{}}
\toprule Circuit and noise & Guaranteed lower bound & Lifetime upper bound & Scalable-memory labels\\\midrule
6D (A), WS, Q & \(2^{\lfloor c_{\rm e}L\rfloor}\) rounds & \(\exp(O_n(L^4))\) rounds by resets of four-dimensional code fibres (Proposition~\ref{v:certification}(i)) & \(\Pi^0_1\)-complete\\\addlinespace[4pt]
4D (B), WS & \(\exp(c_4L^{\alpha_4})\) elementary layers, \(\alpha_4<2\times10^{-4}\) & \(\exp(O_n(L^4))\) elementary layers by full-register reset \eqref{f4:lifetime-upper} & \(\Pi^0_1\)-complete\\\addlinespace[4pt]
3D (C), WS, Q & \(\lfloor cL\rfloor\) rounds & \(<201L+C_0\) rounds for large \(L\) (Theorem~\ref{u:eb}) & \(\Pi^0_1\)-complete\\\addlinespace[4pt]
1D (D), Q & \(\lfloor c_1L\rfloor\) elementary layers & \(\exp(O_n(L))\) elementary layers by full-register replacement (Section~\ref{q1:effective}) & \(\Pi^0_1\)-complete\\\bottomrule
\end{tabular}
\end{table}

\subsection{What is engineered and what is hard}\label{intro:nontrivial}
Each construction uses one fixed local, translation-invariant rule, independent of \(n\). The noise rate is its only variable physical parameter. Given a rate in \(\bigcup_n I_n\), the index \(n\) is found by testing membership in the computable windows. The circuit must sense the rate while subject to the same noise, correct quantum errors correlated with faults in the classical controller, and bound the recoverable fidelity even when the encoding uses the full register. The forgetting bound also applies when the initial state uses a different index or puts information into the controller or work registers.

The windows are computable whether or not \(U(n)\) halts, so their rational midpoints suffice for the reduction. For each fixed rational \(0<\eta<3/8\), the halting-side time to \(F^*\le1/2+\eta\) of the circuits of Theorems~\ref{thm:exponential}--\ref{thm:main} has no total computable bound in \(n\) (Corollary~\ref{a:time}, Proposition~\ref{v:certification}(iii) and the end of Section~\ref{f4:forgetting}). The limit on \(\eta\) puts the forgetting bound strictly below the lifetime fidelity \(7/8\).

\subsection{How it works}\label{intro:mechanism}
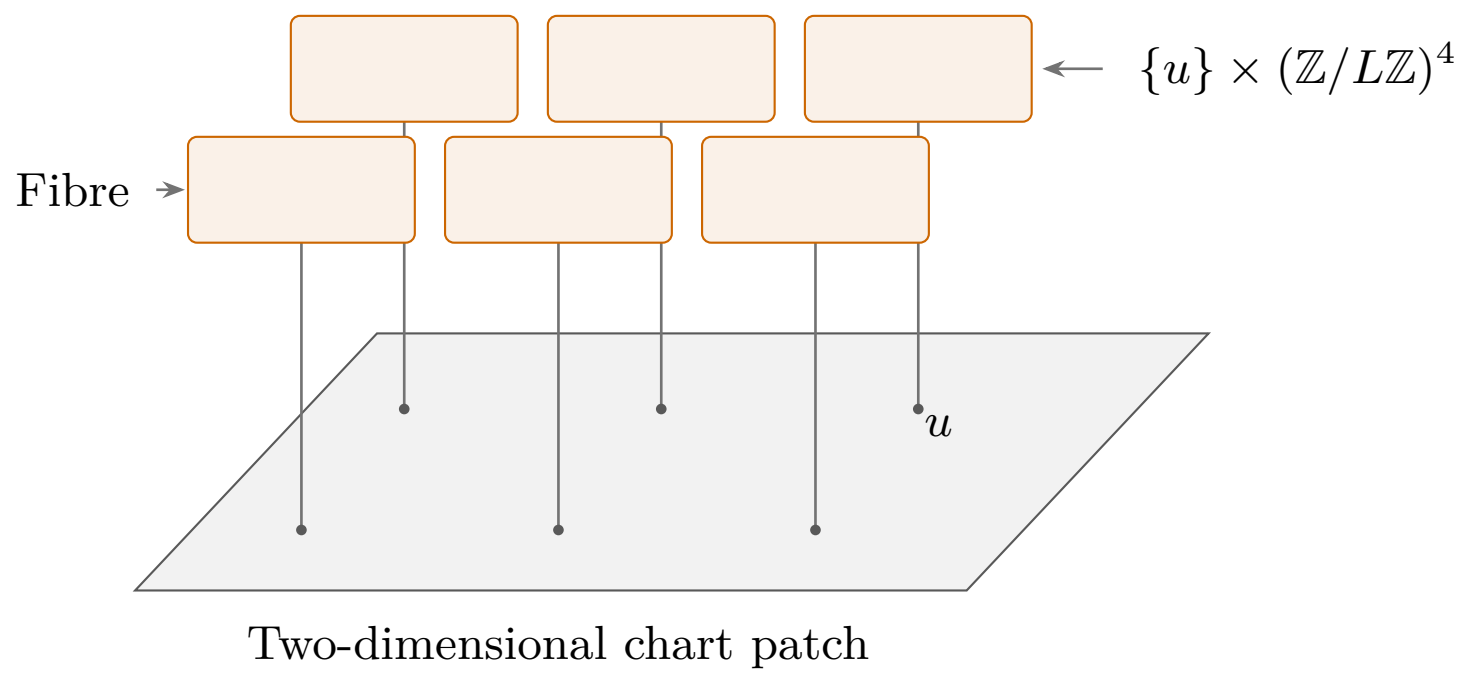
\begin{figure}[!tb]
\centering\begin{tikzpicture}[x=1mm,y=1mm,font=\footnotesize,
  fibre/.style={draw=rxaccent,fill=rxaccent!9,rounded corners=.6mm,minimum width=15mm,minimum height=7mm},
  edge/.style={black!55,line width=.5pt}]
\filldraw[fill=black!5,draw=black!65] (-4,-4)--(51,-4)--(67,13)--(12,13)--cycle;
\foreach \x in {7,24,41} {
  \foreach \y in {8,0} {
    \coordinate (base) at ({\x+.85*\y},\y);
    \draw[edge] (base)--++(0,19);
    \fill[black!65] (base) circle (1pt);
    \node[fibre] at ($(base)+(0,22.5)$) {};
  }
}
\node[anchor=east] at (-3,22.5) {Fibre};
\draw[edge,-{Stealth[length=1.5mm]}] (-2.6,22.5)--(-0.7,22.5);
\node[anchor=north] at (24,-5) {Two-dimensional chart patch};
\node[anchor=north west,inner sep=1pt] at (47.8,8) {$u$};
\node[anchor=west] at (61,30.5) {$\{u\}\times(\Z/L\Z)^4$};
\draw[edge,-{Stealth[length=1.5mm]}] (60,30.5)--(56,30.5);
\end{tikzpicture}
\caption{Geometry of the six-dimensional circuit. Two coordinates locate the classical chart patch. Holding a chart point \(u\) fixed leaves a four-dimensional toric-code fibre \(\{u\}\times(\Z/L\Z)^4\). The boxes represent fibres schematically.}\label{fig:chart-fibres}
\end{figure}

\begin{figure}[!tb]
\centering\begin{tikzpicture}[x=1mm,y=1mm,font=\footnotesize,
box/.style={draw=black!65,fill=black!6,rounded corners=.8mm,align=center,minimum height=10mm},
arr/.style={-{Stealth[length=2mm]},line width=.55pt}]
\node[box,text width=28mm] (rate) at (15,20) {Noise samples\\from sampling\\registers};
\node[box,text width=32mm] (input) at (56,20) {Tests check the\\rate scale for $n$};
\node[box,text width=38mm] (run) at (103,20) {Check the finite static\\chart for $U(n)$};
\node[box,text width=33mm] (defect) at (48,0) {Failed rate test,\\local chart defect\\or counter expiry};
\node[box,text width=25mm,draw=black!65,fill=black!6] (erase) at (48,-19) {Erase\\quantum data};
\node[box,text width=42mm,draw=rxaccent,fill=rxaccent!12] (store) at (110,0) {Local correction protects\\the qubit through the\\guaranteed horizon};
\draw[arr] (rate)--(input);
\draw[arr] (input)--(run);
\draw[arr] (run.south west)--(defect.north east);
\draw[arr] (input.south)--(defect.north);
\draw[arr,draw=rxaccent] (run.south)--(store.north);
\draw[arr] (defect.south)--(erase.north);
\end{tikzpicture}
\caption{One round of the six-dimensional circuit. Noise samples from sampling registers test the rate scale, and the circuit checks a finite static chart for \(U(n)\) prepared by the encoder. A failed rate test, a local chart defect or counter expiry can trigger erasure. On a nonhalting window, local correction protects the qubit through the guaranteed horizon (orange branch). Noise follows every elementary layer, including those inside each box.}\label{fig:round}
\end{figure}
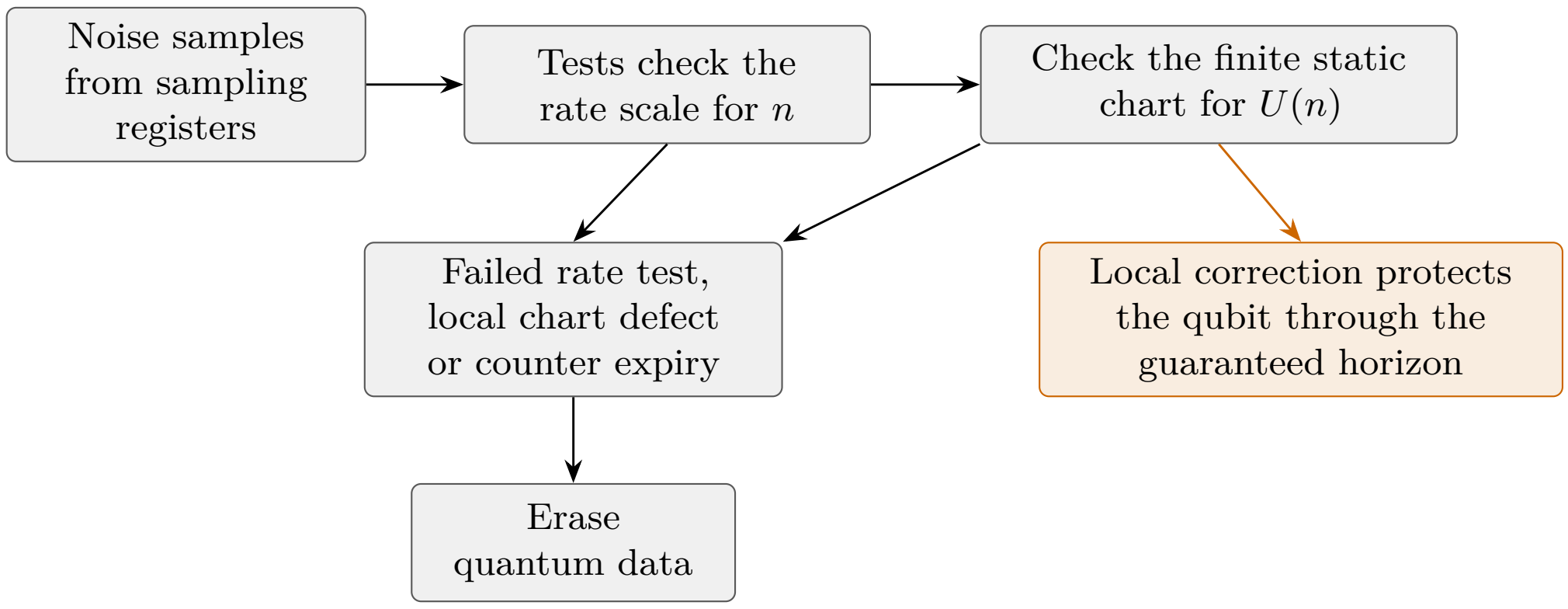

The three- and six-dimensional constructions use the same rate-selection strategy; the ring construction is described at the end of this subsection. The computation chart is a set of local records of a simulated run. It occupies space-time in three dimensions and a static square in six dimensions (Figure~\ref{fig:chart-fibres}). Table~\ref{tab:mechanism} follows the two implementations. Figure~\ref{fig:round} shows one round in six dimensions.

\begin{table}[!tb]
\centering\small
\caption{How the three- and six-dimensional circuits select, protect and erase.}\label{tab:mechanism}
\begin{tabular}{@{}>{\raggedright\arraybackslash}p{.14\linewidth}>{\raggedright\arraybackslash}p{.4\linewidth}>{\raggedright\arraybackslash}p{.4\linewidth}@{}}
\toprule Step & Three dimensions & Six dimensions\\\midrule
Sample noise & Copy sampling registers along rows on a space-time chart (Section~\ref{c:chart}). & Copy sampling registers along rows on a static chart (Section~\ref{v:classical-rule}).\\\addlinespace[3pt]
Select a computation & Local tests keep the depth selected by the rate. On \(I_n\) that depth runs \(U(n)\) (Section~\ref{sec:controller}). & The encoder supplies a candidate chart for \(U(n)\). The circuit checks it with local rate tests that can reject a chart at the wrong depth, even if the encoder supplied it (Sections~\ref{v:classical-rule} and~\ref{v:parameters}).\\\addlinespace[3pt]
Protect the qubit & A protected classical clock schedules local correction on a concatenated-code chain (Section~\ref{sec:quantum}). & At each chart point, local correction acts on a four-dimensional toric-code fibre. The reference square has side \(N_{\rm cap}=\lfloor L/4\rfloor\). Its diagonal counter keeps storage enabled for at least \(2^{N_{\rm cap}}-1\) clean rounds (Sections~\ref{v:classical-rule} and~\ref{v:positive}).\\\addlinespace[3pt]
Erase on halting & A chart inconsistency or failed test propagates and triggers conditional resets (Sections~\ref{sec:controller} and~\ref{s:terminal-erasure}). & A defect, failed rate test or counter expiry triggers conditional resets. A surviving square of side \(N_{\rm cap}\) has clean expiry within \(1+2^{N_{\rm cap}+1}\) rounds, and on a halting window its side is bounded independently of \(L\) (Sections~\ref{v:classical-rule} and~\ref{v:forgetting}).\\\addlinespace[3pt]
Bound all recoverable fidelity & A coupling uses disjoint input blocks to give equal channel marginals. The cloning bound controls every encoding (Section~\ref{sec:forgetting}). & The same coupling and cloning argument applies after bounding the survival of dependence on each input site (Section~\ref{v:forgetting}).\\
\bottomrule
\end{tabular}
\end{table}

\paragraph{How the rate picks a depth.}
A sampling lane starts low and can be refreshed by noise as it is copied along a row. At depth \(r\) the row length is \(\ell_r=2\cdot4^r-1\), so increasing \(r\) by one increases the number of refresh opportunities by about four. For example, in the selector's comparison model with aggregate refresh \(\mu=1/1000\) (illustrative values; the actual windows lie far below), rows of lengths \(31\) and \(127\) have probabilities \(1-(1-\mu)^{31}\) and \(1-(1-\mu)^{127}\) of seeing a refresh, about \(3\%\) and \(12\%\). A count across many lanes distinguishes short from long rows. The actual tests include the refresh bias, finite-copy effects and noise margins and agree on the selected depth throughout \(I_n\) (Section~\ref{c:chart} and Proposition~\ref{c:selector-import}).

In three dimensions a live chart must be refreshed along successive time diagonals. On a torus of side \(L\), a sufficiently late live reading with no marked error would force a wide, locally legal diagonal strip periodic by \((L,-L)\). The tiling obstruction rules out that strip, and the probability that faults sustain a live reading instead gives the order-\(L\) upper bound (Section~\ref{sec:ceiling}). In six dimensions the static chart avoids this obstruction. The positive estimates count connected configurations of distinct faults to control correlated unwanted resets (Section~\ref{v:positive}). The four-dimensional circuit uses a static chart with a folded concatenated-code chain scheduled by a protected classical clock (Section~\ref{sec:four-dimensional}).

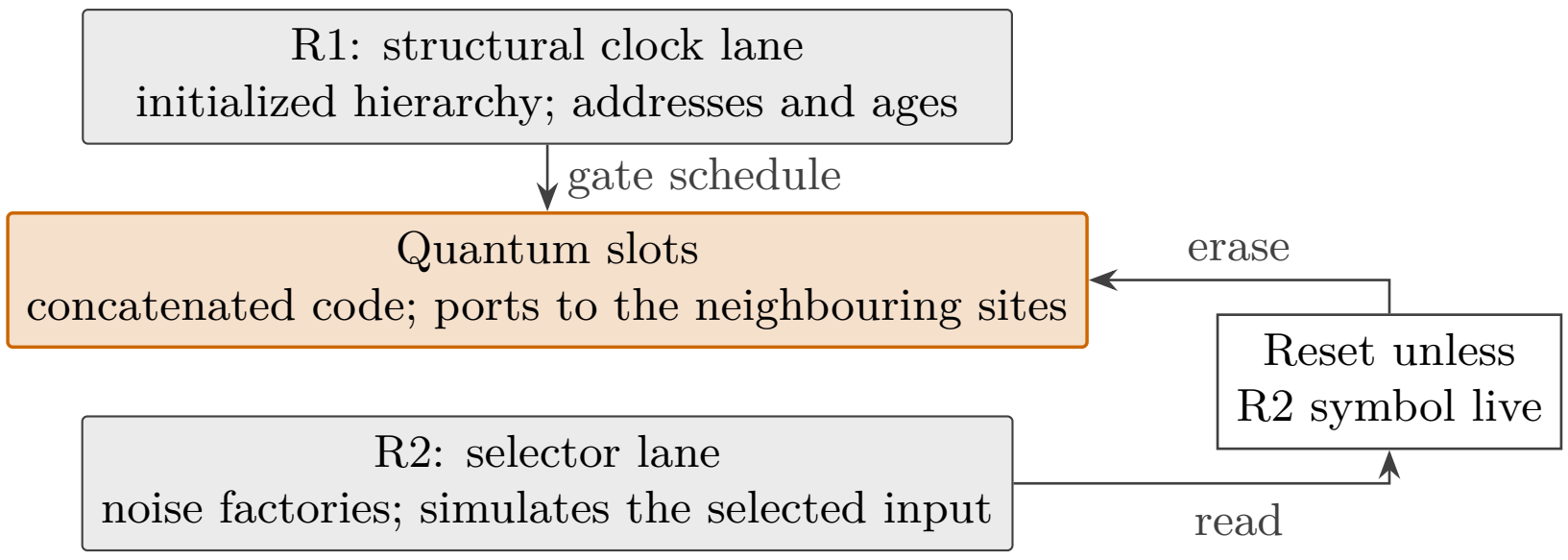
\begin{figure}[!tb]
\centering\begin{tikzpicture}[x=1mm,y=1mm,font=\footnotesize,
 lane/.style={draw=black!75,fill=black!8,line width=.4pt,rounded corners=1pt,minimum width=64mm,minimum height=9mm,align=center},
 qlane/.style={draw=rxaccent,fill=rxaccent!20,line width=.65pt,rounded corners=1pt,minimum width=64mm,minimum height=9mm,align=center},
 gate/.style={draw=black!75,fill=white,line width=.5pt,minimum width=22mm,minimum height=9mm,align=center},
 arr/.style={-{Stealth[length=2mm]},black!75,line width=.5pt}]
\node[lane] (r1) at (0,28) {R1: structural clock lane\\initialized hierarchy; addresses and ages};
\node[qlane] (q) at (0,14) {Quantum slots\\concatenated code; ports to the neighbouring sites};
\node[lane] (r2) at (0,0) {R2: selector lane\\noise factories; simulates the selected input};
\node[gate] (e) at (58,7) {Reset unless\\R2 symbol live};
\draw[arr] (r1.south) -- node[right,align=left]{gate schedule} (q.north);
\draw[arr] (r2.east) -| node[pos=.3,below]{read} (e.south);
\draw[arr] (e.north) |- node[pos=.75,above]{erase} (q.east);
\end{tikzpicture}
\caption{One site of the ring circuit of Theorem~\ref{q1:main}, schematically. The structural lane R1 schedules the gates on the quantum slots; the selector lane R2 decides each round whether the slots are reset. Depolarization follows every elementary layer on every wire, including both classical lanes.}\label{fig:ring}
\end{figure}

\paragraph{How it works on rings.}
Each site carries two classical lanes and quantum slots, all under single-qubit depolarization (Figure~\ref{fig:ring}). The structural lane R1 is a one-dimensional hierarchical clock: it runs the classical construction of \cite{DMNoise1D2026} on a fixed never-halting input, is initialized by the encoder, and its addresses and ages tell each site which gate of a concatenated-code schedule to apply at every level of the hierarchy (Sections~\ref{q1:clock} and~\ref{q1:compiler}). The selector lane R2 runs the same classical construction on the input selected by the rate. Redundant circuits turn the physical depolarization into the exact random replacements that construction requires, and on \(I_n\) the replacement rate lies in the native bin for which it simulates \(U(n)\) (Section~\ref{q1:selector}). Each round, a site resets all its quantum slots unless it reads a live symbol of the selected computation. If \(U(n)\) halts, live symbols survive a burn-in independent of size and rate only inside rare marked balls, and the coupling and cloning argument of Section~\ref{sec:forgetting} gives forgetting for every encoding (Section~\ref{q1:forgetting}). If \(U(n)\) runs forever, the code is protected through one tick of the top level of the hierarchy that fits in the ring. That tick lasts order \(L\) layers, which is why the lower bound is linear (Section~\ref{q1:storage}). The decoder completes the current top tick by simulating the remaining layers at a fixed nominal rate inside the window, so that the whole run has a single change of rate within the window, which every estimate allows; it then applies the fixed hierarchical recovery. Replacing the R2 operations and the conditional erasure block by identity layers gives Corollary~\ref{q1:memory}.

\subsection{No threshold and undecidability on rational rates}\label{intro:threshold}
\begin{corollary}[No positive noise threshold]\label{cor:no-threshold}
For each circuit of Theorems~\ref{thm:exponential}--\ref{q1:main}, below every positive noise rate there are a nonhalting window, on which storage grows without bound in \(L\), and a halting window, on which every encoding forgets after a time independent of size and rate. Thus no positive interval adjacent to zero consists entirely of scalable-memory rates, those with \(\sup_L\tau_{L,\eps}=\infty\) (Definitions~\ref{def:lifetime} and~\ref{def:windows}).
\end{corollary}
\begin{proof}
The windows tend to zero: for Theorem~\ref{thm:exponential} by Section~\ref{v:parameters}, for Theorem~\ref{thm:four-dimensional} by Lemma~\ref{f4:encoder}, for Theorem~\ref{thm:main} by \eqref{eq:windows}, and for Theorem~\ref{q1:main} because its windows lie in monotone preimages of native bins that tend to zero (Section~\ref{q1:parameters}). The halting set of the universal machine \(U\) is undecidable. A finite or cofinite halting set would be decidable, so both sets of labels are unbounded. For any positive rate, choose a label of each kind sufficiently far along the sequence. Parts~(i) and~(ii) of Theorems~\ref{v:detailed}, \ref{f4:detailed}, \ref{thm:detailed} and~\ref{q1:main} give the two conclusions.
\end{proof}
Figure~\ref{fig:rates} illustrates the interleaving of the windows. The storage guarantees have the scales in Table~\ref{tab:results}. The absence of a positive threshold follows from the interleaving alone. The next corollary states the undecidability claim for each higher-dimensional circuit; Theorem~\ref{q1:main} states it on rings.

\begin{corollary}[Undecidability on rational rates]\label{cor:window-undecidable}
For each circuit of Theorems~\ref{thm:exponential}, \ref{thm:four-dimensional}
and~\ref{thm:main}, the labels \(n\) of scalable-memory windows (Definition~\ref{def:windows}),
those with \(\sup_L\tau_{L,\eps}=\infty\) for every \(\eps\in I_n\),
are exactly the nonhalting inputs of \(U\), so this set is \(\Pi^0_1\)-complete.
The lifetime counts rounds in Theorems~\ref{thm:exponential} and~\ref{thm:main} and elementary layers in Theorem~\ref{thm:four-dimensional}
(Definitions~\ref{def:round} and~\ref{def:lifetime}). For the promise problem with rational input
\(q\in\bigcup_n I_n\), no algorithm decides whether \(\sup_L\tau_{L,q}=\infty\).
\end{corollary}
\begin{proof}
Theorems~\ref{v:detailed}, \ref{f4:detailed} and~\ref{thm:detailed}
give unbounded lifetimes on nonhalting windows and, with \(\eta=1/16\),
uniformly bounded lifetimes on halting windows. Thus the set of scalable-memory window labels
is the nonhalting set of \(U\), as in Theorem~\ref{a:complete}
and Proposition~\ref{v:certification}. The rational midpoint of \(I_n\)
is computable from \(n\), giving the reduction for rational rates.
\end{proof}
Every storing window in Theorems~\ref{thm:exponential}, \ref{thm:four-dimensional} and~\ref{thm:main} has a horizon diverging with \(L\), so it is a scalable-memory window. Scalability concerns the optimal lifetime over all encodings. For the explicit maps in Theorems~\ref{thm:exponential} and~\ref{thm:main}, a separate \(\Pi^0_1\)-complete predicate requires every prescribed exact finite test to pass (Proposition~\ref{v:certification}(ii) and Theorem~\ref{a:complete}). On a halting window, enumerating those tests eventually finds a violation.

\subsection{Scope}\label{intro:scope}
In the affine layout of Section~\ref{sec:exponential}, the four independent code directions must leave the two chart coordinates fixed. That requires at least \(4+2=6\) lattice directions in this layout. It is open whether two-dimensional decoders, including the noisy-control constructions \cite{BDL2026,DST2026} and the homogeneous decoders with noiseless classical processing \cite{Lake2025,SBL2026}, can support rate-selected storage under the controller-correlated faults of our model (Section~\ref{a:discussion}).

The number of qubits per site is fixed and very large, as is the local dimension in spectral-gap undecidability constructions \cite{CPGW2015}. The interaction ranges are also fixed and very large. In the six-dimensional estimates, the probability per site and round that noise resamples the sampling registers is bounded above by a number of order \(10^{-578\,792}\). Lemma~\ref{v:density} and \eqref{v:cap-list} give the bound \(\mu_*\le p_{\rm dens}/2=B_{\rm raw}^{-67440}/128\); Section~\ref{v:parameters} relates it to the physical noise rate. We give no order-of-magnitude estimate for the register size or range. The four-dimensional stretched-exponential lower bound eventually exceeds the three-dimensional linear lower bound, but slowly: since \(\alpha_4<2\times10^{-4}\), \(L^{\alpha_4}\ge2\) requires \(L>2^{5000}\). A crossover size depends on the constants and size cutoffs. The four-dimensional upper bound is a full-register replacement bound, and a matching upper bound remains open. The six-dimensional upper bound likewise comes from full resets, applied separately to the four-dimensional fibres.

Rates outside the selected windows remain unclassified. Appendix~\ref{sec:coexistence} gives a separate four-dimensional circuit for independent single-qubit noise with one reserved, explicitly identified exponential-storage window and a tail family on which linear storage versus forgetting is undecidable (Proposition~\ref{p:coexistence}).
\subsection{Related work}\label{intro:related}
\begin{table}[!tb]
\caption{Closely related results and the observable studied here.}
\label{tab:landscape}
\small\centering
\begin{tabular}{@{}>{\raggedright\arraybackslash}p{0.29\textwidth}>{\raggedright\arraybackslash}p{0.29\textwidth}>{\raggedright\arraybackslash}p{0.34\textwidth}@{}}
\toprule
Work & Setting & Observable\\
\midrule
Cubitt--P\'erez-Garc\'ia--Wolf; Bausch--Cubitt--Watson; Purcell--Li--Cubitt \cite{BCW2021,CPGW2015,PLC2024}
 & Local Hamiltonians; interaction parameter & Gap and ground-state phase\\\addlinespace[3pt]
Wolf--Cubitt--P\'erez-Garc\'ia~\cite{WCPG2011} & Finite channel collections & Fidelity reached by some channel word\\\addlinespace[3pt]
Balasubramanian--Davydova--Lake; Drechsler--Oreg--Aharonov; D\"unnweber--Styliaris--Trivedi~\cite{BDL2026,DOA2026,DST2026}
 & Local noisy quantum memories & Positive storage guarantees, including the four-dimensional Toom bound described below\\\addlinespace[3pt]
Aharonov--Ben-Or; G\'acs; G\'acs--Reif; Gottesman; Harrington~\cite{ABO2008,Gacs1986,Gacs2001,GacsReif1988,Gottesman2000,Harrington2004}
 & Hierarchical reliable computation and quantum correction & Protected schedules and encoded gates\\\addlinespace[3pt]
Our prior work (Douglas--Mghirbi)~\cite{DMFixed2026,DMNoise1D2026,DMNoise2026,DMPhase2026} & Noisy classical local rules & Ergodicity and computable noise-window selection\\\addlinespace[3pt]
This paper & Fixed noisy quantum circuits; physical noise rate & Storage on computable windows, forgetting for all encodings, exact certification for the circuits of Theorems~\ref{thm:exponential}, \ref{thm:main} and~\ref{q1:main} (Theorem~\ref{a:complete}, Proposition~\ref{v:certification} and Section~\ref{q1:effective})\\
\bottomrule
\end{tabular}
\end{table}

\paragraph{Closest prior work.}
Bausch, Cubitt and Watson construct a continuous one-parameter Hamiltonian family in which a Turing machine controls the ground-state phase of each parameter region \cite[Theorem~2.1]{BCW2021}. Their theorem gives a specified nonhalting interval and an interval of positive extent for a halting input. In our circuit the parameter is a physical noise rate, the observable is storage lifetime, and the same procedure computes \(I_n\) for either halting outcome. For the designated four-dimensional toric-code encoding with Toom correction, Drechsler, Oreg and Aharonov \cite[Section~3.2]{DOA2026} bound the worst-case decoded infidelity by \(2Cme^{-cL}\) at every round through \(m\), below a nonzero noise threshold. Their synchronous noise is a product of single-qubit Pauli channels applied between correction updates. Here depolarization acts after every elementary layer, including layers within a correction update, and faulty controller decisions can cause correlated erasures. Relative to our classical noise diagram \cite{DMNoise2026}, the additional ingredients are quantum recovery under those faults and forgetting for every encoding. Table~\ref{tab:landscape} compares the observables.

Noise-rate phase diagrams for quantum information predate this computability question. Aharonov studied a quantum-to-classical transition through entanglement length \cite{Aharonov2000}. Upper bounds on tolerable noise and storage time were developed by Razborov, Kempe--Regev--Unger--de Wolf, and Harrow--Nielsen \cite{HarrowNielsen2003,KRWdW2010,Razborov2004}. The need for fresh low-entropy resources is central to noisy reversible computation and quantum refrigeration \cite{ABIN1996,BGH2013}; reset supplies such resources here. Baumann, Stein and Linnhoff-Popien give an all-encoding memory ceiling in terms of fresh qubits per step \cite{BSL2026}. Pastawski, Kay, Schuch and Cirac give an upper bound under depolarization without fresh resources \cite{PKSC2009}. Our circuits reset a number of qubits proportional to their volume, so that resource is not fixed as size grows.

Balasubramanian, Davydova and Lake give a strictly local circuit for the two-dimensional toric code that preserves a logical state for exponential time under circuit-level noise. Its schedule varies in space and time, and its time-translation-invariant variant uses three-dimensional stacks \cite{BDL2026}. Lake's homogeneous local decoder achieves a stretched-exponential lifetime using \(O(\operatorname{polylog}L)\) noiseless classical bits per noisy qubit \cite{Lake2025}. The time-translation-invariant surface-code decoder of Selub, Bhardwaj and Lake also achieves a stretched-exponential lifetime under phenomenological noise and assumes noiseless classical operations. Its streaming version is spatially translation-invariant, while its constant-density construction uses a spatially varying hierarchy \cite{SBL2026}. Section~\ref{a:discussion} describes the remaining question for these decoders. The classical-to-quantum correspondence of Drechsler, Oreg and Aharonov and the sweep construction of Ponarovsky are close comparisons \cite{DOA2026,Ponarovsky2026}. The memory tools used here follow G\'acs and G\'acs--Reif's reliable computation, Harrington's hierarchy, Aharonov--Ben-Or's coherent Calderbank--Shor--Steane (CSS) gadgets, and the toric-code and sweep-decoder work of Dennis--Kitaev--Landahl--Preskill, Ahn, Kubica--Preskill and Vasmer--Browne--Kubica \cite{ABO2008,Ahn2004,DKLP2002,Gacs2001,GacsReif1988,Harrington2004,KubicaPreskill2019,VBK2021}. Level reduction is in the extended-rectangle tradition of Aliferis, Gottesman and Preskill \cite{AGP2006}. Breuckmann, Duivenvoorden, Michels and Terhal study local decoders for the two- and four-dimensional toric codes \cite{BDMT2017}. Measurement-free correction and local-gate fault tolerance provide further context \cite{Gottesman2000,Heussen2024,PazSilva2010}.

\paragraph{One-dimensional memories.}
Aharonov and Ben-Or proved a one-dimensional nearest-neighbour threshold without measurement or classical processing during computation \cite[Theorem~13]{ABO2008}; Gottesman proved a one-dimensional local-gate threshold with next-to-nearest-neighbour gates and suggested interspersing G\'acs-type classical fault-tolerant regions with quantum data \cite{Gottesman2000}, and his textbook discusses replacing classical processing by fault-tolerant quantum circuits \cite[Section~15.3.1]{Gottesman2024}. These are compiled spacetime schedules, neither translation-invariant nor periodic with a size-independent period. Shtanko and Sharma give a further non-uniform, measurement-free construction under nonunital noise \cite{SS2025}, building on quantum refrigeration \cite{BGH2013}; here the noise is unital depolarization and entropy is removed by explicit resets. Fitzsimons and Twamley and Kay obtained one-dimensional global-control thresholds with externally supplied pulse schedules \cite{FT2007,Kay2007,Kay2005}, and Gidney and Bergamaschi prove a constant-rate fault-tolerant computer on a line with noiseless nonlocal classical operations and name translation invariance as a further goal \cite{GB2025}. Balasubramanian, Davydova and Lake pose a local quantum analogue of G\'acs's automaton as an open problem \cite{BDL2026}; Drechsler, Oreg and Aharonov note that the one-dimensional fault-tolerant constructions break translation invariance and ask about autonomous synchronous quantum memories in one dimension \cite[Sections~1.2 and~1.8]{DOA2026}; Selub, Bhardwaj and Lake identify removal of noiseless classical processing as an open direction and suggest G\'acs's techniques \cite[Section~1.3]{SBL2026}. Earlier questions about fault tolerance in quantum cellular automata appear in \cite[Section~IV]{Kay2007} and \cite[Section~10.1.2]{PerezDelgado2007}. The classical foundations are G\'acs's one-dimensional reliable computation, from an initialized hierarchy \cite{Gacs1986} and later with self-organization \cite{Gacs2001}, Cirel'son's hierarchical storage scheme \cite{Cirelson1978}, G\'acs's two-dimensional arrays \cite{Gacs1989} and the three-dimensional construction of G\'acs and Reif \cite{GacsReif1988}; Gray gives a guide to G\'acs's work \cite{Gray2001}, and Marsan, Sablik and T\"orm\"a give a one-dimensional automaton whose ergodicity changes twice as the noise rate grows \cite{Marsan2025}. We are not aware of a previous proof of Corollary~\ref{q1:memory}. It is synchronous, a PCA-type model in the terminology of Drechsler, Oreg and Aharonov, and its initialization and unrestricted encoder and decoder are part of its hypotheses. Its linear lifetime is weaker than the guarantees of the non-uniform constructions above; the new feature is the model. Its resets supply fresh low-entropy qubits \cite{ABIN1996}. Its use of the non-Clifford Toffoli gate places it outside the Clifford-only memory bound of Nelson, Rajakumar and Gullans \cite{NRG2026}. Its active dynamics lie outside the hypotheses of the thermal and Gibbs-sampler results for one-dimensional chains \cite{Bardet2023,BC2025,CD2026}, and its concatenated code has nonlocal stabilizers, unlike the geometrically local stabilizer Hamiltonians and local commuting constraints of \cite{BravyiTerhal2009,BPT2010}.

Thermal and dissipative memories concern different dynamics \cite{Alicki2010,BDLin2026,KumarWeimer2026,PCC2011}; dimensional no-go results for those models do not constrain the dimension of our circuits \cite{BravyiTerhal2009,Brown2016,Terhal2015}. Hamiltonian protection \cite{PKSC2009} and space-overhead bounds \cite{FMSh2022} address complementary resources. Xu and Xu's computable sufficient criterion for a restricted class of three-dimensional stabilizer Hamiltonians is a useful contrast \cite{XuXu2026}; our undecidability statement concerns fixed noisy circuits.

For undecidability in physics, see Perales-Eceiza and coauthors \cite{UndecidabilityReview2025}. Spectral gaps and phase diagrams, quantum-channel reachability, memory-channel capacity, measurement occurrence, tensor-network properties and thermalization have undecidable instances \cite{BCW1D2020,BCW2021,CPGW2015,CPGW2022,DeLasCuevas2016,EMG2012,Elkouss2018,KGE2014,PLC2024,ShiraishiMatsumoto2021,WCPG2011}. Bhattacharyya, Mehta and Zhao show hardness of approximating quantum capacity and uncomputability of a restricted entanglement-assisted zero-error capacity \cite{BMZ2026}. Xu and coauthors give convergent certified upper bounds on the bulk spectral gap but rule out general convergent lower bounds \cite{XuEtAl2026}. Matsuura studies fidelity hitting times in fixed periodically driven unitary dynamics \cite{Matsuura2026}; here the parameter is a physical noise rate and the conclusion quantifies over every encoding. The open rate windows give robustness in that parameter. They do not assert robustness under perturbations of the gates, a distinction also relevant to unstable undecidable spectral-gap families \cite{CastillaLucia2026,XuEtAl2026}.

The classical lineage includes Kurdyumov and Toom's algorithmic ergodicity questions, probabilistic cellular automata and undecidable unique ergodicity \cite{BMM2013,Kurdyumov1978,Toom2000,ToomContinuous2000,Torma2015}. Marsan and Sablik show that stability of a given cellular automaton invariant measure under uniform perturbation is undecidable \cite{MarsanSablik2024}. Braverman, Grigo and Rojas study computability of invariant measures after perturbation of a fixed classical dynamical system \cite{BravermanGrigoRojas2012}, whereas the undecidable observable here is the growth of a transient storage lifetime across system sizes. Noise-driven ergodicity and nonmonotone transitions are studied in \cite{MST2019,MarcoviciTaati2026,Marsan2025}; see also Gray's guide to G\'acs's positive-rates construction \cite{Gray2001} and Taati's theorem on rapid forgetting by reversible cellular automata under positive additive noise \cite{Taati2021}. Our classical interfaces are drawn from our preprints \cite{DMFixed2026,DMNoise2026,DMPhase2026}, in the tiling and reliable-computation tradition of Robinson, Kari, Toom and Berman--Simon \cite{BermanSimon1988,Kari1992,Robinson1971,Toom1980}. Appendix~\ref{app:imports} lists the source statements and the proofs adapted here.

\subsection{Organization and reading guide}\label{intro:organization}
Readers who want only the statements can stop after this section and Section~\ref{s:model}, which fixes the model, the fidelity and the lifetime. Fault-tolerance readers will find the quantum arguments in Section~\ref{sec:quantum} (storage under controller-correlated faults), Section~\ref{sec:forgetting} (the forgetting bound for every encoding, which stands alone) and Section~\ref{q1:storage} (the memory on rings). Readers interested in the undecidability can read Sections~\ref{sec:controller} and~\ref{a:certification}; for the ring construction, Section~\ref{q1:model} and the statements of Lemmas~\ref{q1:l18}, \ref{q1:l13}, \ref{q1:l14} and~\ref{q1:l15} give the logic. Section~\ref{sec:compilation}, Sections~\ref{q1:clock}--\ref{q1:parameters} and Appendices~\ref{app:code-estimates}--\ref{app:f4-sources}, \ref{sec:fold-controls} and~\ref{sec:folded} contain compilation, probability estimates and constant bookkeeping, and can be treated as black boxes on a first reading. The body proves Theorem~\ref{thm:main} first because Theorems~\ref{thm:exponential} and~\ref{thm:four-dimensional} reuse its controller, compiler and whole-channel forgetting argument; the four-dimensional proof also reuses the six-dimensional static chart. Sections~\ref{q1:model}--\ref{q1:effective} then prove Theorem~\ref{q1:main} and Corollary~\ref{q1:memory}. Table~\ref{tab:where} gives the proof locations, and Tables~\ref{tab:terms} and~\ref{tab:terms-ring} the recurring terminology. Lemmas~\ref{f:iteration}, \ref{f:bell}, \ref{q1:l1} and~\ref{q1:l1.1}, conditional forms of Theorem~\ref{f:forgetting} and of the fidelity bound of Lemma~\ref{q1:l18}, and Theorem~\ref{lean:conditional} are also checked in Lean (Appendix~\ref{app:lean}); Theorems~\ref{thm:exponential}--\ref{q1:main} are not.

\begin{table}[!tb]
\centering\small
\setlength{\tabcolsep}{5pt}\caption{Proof locations for storage, forgetting, optimal lifetime and finite-test certification.}\label{tab:where}
\begin{tabular}{@{}>{\raggedright\arraybackslash}p{.14\linewidth}>{\raggedright\arraybackslash}p{.19\linewidth}>{\raggedright\arraybackslash}p{.19\linewidth}>{\raggedright\arraybackslash}p{.17\linewidth}>{\raggedright\arraybackslash}p{.21\linewidth}@{}}
\toprule Result & Storage & Forgetting & Lifetime ceiling & Finite-test certificate\\\midrule
\ref{thm:main}: Theorem~\ref{thm:detailed} & Proposition~\ref{d:storage} & Theorem~\ref{f:forgetting} & Theorem~\ref{u:eb} & Theorem~\ref{a:complete}; Corollary~\ref{a:time} for time bounds\\\addlinespace[3pt]
\ref{thm:exponential}: Theorem~\ref{v:detailed} & Section~\ref{v:positive} & Section~\ref{v:forgetting} & Proposition~\ref{v:certification}(i) & Proposition~\ref{v:certification} \mbox{(ii)--(iii)}\\\addlinespace[3pt]
\ref{thm:four-dimensional}: Theorem~\ref{f4:detailed} & Section~\ref{f4:positive} & Section~\ref{f4:forgetting} & Bound~\eqref{f4:lifetime-upper} & ---\\\addlinespace[3pt]
\ref{q1:main}: Section~\ref{q1:effective} & Section~\ref{q1:storage} & Section~\ref{q1:forgetting} & Section~\ref{q1:effective} & Section~\ref{q1:effective}\\\addlinespace[3pt]
Coexistence & \multicolumn{4}{l}{Proposition~\ref{p:coexistence}, with Lemma~\ref{x:exponential} and Proposition~\ref{q:linear}}\\\addlinespace[3pt]
Window corollaries & \multicolumn{4}{l}{Section~\ref{intro:threshold}, from Theorems~\ref{thm:exponential}--\ref{thm:main}; Theorem~\ref{q1:main} states its own}\\
\bottomrule
\end{tabular}
\end{table}

Appendices~\ref{app:reproducibility}, \ref{app:imports} and~\ref{app:lean} document reproducibility, imported results and formalization.

\section{Model, fidelity and detailed three-dimensional theorem}\label{s:model}
This section fixes the local circuit model, the two noise models and the fidelity and lifetime used throughout. It states Theorem~\ref{thm:detailed}, the detailed form of Theorem~\ref{thm:main}.

Universality supplies a total computable compiler from program codes \(e\) to inputs \(n\) such that \(U(n)\) halts exactly when program \(e\) halts on input \(0\) (Section~\ref{a:certification}). Each construction has nonempty open windows \(I_n\) with rational endpoints. The total computable noiseless encoder depends on \((n,L)\), and the total computable noiseless decoder on \((n,L,t)\). On a halting window the forgetting bound holds for all encoders and decoders on the entire register, for every \(L\), every rate in \(I_n\), and every integer \(t\) at or above the cutoff \(C_{n,\eta}\) in the corresponding theorem. Time counts rounds in Theorems~\ref{thm:exponential} and~\ref{thm:main} and elementary layers in Theorems~\ref{thm:four-dimensional} and~\ref{q1:main}.

\subsection{Terminology}
We call the registers of the rule before physical compilation \emph{semantic registers}. A site can hold several such registers and the extra wires used to implement them.

\begin{table}[!tb]
\caption{Terminology used in the construction and the proofs. The source vocabulary (bin, star, seed, witness, certificate, charge, image, catalogue and overlay) is explained in Section~\ref{c:named-sources}. Table~\ref{tab:terms-ring} lists the terms used on rings.}\label{tab:terms}
\small\centering
\begin{tabular}{@{}>{\raggedright\arraybackslash}p{.23\linewidth}>{\raggedright\arraybackslash}p{.51\linewidth}>{\raggedright\arraybackslash}p{.19\linewidth}@{}}
\toprule Term & Meaning & Defined in\\\midrule
Selector and chart & Rate-selecting classical controller and its locally checked space-time diagram. In six and four dimensions the chart is a static patch. A live letter is a nonbottom symbol. & Sections~\ref{c:chart}, \ref{sec:exponential} and~\ref{sec:four-dimensional}\\\addlinespace[3pt]
Cap, profile and death flag & Finite static computation patch with a diagonal overflow counter. Its per-site profile is \(\mathsf{pr}\). The flag \(\mathsf d\) records whether erasure has been triggered at or has propagated to a site. & Section~\ref{sec:exponential}\\\addlinespace[3pt]
Replica and sampling lane & A replica is one transverse displacement supplying one vote. A sampling lane follows a row by copying a register that samples noise. One such copying interval is called a ride. & Sections~\ref{c:chart} and~\ref{v:parameters}\\\addlinespace[3pt]
Censor & Local statistical test that can replace a computation symbol by \(\bot\). & Section~\ref{sec:exponential}\\\addlinespace[3pt]
Storage-enable bit and cut & In six and four dimensions the storage-enable bit controls conditional reset. A cut removes incoming dependence through that reset. & Sections~\ref{sec:exponential}, \ref{sec:four-dimensional}\\\addlinespace[3pt]
Conditional reset and mixed ancilla & Controlled swap with a fresh maximally mixed ancilla, then reset of that ancilla. In the clean circuit a triggered cut replaces each data qubit by \(I/2\). & Section~\ref{sec:compilation}\\\addlinespace[3pt]
Hidden cover, realization and rate bound & A hidden cover adds letters to projected chart geometry so legality is local. A realization compiles the rule for one noise model. A rate bound is an upper bound on a fault or refresh probability. & Sections~\ref{c:chart} and~\ref{sec:compilation}\\\addlinespace[3pt]
Dirty location & Circuit location touched by a fault or by a wrong controller output. & Section~\ref{d:dispatch}\\\addlinespace[3pt]
Sheet and dispatch & A sheet is one affine branch of a folded code. Dispatch is the selection and execution of a local gate through its ports. & Section~\ref{d:dispatch}, Appendices~\ref{app:f4-dispatch} and~\ref{sec:folded}\\\addlinespace[3pt]
Clock and exclusive dispatcher & The clock is a classical scheduling field. The dispatcher uses exclusive ports to implement the chosen gate. & Section~\ref{d:calendar-section}, Appendices~\ref{app:f4-calendar} and~\ref{app:f4-dispatch}\\
\bottomrule
\end{tabular}
\end{table}
Table~\ref{tab:geometry} lists the coordinate roles. In the six-dimensional construction \(u\) denotes the patch coordinates and \(P\) the four code coordinates. In the four-dimensional construction \(\mathsf C\) is the computation field and \(\mathsf S\) the clock field. The chain coordinate is \(\sigma\), the clock block length is \(P_{\rm cal}\), the simulation period is \(T_{\rm cal}\), and \(k\) is the concatenation depth. The window bound \(w_n\) is on bin height and \(\alpha_4\) is the storage exponent. Names are local to each construction (the three-dimensional controller and its six-dimensional reuse; the four-dimensional calendar; the rings) unless a section says otherwise.

\begin{table}[!tb]
\caption{Coordinate roles. The second coordinate \(t-i\) of the three-dimensional computation is a space-time coordinate.}\label{tab:geometry}
\centering\small
\begin{tabular}{@{}l>{\raggedright\arraybackslash}p{.32\linewidth}>{\raggedright\arraybackslash}p{.49\linewidth}@{}}
\toprule Circuit & Computation & Quantum data\\\midrule
6D (A) & Static patch in \((u_x,u_y)\) & Four-dimensional toric code in \((P_1,P_2,P_3,P_4)\)\\\addlinespace[3pt]
4D (B) & Static patch in \((x,y)\) & Chain along \(\sigma\), folded into two internal roles\\\addlinespace[3pt]
3D (C) & Space-time diagram in \((i,t-i)\) & Chain along \(\sigma\)\\\bottomrule
\end{tabular}
\end{table}

The separate four-dimensional circuit of Appendix~\ref{sec:coexistence}
uses the four-sheet geometry of Appendix~\ref{sec:folded} and a separate
closed-code track. The recurring constants are explained in
Appendix~\ref{app:code-estimates}.

\subsection{Local circuits and two noise models}\label{s:local-circuits}
Each construction has a fixed local register and a fixed repeating schedule. We begin with the three-dimensional register convention; the later theorems specify their other lattice dimensions.
Let \(\mathbb T_L^3=(\Z/L\Z)^3\). One physical site carries
\(\mathcal H_x=(\mathbb C^2)^{\otimes N}\), where \(N\) is one fixed
finite integer. It includes every data, controller, sensor, transport
and work qubit. A gate layer is a translated product of gates on
disjoint internal tensor factors. Several such gates can meet the same
physical product site on different factors. Every gate has at most
three input qubits and uses one of finitely many displacement templates, independent of \(n,L,\eps\).

\begin{definition}[Whole-site depolarizing noise, WS]\label{def:noise-w}
Following each gate layer, independently at each physical site, apply
\begin{equation}\label{eq:site-noise}
 \mathcal D_{\eps,x}(\rho)
 =(1-\eps)\rho+\eps\,\frac{I_x}{2^N}\otimes\Tr_x\rho.
\end{equation}
\end{definition}
\begin{definition}[Independent single-qubit depolarizing noise, Q]\label{def:noise-q}
In realization \(\mathrm Q\), independently on every wire after every layer, apply
the one-wire channel \(\rho\mapsto(1-\eps)\rho+\eps\Tr(\rho)\,I/2\), tensored with the identity on all other wires.
\end{definition}
In both realizations the same
physical rate follows preparation, dephasing, reset, and idle layers.
\(\mathrm{WS}\) and \(\mathrm Q\) are separately compiled fixed circuits; their dimensions,
schedules and physical windows may differ. Gates in a layer are disjoint
on qubits and may share physical sites. All spatial gates have fixed
finite range, which need not be one lattice spacing. The gate list is
\[
 \{X,H,\mathrm{CNOT},\mathrm{Toffoli},R_0,D_Z,I\}.
\]
The unitary subset contains the computationally universal pair Toffoli and Hadamard \cite{Aharonov2003,Shi2003} (universality here is the real-amplitude simulation of complex computation with one extra qubit, in the sense of \cite[Definition~3 and Theorem~2]{Aharonov2003}). Here \(R_0(\rho)=|0\rangle\langle0|\Tr\rho\) and \(D_Z\) is complete
computational-basis dephasing; a measurement is a \(D_Z\) whose outcome
stays on a basis-valued wire.
\Needspace{7\baselineskip}
\begin{definition}[Round channel]\label{def:round}
Repeating the fixed circuit defines a round channel \(\Lambda_{L,\eps}\).
One application is one round. If the period is \(J_{\rm per}\) elementary layers,
then at aligned times \mbox{\(t_{\rm layer}=J_{\rm per}t_{\rm round}\)}.
For Theorem~\ref{thm:four-dimensional}, including Theorem~\ref{f4:detailed},
bound~\eqref{f4:lifetime-upper}, Table~\ref{tab:results} and
Corollary~\ref{cor:window-undecidable}, fidelity and lifetime are measured
in elementary layers. In this convention use \(\Lambda^{(t)}_{L,\eps}\),
the channel of the first \(t\) noisy layers starting at a round boundary,
in Definitions~\ref{def:fidelity} and~\ref{def:lifetime}. Convert round
quantities in the proof at aligned times using the period.
Unless elementary layers are specified, \(t\) counts rounds.
\end{definition}

\subsection{Storage fidelity}
CPTP means completely positive and trace preserving.
\begin{definition}[Entanglement fidelity]\label{def:fidelity}
Let \(\beta=|\beta\rangle\langle\beta|\), with
\(|\beta\rangle=(|00\rangle+|11\rangle)/\sqrt2\). The reference qubit
is never acted upon. For CPTP encoding and decoding maps \(\cE,\cD\), put
\begin{equation}\label{eq:fidelity}
 F(\cE,\cD;L,t,\eps)=
 \langle\beta|[\mathrm{id}\otimes
 (\cD\Lambda_{L,\eps}^t\cE)](\beta)|\beta\rangle,
 \qquad F^*_{L,\eps}(t)=\sup_{\cE,\cD}F(\cE,\cD;L,t,\eps).
\end{equation}
This is entanglement fidelity for the maximally mixed logical qubit.
\end{definition}
The supremum allows arbitrary encodings into the entire physical
algebra, including work and controller registers, with no persistent
outside quantum memory. \begin{definition}[Memory lifetime, or storage time]\label{def:lifetime}
Let
\[
 \tau_{L,\eps}=\sup\{t\in\N:F^*_{L,\eps}(t)\ge7/8\}.
\]
\end{definition}
Optimal fidelity is nonincreasing in \(t\). A later decoder composed with the intervening channel is a decoder at the earlier time. The passing layer times
therefore form an initial segment, and the passing round times are
precisely the aligned times in that segment. For a period of \(J_{\rm per}\)
layers, the optimized integer lifetimes satisfy
\(\tau_{\rm round}=\lfloor\tau_{\rm layer}/J_{\rm per}\rfloor\) when finite,
and are both infinite otherwise. Our explicit decoder can depend
on \(t\), as well as \(n,L\), and remains independent of the rate.

\begin{definition}[Storing and forgetting windows]\label{def:windows}
A \emph{rate window} is a nonempty open interval with rational endpoints. For a specified encoder family, decoder family, horizon and size cutoff, a \emph{storing window} is one on which the stated fidelity guarantee holds at every rate and every indicated size and time. A \emph{forgetting window} is one on which, for every \(\eta>0\), the optimal fidelity is at most \(1/2+\eta\) after a finite time uniform in size and rate. A \emph{scalable-memory rate} is one with \(\sup_L\tau_{L,\eps}=\infty\). A \emph{scalable-memory window} consists entirely of such rates. Nonhalting and halting windows are indexed by \(n\) with \(U(n)\uparrow\) and \(U(n)\downarrow\), respectively. The theorems give storing windows with fidelity \(15/16\); certification tests the weaker bound \(7/8\). The limiting benchmark \(1/2\) is the largest entanglement fidelity attainable by a measure-and-prepare qubit channel.
\end{definition}

\begin{theorem}[Detailed form of Theorem~\ref{thm:main}]\label{thm:detailed}
For each of \(\mathrm{WS}\) and \(\mathrm Q\) there are a computably specified,
translation-invariant, time-periodic circuit as above, a computable sequence
of disjoint nonempty rational open intervals
\(I_n\subset(0,1)\), a rational \(c>0\), and computable size cutoffs
\(L_0(n)\ge1\), with \(c\) independent of \(n\), and the following properties.
\begin{enumerate}
\item There are total computable maps \(\cE_{n,L}\) and
\(\cD_{n,L,t}\), independent of \(\eps\) and of whether \(U(n)\)
halts. If \(U(n)\uparrow\), then for every \(\eps\in I_n\),
\(L\ge L_0(n)\), and every integer \(0\le t\le\lfloor cL\rfloor\),
\(F(\cE_{n,L},\cD_{n,L,t};L,t,\eps)\ge15/16\).
\item If \(U(n)\downarrow\), then for every \(\eta>0\) there is
\(C_{n,\eta}<\infty\) such that
\(F^*_{L,\eps}(t)\le1/2+\eta\) for every \(L\),
\(\eps\in I_n\), and \(t\ge C_{n,\eta}\).
\item There are computable constants \(A,a>0\), \(b\in\N\), an integer \(C_0\ge2\)
and \(L_{\rm EB}\ge C_0/2\), all independent of \(n\), such that for every \(n\), every
\(\eps\in I_n\) and every \(L\ge L_{\rm EB}\), with \(s_L=201L+C_0\),
\[
 \Lambda_{L,\eps}^{s_L}=(1-r_L)\Phi^{\rm EB}_{L,\eps}+r_L\mathcal B_{L,\eps},
 \qquad r_L\le q_L\le AL^be^{-aL},
\]
where \(\Phi^{\rm EB}_{L,\eps}\) is entanglement breaking (EB) and
\(\mathcal B_{L,\eps}\) is a channel. In particular
\mbox{\(\|\Lambda_{L,\eps}^{s_L}-\Phi^{\rm EB}_{L,\eps}\|_\diamond\le2q_L\)} and
\(F^*_{L,\eps}(t)\le1/2+q_L/2\) for every \(t\ge s_L\).
\item The predicate requiring the explicit fidelity to be at least
\(7/8\) for every integer \(L\ge L_0(n)\), integer
\(0\le t\le\lfloor cL\rfloor\), and rational \(\eps\in I_n\)
holds exactly when \(U(n)\uparrow\) and is \(\Pi^0_1\)-complete.
Every finite test is an exact rational comparison.
\end{enumerate}
\end{theorem}

If \(U(n)\) never halts, these bounds give
\(\lfloor cL\rfloor\le\tau_{L,\eps}<201L+C_0\) for every \(\eps\in I_n\)
and every \(L\ge L_C(n):=\max\{L_0(n),L_{\rm EB},L_2\}\). Here \(L_2\) is a
computable size beyond which \(AL^be^{-aL}<3/4\). One can search beyond
\(b/a\), where this bound decreases, until a certified upper estimate
is strictly below \(3/4\). Thus
\(\tau_{L,\eps}=\Theta(L)\), uniformly across the window. On halting windows the lifetime is
uniformly bounded in \(L\). These halting-side times are finite but not
computable from \(n\): for each fixed rational \(0<\eta<3/8\), no total
computable function of \(n\) can serve as \(C_{n,\eta}\) in part (ii)
(Corollary~\ref{a:time}). Computable windows and their rational midpoints
already suffice for the undecidability assertion. The scope is as in Section~\ref{intro:scope}.

For fixed \(L\) and \(\eps>0\), complete replacement of every physical
factor at the last layer of a round has probability
\(p_{L,\eps}=\eps^{L^3}\) in \(\mathrm{WS}\) and
\(p_{L,\eps}=\eps^{NL^3}\) in \(\mathrm Q\). If it has occurred, the output is
independent of the input. Thus
\begin{equation}\label{eq:finite-reset}
 F^*_{L,\eps}(t)\le\frac14+\frac34(1-p_{L,\eps})^t.
\end{equation}
This also controls the finitely many small volumes omitted from
asymptotic estimates.

Sections~\ref{sec:controller}--\ref{a:certification} prove Theorem~\ref{thm:detailed}. Section~\ref{sec:controller} constructs the classical selector and Section~\ref{sec:compilation} compiles it and fixes the constants. Section~\ref{sec:quantum} proves (i), Section~\ref{sec:forgetting} proves (ii), Section~\ref{sec:ceiling} proves (iii), and Section~\ref{a:certification} proves (iv).

\section{A classical computation that controls erasure}\label{sec:controller}

The classical processor decides when to reset local quantum data. Its noise samples select a computation. We state the imported classical interface, then prove the estimates that connect it to quantum storage and to loss of quantum dependence.

We call the rate-selecting controller the selector. It updates a space-time diagram on a hierarchical tiling. The noise tests can set a symbol to \(\bot\), which spreads under the fault-free update.

\subsection{The selector and its computational chart}\label{c:chart}
For the three-dimensional construction use coordinates
\({\bf z}=(i,P_1,P_2)=(i,\sigma,z_3)\in\Z^3\).
Here the two components of \(P\) are the transverse voting coordinates.
The selector \(C_{\rm sel}\) has chart coordinates
\[
 (x,y)=(i,t-i)
\]
at round \(t\).
In Section~\ref{sec:exponential}, \(P\) has four components. All carry the code, and \(P_1,P_2\) are the voting coordinates. The four-dimensional construction of Appendix~\ref{sec:fold-controls} adds a copied coordinate \(Q\) and four signed orientations. Formulas involving \(Q\) omit that coordinate in three dimensions.
\par\emph{The chart.} A chart alphabet \(\mathcal A\) is extended by a symbol \(\bot\); a chart letter is live when it differs from \(\bot\). The west and south chart neighbours are the actual previous-time sites \((i-1,t-1)\) and \((i,t-1)\), respectively. A partial completion map \(\operatorname{cmp}\) sends a west letter and a south letter to a new letter. It is functional, so there is at most one completion, and bottom-spreading, so a live completion requires two live inputs. The chart history has fixed depth \(T_{\rm hist}=1024\), and its local history consistency tests are included in the ordinary completion map.

\emph{Replicas and the ordinary proposal.} Set
\begin{equation}\label{eq:c-boxes}
 S_0=[0,15]^2,\qquad S_1=160e_1+S_0,\qquad S_2=160e_2+S_0,\qquad K=256,\quad h=512,
\end{equation}
where \([0,15]\) is an interval of integers, so each box holds \(K\) replicas. A replica uses both logical inputs at its own transverse displacement. The ordinary proposal at \((i,P,t)\) is the unique letter proposed by at least \(h\) of the \(3K\) replicas in \(S_0\cup S_1\cup S_2\); if there is none, it is \(\bot\). Since \(2h>3K\), two different letters cannot both receive \(h\) votes.

\emph{Two coordinates and the rate tests.} The second chart coordinate records the ordinary proposal. The first, retained coordinate equals the ordinary proposal unless additional sampling-register coordinates, which test the scale of the noise, replace it by \(\bot\). A \emph{ride} is a sampling-register copying interval along a chart row. The tests compare short rides with recent rides, at the scales
\begin{equation}\label{eq:c-scales}
 \ell_r=2\cdot4^r-1,\qquad W=\lfloor C_h/\mu\rfloor.
\end{equation}
Here \(\mu\) is the aggregate refresh rate and \(C_h\) is a fixed rational constant selected later. The physical rule reads neither number; \(W\) is used only to analyse the experiments. A fixed bank of \(B\) sampling lanes supplies the test votes. Each binary-coded sampling register has one low code, one unknown code and 126 high codes.

We use the chart and machine construction of \mbox{\cite[Proposition~4.3]{DMNoise2026}} and Lemmas~4.1\nobreakdash--4.2 and~11.2\nobreakdash--11.3 of that preprint, its
finite-history boundary-path estimate
\mbox{\cite[Lemma~11.6]{DMNoise2026}}, its intrinsic experiment and source bounds
\cite[Lemmas~6.1--6.3 and~10.4]{DMNoise2026}, and the scaffold parent and
soundness lemmas \cite[Lemmas~A.3 and~A.2]{DMPhase2026}. We use the replica-majority construction of \cite[Theorem~2.1]{DMFixed2026}, where replicas are called providers, in the tradition of G\'acs--Reif and Toom \cite{GacsReif1988,Toom1980}. The sampling registers are the helpers of \cite{DMNoise2026}. A computation board
is a finite grid carrying successive configurations of the fixed machine;
legality means its local transition and boundary constraints hold. Its
row flag is a binary chart label. A critical gate is the sampling-register test at
the bottom end of a row at one of the two depths specified below.

The rate selects a cutoff depth by separating short rows from long rows. Along short rows few lanes have refreshed, and along recent segments of long rows enough have refreshed for the test to distinguish them. The chart marks which side of this distinction its rows should satisfy. A \emph{cover} decorates its projected geometry with finite-state letters so that legality can be tested locally. The cutoff and the guarded tests at the ends of the critical rows tie this legal geometry to the machine input.

\begin{proposition}[Classical chart interface]\label{c:selector-import}
There is an effective finite chart alphabet and a fixed finite sampling-register rule
with the following properties. For a universal machine \(U\), choose
\(a=n+n_0+1\), with \(n_0\) the fixed offset of
Section~\ref{s:parameters}, item~4, and the nominal window for the
aggregate refresh rate \(\mu\)
\begin{equation}\label{eq:c-window}
 \mathcal J_n=\left(\frac{2C_h}{\ell_a},\frac{21C_h}{10\ell_a}\right).
\end{equation}
For every \(\mu\in\mathcal J_n\), \(W=\lfloor C_h/\mu\rfloor\)
satisfies \(\ell_{a-1}<W\), \(2W\le\ell_a\), and \(W\le4^a-1\). If \(U(n)\) does not halt, the chart has a canonical
projected geometry with cutoff \(a\) with a legal finite-state cover and legal
finite computation boards. If \(U(n)\) halts, every full chart has a wrong
critical gate of one of two types: a short depth-\((a-1)\) gate with flag
zero, or a long depth-\(a\) gate with flag one.

The canonical projected geometry is computable on signed coordinates.
Its sampling-register copy characteristic preserves the base chart row of each lane.
For a fixed candidate critical gate and a deterministic finite footprint,
the intrinsic experiment estimates, namely the bounds \(u_B,z_B,w_B\) of
\eqref{eq:c-eta}, and the finite-history boundary-path estimates
\cite[Lemma~11.6]{DMNoise2026} hold with constants uniform on \(\mathcal J_n\). On halting, the finite
motif radius needed for these estimates is found by a terminating search
through finite locally compatible patches.
\end{proposition}

\emph{Termination of the motif search.} The finite motif search rejects a patch when it contains either of
the two wrong-gate rides. If arbitrarily large survivors remained,
compactness would give a full chart with neither ride, contrary to the
machine construction. This explains both termination on halting and why
the search is not a halting decision procedure.

\emph{The chart cover.} The effective chart cover
uses the northeast (NE) presentation of Le Gloannec and Ollinger
\cite[Theorem~3]{LGO2012}. The limit set of a \(2\times2\) substitution has
an effective finite cover whose full west and south neighbours determine
at most one tile. Their Theorem~4 gives the stronger four-way presentation. The limit set here consists of full-plane colourings
admitting arbitrarily deep compatible desubstitutions. The canonical
projected geometry belongs to it. One can see this by using its finite
output automaton as a state substitution, translating state quadrants
far into the positive quadrant, and taking compatible limits at every
substitution depth. The cover completion and incoming edges fix the
geometry. The flags and the chart bit are fixed by the west and south
inputs, with the horizontal row flag coming from the west and the
vertical flag from the south. The carrier, boundary and interior rules
of \cite[Lemma~11.3 and its preceding update table]{DMNoise2026} set the
computation fields, with the phase read from the geometry letter.

\emph{History slots and rejection.} For west and south history tuples \(w=(w_j)\) and \(s=(s_j)\), slot zero
is the completion of \(w_0,s_0\). The remaining slots copy \(s_{j-1}\)
for \(1\le j<T_{\rm hist}\), as in \cite[Section~11.4]{DMNoise2026}.
Collar overlaps, cutoff compatibility, forbidden halting symbols and
vertical and horizontal history matches only reject the candidate fixed
by the two inputs.

The offset in the machine input makes the computation at the selected depth \(U(n)\).
Let \(V(m)=U(m-1)\), \(m\ge1\). If the scan's first zero is
\(j_*\), it accepts \(j_*+2=2^e\), writes \(e\) tokens and
consumes \(n_0+1\) of them before calling \(V\) on
\(m=e-n_0-1\). At cutoff rank \(a\),
\(j_*=2^{a+1}-2\), so \(m=a-n_0=n+1\) and the simulated
computation is exactly \(U(n)\). The finite scan machine is built
after \(n_0\) is fixed.

\subsection{Total finite initialization}\label{subsec:c-initialization}
A global choice of hidden cover letters need not be computable. Only a finite legal
choice is required by the encoder.

\begin{lemma}[Finite controller initialization]\label{c:initialization}
From \((n,L)\) and a declared finite time horizon one can compute an initial
controller word, without the noise rate or a halting answer.
Suppose that the space-time cone of the stored data through that horizon,
with every initial history slot and all fixed lane and collar offsets,
injects into one chart of the side-\(L\) torus whose seams lie outside it.
For the extended storage horizon used in the proof of
Proposition~\ref{d:storage}, this condition is guaranteed by its size
cutoff \(L_0(n)\). Then on a nonhalting input the word agrees with a
correct-cutoff reference on that entire cone.
Its sampling-register initialization
satisfies the same short upper and recent-long lower experiment bounds as
the classical interface.
\end{lemma}
\par\noindent\emph{Proof idea.} Compute and search only the finite space-time region needed for the requested horizon. The legal nonhalting extension supplies a candidate, and a fixed fallback makes the search total.

\begin{proof}
Compute the canonical projection, cutoff and finite payload on a
square containing the
whole cone, every initial history slot and all fixed lane and collar
offsets. Exhaust the finite list of cover-letter assignments on that
square, imposing the computed projection and every local constraint.
Return the first successful assignment, with a fixed fallback if the list
is empty. The algorithm always terminates. On nonhalting, a global cover
of the projected reference supplies a successful finite assignment.
Functionality forces its future on the smaller cone. Assignments for
different sizes need not be restrictions of a common hidden cover.

Initialize sampling registers deterministically: use low on reference flag-one rows
and high on flag-zero rows. Each sampling-register field must use its own base row.
The copy characteristic preserves that row, even when the physical field
index changes its offset. On a short ride, a low initial value can only
remove a contribution until a genuine future mark is seen; hence its
upper bound is preserved. On a long ride, a high initial value dominates
the recent-\(W\) comparison until \(W\) genuine future atoms have become
available. Thereafter the latest genuine mark gives the usual comparison.
This argument also covers an experiment crossing time zero.
\end{proof}

\subsection{Named sources under layer noise}\label{c:named-sources}
Section~\ref{sec:compilation} compiles the controller so that all microscopic faults at one site in one round have a single source name (Section~\ref{s:source-name}). We record the resulting source field.

\begin{definition}[Raw sources, the marked field and source labels]\label{c:sources-def}
For a site \({\bf z}\) and round \(t\), let \(A_{{\bf z},t}\) record any internal fault at \({\bf z}\) in round \(t\) and let \(E_{{\bf z},t}\) be its independent aggregate refresh (Sections~\ref{s:ws-realization} and~\ref{s:probe}), and put
\[
 H_{{\bf z},t}=A_{{\bf z},t}\lor E_{{\bf z},t},\qquad p=\Pr(H_{{\bf z},t}=1)\le2\mu.
\]
The family \(H\) is independent on physical site-round names. One name can corrupt arbitrary outgoing copies of that site's old word, which may disagree with one another, and its own new output. Accordingly the single name \(({\bf z},t)\) seeds both formal occurrences \(({\bf z},t-1)\) and \(({\bf z},t)\); nonexistent pre-zero sources are omitted. The marked field \(M^0\) is the least finite-start field that contains these seeds and marks a nonfresh vertex whenever marked ordinary parents occur in two distinct replica boxes. For the positive source estimate, use base characteristics and bins of length \(W\); a \emph{source label} is
\[
 \lambda=(i-t,\;P+58t(1,1),\;Q,\;\lfloor t/W\rfloor).
\]
\end{definition}
Source labels do not count independent microscopic faults: an intrinsic test can involve many physical atoms. All occurrences with the same label are kept together, and geometric roles annotate a witness without splitting a label.

A \emph{bin} is an integer block of rounds. A \emph{star} joins appearances of one source, and a \emph{seed} initiates the marked field. A \emph{witness} is a connected derivation or contour for an error. A \emph{certificate} is a finite configuration counted in an error estimate. A group collects appearances with a common label or a common covering ball, as specified in the argument. A \emph{charge} indexes a potential, and a source's \emph{image} projects it into the coordinates used by the code. A catalogue lists local templates by type. Overlay faults occur outside the nominal replacement experiment. In the counting arguments, the cost of a configuration is its probability factor, and its entropy refers to the number of possible descriptions.

\begin{proposition}[Joint source law]\label{c:source-law}
There are fixed finite support lists and computable overlap divisors
\(D_{\rm bin},D_{\rm hist}\), independent of \(W,n\), such that the
following holds for every \(n\) and every aggregate refresh rate
\(\mu\in\mathcal J_n\). Let \(O_\lambda\) mean an internal fault in the complete
hardware support of an experiment in label \(\lambda\), and suppose
\(\Pr(O_\lambda)\le\alpha_{\rm bin}\) for every label \(\lambda\). For any \(g\) distinct source labels,
\begin{equation}\label{eq:c-gamma}
\begin{gathered}
 \Pr(\textup{all labels have a raw or intrinsic exception})\le\gamma_{\rm src}^g,\\
 \gamma_{\rm src}=\sqrt{2C_h}+\eta_B^{1/6}+\alpha_{\rm bin}^{1/(2D_{\rm bin})},
\end{gathered}
\end{equation}
where
\begin{align}\label{eq:c-eta}
 u_B&=2^{-\lfloor3BC_h/100\rfloor},&
 z_B&=2^{-\lfloor BC_h/3\rfloor},&
 w_B&=2^{-\lfloor BC_h^2/50\rfloor},\notag\\
 \eta_B&=\frac{3u_B}{1-4z_B}+2w_B,
 &4z_B&<1.
\end{align}
For \(m\) distinct historical fixed-gate tests the augmented intrinsic
intersection has probability at most \(\theta_{\rm pkt}^m\), where
\begin{equation}\label{eq:c-theta}
 \theta_{\rm pkt}=\eta_B+\alpha_{\rm bin}^{1/D_{\rm hist}}.
\end{equation}
A larger common divisor can be used for boundary tests.

Every erroneous initialized controller reading has a finite acyclic
backward derivation through ordinary two-box supports, sampling-register corridors,
or these source labels. In characteristic coordinates sampling-register corridors
preserve all five potentials used by the quantum proof. Collapsing them
leaves a finite set of roles and bounded bin shifts.
\end{proposition}
\par\noindent\emph{Proof idea.} Bound the statistical tests using their sampling means and tails. Colour the overlapping supports and combine the independent sampling and hardware events to obtain the joint bound.

\begin{proof}
The binary sampling-register coding changes the upper fresh-mark prefactor to
\(127/128\) and the lower to \(126/128\ge 0.98\). On a short depth
\(a-1-r\) ride the mean is less than \(0.53C_h\cdot4^{-r}\); the voting
threshold is at least \(0.75C_h\). On a recent long ride the mean exceeds
\(0.9C_h\), while the threshold is at most \(0.8C_h\). The exponential
moment bounds for these two Bernoulli comparisons give \(u_B,z_B,w_B\);
summing short depths gives the geometric denominator in
\eqref{eq:c-eta}. These are the intrinsic estimates in the classical
interface, with conservative binary-coding prefactors.

Raw \(H\) bins have disjoint physical atoms and probability at most
\(pW\le2C_h\). Baseline intrinsic tests use only \(E\) and fresh baseline mixed ancilla
marks. On \(A=0\) these are the actual marks; dirty outputs on \(A=1\)
can have arbitrary values, so we count them as internal hardware faults. Only
\(E\cap\{A=0\}\) is used to infer an actual uniform product. Distinct characteristics own disjoint lane characteristics, and
every third bin has disjoint intrinsic support; hence \(g\) specified
baseline tests cost at most \(\eta_B^{g/3}\).

An internal exception includes the full support of every short ride and
recent experiment in the bin. Its atoms lie on a fixed characteristic
translate list and in a fixed number of adjacent bins, so there are at
most \(h_0W\) atoms. Its probability is at most \(h_0W\nu\), where
\(\nu=\Pr(A=1)\). Supports can overlap only at differences from a fixed
finite list. A proper colouring of that overlap graph, with at least
one more colour than its degree, yields
\(\Pr(\bigcap O_\lambda)\le\alpha_{\rm bin}^{g/D_{\rm bin}}\).
The \(E\) and \(A\) families are independent. Combine their intrinsic
bounds, then use Cauchy--Schwarz against the raw family, which overlaps
both. Expanding the possible source types gives exactly
\eqref{eq:c-gamma}. Different \(Q\) copies within one selector have disjoint physical
innovations; their copy names remain part of the label.

For historical tests, write an atom of a ride starting at chart endpoint
\(g=(g_x,g_y)\), time \(b\) and replica \(P_b\) as
\[
 x=g_x-\ell+\delta_x,\quad y=g_y+j+\delta_y,
 \quad P=P_b+58\ell(1,1)+\delta_P,
 \quad s=b-\ell+\delta_t.
\]
The offset \(\delta\) ranges over a fixed support list. Equality of a
physical atom between two tests implies
\[
 b-b'=(\delta_x-\delta_t)-(\delta'_x-\delta'_t).
\]
Thus the difference of their start times lies in a fixed finite list,
independently of \(W\) and of the replica path. Colouring gives
\(D_{\rm hist}\), and independence of the baseline and internal
families gives \eqref{eq:c-theta}. This remains a fixed-support statement
when a later union bound chooses paths after seeing the noise.

A historical fixed-gate test is one physical sampling experiment,
given by a gate endpoint \(g\), start time \(b\) and transverse base
\(P_b\). Different annotations or roles of one experiment are the
same test. Its baseline atoms are the refresh coins of its own \(B\)
lanes on its own ride, the atoms with \(\delta=0\) above. On one
gate such an atom determines \(\ell=g_x-x\), \(b=s+\ell\) and
\(P_b=P-58\ell(1,1)\). Across gates, ruler and lane ownership
\cite[Lemmas~10.4 and~11.4]{DMNoise2026} assigns it one owner.
Hence distinct tests, including tests with \(b=b'\), have disjoint
baseline atoms. In Proposition~\ref{c:late-live} they are the
experiments of one candidate gate at distinct transverse bases
\cite[Lemma~11.6]{DMNoise2026}, on distinct copy characteristics.
So the baseline family contributes \(\eta_B^m\) with divisor one,
as in Section~\ref{v:parameters}. Only internal-fault envelopes
overlap. Equality of one envelope atom forces \(b-b'\) and
\(P_b-P'_b\) into fixed finite lists, read off the \(x\)-, \(s\)-
and \(P\)-coordinates above, independent of \(B,W,n\). This bounds
the overlap degree and gives \(D_{\rm hist}\).

Finally, a wrong clean ordinary output has more than \(K\) failing
replicas, so wrong inputs occupy two boxes. A sampling-register error either has
an intrinsic exception or follows its unique copy characteristic to a
wrong driver. That corridor preserves its characteristic potentials;
a driver step then makes positive progress. Expanding the fixed-depth
query and ordinary rule strictly decreases the formal history order,
ending at a named source or the ideal initial reference. This proves
finiteness and acyclicity. The projected role list is independent of the
lane length because its fixed offsets cancel along a complete corridor.
\end{proof}

\subsection{All-start halting estimates}
The estimates so far assume the intended initialization. We now work in one selector marginal and define events that dominate the corresponding events for all initial words at once.

\begin{lemma}[Finite-age coherence]\label{c:coherence}
Let \(\mathrm{MP}({\bf z},t)\) mean that an ordinary parent of \(({\bf z},t)\)
belongs to \(M^0\), and let \(D_s({\bf z},t)\) be the event that a backward
ordinary path of length \(s\) has at least \(\delta s\) such positions,
where \(\delta=1/1404\). Outside the corresponding two \(D_s\) events,
two unmarked live readings at the same logical position and round \(t\),
at replica distance \(d\), with \(t\ge s\ge\max(1,2d)\), agree. A clean
child check and this equality certify local west/south compatibility.
There are a computable source-rate cap and constants \(A,a>0\),
independent of \(s\), \(({\bf z},t)\), \(L\) and \(n\), such that below
the cap \(\Pr(D_s({\bf z},t))\le A\exp(-as)\) for every \(s\ge1\)
whenever the depth-\(s\) backward cone of \(({\bf z},t)\) unwraps
injectively. The age of a reading is its round index \(t\).
\end{lemma}
\par\noindent\emph{Proof idea.} Pair clean predecessor readings to compare the two proposals. A discrepancy needs many marked predecessors, whose complete contours can be counted by disjoint source slabs.

\begin{proof}
The recurring constants in this proof are derived in Appendix~\ref{q:constants-origin}.
At an unmarked root at most one replica box contains marked ordinary
inputs. Therefore at least \(h-K=256\) successful replicas read actual
unmarked words. When neither comparison root has \(\mathrm{MP}\), pair
replicas by a unit shift reducing their signed \(\ell^1\) distance.
There are \(3\cdot16\cdot15=720\) pairs. At most
\(2(3K-h)=512\) fail one of the two checks, leaving a reducing pair.
Functionality selects a common conflicting west or south input. At an
exceptional step choose any clean successful replica at each root;
functionality still selects a conflicting logical input, and replica
distance increases by at most 350. After one step both readings are
actual first-coordinate words. Distance-zero conflict is then impossible,
including at time zero. With fewer than \(2\delta s\) exceptional
steps the final distance is strictly less than
\[
 d-s+351(2\delta s)=d-s/2,
\]
a contradiction. For compatibility use one clean child replica, and
compare its two actual old inputs with the chosen reference inputs.

For the probability bound, use potentials
\(3(\chi_1+\chi_2),-3\chi_1,-3\chi_2\), where
\(\chi=P+58t(1,1)\). Good gains are 129, the largest loss is 351,
and a grouped old/new name has total oscillation 696. The balanced
terminal-arm count of Lemma~\ref{q:frontier} gives fewer than \(17f\) positive edges for \(f\)
distinct names; the conservative budget \(64f\) also covers the group
stars. Attach complete contours at the marked parents along the selected
path. Retain each complete contour and restart at least two times before
its earliest physical name. Disjoint source slabs then cover at most
\(68F\) path positions with \(F\) distinct names. Thus
\(F\ge\lceil\delta s/68\rceil\), and the path and contours form a
connected graph of \(O(F/\delta)\) edges. A fixed finite-degree graph has
at most \(K_\delta^F\) such rooted annotated graphs. Summing their
probabilities \(p^F\) proves the estimate at a small enough fixed rate.
\end{proof}

The constants \(A_n,a_n\) in Proposition~\ref{c:late-live} are decay
constants local to this section; the window endpoints \(a_n,b_n\) of
Section~\ref{s:parameters} are different.

\begin{proposition}[Late unmarked liveness]\label{c:late-live}
If \(U(n)\) halts, there are finite \(A_n,a_n>0\), uniform for
\(\mu\in \mathcal J_n\) and independent of \(L\), \({\bf z}\) and \(t\),
such that whenever the depth-\(t\) determining cone of \(({\bf z},t)\)
unwraps injectively on the physical torus,
\begin{equation}\label{eq:c-late}
 \Pr(\textup{some initial word gives an unmarked live reading at }({\bf z},t))
 \le A_n(1+t)^b e^{-a_nt}.
\end{equation}
The same conclusion applies to ordinary-proposal readings. Here the
exponent \(b\) is fixed by the finite cone geometry.
\end{proposition}
\par\noindent\emph{Proof idea.} Use finite-age coherence to expose a wrong critical test in a late valid chart. Count the complete boundary explanations that could prevent its error from reaching the target.

\begin{proof}
Use Lemma~\ref{c:coherence} at the deterministic roots of a late box.
The finite-history construction in the classical interface chooses that
box so its minimum age dominates twice its replica diameter. Its
exception has a polynomial-times-exponential bound. On the complement,
a live unmarked target determines a coherent same-history reference
patch. The halting motif supplies a wrong critical gate \(\Gamma\) at
time \(b_\Gamma\), with
\(t-b_\Gamma\ge c_0t\,-\,O_n(1)\) for a fixed \(c_0>0\) given by the
halting motif. Candidate gate positions, types, finite
phase residues and footprints range over a deterministic polynomial
list. The reference labels of a selected history are not included in
that list.

For each candidate let \(F_\Gamma\) be its deterministic slab footprint:
the routed vertices \((i_\Gamma,P,b_\Gamma)\), \((i_\Gamma,P,b_\Gamma+1)\)
and \((i_\Gamma+1,P,b_\Gamma+1)\), with \(i_\Gamma\) the candidate's first northeast
position (as below) and \(P\) in the candidate's
transverse footprint. Let \(S_\Gamma\subset F_\Gamma\) be the primitive
boundary exception set: every vertex of \(F_\Gamma\) carrying a selected
complete raw or deep certificate whose flags have physical time at most
\(b_\Gamma+1\), or a failed intrinsic test, enlarged by the packed balls
of \cite[Lemma~11.6]{DMNoise2026}, taken in characteristic distance and
intersected with \(F_\Gamma\), so that each ball covers both slab rows.
A flag of a boundary name at one of its formal occurrences in
\(F_\Gamma\) is a complete raw certificate there, with \(f=1\). Let
\(R_+\) contain both formal occurrences of every flagged raw physical
name of time at least \(b_\Gamma+2\).
Define \(M_\Gamma\) as the least field containing \(S_\Gamma\cup R_+\)
and closed under the two-box ordinary-parent rule at every child time
\(s\ge b_\Gamma+1\), including the upper row of the slab.
There are no other births. Deep queries, sampling
registers and the censor only replace a live proposal by \(\bot\)
(Section~\ref{c:chart}), so none of them is a birth. These definitions use only noise and the
candidate geometry. Minimize flag count and then a finite lexicographic
code to select certificates. The ordinary entropy count already includes
all these types. The envelope can overinclude actual errors, which only
strengthens domination of the target.

Reserve all boundary raw physical names through \(b_\Gamma+1\). The
formal slab is \(\{b_\Gamma,b_\Gamma+1\}\), and includes new seeds at
both first northeast positions \(i_\Gamma\) and \(i_\Gamma+1\).
Future raw names start at \(b_\Gamma+2\). Use west when
\(x>i_\Gamma\), and south on \(x=i_\Gamma\); this stays in the
northeast cone. Full supports, rather than their realized flagged subsets,
determine all restart cuts. These conventions separate boundary and
future raw names exactly.

The finite-history boundary-path estimate of the classical interface
extends to these grouped sources as follows. A genuine wrong live deep
query traverses all \(T_{\rm hist}=1024\) levels, so its \(64f\) budget requires
at least sixteen distinct names. A faulty current lookup is a direct
local exception instead. Pack complete raw or deep witnesses in disjoint
physical source slabs. Historical intrinsic tests obey
\eqref{eq:c-theta}; Hölder gives raw weight \((2\mu)^{3f/4}\) and
intrinsic weight \(\theta_{\rm pkt}^{m/4}\). The lag, gap and endpoint choices
cost at most \(W^{6f/16}\). Since \(W\le C_h/\mu\), the resulting raw
power per name is \(\mu^{3/8}\). Expand any propagated upper-row boundary terminal through this one
ordinary level to its primitive seeds, adding only fixed-range support
edges and fixed roles, with no new intrinsic experiment or lag choice.
Keep the expanded boundary contour whole, including old/new raw
occurrences, before making restart cuts. The boundary and future
physical-name sets remain disjoint by the preceding time partition.
Enlarge the fixed contour and entropy constants for this level before
choosing the rate cap. Consequently the strengthened boundary field
satisfies
\[
 \Pr(\text{a backward }N\text{-path at times }s\ge b_\Gamma+1\text{ has at least}
              \ N/2\text{ vertices in }M_\Gamma)
 \le A_\partial e^{-a_\partial N}.
\]
Choosing \(B\) makes the intrinsic base small, padding makes its
internal overlay small, and the nominal rate bound makes the remaining
raw base small. These choices are uniform on each selected window.

Boundary domination places the target in \(M_\Gamma\).
On the gate row let a routed footprint vertex be read live by a routed
child on row \(b_\Gamma+1\). If the copy read is flagged, its name has
physical time \(b_\Gamma+1\) and is a one-flag certificate at that
gate-row vertex, even when its actual output is \(\bot\). Otherwise its
actual output is live, and the intrinsic/raw/deep alternative (item~3
of the halting proof in \cite[Section~11.7]{DMNoise2026}) applies.
Either way the vertex is in \(S_\Gamma\). On the next row, a routed
child whose own output is flagged carries a one-flag certificate by its
boundary name of time \(b_\Gamma+1\), and so lies in \(S_\Gamma\).
Otherwise a live ordinary completion has \(h>K\) successful replicas
with live selected-index inputs on the gate row, by bottom-spreading.
These inputs are in \(S_\Gamma\) and occupy two boxes, so closure at
\(s=b_\Gamma+1\) marks the child.
The same induction continues at every later row and includes ordinary
proposals before censoring
\cite[Lemma~11.6, domination paragraph]{DMNoise2026}.
Put \(U_\Gamma=M_\Gamma\setminus M^0\) at times
\(s\ge b_\Gamma+1\). A vertex of \(U_\Gamma\) is not a future raw
seed, since that name also seeds \(M^0\). Strictly above the slab,
membership therefore comes from a closure step with selected marked
parents in two boxes. If all those selected parents belonged to
\(M^0\), its own two-box closure would mark the child.
Hence a selected parent belongs to \(U_\Gamma\). Following such a
parent strictly decreases time until it reaches the slab.
The result is an actual Boolean path of length at least
\(t-b_\Gamma-1\), entirely in \(M_\Gamma\) at times
\(s\ge b_\Gamma+1\). Apply the
boundary-path estimate, union over the candidate list, and add the
coherence exception. This proves \eqref{eq:c-late} simultaneously over
initial words.
\end{proof}

\begin{lemma}[Rare live proposals]\label{c:rare}
Fix \((i,Q)\) and \(t_0\ge0\). With \(\Theta_h\) denoting the
replica threshold (for a transverse field \(V\), \(\Theta_h(V)(P)=1\)
when \(V(P+\delta)=1\) for at least \(h\) of the \(3K\) offsets
\(\delta\in S_0\cup S_1\cup S_2\)) and \(B^{\rm fr}_s=H_s\lor H_{s+1}\) at \((i,P,Q)\), put
\[
 \mathcal W_{t_0-1}\equiv0,\qquad
 \mathcal W_s=B^{\rm fr}_s\lor\Theta_h(\mathcal W_{s-1}),\qquad
 N_s=\Theta_h(\mathcal W_s)\quad(s\ge t_0).
\]
For every initial word and every \(t>t_0\), if the ordinary proposal
at \((i,P,Q,t)\) is live, then (a) a direct destination exception
occurs: \(H\) at \((i,P,Q)\) at one of the times \(t-1\), \(t\), \(t+1\), or (b) \(N_{t-1}(P)=1\), or (c) there is a
backward path of live readings without \(B^{\rm fr}\) from \((i,P,Q,t)\)
to time \(t_0\), through south inputs of successful replicas.
Every new \(N\)-birth either has old support in two boxes or uses
at least 256 distinct fresh replica positions. Tile the transverse
plane by cubes \(C_c=176c+[0,176)^2\), \(c\in\Z^2\). For
\(P\in C_c\),
\(P+(S_0\cup S_1\cup S_2)\subset D_c=176c+[0,352)^2\).
The seed group \(\Sigma_{c,s}\), meaning that some \(P\in C_c\)
has a fresh alternative at time \(s\), depends only on \(H\) on
\(D_c\times\{s,s+1\}\) and has probability at most
\(N_{\rm seed}p^{256}\), where
\mbox{\(N_{\rm seed}=\tbinom{2\cdot352^2}{256}\)}.
Within each of the eight classes \((c\bmod2,s\bmod2)\) these supports
are pairwise disjoint. Thus \(g\) distinct specified seed groups
have joint probability at most
\begin{equation}\label{eq:c-rare}
 (N_{\rm seed}p^{256})^{g/8}.
\end{equation}
\end{lemma}
\par\noindent\emph{Proof idea.} Follow live south inputs backward until a fresh source or the initial row is reached, then count fresh source groups on disjoint doubled cubes.
\begin{proof}
A live completion needs both inputs, so selecting south keeps
\(i,Q\) fixed. If (a) and (c) fail, every backward chain of live
readings through successful replicas ends at a vertex with \(B^{\rm fr}\)
at time at least \(t_0\). A non-\(B^{\rm fr}\) live reading equals its live
ordinary proposal and has \(h\) successful replicas with live south
inputs. Induction up these finite chains puts every chain vertex
in \(\mathcal W\) and the root in \(N_{t-1}\), proving (b).
If fewer than \(h-K=256\) replicas are fresh, more than \(K\)
old replicas remain and occupy two boxes. Otherwise there are 256
distinct spatial positions, hence 256 distinct physical names, even
though either of two times can supply each position. The replica
boxes lie in \([0,175]^2\), which proves the displayed doubled-cube
inclusion. A union over subsets of 256 of the \(2\cdot352^2\)
physical names gives the seed bound. The four spatial cube parities
and two time parities give eight disjoint-support classes. One class
contains at least \(g/8\) of any specified collection, proving
\eqref{eq:c-rare} by independence of \(H\).

On a halting window take \(t_0=\lfloor t/2\rfloor\). Alternative
(c) gives either an unmarked live reading of age at least \(t_0\)
in the depth-\(t\) cone, excluded by Proposition~\ref{c:late-live}
with a polynomial union, or an all-\(M^0\) backward path of length
\(\lceil t/2\rceil\), excluded by the raw-path tail. These are the
events unioned in Propositions~\ref{c:path-tail} and~\ref{f:ancestor}.
\end{proof}

\begin{proposition}[Classical control of arbitrary data paths]\label{c:path-tail}
For each fixed finite catalogue of quantum dependency templates, the
rate bound can be chosen so that, on a halting window, the probability of
an \(s\)-step permitted data path through a specified physical initial
site is at most \(A'_n(1+s)^{b'}e^{-a'_ns}\), where \(A'_n,a'_n>0\)
depend only on \(n\) (uniformly over \(\eps\in I_n\), the size \(L\), the
site and \(s\)) and \(b'\) only on the catalogue. For the folded appendix, a path may switch among its four sheets
arbitrarily. It must cross at each step either its assigned live
selector or a named local fault at the conditional-reset block. The bound is
simultaneous over all initial combined classical words and holds in
finite cones that unwrap injectively on the physical torus.
\end{proposition}
\par\noindent\emph{Proof idea.} Combine the late valid-reading bound with the raw path bound in the data cone. Any residual quantum dependence must repeatedly pay for an exceptional classical reading or a hardware fault.

\begin{proof}
Union Proposition~\ref{c:late-live} and the marginal raw-path tails
over the selector collection and the polynomially many roots in the late actual
physical data cone. This union uses no independence between selectors.
For a late reading at time \(t\), take \(t_0=\lfloor t/2\rfloor\)
in Lemma~\ref{c:rare}. The union excludes alternative~(c), leaving
rare-field ancestry or a direct destination exception. Attach complete rare contours in
their own orientations and direct exceptions to the physical
path. Keep entire deterministic doubled-time cube supports and restart
below their earliest full physical time minus two. The union of the
finite controller collection and all possible programmed quantum edges is
a finite-degree graph, irrespective of the initial profile.

Let \(g\) count distinct typed rare groups and \(k_{\rm dir}\) distinct direct
physical names. With \(n_{\rm sel}=1\) in three dimensions and
\(n_{\rm sel}=4\) in the folded appendix, marginal bound \eqref{eq:c-rare} and
Hölder across selector families give seed intersection weight at most
\((N_{\rm seed}p^{256})^{g/(8n_{\rm sel})}\). Direct tests have weight
\(p^{k_{\rm dir}}\); they can overlap seed supports. Cauchy--Schwarz between
these two intersections therefore gives the combined bound
\begin{equation}\label{eq:c-mixed}
 (N_{\rm seed}p^{256})^{g/(16n_{\rm sel})}p^{k_{\rm dir}/2}.
\end{equation}
The rare power is \(256/(16n_{\rm sel})\ge4>0\). Temporal coverage and connected
edge budgets are fixed maxima over the four orientations. A length-\(s\)
path requires \(g+k_{\rm dir}\ge s/B_{\rm rare}\), while its annotated graph
count is at most \(K_{\rm rare}^{g+k_{\rm dir}}\). Choose the architecture's
cap so both entropy-weighted bases in \eqref{eq:c-mixed} are at most
\(1/64\). The resulting geometric sum decays exponentially. Polynomial
unions over accessible final roots and the earlier finite-cone
exceptions preserve that decay for the full joint physical history.
\end{proof}

\section{Elementary implementation and a finite parameter choice}\label{s:compile}\label{sec:compilation}
This section compiles the controller into the seven physical gates and gives a terminating order for choosing the constants. Sections~\ref{s:binary-alphabets} and~\ref{s:source-name} show that the finite controller alphabet affects only on-site depth and that all faults at one site in one round have one source name. Section~\ref{d:dispatch} compiles quantum gates through private ports. Section~\ref{s:terminal-erasure} builds the conditional reset that cuts quantum dependence. Sections~\ref{s:ws-realization} and~\ref{s:probe} realize the two noise models and derive the rates \(\mu,\nu,p\). Section~\ref{s:parameters} fixes the constants and windows.

\subsection{Binary alphabets and a streamed Boolean circuit}\label{s:binary-alphabets}
Encode a chart word in \(r=\lceil\log_2|\mathcal A|\rceil\) bits.
Unused codes mean the bottom symbol. Dephase every retained controller
word before using it, decode it and write canonical output codes. This
makes bottom aliases harmless even for an arbitrary initial state.
For sampling-register fields use seven bits: one low code, one unknown code and
126 high codes. Uniform whole-site replacement supplies uniform bits
in these registers. It leaves the needed upper and lower sampling-register laws
\begin{equation}\label{eq:helper-law}
 u_d=\frac{127}{128}\bigl[1-(1-\mu)^{\ell_d}\bigr],\qquad
 l_{\rm long}=\frac{126}{128}\bigl[1-(1-\mu)^W\bigr].
\end{equation}
Here \(\mu\) is the aggregate semantic refresh rate, the probability per site and round of the refresh event \(E\), which replaces every semantic register. The short experiment uses
only \(u_d\le\mu\ell_d\); the long-ride mean \(l_{\rm long}\) uses
\(126/128\ge 0.98\), so \(l_{\rm long}>0.9C_h\) in the parameter range below.

For a classical function \(f:\{0,1\}^m\to\{0,1\}^r\), start with the new output word and scratch zeroed, then enumerate
inputs \(w\). Negate wires corresponding to negative literals of \(w\),
compute their conjunction in zeroed scratch wires by Toffoli gates,
copy the resulting indicator into each output bit with \(f(w)_j=1\),
and uncompute. The cases \(m=0,1\) use constants and CNOT. A valid bound is
\[
 G_f\le2^m(8m+2r+20)+10(m+r+1).
\]
There are \(2^m\) rows and \(r2^m\) truth-table bits; neither need be
stored simultaneously. Computing \(f(w)\) from the finite semantic rule
and writing the corresponding gates is a terminating finite algorithm.
It uses no answer to a halting question. All its gates are on site.
The same construction computes finite matching rules for the quantum
processor. Scratch and transport registers reset before first use and
before the end of each round.

\subsection{Classical transport and the source name}\label{s:source-name}
\begin{lemma}[One source name per site and round]\label{d:transport-source}
All microscopic flags at a site in one round have one spatial source name, with old and new appearances at the two adjacent formal times.
\end{lemma}
\begin{proof}

For each displacement and bit allocate a distinct sender and a distinct
zeroed receiver. Fan out the old dephased word locally using CNOTs.
In one tensor-disjoint layer send each sender exactly once into its
receiver by CNOT. No received word is subsequently forwarded to another
site. The receiver then participates only in local computation.

A flag before transport at a sender corrupts copies of that site's
old word. A flagged zero receiver affects only that destination's new
computation. A flag after transport likewise affects only the new
computation at its site. Consequently all microscopic flags at one
site in one round have one spatial source name, with old and new
appearances at the two adjacent formal times. Reusing the same received
word for all roles reading that old site is essential. A spatial SWAP
implemented by three noisy CNOTs would allow a middle flag to be
forwarded. The one-way transport supplies the attribution used here.
\end{proof}

All information-moving quantum gates occur before the long on-site
Boolean calculation. The fixed native candidate-port
hardware has a bounded propagation cone. A fault during that hardware
is an arbitrary channel on a bounded semantic support. During the later
Boolean calculation quantum data is idle until the final erasure.
The spatial quantum-fault cone therefore does not grow with Boolean
depth or sampling lane length. Remote controller errors are charged by
the controller's derivation, rather than by a new quantum fault at
every induced erasure.

\subsection{Unconditional native operations on private ports}\label{d:dispatch}
For every native gate type in \eqref{d:native} and every translated
operand tuple, allocate private type and position ports. Reset them,
route matching semantic operands into their ports by metadata-controlled
local swaps, apply the native gate unconditionally to every translated
role tuple, reverse the routes and reset the ports. On the clean
circuit, either all operands of a matching active tuple are routed or
none are. Swap conjugation implements the native channel on data and
reference; an inactive gate acts only on disposable ports.
A fault inside an inactive route can affect its data cell and is counted
as a local dirty location.

Position role \(j\) at site \(u\) belongs only to tuple base
\(u-\delta_j\). Therefore all translated gates are tensor-disjoint,
even on small tori where sites coincide. Serialize types, incompatible
routes and elementary decompositions. Malformed masks still define a
valid circuit on the same finite neighbourhood. One semantic qubit per
chain cell suffices; code-sized work is distributed over additional
cells rather than added to this fixed role catalogue.

The controlled route is the Fredkin word
\[
\mathrm{CNOT}(b\to a),\quad
\mathrm{Toffoli}(c,a\to b),\quad
\mathrm{CNOT}(b\to a).
\]
For \(c=0\) the CNOTs cancel; for \(c=1\) it swaps the two factors.
The other gate words are
\[
 \mathrm{SWAP}=\mathrm{CNOT}_{ab}\,\mathrm{CNOT}_{ba}\,\mathrm{CNOT}_{ab},
 \quad \mathrm{CZ}=H_b\,\mathrm{CNOT}_{ab}\,H_b,\quad Z=HXH.
\]
Discard applies \(R_0\) to its retained slot.
Consequently every physical instruction belongs exactly to the seven-gate inventory in Section~\ref{s:model}. Every sublayer in these words
receives noise. The real closed native subsystem of
Section~\ref{sec:quantum} supplies every intended opcode.

Let \(\mathcal M\) denote all autonomous classical fields. Initially dephase
\(\mathcal M\); its own gates are Boolean basis permutations, resets and
dephasings. In an interaction with quantum data, \(\mathcal M\) is only a
control and never a target, at every elementary sublayer. Depolarization
preserves block diagonality in \(\mathcal M\). Induction shows that its noisy
history law is independent of the quantum input even for malformed
initial words and intermediate faults. Quantum-derived records stay
in the quantum track. Old clock symbols control the current port;
old independent copies compute the next classical symbols.

\subsection{A terminal erasure that cuts dependence}\label{s:terminal-erasure}
\begin{lemma}[Complete cut]\label{d:complete-cut}
A clean terminal conditional-reset block with erase bit \(c=1\) replaces every retained quantum factor by \(I/2\) and discards its previous reference correlations. Every path from retained input to retained output quantum factors in a round crosses that round's final conditional-reset block. A clean bottom trigger cuts these paths, including those through temporary hardware.
\end{lemma}
\begin{proof}

For each data qubit \(a\), prepare a mixed ancilla \(b\) by \(R_0,H,D_Z\),
forgetting the measurement outcome. Conditional on every intervening
flag history the result is exactly \(I/2\). With no flag this follows
from the three gates; after a flag the remaining operations are unital
on an already maximally mixed local quantum register.
Let \(c\) be the classical erase bit. The Fredkin word in Section~\ref{d:dispatch}
is identity on \(a,b\) when \(c=0\), and swaps them when \(c=1\).
Apply it to every retained quantum factor, including native syndrome,
ancilla and work factors, and then reset the mixed ancillas.
A clean block with \(c=1\) replaces the data by \(I/2\) and discards
its entire previous reference correlation. In \(\mathrm{WS}\), a site flag within the
block replaces all local quantum factors; the remaining controlled
unitaries preserve their maximally mixed state. The clean replacement identity is a complete channel identity,
including on inputs entangled with a reference.

For arbitrary-input forgetting we use the direct dependence statement:
in every round, every path from retained input quantum factors to
retained output quantum factors crosses that round's final conditional-reset
block. A clean bottom trigger cuts the path, including paths through
temporary hardware. Only preparation, storage, noise and final
work discard or reset follow the block. This statement is stronger than
resetting mixed ancillas at endpoints: a swap into a mixed ancilla followed by a
swap back after erasure would violate it. The actual schedule satisfies
the per-round condition.
\end{proof}

\subsection{Whole-site realization WS}\label{s:ws-realization}
Let \(J\) be the number of internal layers, including their following
noise. Add \(P_{\rm pad}\) identity/noise layers at the end of each round. At
site \(x\) and round \(t\), let \(A_{x,t}\) mean at least one internal
replacement and \(E_{x,t}\) at least one padding replacement. Then
\begin{equation}\label{eq:macro-rates}
\begin{gathered}
 \nu=1-(1-\eps)^J,\qquad \mu=1-(1-\eps)^{P_{\rm pad}},\\
 H=A\lor E,\qquad p=1-(1-\nu)(1-\mu)\le2\mu\quad(P_{\rm pad}\ge J).
\end{gathered}
\end{equation}
These are independent families across physical
site-round names; \(A\) and \(E\) are independent. Padding can be
represented by one replacement mark, the last one, when \(E=1\).
Constituent marks are independent uniform variables conditional on that
one site flag. Raw source tests may use \(H\); baseline sampling-register tests use
only padding flags and marks.

For \(\mu\le1/2\),
\[
 \nu\le J\eps\le2J\mu/P_{\rm pad}.
\]
A source bin using at most \(h_0W\) site-round source names, with \(W\le C_h/\mu\),
is spoiled by internal flags with probability at most
\begin{equation}\label{eq:overlay}
 \alpha_{\rm bin}\le 2h_0JC_h/P_{\rm pad}.
\end{equation}
The overlap graph has fixed finite colour divisors \(D_{\rm bin}\)
and \(D_{\rm hist}\), chosen at least one above its maximum degree.
They depend on the spatial supports within one round, not on \(J,W,n\).
Independence of \(A\) and padding experiments, colouring, and
Cauchy--Schwarz give the augmented estimates
\begin{equation}\label{eq:gamma}
 \gamma_{\rm src}\le\sqrt{2C_h}+\eta_B^{1/6}
       +\alpha_{\rm bin}^{1/(2D_{\rm bin})},\qquad
 \theta_{\rm pkt}=\eta_B+\alpha_{\rm bin}^{1/D_{\rm hist}}.
\end{equation}
Distinct labels within one selector obey the joint bound \(\gamma_{\rm src}^g\), as used in
Lemma~\ref{d:sources}. The historical packet bound is \(\theta_{\rm pkt}\). Labels,
rather than controller-induced erasures, carry this bound.

\subsection{Independent single-qubit realization Q}\label{s:probe}

\begin{lemma}[Refresh law in Q]\label{d:qubit-refresh}
Let \(M\), \(K_{\rm samp}\), \(N_{\rm np}\), \(J\), the internal-fault family \(A\) and \(\nu=\Pr(A)\) be as constructed in the proof below.
The aggregate refresh event \(E\) is independent of the semantic input and reference, with probability
\[
 \mu(\eps)=1-\left[\frac{1+(1-\eps)^{K_{\rm samp}}}{2}\right]^M.
\]
Its coin family is independent of the internal-fault family \(A\). The latter has probability \(\nu\) bounded by \eqref{eq:q-overlay}. On \(E\land\neg A\) the semantic output is the independent product of \(I/2\) on every factor, and on \(\neg E\land\neg A\) it equals the clean core output.
\end{lemma}
\begin{proof}
The \(\mathrm Q\) realization supplies the same round interface by a separate
physical circuit. It adds a local sampling experiment and controlled
refresh of every semantic register, including all quantum, clock,
selector and sampling-register fields (and the profile in the folded appendix). Denote by \(N_{\rm np}\) the number of
nonprobe wires per site: the core, one mixed ancilla per semantic wire,
two OR scratch bits and all fixed work. This count is independent of
the number \(M\) of probes.

To see this, compute OR serially. Given a basis-valued accumulator
\(a\), probe \(q_{\rm bit}\) and target \(b=0\), apply \(\mathrm{CNOT}(a\to b)\), \(\mathrm{CNOT}(q_{\rm bit}\to b)\) and
\(\mathrm{Toffoli}(a,q_{\rm bit}\to b)\). Then
\mbox{\(b=a\oplus q_{\rm bit}\oplus aq_{\rm bit}=a\lor q_{\rm bit}\)}.
Reset \(a\), copy \(b\to a\) and reset \(b\). Repeating this
six-gate pattern uses two scratch wires.

After the core and its terminal quantum cut, reset all \(M\) probes to
zero. Including the noise following that reset as the first layer,
let them undergo \(K_{\rm samp}\) sampling noise layers, with identities in the
remaining layers. Every nonprobe wire idles and is noisy as well.
At the endpoint, each probe has independent probability
\(q_{\rm end}=[1-(1-\eps)^{K_{\rm samp}}]/2\) of value one. All its sampling states are
diagonal. Define the exogenous event \(E\) as the OR of these endpoint
values, before any subsequent noisy OR computation. Its exact rate is
\begin{equation}\label{eq:q-refresh}
 \mu(\eps)=1-\left[\frac{1+(1-\eps)^{K_{\rm samp}}}{2}\right]^M.
\end{equation}
It is independent of the semantic input and reference because of the
initial probe reset.

Compute OR, prepare private mixed ancillas by \(R_0,H,D_Z\) and use
the controlled-swap gadget to refresh every semantic factor when the
computed control is one. Reset all swapped-out contents, probes and
scratch before declaring the round's outputs. No discarded quantum
state is recovered afterwards. Let \(J\) count all nonsampling layers,
including the core and this finite OR/refresh/discard circuit. We fix \(J\) after \(M\) and before \(K_{\rm samp}\).

Define \(A\) as the union of all flags on nonprobe wires during
sampling and all flags on every wire in the nonsampling layers.
The primitive coins defining \(A\) are disjoint from all probe
sampling flags and endpoint marks defining \(E\). Hence \(A\) and \(E\) are independent, and each is an iid site-round field, with
\begin{equation}\label{eq:q-overlay}
 \nu\le[N_{\rm np}K_{\rm samp}+(M+N_{\rm np})J]\eps.
\end{equation}
Post-sampling probe faults, including an incorrect OR, belong to
\(A\). On \(E\land\neg A\) the semantic output is exactly the
independent product of \(I/2\) on every factor; on
\(\neg E\land\neg A\) it equals the clean core output. We use only
this clean refresh implication in \(\mathrm Q\). Actual output on \(A\) may be
arbitrary; in particular, \(E\) alone does not imply uniform output.
Counterfactual baseline mixed ancilla marks can be taken independent and
uniform on dirty rounds, agreeing with the actual refresh on clean
rounds. These marks support the original sampling-register experiments, while
\(A\) enlarges every dirty experiment to an intrinsic exception.

For \(MK_{\rm samp}\eps\le1/2\), elementary union and product bounds give
\[
 \frac{K_{\rm samp}\eps}{2}\le1-(1-\eps)^{K_{\rm samp}}\le K_{\rm samp}\eps,
 \qquad
 \frac{MK_{\rm samp}\eps}{8}\le\mu(\eps)\le\frac{MK_{\rm samp}\eps}{2}.
\]
Consequently
\begin{equation}\label{eq:q-ratio}
 \frac\nu\mu\le
 \frac{8N_{\rm np}}M+
 8\left(1+\frac{N_{\rm np}}M\right)\frac{J}{K_{\rm samp}}.
\end{equation}
Given an overlay target \(0<\delta_{\rm ov}\le1\), choose
\(M\ge\max(N_{\rm np},32N_{\rm np}/\delta_{\rm ov})\), compile the
finite nonsampling circuit and its \(J\), and then choose
\(K_{\rm samp}\ge128J/\delta_{\rm ov}\). The ratio is at most
\(3\delta_{\rm ov}/8\). At \(\eps_0=1/(2MK_{\rm samp})\), the refresh rate is at
least \(1/16\), and it is strictly increasing from zero. Thus every
nominal interval below \(1/100\) has a physical preimage strictly
below \(\eps_0\). All rate comparisons in the infinite list stay in
the proven range of~\eqref{eq:q-ratio}.

An intrinsic test sees only \(O(W)\) site-round source names in fixed characteristic
translates. Since \(W\mu\le C_h\), its overlay probability is at most
\(c_{\rm ov}C_h\delta_{\rm ov}\), with a fixed support count
\(c_{\rm ov}\). The overlap colours and old/new attribution are
unchanged: sampling and refresh add only on-site depth. This replaces
\eqref{eq:overlay} in~\eqref{eq:gamma}. Also
\(p=\Pr(A\lor E)\le(1+\delta_{\rm ov})\mu\le2\mu\).
The \(\mathrm Q\) site has \(M+N_{\rm np}\) wires, all noisy after every layer.
Arbitrary initial probe, mixed ancilla and scratch states are reset before
use. Every outside data slot retains its own selector-controlled reset.
Thus neither the new hardware nor the refresh can bypass the original
per-round complete cut.
\end{proof}

\subsection{A terminating choice of all constants}\label{s:parameters}
The choices specify a fixed circuit by terminating algorithms in the following order.
\begin{enumerate}[label=\arabic*.]
\item \emph{Quantum layout.} First
fix the semantic real-gate inventory, roles, autonomy and complete-cut
templates. The semantic width, halo and participation bounds determine
\(a_{\rm sep}\) and \(b_{\rm FT}\) in \eqref{d:budgets}.
Choose \(r_{\rm FT}=8b_{\rm FT}\), the least code parameter
in \eqref{d:capacity}, the hardened finite quantum table, and then
\(P_{\rm cal},T_{\rm cal}\) and its classical interpreter.
This fixes the full witness geometry and entropy used by the positive
threshold. The budget \(b_{\rm FT}\) already includes the physical conversion multiplicity.

\item \emph{Sampling scale and lane count.} Choose rational \(0<C_h<0.01\) small enough for the strict source
inequalities. Search powers \(B=4^{k_B}\), with \(B\ge20/C_h\), until
\(4z_B<1\) and every intrinsic source base is below its required
bound. The quantities in \eqref{eq:c-eta} decay exponentially in
\(B\), while their support and entropy prefactors grow polynomially.
This search terminates. With \(s_{\rm vote}=\lceil3C_hB/4\rceil\), it also gives
\(s_{\rm vote}/B\le0.8C_h\), below the mean of the long sampling experiment.

\item \emph{Rate bound.} After \(B\), choose support colours, overlay margins and a positive
macro-rate bound. Decrease it to meet the source-certificate recursion
of Lemma~\ref{d:recursion}, the historical/boundary and rare-path
bounds of Section~\ref{sec:controller}, and the wrapped bound of
Lemma~\ref{u:wrapped}. Deep queries require at least sixteen raw names:
their lag entropy \(W^{6f/16}\) is beaten by raw weight
\((2\mu)^{3f/4}\), leaving the positive power \(\mu^{3f/8}\).
Together with \(W\mu\le C_h\), this makes the smallness searches
terminate. The rate bound is chosen for this fixed processor catalogue.

\item \emph{Window offset.} Choose \(\mu_*\le\min(C_h/100,1/100,p_*/2)\), with \(p_*\) from \eqref{eq:u-cap}, inside all these margins
and then \(n_0\ge k_B+2\) with
\(21C_h/[10\ell_{n_0+1}]<\mu_*\).
\item \emph{Chart and hardware.} Build the fixed scan machine,
its chart alphabet and binary implementation only now. Full source
geometry and rate bounds do not depend on the number of palette codes.
In \(\mathrm{WS}\), count every physical internal layer \(J\) and choose
independent padding \(P_{\rm pad}\) large enough for
\eqref{eq:overlay}. In \(\mathrm Q\), choose the overlay target,
count all nonprobe wires, choose \(M\), count the physical nonsampling
\(J\), and then choose \(K_{\rm samp}\) as above. Every port, idle,
reset and decomposition layer is included. Long on-site depth changes
these counts without adding a spatial quantum stencil.

\item \emph{Windows.} For \(n\ge0\), define
\begin{equation}\label{eq:windows}
 \mathcal J_n=\left(\frac{2C_h}{\ell_{n+n_0+1}},
            \frac{21C_h}{10\ell_{n+n_0+1}}\right).
\end{equation}
Use the separately chosen strictly increasing rational refresh polynomial
\(\mu_{\mathrm{WS}}\) or \(\mu_{\mathrm Q}\). Enumerate rational pairs
\(0<a_n<b_n\) in the overlay-valid regime until
\[
 \inf \mathcal J_n<\mu(a_n)<\mu(b_n)<\sup \mathcal J_n,
 \qquad I_n=(a_n,b_n).
\]
Exact rational polynomial comparisons suffice, and density guarantees
termination. These are rational inner intervals; their closures lie
strictly inside the full preimages, whose endpoints can be algebraic.
They are nonempty, disjoint and tend to zero. The offset calculation
following Proposition~\ref{c:selector-import} identifies their exact
machine input.
\item \emph{Size and time.} Choose the rational storage coefficient \(c\) after
the final speed and source footprint, and the total size cutoff \(L_0(n)\)
after all finite label margins, as in Proposition~\ref{d:storage}.
\end{enumerate}

\section{Three-dimensional storage with a protected classical clock}\label{sec:quantum}

This section proves Theorem~\ref{thm:detailed}(i). It develops the concatenated-code and coherent-gadget estimates (Lemmas~\ref{d:code}--\ref{d:reduction}), constructs the classical clock and its source law (Lemma~\ref{d:sources}), and proves the renormalization bound and storage at every requested time (Lemma~\ref{d:recursion} and Proposition~\ref{d:storage}). The new source estimates account for correlated errors in the protected clock.

Semantic registers are the registers of the scheduled circuit before
compilation. Write the physical coordinates as \((i,\sigma,z)\).
Quantum data move only along \(\sigma\), at fixed \((i,z)\). One semantic qubit per chain cell suffices. Code coordinates and gadget work occupy additional cells. Two autonomous classical fields have
distinct jobs. The selector
\(C_{\rm sel}\) supplies the compulsory final cut; a field \(S\)
schedules quantum correction. Every syndrome and every bit computed
from quantum data belongs to the quantum track, including when it is
basis-valued.

The hierarchy follows the reliable-computation tradition of G\'acs
and G\'acs--Reif \cite{Gacs2001,GacsReif1988}; hierarchical local quantum
decoding has an early precedent in Harrington \cite{Harrington2004}.
The closest quantum precedent is the two-dimensional translation-invariant automaton of D\"unnweber, Styliaris and Trivedi \cite{DST2026}, combining a G\'acs-type self-simulating classical hierarchy with a measurement-free concatenated code; here the clock must also tolerate noise after every elementary layer and coexist with controller-triggered erasures.
We use the coherent CSS gadget designs of Aharonov and Ben-Or
\cite{ABO2008}. The support bound needed here is proved below
for the specified real-gate subsystem.

\subsection{An explicit binary code and one recovery}

\begin{lemma}[Binary code family]\label{d:code}
For every integer \(s\ge1\) there is an effective CSS code with parameters
\[
 [[n_s,1,d_s]]=[[2^{2s+1}-1,1,2^{s+1}-1]],
 \qquad q_s=2^s-1
\]
and one stabilizer recovery correcting the union span of all errors of
weight at most \(q_s\). Encoding, recovery and the total classical
functions used by the gadgets have finite exact real circuits.
\end{lemma}
\par\noindent\emph{Proof idea.} Use the orthogonality and distance of the binary evaluation codes to obtain one CSS logical qubit. Finite elimination and syndrome recovery give exact encoding and correction procedures.

\begin{proof}
Let \(\mathcal R_s=\mathrm{RM}(s,2s+1)\), where \(\mathrm{RM}(s,m)\)
is the binary Reed--Muller code of evaluations of multilinear polynomials
of degree at most \(s\) on \(\mathbb F_2^m\).
Products of two generating monomials have degree below \(2s+1\),
so their evaluation inner product is even. Also
\(\dim\mathcal R_s=\sum_{j=0}^s\binom{2s+1}{j}=2^{2s}\), half the
length. Thus \(\mathcal R_s\) is self-dual. Each generator has weight
divisible by four. Self-orthogonality makes every pairwise intersection
even, and \(\mathrm{wt}(u+v)=\mathrm{wt}(u)+\mathrm{wt}(v)
-2|\supp u\cap\supp v|\) proves double evenness for every word.

The usual polynomial splitting \(f=f_0+x_m f_1\), with \(f_0,f_1\) local to this argument, proves the distance of
\(\mathrm{RM}(r,m)\) to be \(2^{m-r}\): if \(f_1=0\), the weight
is twice that of \(f_0\); otherwise the weights of \(f_0\) and
\(f_0+f_1\) sum to at least the weight of \(f_1\). Induction gives
the bound, and a product of \(r\) coordinate literals attains it.
Puncture at zero to obtain \(\mathcal C_s\). Its dual
\(\mathcal D_s=\mathcal C_s^\perp\) is the shortened code, contained
in \(\mathcal C_s\) with codimension one. These are the CSS codes of
punctured doubly even self-dual codes used by Shor
\cite[Sections~5 and~7]{Shor1996}, here from the self-dual Reed--Muller
codes; \(s=1\) gives the \([[7,1,3]]\) Steane code. The logical words are
\[
 |a_L\rangle=|\mathcal D_s|^{-1/2}
       \sum_{w\in\mathcal D_s}|w+a\one\rangle,
       \qquad a\in\mathbb F_2.
\]
Puncturing lowers distance by at most one, and
\(\prod_{j=1}^s(1+x_j)\) has value one at the removed point and
attains weight \(2^{s+1}-1\) in the nontrivial coset. This gives
\(d_s\) and \(q_s\). Products of two weight-\(q_s\) Pauli errors
have weight below \(d_s\); their compression to the code is scalar
or zero. The Knill--Laflamme equations \cite{KnillLaflamme1997} give one recovery
for the entire error span, including coherent cross terms.

Choose generators of \(\mathcal D_s\), complete them to an invertible
binary map, prepare the generator variables with Hadamards and the
remaining variables in zero, and apply that map by CNOT and SWAP.
This prepares the zero word; its transversal Hadamard gives the plus
word. Finite syndrome tables with a fixed choice outside the decoding
radius give total recovery circuits. Boolean functions use the
compute, copy and uncompute construction of
Section~\ref{sec:compilation}. All these prescriptions terminate for
each fixed \(s\). The code size and the running time may be enormous.
\end{proof}

\subsection{Real native procedures and the sign bit}
In the gadget bound below, \(j\) counts faults; in coordinatewise
operations and Table~\ref{tab:toffoli}, \(j\) indexes a protected
coordinate. Preparation depth \(t\) is local to the definition of \(V_t\).
The clock issues semantic gates, which Section~\ref{d:dispatch}
compiles into physical gates. Use the semantic inventory
\begin{equation}\label{d:native}
 \mathcal G_{\mathbb R}=
 \{I,X,H,\mathrm{CNOT},\mathrm{CZ},\mathrm{Toffoli},
                      \mathrm{SWAP},R_0,\text{discard}\}.
\end{equation}
Discard on an existing physical slot is reset. The transversal logical
identities follow directly from
\((a\one+w)\cdot(b\one+w')=ab\pmod2\) for
\(w,w'\in\mathcal D_s\), and character orthogonality for the Hadamard.
They give logical \(X,Z,H,\mathrm{CNOT},\mathrm{CZ}\). For \(Z\), use
the identity with \(b=1\) and \(w'=0\). SWAP exchanges blocks. Toffoli is implemented coherently with a resource state below.

For a basis word \(w\in\mathbb F_2^{n_s}\), define \(a(w)\) by
bounded-distance decoding: if \(w\) is within Hamming distance \(q_s\)
of \(\mathcal C_s\), choose its unique nearest codeword and return
its coset in \(\mathcal C_s/\mathcal D_s\). Otherwise set \(a(w)=0\).
Uniqueness follows from \(d(\mathcal C_s)\ge2q_s+1\). In particular,
any at most \(q_s\) bit errors on a word in either logical coset
leave \(a(w)\) unchanged. Every decoded logical bit below, of a cat
block or of a consumed resource, data or target block, uses this decoder.

The sign operation at protected coordinate \(j\) is
\begin{equation}\label{d:sign}
 |w,b_j,c_j,d_j\rangle\longmapsto
 (-1)^{a(w)(b_jc_j+d_j)}|w,b_j,c_j,d_j\rangle.
\end{equation}
Compute \(a(w)\) in private scratch, apply
\(\mathrm{CCZ}(a,b_j,c_j)\mathrm{CZ}(a,d_j)\), then uncompute.
Here
\[
 Z=HXH,\quad \mathrm{CZ}=H_b\,\mathrm{CNOT}_{ab}\,H_b,
 \quad \mathrm{CCZ}=H_{\rm target}\,\mathrm{Toffoli}\,H_{\rm target}.
\]
The module touches one expendable cat block and only coordinate \(j\)
of each protected block. Its possibly large Boolean computation has
no other data target.

\begin{lemma}[Complete routed gadget bound]\label{d:gadgets}
The code family admits coherent error correction, deterministic verified
zero preparation, and coherent encoded bodies for
\(\mathcal G_{\mathbb R}\), with one-dimensional routing. A fault-free
complete error correction maps every input into the code. For a
correctable input deviating on at most \(f\) coordinates per block,
a complete routed error-correction-plus-body rectangle with \(j\)
faulty locations has output in the operator span of the ideal output
with at most \(8j\) erroneous coordinates per block, provided
\begin{equation}\label{d:margin}
 8(f+j)\le q_s-8.
\end{equation}
The statement includes arbitrary environment and reference coefficients.
Physical decomposition multiplicity is counted in the location budget,
rather than in the constant eight.
\end{lemma}
\par\noindent\emph{Proof idea.} Count touching faults for each routed primitive. Verified preparation, repeated recovery checks and separate port work keep the residual error within the correction bound.

\begin{proof}
The proof treats verified preparation, coherent error correction, coherent Toffoli and routing in turn. A native location is a gate, wait
or reset of the circuit before routing. All records and their reversible
processing are quantum work. A private correction decision for coordinate
\(i\) controls only that coordinate. Record copying is done before
these private computations. Thresholds and repetitions below exceed
the admitted fault budget.

\smallskip\noindent
\emph{Deterministic verified preparation.}
Here \(V_t\) denotes verified preparation. Let \(V_0\) be the exact encoder for \(|0_L\rangle\).
For \(t\ge1\), \(V_t\) uses five \(V_{t-1}\) blocks: a candidate
\(B_1\), testers \(B_2,B_3,B_4\) and spare \(B_5\).
Apply transversal CNOTs \(B_1\to B_2\), then \(B_3\to B_4\) to
check \(B_3\)'s bit errors before it interacts with the candidate.
Use the resulting basis words of \(B_2,B_4\) as the two raw bit
records. Then apply transversal CNOT \(B_3\to B_1\) and transversal
\(H\) to \(B_3\), whose basis word is now the raw phase record.
Copy each raw record in the computational basis for each coordinate
before any decoding. On its own copies, each coordinate computes
minimum-weight representatives modulo \(\mathcal D_s\) for both bit
records, so the check includes logical \(X\), and modulo
\(\mathcal C_s\) for the phase record. These are total functions with
a fixed lexicographic tie-break. For input Pauli components
\(X^{x_i}Z^{z_i}\) on \(B_i\), clean checks give the bit shifts
\(x_1+x_2\) and \(x_3+x_4\) modulo \(\mathcal D_s\) and the phase shift
\(z_1+z_2+z_3+z_4\) modulo \(\mathcal C_s\). The candidate carries
\(X^{x_1+x_3}Z^{z_1+z_2}\) before selection. These identities follow by
propagating \(X\) forward and \(Z\) backward through the CNOTs. Shifts
by \(\mathcal D_s\) and phases from \(\mathcal C_s=\mathcal D_s^\perp\)
fix \(|0_L\rangle\).
Each private decision tests whether the union of the three supports
has weight at most \(t\), and controls only its own Fredkin between
\(B_{1,i}\) and \(B_{5,i}\), swapping on rejection. Dispose of the work
at its declared end. This is a predetermined replacement circuit.

\emph{Claim.} If \(V_t\) contains \(k\le t\) faults, its output lies in the span of \(F|0_L\rangle\) with \(\mathrm{wt}(F)\le k\). We prove this by induction on \(t\), in two cases.

\emph{Case 1: every recursive subtree has at most \(t-1\) faults.} By induction, each subtree's deviation is at most its own fault count. On each Pauli term, a fault in a transversal check can affect both blocks at its coordinate, and a raw fan-out fault can change the bit at its record index in the copies it reaches, including a proper subset of them; both effects add at most one index per fault to the union of possible record shifts. Raw phase factors act only on the consumed records. Since \(k\le t\le q_s\), every copy reaching a clean private decoder has a unique minimum-weight representative within this support bound. The union used by each such decision therefore has weight at most \(k\), and the decision keeps the candidate. The residual is supported on the charged indices that reach the candidate, together with the targets of faulty private computations or Fredkins; each such fault affects only its own output coordinate.

\emph{Case 2: some subtree has at least \(t\) faults.} It consumes the entire budget: \(k=t\), and the final checks and all other blocks are clean, so all private decisions agree within each coherent raw-record branch. A bad candidate is either replaced by the clean spare or accepted with union pattern of weight at most \(t\). A bad \(B_2\)'s transferred phase errors are detected by the exact phase check; a bad \(B_3\)'s dangerous bit pattern is checked by \(B_4\) before coupling. If only \(B_4\) is bad, rejection selects the clean spare; if only \(B_5\) is bad, the clean candidate passes. Each case gives the same bound.

In neither case can the output be a coordinatewise splice of two arbitrary codewords: in Case~1 every uncorrupted decision keeps the candidate, and in Case~2 all decisions agree. Use depth \(t=q_s\); it is finite and at least the native rectangle budget \(2b_{\rm FT}\) selected below.

\smallskip\noindent
\emph{Coherent error correction.}
For bit correction couple data transversally into a verified plus
block \cite{Steane1997}. For each eventual data coordinate make a
private record and compute its nearest-codeword error vector modulo
\(\mathcal C_s\). Apply only that vector's corresponding correction bit.
A data-record fan-out fault at index \(i\) can flip that bit in any
copy it reaches, even when some copies have already been made. Inside
the decoding margin the nearest codeword stays the same, so the
inferred error changes only at \(i\), and only the decision for
data coordinate \(i\) can apply a different correction. Phases on
the copied words act on consumed work. A fault inside a private
computation reaches only its own data target. Phase correction repeats
the argument in the Hadamard basis, with bit and phase roles
interchanged. The transversal coupling transmits at most one
coordinate per fault, even if a fault affects both operands. The preparation bound just proved charges
those coordinates to their faults. Applying the two stages
removes the incoming correctable error and leaves at most one
coordinate per native fault in each output block.

For normalization, also compute the radius indicator privately.
In an exact syndrome branch outside the prescribed decoding region,
exchange every data coordinate with the corresponding coordinate of
a verified zero block. In a branch inside it, correct the error.
In the fault-free circuit all private indicators agree in each
syndrome branch. Correction maps the admitted syndrome to the code,
and replacement returns the verified zero block. Thus the entire
output has code support for an arbitrary input, including a
purification. Under the margin used here all copies reaching clean
private decoders stay inside the replacement threshold. A faulty
private decision or swap can affect only its own output coordinate.

\smallskip\noindent
\emph{Coherent Toffoli.}
We use a coherent form of Shor's Toffoli construction
\cite[Section~6]{Shor1996}, including the resource state, its preparation
by cat sign tests and its conversion to the gate. The processing is
measurement-free as in \cite{ABO2008}. Put
\[
 |A_0\rangle=\frac12\sum_{a,b}|a_L,b_L,(ab)_L\rangle,
 \qquad |A_1\rangle=X_{L,3}|A_0\rangle.
\]
Their equal superposition is three encoded plus blocks. Prepare each
by \(V_{q_s}\) followed by transversal \(H\). For each of
\(R_{\rm cat}=2\lfloor q_s/2\rfloor+1\) sign tests, prepare a fresh
encoded cat by fanning out a verified plus block via transversal CNOT
into verified zero blocks, all made using \(V_{q_s}\). Use one expendable
cat block per protected coordinate in \eqref{d:sign}. Each test uses
its own cat. On ideal resource words with logical bits \(x,y,z\),
the sum of \(b_jc_j+d_j\) over the protected coordinates is \(xy+z\)
modulo two, by the transversal identities. A clean test in sector
\(xy+z\) writes that sector's parity on the cat and leaves the three
protected logical qubits unchanged.

Apply complete error correction (EC) to the resource blocks between
tests. Apply transversal Hadamards to the used cats. Each protected
coordinate privately decodes all cat blocks by bounded-distance coset
decoding, takes the parity for each test and computes the majority.
Each private branch decision controls one coordinate of \(X\) on the
third resource block.

The preparation bound charges each residual coordinate of the initial
resource and cat blocks to its originating fault. Transversal fan-out
copies at most one coordinate per fault into each cat block, leaving
its decoded coset unchanged within the margin. Each sign computation
can spoil its own expendable cat block and touch only one coordinate
of each resource block. In each sign sector, expand a disturbed test
in operators on the affected cat and resource coordinates. Propagating
a resource-coordinate error through its controlled sign module adds
only cat operators and errors at that same resource index. The next
EC removes the resource error into a syndrome factor, so the ideal
logical words in the sector are unchanged, with all exterior systems
retained as coefficients. A residual left by the initial preparation
or by one EC can affect only the next test, since the following EC
removes it. This description iterates with earlier cats and syndromes
retained in the exterior coefficients. A fault on a used cat during
its Hadamards or record copying reaches only that test's record.
Each native fault can therefore spoil at most one repeated test for
the undisturbed private decisions. Their majority is correct for at
most \(\lfloor q_s/2\rfloor\) native faults. Faults in a private parity,
majority or branch correction affect only the corresponding target
coordinate. The leading EC of the conversion removes the remaining
resource errors before data contact. With at most \(2j\) native
faults from \(j\) routed faults, the margin in \eqref{d:margin}
covers both the majority and this EC.

The resource-to-gate conversion uses transversal CNOT and Hadamard,
private decoding of consumed blocks and private real corrections.
Its disposal preserves an unknown complex qubit. For half-Toffoli on
logical data bits \(c,d\), let the resource carry \((x,y,xy)\); CNOTs
from its first two blocks into the data leave consumed outcomes
\(u=c\,+\,x\), \(v=d\,+\,y\). Add
\(xv\,+\,yu\,+\,uv\) to the third register and apply \(X^u,X^v\) to
the first two. Since \(xy+xv+yu+uv=(x+u)(y+v)=cd\), the
retained registers become \((c,d,cd)\) in every branch;
\((u,v)\) is uniform independently of \((c,d)\), and all decoder
garbage depends on \((u,v)\) alone. For full Toffoli on data bits
\((a,b,c)\), apply this to \((a,b)\) and add the old target to the
third resource block by CNOT, so the new target is \(c+ab\). The old
target, after Hadamard, decodes to a bit \(h\) and contributes phase
\((-1)^{hc}\); the output correction, \(Z^h\) on the new target and a
\(\mathrm{CZ}^h\) on the two retained control blocks, contributes
\[
 (-1)^{h(ab+(c+ab))}=(-1)^{hc}.
\]
They cancel, leaving the same garbage vector for every \((a,b,c)\).
Consequently disposal gives the correct gate for arbitrary complex
amplitudes and a reference.

Table~\ref{tab:toffoli} gives the conversion before routing.
The resource blocks \(1,2,3\) are retained; the data blocks of the
controls, and for full Toffoli of the old target, are consumed. For
each coordinate \(j\), a private register \(R_j\) receives copies of
the consumed blocks and holds the decoded bits \(u_j,v_j\) (and
\(h_j\)); its outputs control coordinate \(j\) only.

\begin{table}[!tb]
\caption{Native schedule of the resource-to-gate conversion. Steps
marked (F) occur only for full Toffoli. Outside EC and record preparation,
the gates act coordinatewise. A superscript gives the private control bit.
In step 5, the controlled CNOTs and the doubly controlled \(X\) are
Toffolis. In step 6, \(Z^{h_j}\) is a Hadamard-conjugated CNOT and
\(\mathrm{CZ}^{h_j}\) a Hadamard-conjugated Toffoli.}\label{tab:toffoli}
\small\centering
\begin{tabular}{@{}l>{\raggedright\arraybackslash}p{0.50\textwidth}>{\raggedright\arraybackslash}p{0.33\textwidth}@{}}
\toprule
Step & Operation & Retained coordinates reached by one native fault before routing\\
\midrule
1 & Complete EC on every block. & As in the EC bound.\\\addlinespace[3pt]
2 & CNOT from resource blocks 1 and 2 into the two control data blocks. & At most the fault's coordinate in each retained block, including propagation to block 3 in step 5.\\\addlinespace[3pt]
3 (F) & CNOT from the old-target block into resource block 3, then \(H\) on the old-target block. & At most the fault's coordinate in each retained block, including propagation from block 3 to blocks 1 and 2 in step 5.\\\addlinespace[3pt]
4 & For each \(j\), fan out the consumed blocks by CNOTs into \(R_j\) and decode \(u_j,v_j\) (and \(h_j\)) by bounded-distance coset decoding. & A raw-coordinate fault leaves all clean decoder outputs unchanged within the margin. A private decoder fault can reach coordinate \(j\) in each retained block in steps 5--6.\\\addlinespace[3pt]
5 & On coordinate \(j\): \(\mathrm{CNOT}^{v_j}_{1\to3}\), \(\mathrm{CNOT}^{u_j}_{2\to3}\), \(X_3^{u_jv_j}\), \(X_1^{u_j}\), \(X_2^{v_j}\), in this order. & Coordinate \(j\) of blocks 1, 2 and 3.\\\addlinespace[3pt]
6 (F) & On coordinate \(j\): \(Z_3^{h_j}\), then \(H_2\,\mathrm{CNOT}^{h_j}_{1\to2}\,H_2\), which is \(\mathrm{CZ}^{h_j}\) on blocks 1 and 2. & Coordinate \(j\) of blocks 1, 2 and 3.\\\addlinespace[3pt]
7 & Reset or discard the consumed blocks and every~\(R_j\). & None.\\
\bottomrule
\end{tabular}
\end{table}

Steps 2, 3, 5 and 6 act coordinatewise, so an error at index \(i\)
can reach only index \(i\) of each retained block. A disturbed \(R_j\)
changes only the coordinate-\(j\) gates in steps 5 and 6. A bit error
on a consumed block can be copied to that same coordinate of later
copies, which stay in the same decoded coset within the margin. From
step 4 onward a consumed block is only a fan-out control and is then
discarded.
A phase error on it, including one kicked back from a copy, reaches
no retained block. A phase error before the old target's Hadamard
becomes a bit error at the same index and is covered by the coset margin.
Step 7 acts only on disposed registers. Each native fault thus touches
at most one protected coordinate per retained block. One routed fault
can give at most two native faults by the routing argument below.
The complete EC removes the incoming correctable errors. The EC and
body together obey the stated support bound, as for a transversal body.

\smallskip\noindent
\emph{Routing and the numerical envelope.}
Route the already complete EC-plus-body circuit along a line. Bring
its at most three operand tokens together by a fixed adjacent-SWAP
word, perform the native gate, and undo the permutation. Include
identity locations on waiting tokens. Reset a work lifetime before
its first use and retain it until its declared reset or disposal.
An ideal SWAP permutes errors. Expand a faulty two-token SWAP in a
product operator basis and commute its factors through the remaining
permutations. The operand factor is an arbitrary fault at the next
native gate; a bypass token's factor is a fault at its native idle.
Return routes have the same description. Thus one routed fault gives
at most two native faults, with possibly correlated coefficients,
independently of route length and \(s\). Waits and resets give one.

The premise \eqref{d:margin} implies
\[
 f+2(j+1)\le2(f+j+1)\le q_s/4\le\lfloor q_s/2\rfloor.
\]
All unique-decoding and cat-majority arguments are therefore inside
their stated margins, including one additional insertion. The routed
total deviation is at most \(2j\), hence at most \(8j\).
The complete table counts all EC and Toffoli resource preparations,
including the \(5^{q_s}\) leaves of each verified zero preparation for
the initial plus blocks and for the fresh cats. These preparations,
route length and private table size affect location count and time.
The contact bound is per fault.

Finally these are operator statements. Expand arbitrary faulty Kraus
operators in local matrix or Pauli bases, retaining all exterior
systems as coefficients. In EC a coset-uniform record
\(\sum_{v\in\mathcal C_s}|v+x\rangle\) depends on the error coset,
not the logical amplitude. The two Toffoli cancellations show the
same independence for its consumed records. On the correctable span
each term is an ideal logical map times an input-independent
syndrome/environment coefficient. The support analysis applies
coefficientwise, so coherent sums, cross terms and the full reference
are covered by the one recovery of Lemma~\ref{d:code}.
\end{proof}

All procedures in this proof use \(\mathcal G_{\mathbb R}\).
An encoded Toffoli table contains physical Toffolis; replacing them
at the next level is finite circuit substitution. Induction on the
number of levels proves that every finite concatenated identity
circuit remains in this inventory. Invalid opcodes select identity.
Section~\ref{d:dispatch} reduces this inventory to the seven physical gates.

\subsection{Repeated error correction before each gadget}
Level reduction follows the extended-rectangle tradition of Aliferis, Gottesman and Preskill \cite{AGP2006}. Each rectangle here starts with repeated full error correction before the encoded body. The rectangles do not overlap or include trailing error correction. The proof treats the correlated faults generated by the controller.

Fix the constant semantic width, candidate-port geometry and complete
boundary halo before selecting a code. Let \(\kappa_0\) count physical
conversion into dirty semantic locations, \(\kappa_1\) bound child-box
participation, and \(v_0\) bound parent boxes meeting a rectangle.
These are incidence bounds per cell and time; extra ancillas use more
cells and extra correction passes use more time. With complete halo
\(h_{\rm halo}\) (Section~\ref{d:threshold-section}) and a fixed image gap \(\rho\), choose
\begin{gather}
 a_{\rm sep}>(4h_{\rm halo}+2\lceil\rho\rceil+5)^2,
 \qquad b_{\rm FT}=\max\{1,\kappa_0\kappa_1v_0(a_{\rm sep}-1)\},\label{d:budgets}\\*
 r_{\rm FT}=8b_{\rm FT},\quad
 s=\min\{s\ge1:2^s-1\ge40\,960(b_{\rm FT}+1)\}.\label{d:capacity}
\end{gather}
The bound \(q_s\ge72b_{\rm FT}+8\) suffices for the
fault-tolerance margin here. We use the larger constant shared with
Section~\ref{sec:four-dimensional}.
The constant \(\kappa_0\), and hence \(b_{\rm FT}\), counts every fault in the Fredkin, CZ, SWAP, CCZ and \(Z\) decompositions. The preparation depth is
\(q_s\ge2b_{\rm FT}\). The integer \(s\) is fixed for the circuit.

Let \(B_g\) be one full error-correction pass followed by an
encoded body. Precede every \(B_g\) by \(b_{\rm FT}+1\) time-disjoint
full error-correction passes on each input block, with fresh private work. Synchronize
these normalizing phases before interblock bodies. Every idle block
normalizes too.

\begin{lemma}[Level reduction]\label{d:reduction}
Let the code member satisfy \(q_s\ge72b_{\rm FT}+8\), as \eqref{d:capacity} does,
and let every \(B_g\) be preceded by \(b_{\rm FT}+1\) normalizing passes as above.
If a parent rectangle contains at most \(b_{\rm FT}\) bad child
circuit locations, its output is correctable by one fixed recursive
recovery. Good logical subsystems are preserved jointly, including
when entangled with bad blocks and an arbitrary reference. Effective
faults at the next level have support only on the union of the named
bad child locations' outgoing wires.
\end{lemma}
\par\noindent\emph{Proof idea.} A good block has a clean error-correction pass after its last arbitrary input. Track the remaining child faults through the body and decode the block to its intended operation.

\begin{proof}
Each good block's normalization has a last clean EC pass, because
there are more full passes than touching faults. That pass maps any
input into the code. Every later full EC pass and \(B_g\) has input
deviation at most \(r_{\rm FT}\) and at most \(b_{\rm FT}\) faulty
locations, and
\[
 8(r_{\rm FT}+b_{\rm FT})=72b_{\rm FT}\le q_s-8.
\]
Lemma~\ref{d:gadgets}, applied afresh at each full EC and at \(B_g\),
leaves at most \(r_{\rm FT}\) residual coordinates. A correctable
input retains its original logical subsystem. An arbitrary input has
output support in
\[
 \mathcal C_{r}=\operatorname{span}
       \{FV|\psi\rangle:\mathrm{wt}(F)\le r_{\rm FT}\}.
\]
Different good blocks can have their last clean pass at different times.

Actual faults can couple blocks. For each good block expand shared
operators on its coordinates, with all other systems as coefficients.
The preceding argument puts each output in \(\mathcal C_r\).
Positivity then puts the joint output in the tensor product of these
spaces: every complementary projection has expectation zero. In fixed
recovery coordinates \(\mathcal C_r\simeq\mathcal H_{\rm log}
\otimes\mathcal H_{\rm syn}\), the operator identity in
Lemma~\ref{d:gadgets} fixes the whole originally good logical algebra.
Its Stinespring map is
\[
 I_{\text{all good logicals}}\otimes
       K_{\text{bad inputs, syndromes, memory}}.
\]
The coefficient is the same for every good logical amplitude, including
off-diagonals; the factor on the remaining systems may be correlated.

At a common coarse cut, compose the next normalizations with the
preceding bad outgoing wires before crossing any interblock body.
In recovery coordinates their joint action is identity on the good
logical wires and a joint map on the union of those bad wires and
syndrome memory. Expand that map in tensor matrix bases on the named
bad wires. It adds neither a wire nor a bad location. Consecutive bad
rectangles can remain distinct; their common environment coefficients
are allowed. Good bodies now implement their ideal logical gates.
The finite directed circuit permits this reduction successively at
every level. Full scheduled normalizers remain covered at their own
scale; their circuit is never truncated to one fine time step.

The decoder is fixed bottom-up. At a parent there
are at most \(r_{\rm FT}\) arbitrary child-block operators; all other
children are recursively correctable. Lower recovery turns the former
into arbitrary operators on at most \(r_{\rm FT}\) parent coordinates,
with exterior coefficients. The base recovery corrects their entire
union span. Induction corrects all recursively allowed supports and
their coherent combinations. The same decoder serves every history.
\end{proof}

\begin{lemma}[Clean schedule and ownership]\label{d:ownership}
In the clean reference under an idle top, at every level \(\ell\)
and every level-\(\ell\) step the requests form a matching of native
tuples with consistent types, roles and offsets. Each level-\(\ell\)
location (gate, routing SWAP, waiting-token idle, reset, discard or
EC-pass gate) is a step of exactly one level-\((\ell+1)\) table share.
It is executed within one level-\((\ell+1)\) period on child cells
of the at most three operand blocks of that parent location.
Consequently its complete support lies in at most five consecutive
level-\(\ell\) cells during one level-\(\ell\) period, enlarged by
the halo of its descendants. Each level-\(\ell\) cell-time belongs
to exactly one level-\(\ell\) location, and no token is in flight
across a level-\((\ell+1)\) period boundary.
\end{lemma}
\begin{proof}
Use downward induction from the idle top. At level \(k\) all cells
are idle, a matching of unary tuples. Suppose the level-\((\ell+1)\)
requests form a consistent matching. A tuple of arity \(a\) occupies
\(a\) operand blocks, each executing its share at the common age,
a function of time in the clean reference. By the multi-block
specification in Section~\ref{d:calendar-section}, the shares restrict
one table whose every layer is a matching on its \(aP_{\rm cal}\)
child cells. Thus their child requests form a consistent matching,
also at block seams inside the tuple. Distinct parent tuples use
disjoint blocks. Each child step belongs to the table issuing it,
including resets, discards, EC gates and padded idles; every table
returns all tokens home. The two-cell operand offsets give the
five-cell window, and the complete descendant halo is the geometric
series in Section~\ref{d:threshold-section}. No tuple crosses a
top-level block boundary because the top is idle. The cyclic seam
of Lemma~\ref{f4:periodic} is such a boundary; for Theorem~\ref{thm:main}
every seam is outside the extended cone of Proposition~\ref{d:storage}.

Thus \(\kappa_0\) is the fixed dispatch conversion bound of
Section~\ref{d:dispatch} and Lemmas~\ref{f4:dispatch}--\ref{f4:dispatch-partial}.
Every resulting primitive location has exactly one ancestor at each
level, so a primitive station is charged to at most \(\kappa_0\)
locations at every level. The constants \(\kappa_1\) and \(v_0\)
are finite maxima of child participation and parent-box intersection
over the five-cell window, one period and the fixed child-index collar.
These normalized geometric maxima depend only on the arity bound,
that window and \(h_{\rm halo}\), uniformly in level and code size.
This derives the incidence constants used in \eqref{d:budgets} and
Lemma~\ref{f4:geometry}.
\end{proof}

\subsection{A finite classical interpreter for the schedule}\label{d:calendar-section}
The classical field \(S\) issues the schedule through a finite tower of block simulations. A block of \(P_{\rm cal}\) cells simulates one cell at the next level, and the top level is idle.

Choose a sufficiently large even block length \(P_{\rm cal}\) and a
period \(T_{\rm cal}=P_{\rm cal}^4\), or one larger fixed polynomial.
The radius-one classical field \(S\) has finite address and age,
a mode, distributed old/next information bits, read-only program bits,
and finite tape/head and broadcast tracks. A block of
\(P_{\rm cal}\) cells stores one virtual state as a binary string of
length \(O(\log P_{\rm cal}+\log T_{\rm cal})\), with blank padding.

During a simulation period it copies its old string and those of its
two neighbours, interprets one local transition and its old quantum
instruction, broadcasts the instruction, executes its padded encoded
quantum table, and commits the next string. Communication and broadcast
cost \(O(P_{\rm cal})\); one virtual transition uses polynomial time
and space in the logarithms of the parameters. The algorithm computes
one transition, rather than running the entire virtual period.
For the already fixed quantum table, padding and line routing take
at most \(C_QP_{\rm cal}^2\) steps. Include quantum idles during
classical work, the hardened leading EC and every reset. Thus a
terminating search finds \(P_{\rm cal}\) large enough for all work
to fit. Deadlines and fixed conflict priorities make malformed
classical words total.

\emph{Multi-block instructions.} A virtual instruction is a triple
\((\tau,r,\kappa)\): a native type \(\tau\in\mathcal G_{\mathbb R}\)
of arity \(a\le3\), a role \(r\in\{1,\ldots,a\}\), and the offsets
\(\kappa\) of the other operands, each within two virtual cells. For
each \(\tau\) fix one routed encoded table \(E_{\mathbb R}(\tau)\)
on the \(aP_{\rm cal}\) child cells of the \(a\) operand blocks, in
operand order. It consists of the normalizing EC passes, the body and
the line routing of Lemma~\ref{d:gadgets}, acts only on those child
cells, and returns every token to its home cell at its end. Its share
\(E_{\mathbb R}(\tau,r)\) lists, for each child age and child address
in block \(r\), that child's instruction. A block with old instruction
\((\tau,r,\kappa)\) broadcasts this triple to its cells, and each
cell's current request is
\(E_{\mathbb R}(\tau,r)(\text{age},\text{address})\)
(Table~\ref{f4:fields}). No cell decides a tuple from its
neighbours' words. The role field adds a fixed number of bits and
does not change the period budget.

The local algorithm is defined first. The encoder writes its finite
description as ordinary program data. The interpreter computes that
algorithm on three virtual states, including their program-bit fields;
it does not require a self-description fixed point. In the idle mode
the virtual instruction is identity. A simulation block has the exact
identities
\[
 \begin{aligned}
 \text{new virtual state}&=\text{one classical transition},\\
 \text{quantum channel}&=E_{\mathbb R}(\text{old virtual instruction}).
 \end{aligned}
\]
Induction gives a finite tower of \(k\) simulation levels under a top
idle state, executing \(E_{\mathbb R}^k(I)\). Higher clocks and
programs stay classical. The encoder computes every finite initial symbol by \(k\) block decompositions and finite encodings.

\subsection{Protecting the entire clock symbol and counting sources}
Clone the intended \(S\) trajectory over the transverse \((i,z)\)
plane. At every physical update predict its next whole symbol from
its three old \(\sigma\)-neighbours at every replica in
\eqref{eq:c-boxes}; retain a symbol with at least 512 votes, or a
fixed default. A clean wrong output needs at least 257 wrong
predictions, occupying two replica boxes. Each wrong prediction
has a wrong old input. Use private copies of old \(S\) for the
current quantum instruction, before committing its next symbol.
A later correct retained symbol does not undo a wrong current use.

\begin{lemma}[Combined source certificates]\label{d:sources}
There are constants \(C_{\rm w},K_{\rm w}\), independent of \(n\), \(W\),
\(L\) and the hierarchy level, such that every primitive dirty quantum
location has a connected full witness with \(g\) distinct canonical source
labels and at most \(C_{\rm w}g\) edges, and its anchored description
count is at most \(K_{\rm w}^g\). For \(\mu\in\mathcal J_n\), under the
joint physical law of Proposition~\ref{c:source-law}, any requested \(g\) labels have
intersection probability at most \(\gamma_{\rm eff}^g\), where
\[
 \gamma_{\rm eff}=\max\{\sqrt{\gamma_{\rm src}},\sqrt p\}.
\]
Each source has one Lipschitz image in chain space-time. Full bin
and role information is retained when witnesses are counted.
\end{lemma}
\par\noindent\emph{Proof idea.} Trace wrong clock and controller outputs to grouped noise sources. Bound the complete support of each explanation in the coordinates used by the concatenation.

\begin{proof}
For the clock field \(S\) (whose replica boxes are \(S_0,S_1,S_2\)) use characteristics \(i+58t,z+58t\) and balanced potentials
\(-3(i+58t)\), \(-3(z+58t)\) and \mbox{\(3(i+z+116t)\)}. The three replica boxes give
ranges
\[
\begin{array}{crrr}
 &F_1&F_2&F_3\\
 S_0 &[129,174]&[129,174]&[-348,-258]\\
 S_1 &[-351,-306]&[129,174]&[132,222]\\
 S_2 &[129,174]&[-351,-306]&[132,222]
\end{array}
\]
for backward gains. Any two boxes provide a positive parent for each
potential, with margin at least 129. The \(\sigma\)-shift affects none
of these potentials. Lemma~\ref{q:frontier}'s grouped-terminal proof
therefore gives \(O(g_H)\) edges for distinct physical raw names
\((i,\sigma,z,t)\). Old/new roles and quantum hardware use the same
raw source label \(H_{i,\sigma,z,t}\). The literal image \((\sigma,t)\)
is Lipschitz; there are no sampling-register contractions in this raw graph.

For \(C_{\rm sel}\), the canonical source label is
\[
 \lambda=(i-t,\sigma+58t,z+58t,\lfloor t/W\rfloor).
\]
Its event is the union of its raw, statistical and overlay exceptions,
with the law \(\gamma_{\rm src}^{g_C}\) in Proposition~\ref{c:source-law}.
These three types annotate one source label. Sampling-register gather/consult links cancel
on the base characteristic; a clean predictor entry traverses all
1024 levels, distributing its fixed debit over those levels.
The selector link calculation and grouped-terminal argument are given
in Table~\ref{q:links} and Lemma~\ref{q:frontier}; deleting the copied
coordinate changes neither. Keep the bin and entry roles in the full
bounded-degree graph. On a quantum line \(i=i_0\) its image is
\[
 t^*=i_0-(i-t),\qquad
 \sigma^*=(\sigma+58t)-58t^*.
\]
It agrees with the root and is constant on each canonical source label.
Contracted sampling-register links have zero image increment; all other typed
links have fixed bounded increments.

Glue the two full graphs at quantum-use roots by the fixed
hardware stars. For \(g_C\) owner source labels and \(g_H\) raw source labels,
Cauchy--Schwarz gives
\[
 \Pr(\text{all})\le\gamma_{\rm src}^{g_C/2}p^{g_H/2}
                  \le\gamma_{\rm eff}^{g_C+g_H}.
\]
This pays every physical overlap. Bounded full degree and the linear
edge budget give exponential anchored enumeration by a doubled-tree
walk with finite annotations. In particular neither source types nor
coincident images create duplicate charges or free bin choices.
\end{proof}

\subsection{Separated complete supports and the threshold}\label{d:threshold-section}
Use regular chain space-time boxes of spatial side
\(P_{\rm cal}^{\ell}\) and temporal side \(T_{\rm cal}^{\ell}\).
Their fixed child-index cover includes a constant collar. Geometric
series bound all complete descendants by a fixed halo \(h_{\rm halo}\).
Let \(N_0\) mean primitive dirtiness, and define \(N_{\ell+1}\)
by at least \(a_{\rm sep}\) bad children in that cover. This is a
monotone geometric predicate on scheduled bad roots. If all parent
boxes covering a rectangle are good, \eqref{d:budgets} bounds its
bad child circuit locations by \(b_{\rm FT}\). Thus
Lemma~\ref{d:reduction} applies.

The packing bound in \eqref{d:budgets} ensures two bad complete child
supports separated by \(\rho P_{\rm cal}^{\ell}\) whenever the
parent is bad. Separation in time can be still larger, because
\(T_{\rm cal}^{\ell}\ge P_{\rm cal}^{\ell}\). There are finitely
many candidate pairs per parent, independently of \(\ell\).

\begin{lemma}[Source-certificate recursion]\label{d:recursion}
If \(\gamma_{\rm eff}\) of Lemma~\ref{d:sources} is at most a computable
bound, chosen after the entropy constants (Section~\ref{s:parameters},
items 2--5), there are constants \(\bar A\ge1\) and \(D=1/(2\bar A)\)
such that, for every level-\(\ell\) box whose complete cone injects into
the torus,
\[
 \Pr(N_\ell)\le z_\ell\le D\,2^{-2^\ell},
\]
uniformly in the hierarchy level and selected nonhalting window.
\end{lemma}
\par\noindent\emph{Proof idea.} Prune the certificate tree at a long primitive witness. Otherwise the proof extracts separated child certificates and uses the joint source law to multiply their weights.

\begin{proof}
Let \(z_\ell\) be the sum of source weights of recursively pruned
fully embedded certificates in one level-\(\ell\) box. At a parent,
if a selected primitive witness has full graph length at least a
small fixed multiple of \(P_{\rm cal}^{\ell}\), replace its entire
repeated subtree by that witness and its root. Lemma~\ref{d:sources}
pays at most
\(B_{\rm tail}P_{\rm cal}^{d_0\ell}
 \exp(-a_{\rm tail}P_{\rm cal}^{\ell})\).
The polynomial counts possible roots, including the anisotropic time
extent. Otherwise every witness is shorter than one third of the
separated-support gap. The two complete child certificates have
disjoint source-label sets, since a label has one image. Their
weights multiply under the joint intersection bound, and enumeration
factors. Hence
\begin{equation}\label{d:sum-recursion}
 z_{\ell+1}\le A_{\rm rec}z_\ell^2+
 B_{\rm tail}P_{\rm cal}^{d_0\ell}\exp(-a_{\rm tail}P_{\rm cal}^{\ell}).
\end{equation}
The long case discards the full subtree; its leaves are not counted
again. This is a recursion for certificate sums, not for independent
bad-location probabilities.

Put \(\bar A=\max(1,A_{\rm rec})\). Choose the source bound so that
\[
 z_0\le\frac1{4\bar A},\qquad
 a_{\rm tail}\ge\max\{0,\log(4\bar A B_{\rm tail})\}
       +\frac{d_0\log P_{\rm cal}}{P_{\rm cal}}+2\log2.
\]
For \(P_{\rm cal}\ge3\), use
\(\ell/P_{\rm cal}^{\ell}\le1/P_{\rm cal}\) for \(\ell\ge1\)
and \(2(2/P_{\rm cal})^\ell\le2\).
The remainder in \eqref{d:sum-recursion} is at most
\((D/2)2^{-2^{\ell+1}}\). If \(z_\ell\le D\,2^{-2^\ell}\), its
squared contribution is at most the same quantity because
\(A_{\rm rec}D^2\le D/2\). The base inequality is
\(z_0\le D/2\). This proves the displayed bound with the fixed
exponent \(\log2\). All entropy constants precede the source bound;
shrinking that cap makes these strict conditions attainable.
\end{proof}

\subsection{A total encoder and every requested time}

\begin{proposition}[Three-dimensional linear storage]\label{d:storage}
There are a rational \(c>0\), independent of \(n\), and total
computable \(L_0(n)\), \(\cE_{n,L}\), \(\cD_{n,L,t}\) satisfying
part (i) of Theorem~\ref{thm:detailed}. One encoder works for every
requested time and every rate in its window.
\end{proposition}
\par\noindent\emph{Proof idea.} Choose a logarithmic block scale so the probability of a bad top-level box is small over the whole linear horizon. One recovery handles the permitted histories at each requested time.

\begin{proof}
Let \(k\) be the least integer \(k\ge1\) with
\(2^k\ge64\lceil\log_2(L+2)\rceil\), and set
\[
 t_{\max}=\lfloor cL\rfloor,\qquad
 t_{\rm ext}=t_{\max}+T_{\rm cal}^{k}.
\]
The hierarchy dimensions \(P_{\rm cal}^k,T_{\rm cal}^k\) are fixed
powers of \(\log(L+2)\), hence \(o(L)\).
Choose \(0<c\le1/[1000(V_*+1)]\) after the final catalogue and full
source speed \(V_*\) are known. A total computable size cutoff
makes the complete extended quantum and source cones inject into
one torus chart, with all seams outside. Fixed label-dependent
initialization and bin margins enter \(L_0(n)\).

Prepare the finite \(k\)-level quantum block and classical simulation
tower under an idle top. Clone intended \(S\) over the required
transverse cone. Lemma~\ref{c:initialization} supplies the finite
selector chart by exhaustive hidden-cover search, with low/high
sampling registers on their base rows. Nonhalting guarantees success. The search
is finite even for halting inputs; failure and small sizes use a fixed
fallback. Neither a noise rate nor a halting answer is read.

At most \(C_*L^4\) complete level-\(k\) boxes meet the extended
cone and its fixed cover. Define \(\mathcal G\) by the absence of
\(N_k\) in all of them. Lemma~\ref{d:recursion} gives
\[
 \Pr(\mathcal G^c)\le C_*L^4D(L+2)^{-64}\le1/16
\]
after a computable threshold enlargement. The event is exogenous
and independent of the stored state. On it the deterministic
reduction restores the ideal top identity on its full logical algebra.

At a requested integer \(t\le t_{\max}\), retain every actual
in-flight quantum data, syndrome, partial correction, ancilla and
routing/work factor. Ignore the received autonomous metadata and
append the intended noiseless future quantum circuit until
\(t'=T_{\rm cal}^k\lceil t/T_{\rm cal}^k\rceil\).
Existing work is reset only at a declared future reset instruction;
new external work is introduced only as genuinely fresh work.
Physical disposable ports already cleaned at the round boundary may
be discarded. This external completion adds neither noise nor
selector cuts. Delete only future primitive bad roots in the comparison.
Monotonicity of the pure \(N_\ell\) predicates preserves goodness,
so Lemma~\ref{d:reduction}'s one bottom-up recovery gives the Bell
state exactly. Averaging gives fidelity at least \(15/16\).
The initial encoder did not depend on the requested time. The
external circuit is total and computable from \((n,L,t)\), independently
of the rate and fault history.
\end{proof}

This proves Theorem~\ref{thm:detailed}(i). Section~\ref{sec:forgetting} proves (ii) for every encoding.

\section{Size-uniform fidelity bound for all encodings}\label{sec:forgetting}

Forgetting in the classical processor does not rule out quantum information elsewhere in the register. We instead bound the initial factors on which the output channel can depend. A small surviving-dependence probability gives a two-output extension with equal full-channel marginals. Iteration and Lemma~\ref{f:bell} then give \(1/2+1/(2k)\), the optimal entanglement fidelity for universal \(1\to k\) qubit cloning \cite{KeylWerner1999,Werner1998}. Lemma~\ref{f:bell} includes a proof of this bound. Such equal-marginal extensions can be symmetrized, making the channel's Choi state \(k\)-extendible \cite{Kaur2019,Kaur2021}; the proof below uses only the displayed marginal equalities. For isotropic-state sharability, see \cite{JohnsonViola2013}.

\subsection{Exact conditional channels and initial supports}

Fix a rate \(\eps\) and write \(\Lambda_L=\Lambda_{L,\eps}\) for one full round on the complete physical register. Dephase the retained classical words at the beginning of
the round. Write \(a\) for their measured initial word and \(\omega\)
for one joint history of autonomous metadata and exogenous
innovations, including compiler marks. The argument below also applies
to the four-dimensional appendix: put \(d_{\rm lat}=3\), \(n_{\rm sel}=1\)
for Theorem~\ref{thm:detailed}, and \(d_{\rm lat}=4\), \(n_{\rm sel}=4\)
for the folded part of the coexistence circuit (Appendices~\ref{sec:coexistence}--\ref{sec:folded}). The latter history includes all selectors and the
profile together. Syndrome measurements and their outcomes stay
inside the quantum CPTP circuit. Conditional on \((a,\omega)\),
the controller trajectory is fixed, and the data evolution over \(s\)
rounds is a CPTP map \(T_{a,\omega}\). All auxiliary hardware is reset before
reading and discarded or reset after the final conditional-reset block.

Draw a conservative dependence graph for this conditional circuit.
A preparation has no incoming quantum edge. An ordinary local CPTP
map has edges from every input factor to every output factor. A clean
bottom-triggered reset, including its discarded swap register,
has no incoming edge to its retained outputs. Each physical replacement
also removes the incoming edges of the factors it replaces. Include temporary quantum work
until its actual discard. Let \(A_s(a,\omega)\) be the set of initial
physical sites supporting any retained quantum factor that has a path to any retained output after \(s\)
rounds.

\begin{lemma}[Initial-support factorization]\label{f:factorization}
For every \(s\ge1\) and every \((a,\omega)\), there is a CPTP map \(G_{a,\omega}\) such
that
\begin{equation}\label{f:factor}
 T_{a,\omega}
 =G_{a,\omega}\circ
   \operatorname{Tr}_{A_s(a,\omega)^c}.
\end{equation}
This identity holds on the complete initial quantum algebra and after
tensoring with an arbitrary reference. The fixed conditional final
classical word and all retained hardware outputs are included.
\end{lemma}
\begin{proof}
Propagate an arbitrary output observable backward in the Heisenberg
picture. A CPTP gate's adjoint is unital, so gates outside the
observable's backward support contribute identities. At a
preparation or complete erasure, backward dependence terminates.
By the graph definition the resulting observable is supported on
\(A_s(a,\omega)\). This is true for every output observable, proving
that the conditional channel depends only on the reduced input on
that initial set. Fix a reference density matrix on its complement
and define \(G_{a,\omega}\) by inserting that reference matrix before
\(T_{a,\omega}\). It is CPTP and gives \eqref{f:factor}.

The statement uses the complete cut in every round. In the
concrete schedule all spatial quantum work precedes the final
controlled-swap and discard block. What follows consists of controller
storage, preparation, reset and noise, so no old quantum state can
be returned from temporary hardware to retained data. Conditional
syndrome processing remains part of a CPTP map throughout this
argument.
\end{proof}

\begin{proposition}[Halting ancestor tail]\label{f:ancestor}
If the computation indexed by \(n\) halts, there are finite positive
constants \(B_n,\kappa_n\) and an integer \(d_n\), uniform over
\(\eps\in I_n\), such that for every measured initial word \(a\)
and initial site \(v\),
\begin{equation}\label{f:ancestor-tail}
 \Pr(v\in A_s(a,\omega))
 \le B_n(1+s)^{d_n}e^{-\kappa_ns}
\end{equation}
on every sufficiently large torus whose depth-\(s\) determining cones
unwrap injectively. The event is determined by innovations in a
spatial cone of radius \(O(s)\) during these \(s\) steps.
\end{proposition}
\par\noindent\emph{Proof idea.} Follow the dependence of the output backward through each conditional reset. Survival requires an exceptional controller reading or a hardware fault at repeatedly spaced steps.

\begin{proof}
A surviving path crosses a retained conditional-reset block in every round.
A fault-free reset triggered by bottom cuts it, so at each crossing either
the ordinary checked proposal is live or there is a direct named
local fault. Proposition~\ref{c:late-live} bounds, simultaneously over all
initial controller words, the possibility that a sufficiently late
live proposal is outside the raw marked field. Only polynomially
many roots of the selector collection occur in a specified depth-\(s\)
quantum cone. Union over these candidates preserves its exponential
bound.

Also union the marginal all-\(M^0\) backward-path tails in this late
cone. For a late crossing at time \(t\), use
\(t_0=\lfloor t/2\rfloor\) in Lemma~\ref{c:rare}; the two unions
exclude its alternative~(c). On the complement each crossing is
covered by \(N_{t-1}\) for its current sheet or a direct destination
exception. The path may change sheet at every round. Within
one selector use the south comparison of Lemma~\ref{c:rare}; its
rare birth needs \(256\) fresh positions and its seed groups have
the eight-class bound. Pack complete per-sheet contours along the
physical path, reserving each seed test's entire deterministic
two-time support and restarting below its minimum time minus two.

For the above \(n_{\rm sel}\), \(g\) prescribed typed rare groups, and \(k_{\rm dir}\) distinct
direct physical source names, H\"older across the sheets followed by
Cauchy--Schwarz between rare and direct events gives
\[
 (N_{\rm seed}p^{256})^{g/(16n_{\rm sel})}p^{k_{\rm dir}/2}.
\]
The rare power is \(256/(16n_{\rm sel})\ge4\), and the direct power is \(1/2\).
Their positive exponents beat the finite entropy of the graph
containing every allowed physical quantum template, as in
Proposition~\ref{c:path-tail}. Union over polynomially many
reachable endpoints and the preceding marginal exceptions proves
\eqref{f:ancestor-tail}, uniformly in the combined measured word.
For the torus statement first use the literal finite determining
cone of the event. When that cone is injective, its innovations have
the same joint law as the corresponding finite region of an
independent lattice field. Extend outside this cone with independent
innovations to apply the preceding infinite-lattice bound. A separate
extension may be used for each individual upper bound. Dependence
between different torus block events will instead be read directly
from their physical determining cones.
\end{proof}

The constants in Proposition~\ref{f:ancestor} may use a finite
halting motif. Their existence does not supply a total algorithm
that produces a cutoff for the all-encoding fidelity bound from every input \(n\).

\subsection{An elementary finite-dependence lemma}

Domination by product measures is the standard way to control finite-dependence fields \cite{LSS1997}. We give the elementary finite version needed to turn the ancestor tail into a channel extension. Independence from nonneighbours means
independence from their entire vector.

\begin{lemma}[Domination by a product measure]\label{f:domination}
Let \((X_j)_{j\in J}\) be \(\{0,1\}\)-valued variables on a finite
dependency graph of maximum degree \(\Delta\). Suppose \(X_j\) is
independent of the vector of all its nonneighbours. Let
\(0<q\le1/2\), put \(r=q/2\), and suppose
\[
 \Pr(X_j=1)\le p\le r^{\Delta+1}
 \quad\hbox{for every }j.
\]
Then their joint law is stochastically dominated by independent
Bernoulli\((q)\) variables.
\end{lemma}
\par\noindent\emph{Proof idea.} Add independent marks to the finite-dependent field. A bound on each conditional mark probability gives a sequential coupling to independent Bernoulli variables.

\begin{proof}
Add independent variables \(Z_j\) of Bernoulli parameter \(r\),
independent of \(X\), and let \(\tilde X_j=X_j\lor Z_j\). We first prove,
for every \(j\) and every positive-probability pattern \(w\) on a
set \(K\) excluding \(j\), that
\begin{equation}\label{f:conditional}
 \Pr(X_j=1\mid \tilde X_K=w)\le r.
\end{equation}
Induct on \(|K|\). The empty case follows from \(p\le r\). Split
\(K=F\sqcup H\) into nonneighbours and neighbours of \(j\), with
\(|H|\le\Delta\). Vector independence gives
\[
 \Pr(X_j=1,\tilde X_K=w)
 \le p\Pr(\tilde X_F=w_F).
\]
Expose the assignments on \(H\) in any order, after those on \(F\).
A prescribed one has conditional probability at least \(r\), since
its own independent \(Z\)-coin forces it. A prescribed zero has
conditional probability at least \((1-r)^2\): the induction
hypothesis bounds the probability of its \(X\)-variable being one
by \(r\), since the conditioning set at that stage has size less
than \(|K|\), and its own \(Z\)-coin is still independent. Since
\(r\le1/4\), we have \((1-r)^2\ge r\). Hence
\[
 \Pr(\tilde X_K=w)\ge r^{|H|}\Pr(\tilde X_F=w_F).
\]
Dividing gives a ratio at most
\(p r^{-\Delta}\le r\), completing the induction.

The coin \(Z_j\) is independent of \((X_j,\tilde X_K)\), so
\[
 \Pr(\tilde X_j=1\mid \tilde X_K)
 \le r+(1-r)r\le q.
\]
Generate the finite vector \(\tilde X\) sequentially from independent
uniform variables, using its conditional probabilities. With the
same uniforms, the tests against \(q\) generate an independent
Bernoulli\((q)\) vector \(Y\) satisfying \(\tilde X\le Y\). Finally sample
\(X\) conditionally on \(\tilde X\) in its original joint law. Then
\(X\le \tilde X\le Y\) and its marginal law is correct. This is the
claimed finite monotone coupling.
\end{proof}

\subsection{Two outputs with the same full-channel marginal}

\begin{proposition}[Two-output extension]\label{f:extension}
For a halting input \(n\), there are finite integers \(s_n,A^{\rm tor}_n\),
uniform throughout \(I_n\), such that on every torus
\(L>A^{\rm tor}_ns_n\) the complete channel \(\Lambda_L^{s_n}\) has a CPTP
extension to two copies of its whole output algebra with both
marginals equal to \(\Lambda_L^{s_n}\).
\end{proposition}
\par\noindent\emph{Proof idea.} Partition the initial sites into blocks and dominate the blocks still used by independent marks of probability one half. Couple two histories with disjoint sets of used blocks, then route each input block into the history that needs it.

\begin{proof}
Partition the initial torus into coordinate blocks whose sides lie
between \(s\) and \(2s\), which is possible because \(L\) exceeds a
fixed multiple of \(s\). Let \(X_I(a,\omega)\) indicate that block
\(I\) contains an initial quantum factor in \(A_s(a,\omega)\).
Proposition~\ref{f:ancestor} and a union over the \(O(s^{d_{\rm lat}})\) sites in a block give the following bound with \(d_n'=d_n+d_{\rm lat}\), increasing the prefactor \(B_n\) to absorb the constant in the block volume:
\[
 \Pr(X_I=1\mid a)\le
 p_s= B_n(1+s)^{d_n'}e^{-\kappa_ns}\longrightarrow0.
\]
Each indicator is the mask of all quantum factors at the initial physical sites.
The joint innovations determining a block lie in an enlarged block of
radius \(O(s)\). Thus these indicators have a dependency graph of
some degree \(\Delta\) independent of \(s,L,a\). Conditional on
\(a\), nonadjacent determining regions use disjoint underlying
innovations, giving the vector independence in
Lemma~\ref{f:domination}.

Choose \(s=s_n\) large enough that
\(p_s\le(1/4)^{\Delta+1}\), uniformly in the rate window. For a
fixed \(a\), its mask law \(\mathbb P_a\) is dominated by the independent
fair-mask law \(\pi\). Choose a finite monotone coupling and write
\(K_a(X\mid Y)\) for the conditional law of its lower mask given
its upper one. Sample a fair mask \(Y\); both \(Y\) and its
complement \(1-Y\) have law \(\pi\). Sample
\[
 X^{(1)}\sim K_a(\,\cdot\mid Y),\qquad
 X^{(2)}\sim K_a(\,\cdot\mid1-Y).
\]
Their supports are disjoint and each has the exact original mask
law \(\mathbb P_a\). Given these masks, sample two joint
innovation histories of every autonomous field and all shared marks with the original conditional law given the
corresponding mask. Each history retains all physical correlations and its exact original marginal
law. Their initial ancestor sets are disjoint, because each
lies within its own lower mask.

Route every quantum factor at those disjoint initial physical sites to the two maps
\(G_{a,\omega_1}\) and \(G_{a,\omega_2}\) from
Lemma~\ref{f:factorization}, and trace unused factors. Applying the
product of these maps is CPTP even on a globally entangled input.
For either output separately, averaging its exact conditional
history law and using \eqref{f:factor} recovers the original
\(s_n\)-step channel in branch \(a\).

To obtain a channel on the full initial algebra, first measure the
initial classical register, broadcast its classical outcome \(a\)
to both simulations, and perform the conditional construction above.
This is precisely the dephasing already present in the physical
channel. It therefore preserves its action also when the encoded
input entangles the classical register, data and reference.
All final controller words and reset hardware outputs belong to
these simulations. This proves the full-algebra extension.
\end{proof}

The extension is a mathematical information bound. Its construction
may use global conditional sampling; the physical processor itself
is unchanged.

For an integer \(k\ge1\), a \(k\)-output extension with equal marginals of a channel \(T\) is a CPTP map from the input algebra of \(T\) into \(k\) copies of its output algebra whose partial trace over all copies but any one equals \(T\).

\begin{lemma}[Iteration on the entire algebra]\label{f:iteration}
If a channel \(T\) has a two-output extension with equal marginals,
then \(T^j\) has a \(2^j\)-output extension with equal marginals
for every positive integer \(j\).
\end{lemma}
\begin{proof}
Apply the extension separately to every output of the previous
extension. Partial trace over the other outputs commutes with these
trace-preserving maps, so each new marginal is \(T\) composed with
the previous marginal. By induction the depth-\(j\) construction
has \(2^j\) outputs and marginal \(T^j\). Correlations between
previous outputs cause no difficulty for tensor products of CPTP
maps.
\end{proof}

For \(T=\Lambda_L^{s_n}\), this gives \(2^j\) outputs at physical
time \(js_n\). Once the extension exists on a fixed torus, its
iteration uses the full finite algebra; the size requirement remains
\(L>A^{\rm tor}_ns_n\).

\subsection{The cloning bound and a size-independent time}

\begingroup\exhyphenpenalty=10000
\begin{lemma}[Cloning bound for equal-marginal extensions]\label{f:bell}
Let \(k\ge1\) and suppose a channel has a \(k\)-output extension with equal marginals.
For every qubit encoder and common qubit decoder, its entanglement
fidelity is at most
\[
 \frac12+\frac{1}{2k}.
\]
\end{lemma}
\endgroup
\begin{proof}
Apply the encoder to the extension's input and the decoder to each
output. Let \(R\) be the reference qubit and \(B_1,\ldots,B_k\)
the decoded qubits. Write \(P_j\) for the normalized Bell projector
on \(RB_j\), tensored with the identity on the other outputs.
Let \(V_j\) insert this Bell state into \(RB_j\); then
\(P_j=V_jV_j^*\). The nonzero spectrum of \(S=\sum_jP_j\) equals
that of its block Gram operator \(G=(V_i^*V_j)_{i,j}\).
The diagonal blocks are identity. Contracting two normalized Bell
tensors shows that for \(i\ne j\), \(V_i^*V_j\) is one half of a
unitary relabelling of the missing-output tensor factors, so its norm
is \(1/2\). For a block vector \(z\),
\begin{align*}
 \langle z,Gz\rangle
 &\le\sum_j\|z_j\|^2+
       \frac12\sum_{i\ne j}\|z_i\|\,\|z_j\|\\
 &\le\frac{k+1}{2}\sum_j\|z_j\|^2.
\end{align*}
Thus \(\|S\|\le(k+1)/2\). The equal decoded marginals have the
same entanglement fidelity \(F\), whence
\(kF=\operatorname{Tr}(\rho S)\le(k+1)/2\).
\end{proof}

\begin{theorem}[Uniform quantum forgetting on a halting window]
\label{f:forgetting}
If the computation indexed by \(n\) halts, then for every \(\eta>0\)
there is a finite \(C_{n,\eta}\), uniform over all torus sides \(L\)
and all \(\eps\in I_n\), such that every qubit encoder and decoder
through \(\Lambda_L^t\) has entanglement fidelity at most
\(1/2+\eta\) for every integer \(t\ge C_{n,\eta}\).
\end{theorem}
\par\noindent\emph{Proof idea.} Iterate the two-output extension and apply the cloning bound of Lemma~\ref{f:bell}. Treat the finitely many smaller tori by complete replacement, with a probability bounded below on the window.

\begin{proof}
It suffices to consider \(0<\eta<1/2\). Choose \(j\) with
\(2^{-j-1}\le\eta\). Proposition~\ref{f:extension} and
Lemmas~\ref{f:iteration}--\ref{f:bell} give the bound at time
\(js_n\) on every \(L>A^{\rm tor}_ns_n\), uniformly in the rate window.

There are finitely many remaining tori. Let \(N_n\) bound their
physical site counts and let \(N\) be the fixed number
of physical qubits per site. At a fixed final elementary noise layer,
simultaneous replacement of every factor has probability at least
\[
 p_n=(\inf I_n)^{R_n}>0,\qquad
 R_n=\begin{cases}N_n,&\text{in }\mathrm{WS},\\
 N N_n,&\text{in }\mathrm Q.\end{cases}
\]
It replaces every physical factor, including controller and work,
and disconnects the output from the reference. Conditional on at
least one such event, the decoded entanglement fidelity is exactly \(1/4\):
the reference marginal is maximally mixed, and the output is
independent of it. The complementary histories have entanglement fidelity at
most one. Independence of the flags across rounds gives
\[
 F^*_{L,\eps}(t)\le\frac14+\frac34(1-p_n)^t.
\]
Choose \(t_n\) such that this is at most \(1/2+\eta\) for all \(t\ge t_n\), and let
\(C_{n,\eta}=\max\{js_n,t_n\}\).

An upper bound valid for every decoder persists under later
evolution: compose any proposed later decoder with that additional
evolution to obtain a decoder at the earlier time. Thus the bound
holds at every time in the statement.
\end{proof}

\begin{remark}[Dimension-independent forgetting argument]\label{rem:forgetting-interface}
The proofs of Proposition~\ref{f:extension} and Theorem~\ref{f:forgetting} apply in any fixed lattice dimension. Their inputs are Lemma~\ref{f:factorization}; a uniform initial-site ancestor tail of the form \eqref{f:ancestor-tail} whose events are determined by innovations in injective determining cones of radius \(O(s)\), the innovations being independent across disjoint space-time regions and independent of the measured word; and a positive full-replacement probability bounded below uniformly over the window for each of the finitely many small tori. Dimension enters only the polynomial number of sites in a block. The dependency degree stays fixed; the same fair-mask coupling, iteration and cloning bound then give a forgetting time uniform in size and rate.
\end{remark}

\section{A linear upper bound from the obstruction to periodic tiling}\label{sec:ceiling}

The lower bound preserves a qubit for a small fixed multiple of
\(L\). We now prove that this scale is optimal for the chosen circuit.
The upper bound concerns its entire channel and therefore applies to
alternative encoders as well as the explicit concatenated-code encoding.

A late live reading without a marked error would force a locally legal scaffold along a wide space-time diagonal strip, periodic by \((L,-L)\). Lemma~\ref{u:strip} rules out that strip.

\subsection{A finite diagonal-strip obstruction}
The classical chart has a locally recognizable dyadic scaffold as an
edge-matching factor. We need a quantitative finite consequence of that
factor, rather than the assertion that a complete tiling is aperiodic.
\begin{proposition}[Finite scaffold interface]\label{u:scaffold-import}
The finite alphabet tables in \cite[Definitions~3.1, 3.2 and~3.8, Lemma~3.9, and Lemmas~A.1--A.3]{DMPhase2026} have the following local
properties. The quarter encoding declares a parent from a 276-letter
alphabet. Its internal edges force the four declared parent letters to
agree, while external edges impose matching of neighbouring parents.
Dropping the two ruler bits gives the 104-letter base scaffold
\(\mathcal O\). Its thin types fix one parity coset. A \(2\times2\)
block together with its four prescribed south/west extension cells has
one admissible \(\mathcal O\)-parent; fine seam matching gives parent
seam matching. The compiled chart letter has a scaffold projection
\(\pi\): the \(\mathbb A^{\mathrm q}\)-component of the output letter
of its slot-zero geometry letter, with the rank bits \(V,H_{\rm row}\)
dropped. If a chart letter completes its west and south letters,
their \(\pi\)-images match on both shared edges as
\(\mathbb A^{\mathrm q}\) tiles.
\end{proposition}
\begin{proof}
The scaffold properties are those of the cited finite tables.
The cover construction of \cite[Lemma~11.2]{DMNoise2026}, here built
on the presentation named in Section~\ref{c:chart}, composes it with
the output map, rejects every violation of the local scaffold,
routing, phase and collar constraints of \cite[Section~3]{DMNoise2026},
and appends projected edge colours to its covering colours. Those
constraints are the quarter Wang rules of the decorated 276-letter
set, whose edge colours contain those of \(\mathbb A^{\mathrm q}\)
\cite[Definition~3.8]{DMPhase2026}. History slots, flags and computation
fields only reject completions. This is a local statement and needs
no extension of the pattern.
\end{proof}
We also use the quarter encoding of \cite[Definition~3.8 and Lemma~3.9]{DMPhase2026}.

\begin{lemma}[Periodic diagonal strips]\label{u:strip}
Let \(L\ge1\), \(k_L=v_2(L)+1\), and \(R_{k_L}=16\cdot2^{k_L}\).
There is no locally legal quarter-scaffold pattern on a diagonal strip
\[
 \{(x,y)\in\Z^2:a\le x+y\le b\}
\]
of width \(b-a\ge4R_{k_L}+4\) that is periodic by \((L,-L)\).
Moreover \(R_{k_L}\le32L\).
\end{lemma}
\par\noindent\emph{Proof idea.} Iterate the quarter-to-parent map on the interior of the strip. Each step forces an extra factor of two in the period, until the required divisibility contradicts the choice of depth.

\begin{proof}
Adjacent quadrant indices increment by the corresponding
coordinate unit vector modulo two. On a connected strip this fixes one
common quadrant coset. Group its complete \(2\times2\) blocks. Their
common declared parents form a legal pattern on an eroded strip. Now
drop both ruler bits; every remaining letter and edge belongs to
\(\mathcal O\). Subsequent parent steps use this base alphabet alone.

For a base parent, the required block and extension cells have relative
coordinate sums in \([-1,2]\). If its origin is \(\delta+2v\), retain
origins satisfying
\[
 a+1\le\delta_1+\delta_2+2(v_1+v_2)\le b-2.
\]
Rounding endpoints and retaining an additional seam margin gives the
conservative width recurrence
\begin{equation}\label{eq:u-width}
 w_{j+1}\ge(w_j-8)/2.
\end{equation}
The parent is admissible by the finite parent lemma, and adjacent
parents match by local soundness. These conclusions need only the
listed cells.

A period \((L,-L)\) must preserve the quarter or thin-type parity
pattern. Thus \(L\) is even. Grouping at the fixed coset then commutes
with this translation and gives period \((L/2,-L/2)\) on the parent
strip. Repeat. Starting with \(w_0\ge64\cdot2^{k_L}\),
\eqref{eq:u-width} gives
\[
 w_j\ge w_0/2^j-8(1-2^{-j})\ge56\qquad(0\le j\le k_L).
\]
Every required connected strip survives. After at most \(k_L\) parity
recognitions the induced period has odd coordinates, which cannot
preserve the parity pattern. This is the contradiction. Finally,
\(2^{k_L}\le2L\), so \(R_{k_L}\le32L\).
\end{proof}

\subsection{Counting physical coins after a periodic lift}
The subscript \(\sigma\) on \(M^0_\sigma\) in this subsection is
a selector-type index; the chain coordinate is separately specified by
the data geometry. In \(D_N\), \(N\) is the path length in the
family \(D_s\) of Lemma~\ref{c:coherence}.
Through a fixed linear horizon, a periodic lift repeats each physical
noise coin only a bounded number of times. This permits the contour
estimates to be used after wrapping.

\begin{lemma}[Wrapped certificate bound]\label{u:wrapped}
Fix a finite controller stencil, a finite data dependency stencil, and a
horizon coefficient \(C_{\rm hor}\). On the side-\(L\) torus, lift the
finite-start fields \(M^0_\sigma\) and the physical exogenous \(H\) field exactly
periodically in space. An all-marked typed path uses \(M^0_\sigma\)
at a vertex of type \(\sigma\). There are computable integers
\(K_{\rm den},K_{\rm path},Q_{\rm den},Q_{\rm path}\), independent of
\(L\), such that the cap
\begin{equation}\label{eq:u-cap}
 p\le p_*:=\bigl(4\max(K_{\rm den},K_{\rm path})\bigr)^{-
                         \max(Q_{\rm den},Q_{\rm path})}
\end{equation}
implies, through \(C_{\rm hor}L\), the root bounds
\begin{align}
 \Pr(D_N)&\le\frac43\cdot4^{-\lceil\delta N/68\rceil},\label{eq:u-root-bounds}\\*
 \Pr(\textup{an all-}M^0\textup{ backward data path of length }N)
 &\le\frac43\cdot4^{-\lceil N/68\rceil}.\label{eq:u-root-path}
\end{align}
Fixed endpoint and type factors can be included by increasing the
integers in \eqref{eq:u-cap}.
\end{lemma}
\par\noindent\emph{Proof idea.} Lift the periodic recurrence to the lattice and use complete contours or paths there. Within the linear horizon a physical coin has at most \(Q_{\rm wrap}\) spatial lifts (defined in \eqref{eq:u-multiplicity}) and at most \(n_{\rm sel}\) types. Thus \(F\) distinct lifted names require at least \(\lceil F/(n_{\rm sel}Q_{\rm wrap})\rceil\) independent coins.

\begin{proof}
Induction from the empty finite past shows that the lifted marked
recurrence equals the lift of the torus recurrence exactly. This is a
deterministic identity, valid for the spatially periodic flag field.
In the lift, complete old/new contours and the restart-minus-two rule
from Lemma~\ref{c:coherence} give connected certificates with
\[
 F\ge\lceil\delta N/68\rceil\quad\hbox{or}\quad
 F\ge\lceil N/68\rceil,
\]
respectively. Here \(F\) counts distinct sheet-typed lifted physical-time names.
In the folded appendix a path may switch its sheet at each round; attach each contour in
that vertex's own selector orientation and connect by the actual
physical data edges. The union graph has fixed finite degree.
The certificates have at most \(A_{\rm den}F\) or
\(A_{\rm path}F\) edges for fixed computable constants. For example,
the all-marked path can use the edge budget \(256F\).
The density certificate also contains the selected path, whose length
is at most \(68F/\delta\). Whole contours, including all their source
stars, are retained, so no physical source can be forgotten at a
restart.

If controller and data ranges are \(r_M,r_D\), every anchor through
\(T\le C_{\rm hor}L\) lies within a box of radius
\[
 r_DT+r_M(T+1)+r_s\le a_{\rm hor}L
\]
for one computable constant \(a_{\rm hor}\); \(r_s\) covers the fixed
source halo. At one time, a physical torus site has at most
\begin{equation}\label{eq:u-multiplicity}
 Q_{\rm wrap}=(2\lceil a_{\rm hor}\rceil+3)^4
\end{equation}
lifts in that box. Consequently \(F\) distinct lifted names represent
at least \(\lceil F/(n_{\rm sel} Q_{\rm wrap})\rceil\) independent physical coins,
where \(n_{\rm sel}=1\) for Theorem~\ref{thm:detailed} and
\(n_{\rm sel}=4\) for the folded appendix. The fourth power overestimates the lift count in three dimensions. One physical coin can
occur with each type and each spatial lift. A specified certificate
therefore costs at most \(p^{F/(n_{\rm sel} Q_{\rm wrap})}\).

In a graph of degree \(\Delta\) a rooted connected
\(j\)-edge graph has a doubled-edge traversal of length at most \(2j\).
Choosing that traversal, its terminal subset and finitely many roles
costs at most an exponential in \(j\). Summing \(j\le AF\) is still
at most \(K^F\), for a computable integer \(K\). Thus the relevant
probability is at most
\[
 \sum_{F\ge F_0}\bigl(Kp^{1/(n_{\rm sel} Q_{\rm wrap})}\bigr)^F.
\]
Choose \(Q_{\rm den},Q_{\rm path}\) in \eqref{eq:u-cap} to include
the factor \(n_{\rm sel}\). Under that cap the base is at most \(1/4\), giving
\eqref{eq:u-root-bounds} and \eqref{eq:u-root-path}. The argument uses a fixed linear horizon, so its
multiplicity is independent of \(L\).
\end{proof}

\subsection{From one live root to an impossible strip}
Take \(C_{\rm hor}=203\) and impose the cap from
Lemma~\ref{u:wrapped}. Let \(\mathcal G_{\rm den}\) be the event that \(D_{2L+1}\) occurs at no physical root of age at least \(2L+1\) within this horizon. The comparison path length is \(2L+1\).
Use one selector for Theorem~\ref{thm:detailed}; use all four for the folded appendix. The event depends only on the
exogenous \(H\) field, not on an initial chart.

\begin{proposition}[No late unmarked valid region]\label{u:mortality}
On \(\mathcal G_{\rm den}\), every actual live selector trigger at age
\(t\ge200L\), within the specified horizon, belongs to its own \(M^0_\sigma\).
This holds simultaneously for every initial controller word and every
copied selector.
\end{proposition}
\par\noindent\emph{Proof idea.} Compare same-history readings across the periodic replica plane. Late coherent readings would produce a periodic scaffold on a diagonal strip wide enough for the obstruction.

\begin{proof}
Periodic replica \(\ell^1\) distance on its two-torus is at most
\(2\lfloor L/2\rfloor\le L\). The unit-shift pairing proof of
Lemma~\ref{c:coherence} works with a shortest periodic representative:
a good step decreases that distance, and an exceptional step increases
it by at most 350. Hence age \(2L+1\) suffices to compare all
unmarked live replicas at equal physical \(i,t\).

Suppose an unmarked live first or second reading exists at
\((i_0,P_0,Q,t)\), with \(t\ge200L\). Select a clean successful
replica. It supplies both unmarked live actual first-coordinate west
and south inputs at time \(t-1\). Repeat down both logical branches.
At depth \(r\), there are witnesses at every
\(i_0-k\), \(0\le k\le r\), at one common time \(t-r\).
For \(r\ge L\) they cover every residue modulo \(L\).

Take every integer \(r\in[L,L+4R_{k_L}+4]\), where \(k_L=v_2(L)+1\).
All resulting witnesses have age at least
\[
 200L-L-128L-4=71L-4\ge2L+1.
\]
Choose one at each logical position and time. Coherence makes its
letter independent of the replica choice. One clean child check
compares its actual parent letters with these selected reference
letters, supplying both west and south chart edges. Therefore their
scaffold projections form a locally legal strip of width
\(4R_{k_L}+4\) in \((x,y)=(i,s-i)\).
Apply \(\pi\) letter by letter. It has radius zero, so no width is
lost. Each internal edge is a west or south edge of a strip letter
at time at least \(a+1\), where \(a\) is the strip's lower time,
checked by that letter's child comparison. The projections therefore
form a legal \(\mathbb A^{\mathrm q}\) pattern on the whole strip,
to which Lemma~\ref{u:strip} applies.

It is periodic by \((L,-L)\): increasing \(i\) by \(L\) at fixed
time \(s\) denotes the same physical torus coordinate, and
coherence removes replica-choice ambiguity. Lemma~\ref{u:strip}
forbids that strip. Thus no such unmarked live reading exists.
For an actual trigger, a local fault directly marks the vertex; in a
clean computation live retained first data equals its live ordinary
proposal. Hence the same conclusion holds for triggers.
\end{proof}

\subsection{Approximate entanglement breaking of the entire channel}
The final erasure block is a complete dependence cut. Conditional on
the measured initial classical word and exogenous innovations, any
surviving initial quantum dependence must cross a conditional-reset block with a live control or
its named hardware exception in every round. In the concrete
schedule all information-moving quantum work precedes that block.
Its discarded work is not retrieved later.

\begin{theorem}[Linear upper bound for the fixed circuit]\label{u:eb}
There are computable constants \(A,a>0\), \(b\in\N\), an integer \(C_0\ge2\)
and \(L_{\rm EB}\ge C_0/2\), all independent of \(n\), such that for every \(n\), every
\(\eps\in I_n\) and every \(L\ge L_{\rm EB}\),
\begin{equation}\label{eq:u-eb}
 \Lambda_L^{201L+C_0}=(1-r_L)\Phi^{\rm EB}_L+r_L\mathcal B_L,
 \qquad r_L\le q_L\le A L^b e^{-aL},
\end{equation}
where \(\Phi^{\rm EB}_L\) is entanglement breaking and
\(\mathcal B_L\) is a channel. In particular,
\(\|\Lambda_L^{201L+C_0}-\Phi^{\rm EB}_L\|_\diamond\le2q_L\). For every qubit
encoder and decoder and every \(t\ge201L+C_0\),
\begin{equation}\label{eq:u-fidelity}
 F_L(t)\le\frac12+\frac{q_L}{2}.
\end{equation}
The same conclusion, with constants computed for that circuit and the
bound \(q_L\) on the exceptional event taken with \(n_{\rm sel}=4\),
holds for the folded part of the circuit of
Appendix~\ref{sec:coexistence} on every window \(I_j\), \(j\ge1\).
\end{theorem}
\par\noindent\emph{Proof idea.} Exclude large wrapped certificates throughout a late time slab. The remaining local erasure cuts remove dependence on the initial quantum state, which gives the lifetime bound.

\begin{proof}
Let \(c_{\rm type}\) bound the fixed number of endpoint and dependency roles per site.
A union of the estimate \eqref{eq:u-root-bounds} over roots,
selector types and times gives the conservative bound
\[
 q_{\rm den}\le n_{\rm sel} c_{\rm type}(203L+3)L^4\cdot\frac43\cdot
                  4^{-\lceil\delta(2L+1)/68\rceil}.
\]
On \(\mathcal G_{\rm den}\), Proposition~\ref{u:mortality} puts every
permitted live crossing after \(200L\) in \(M^0\). Since
\(L\ge L_{\rm EB}\ge C_0/2\), every time up to \(201L+C_0\) lies within
the horizon \(C_{\rm hor}L=203L\) of Lemma~\ref{u:wrapped} and
Proposition~\ref{u:mortality}. A surviving
initial quantum ancestor at \(201L+C_0\) would therefore give an
all-marked data path of at least \(L\) rounds. The path bound \eqref{eq:u-root-path}, unioned over final output positions and types, gives
\[
 q_{\rm path}\le n_{\rm sel} c_{\rm type}L^4\cdot\frac43\cdot
                                      4^{-\lceil L/68\rceil}.
\]
Enlarge a fixed endpoint factor if the dependency convention requires
it, and put \mbox{\(q_L=\min(1,q_{\rm den}+q_{\rm path})\)}.
Both terms are polynomial times exponential in \(L\).

The complementary good event \(\mathcal G_{\rm EB}\) is exogenous \(H\)-only and covers all
initial words at once. In the Q realization this field
also uses probe endpoint marks; its input independence is the property
needed here. On \(\mathcal G_{\rm EB}\) no initial quantum factor reaches any
retained output. Dephase the entire initial controller, profile and program word \(a\), as the
physical circuit does. Conditional on \(a\) and innovations in \(\mathcal G_{\rm EB}\),
all remaining outputs are preparations. Averaging these preparations
produces states \(\rho_a\), and the good channel has the form
\[
 \Phi^{\rm EB}_L(\rho)=
    \sum_a\operatorname{Tr}[(\Pi_a\otimes I)\rho]\,\rho_a.
\]
It is measure-and-prepare on the entire physical input algebra,
including initial controller/data entanglement. Because \(\mathcal G_{\rm EB}\) is
input-independent, the true channel decomposes as
\[
 \Lambda_L^{201L+C_0}=(1-r_L)\Phi^{\rm EB}_L+r_L\mathcal B_L,
 \qquad r_L\le q_L,
\]
with \(\mathcal B_L\) CPTP. Two channels have diamond distance at
most two, giving the stated diamond estimate. An entanglement-breaking
channel has entanglement fidelity at most \(1/2\), while any channel has
entanglement fidelity at most one. Arbitrary encoding and decoding therefore give
\((1-r_L)/2+r_L\le1/2+q_L/2\). Classical reference correlations in
\(a\) are allowed by this bound. Later evolution can be absorbed into
the decoder; equivalently EB channels remain EB under postcomposition.
This proves \eqref{eq:u-fidelity} at all later times.
\end{proof}

\section{Certification on computable windows}\label{a:certification}

This section proves the certification clause of Theorem~\ref{thm:detailed} (Theorem~\ref{a:complete}), shows that no total computable function bounds the halting-side forgetting time (Corollary~\ref{a:time}), and completes the proof of Theorem~\ref{thm:main}.

Fix either of the separately compiled whole-site (\(\mathrm{WS}\)) and independent single-qubit (\(\mathrm Q\))
realizations in Theorem~\ref{thm:detailed}. The storage reduction has an
exact arithmetical formulation because its encoder and decoder are fixed finite circuits at each size and time.
Write
\[
 f_{n,L,t}(\eps)=
 F(\cE_{n,L},\cD_{n,L,t};L,t,\eps)
\]
for their entanglement fidelity. The encoder and decoder depend on \((n,L,t)\); the noise rate enters only
through the round channel \(\Lambda_{L,\eps}\).

\begin{proposition}[Exact finite fidelity]\label{a:rational}
For fixed \(n,L,t\), the function \(f_{n,L,t}\) is a polynomial in \(\eps\)
with rational coefficients, computable by a total finite algorithm.
In particular, its value at a rational rate can be compared exactly with
\(7/8\).
\end{proposition}
\par\noindent\emph{Proof idea.} Construct the encoder and decoder by finite binary elimination. Expand the fixed noisy circuit in its elementary channels to compute each coefficient of the fidelity polynomial.

\begin{proof}
Choose the external CSS encoder by binary Gaussian elimination. If
\(A\) and \(B\) are the binary spaces of \(X\)- and \(Z\)-stabilizers,
then \(A\subseteq B^\perp\). Extend a basis of \(A\) to a basis of
\(B^\perp\), and then to a basis of the full physical binary space.
Prepare the \(A\)-coefficient wires in \(\lvert+\rangle\), put the logical
input and the fixed other logical bits in the complementary wires, the
wires completing a basis of the full space in \(\lvert0\rangle\), and apply the invertible binary basis change. Hadamards and CNOT gates
implement this encoder. Syndrome extraction, recursive stabilizer
recovery, its total syndrome correction table and inverse encoding use the same real
Clifford gates and computational-basis operations; \(Z=HXH\). The same fixed syndrome table is total outside the correctable span.
At a requested time first append the declared intended future circuit
on the in-flight quantum work, as in
Proposition~\ref{d:storage}. Failed finite chart searches use the
fixed fallback. These rules give total finite circuits in every case.

The Bell density matrix, basis preparations and maximally mixed states
have rational entries. The matrices of \(X\), CNOT and Toffoli are
permutations. Conjugation by a Hadamard replaces each relevant matrix
entry by one half of a signed sum of old entries. Reset, dephasing,
partial trace and sums of unnormalized computational-measurement
branches preserve rational entries. Thus these operations have rational
superoperator matrices, including the external circuits. Each noise map
is affine in \(\eps\), with rational coefficients in its replacement part: \(1/2^N\) for a whole
site in \(\mathrm{WS}\), and \(1/2\) for one wire in \(\mathrm Q\). A finite composition followed
by evaluation of entanglement fidelity is therefore a rational-coefficient polynomial. Finite rational matrix
arithmetic computes it and terminates. The algorithm's resource cost
can be enormous.
\end{proof}

In the conditional theorem below, \(X\) is a polynomial indeterminate and \(A(n,L,t,k)\) is a Boolean test; the stabilizer space \(A\) in the preceding proof is local to that proof.
\begin{theorem}[Conditional window certification]
\label{lean:conditional}
Let $u:\mathbb N\rightharpoonup\mathbb N$ be partial recursive. Suppose there
is a total computable compiler $\gamma$ from partial-recursive program codes
to inputs such that $u(\gamma(e))\downarrow$ if and only if $e(0)\downarrow$.
Let $c\in\mathbb Q$, $c>0$, let $L_0:\mathbb N\to\mathbb N$, and let
$I_n=(a_n,b_n)$ be pairwise disjoint open rational windows with
$0<a_n<b_n<1$. More precisely, for $n\ne m$, assume
$b_n\le a_m$ or $b_m\le a_n$.
Let $p_{n,L,t}\in\mathbb Q[X]$ for $n,L,t\in\mathbb N$.
Let $r:\mathbb N\to\mathbb Q$ be a surjection, and assume a total
computable Boolean test $A(n,L,t,k)$ satisfying exactly
\[
 A(n,L,t,k)=\mathrm{true}\quad\Longleftrightarrow\quad
 \bigl(L_0(n)\le L\ \land\ t\le cL\ \land\ a_n<r(k)<b_n
       \ \Longrightarrow\ p_{n,L,t}(r(k))\ge 7/8\bigr).
\]
Here all size and time variables are natural numbers; $t\le cL$ is
rational comparison and hence equivalent to $t\le\lfloor cL\rfloor$.
No separate computability assumptions on $L_0,a,b,p,r$ are needed:
the displayed exact test is the computational hypothesis.
Assume explicitly the following two physical premises of this paper:
\begin{enumerate}
\item (Nonhalting storage.) If $u(n)\uparrow$, then, for every $L,t\in\mathbb N$
and $q\in\mathbb Q$, whenever $L\ge L_0(n)$, $t\le cL$, and $q\in I_n$,
we have $p_{n,L,t}(q)\ge15/16$.
\item (Halting forgetting.) If $u(n)\downarrow$, there is $T\in\mathbb N$
such that for every $L,t\in\mathbb N$ and $q\in I_n\cap\mathbb Q$,
$t\ge T$ implies $p_{n,L,t}(q)\le9/16$.
\end{enumerate}
Define $P_{\mathbb Q}(n)$ to mean
\[
 \forall L,t\in\mathbb N\ \forall q\in\mathbb Q,\quad
 L\ge L_0(n),\ t\le cL,\ a_n<q<b_n
 \ \Longrightarrow\ p_{n,L,t}(q)\ge7/8.
\]
Define $P_{\mathbb R}(n)$ by replacing $q\in\mathbb Q$ by $x\in\mathbb R$,
coercing the endpoints and $7/8$ to $\mathbb R$, and evaluating the
coefficientwise image of $p_{n,L,t}$ in $\mathbb R[X]$ at $x$; keep
$t\le cL$ as the same rational comparison.
Then all four conclusions hold:
\[
 \begin{gathered}
 \forall n,\quad P_{\mathbb Q}(n)\ \Longleftrightarrow\ u(n)\uparrow;\\
 P_{\mathbb Q}\text{ is }\Pi^0_1\text{-complete};\\
 \forall n,\quad P_{\mathbb Q}(n)\ \Longleftrightarrow\ P_{\mathbb R}(n);\\
 P_{\mathbb Q}\text{ is not a computable predicate}.
 \end{gathered}
\]
For precision, $S\subseteq\mathbb N$ is $\Pi^0_1$ here when there exists
a total computable Boolean function $R:\mathbb N\times\mathbb N\to\{0,1\}$
with $S(n)\Longleftrightarrow\forall s\,R(n,s)=1$; completeness means that
$P_{\mathbb Q}$ is $\Pi^0_1$ and every such $S$ many-one reduces to it
by a total computable map $\mathbb N\to\mathbb N$.
\end{theorem}
\begin{proof}
The physical inputs are explicit hypotheses. Enumerating the exact tests over $(L,t,k)$ puts
$P_{\mathbb Q}$ in $\Pi^0_1$. Nonhalting storage gives every required
inequality. If the input halts, take the rational midpoint of its window,
the time $T$ from halting forgetting, and a size $L\ge L_0(n)$ with
$T\le cL$; this violates the certificate since $9/16<7/8$. Thus the
certified set is exactly the nonhalting set of $u$. The computable
compiler, together with the usual partial-recursive search for a failed
Boolean test, gives many-one $\Pi^0_1$-hardness and undecidability.
Finally, continuity of a real polynomial and density of the rationals in
every open real interval show that a strict real violation has a rational
violation; coefficient coercion preserves rational evaluation. This proves
the equivalence of rational and real certification.
\end{proof}

\begin{theorem}[Window certification]\label{a:complete}
Let \(c>0\), \(L_0(n)\), and \(I_n\) be the rational coefficient,
computable size threshold, and computable rational window in
Theorem~\ref{thm:detailed}. The predicate
\begin{equation}\label{a:predicate}
\begin{split}
 P(n):\quad& f_{n,L,t}(q)\geq 7/8\quad\text{for every integer }L\geq L_0(n),\\
 &0\leq t\leq\lfloor cL\rfloor,
 \quad q\in I_n\cap\mathbb Q
\end{split}
\end{equation}
is \(\Pi^0_1\)-complete. It is equivalent to the same statement with every
real rate in \(I_n\), and holds exactly when \(U(n)\) never halts.
\end{theorem}
\begin{proof}
Apply Theorem~\ref{lean:conditional} with \(u=U\), \(p_{n,L,t}=f_{n,L,t}\) for \(L\ge1\) and \(p_{n,0,t}=0\), and the domain in \eqref{a:predicate}; here \(L_0(n)\ge1\). The exact finite test is total by Proposition~\ref{a:rational}, the
explicit definition at \(L=0\), and computability of the cutoff and endpoints. Theorem~\ref{thm:detailed}(i)--(ii), with \(\eta=1/16\), gives the two physical premises. Universality supplies the input compiler.
\end{proof}

A false certificate has a finite exact witness \((L,t,q)\). The theorem
concerns the computably supplied windows. On rational rates, membership
in their union and the unique window label are decidable. For the chosen
realization, first reject \(q\notin(0,1)\). Otherwise compute its positive
rational refresh rate
\[
 \mu_{\mathrm{WS}}(q)=1-(1-q)^{P_{\rm pad}},\qquad
 \mu_{\mathrm Q}(q)=1-\left[\frac{1+(1-q)^{K_{\rm samp}}}{2}\right]^M.
\]
Enumerate the decreasing nominal-window upper endpoints until they fall
below that value. Only finitely many physical windows remain to test by exact rational
comparison. Hence the set of indices \(n\) whose window has linear
storage is the same \(\Pi^0_1\)-complete set.

\begin{corollary}[No computable uniform forgetting bound]\label{a:time}
For any fixed rational \(0<\eta<3/8\), there is no total computable
\(g:\N\to\N\) such that, for every halting \(n\), every \(L\), every
\(\eps\in I_n\) and every \(t\ge g(n)\), \(F^*_{L,\eps}(t)\le1/2+\eta\).
\end{corollary}
\begin{proof}
Suppose \(g(n)\) were such a bound. At the rational midpoint of \(I_n\),
compute the explicit fidelity at time \(t=\max\{1,g(n)\}\) and a
computably chosen size \(L\geq L_0(n)\) with \(t\leq\lfloor cL\rfloor\).
Nonhalting gives fidelity at least \(15/16\); halting gives fidelity at
most \mbox{\(1/2+\eta<7/8\)}. Proposition~\ref{a:rational} would therefore
decide halting by an exact comparison with \(7/8\), a contradiction.
\end{proof}

\begin{proof}[Completion of Theorems~\ref{thm:main} and~\ref{thm:detailed}]
The exact compilation and parameter choices of
Section~\ref{sec:compilation} give the fixed three-dimensional circuit
and rational windows. Proposition~\ref{d:storage} proves storage at
every requested time. Theorem~\ref{f:forgetting} gives the all-encoding
halting bound, and Theorem~\ref{u:eb} gives the linear entanglement-breaking mixture.
Theorem~\ref{a:complete} proves the certification clause. Their fidelity
gap gives precisely the optimal-lifetime alternatives in Theorem~\ref{thm:main}.
\end{proof}

\section{Exponential storage on the nonhalting windows}\label{sec:exponential}
The finite-torus ceiling in Section~\ref{sec:ceiling} arises because a
causal chart must continue across successive time diagonals. We instead
place a static chart in a finite square and attach a finite counter.
Two coordinates carry this square; four transverse coordinates carry a
closed toric code. The classical rule is protected in two of the latter
coordinates and copied in the other two.

\begin{theorem}[Detailed form of Theorem~\ref{thm:exponential}]\label{v:detailed}
There is a rational \(c_{\mathrm e}>0\) such that for each \(\mathsf r\in\{\mathrm{WS},\mathrm Q\}\) there is a separately
compiled, computably specified, translation-invariant, finite-range,
time-periodic circuit on \(\Z^6\), with \(N_{\mathsf r}\) qubits per
site and elementary inventory
\(\{X,H,\mathrm{CNOT},\mathrm{Toffoli},R_0,D_Z,I\}\), of arity at most
three. Noise in realization \(\mathsf r\) acts after every elementary
layer on every factor, including idle, reset, controller, counter, sampling register, probe, port, work and mixed ancilla factors. There are computable pairwise disjoint
nonempty rational open intervals \(I_n^{\mathsf r}\)
and total computable size cutoffs
\(L_{\mathrm e}^{\mathsf r}(n)\), with these properties on
\((\Z/L\Z)^6\), using fidelity \eqref{eq:fidelity}.
\begin{enumerate}
\item Total computable maps \(\cE^{\mathsf r}_{n,L}\) and
\(\cD^{\mathsf r}_{n,L,t}\), independent of \(\eps\) and of the halting
answer, satisfy
\[
 F(\cE^{\mathsf r}_{n,L},\cD^{\mathsf r}_{n,L,t};L,t,\eps)\ge15/16
 \quad\bigl(U(n)\uparrow,\ \eps\in I_n^{\mathsf r},\
 L\ge L_{\mathrm e}^{\mathsf r}(n),\ 0\le t\le2^{\lfloor c_{\mathrm e}L\rfloor}\bigr).
\]
The same encoder serves every indicated integer time and rate.
\item If \(U(n)\downarrow\), then for every \(\eta>0\) there is a finite
\(C^{\mathsf r}_{n,\eta}\), independent of \(L\) and of
\(\eps\in I_n^{\mathsf r}\), such that
\(F^*_{L,\eps}(t)\le1/2+\eta\) for every \(L\), every such rate and
every integer \(t\ge C^{\mathsf r}_{n,\eta}\). The supremum includes
arbitrary encodings into the entire physical algebra and arbitrary
initial controller, work, intersite and reference entanglement.
\end{enumerate}
The closures of the \(\mathrm{WS}\) windows lie in \((0,1)\); those of the \(\mathrm Q\) windows
lie in \((0,1/(2MK_{\rm samp}))\), where \(M,K_{\rm samp}\) are the fixed probe count and
sampling depth chosen below. For each fixed rational \(0<\eta<3/8\),
no total computable bound on \(C^{\mathsf r}_{n,\eta}\) as a function
of \(n\) exists, by Proposition~\ref{v:certification}(iii).
\end{theorem}

Proposition~\ref{v:certification} gives the lifetime upper bound and the finite-test certification for this circuit; the scalable-memory labels in Table~\ref{tab:results} are classified by Corollary~\ref{cor:window-undecidable}; the scope is in Section~\ref{intro:scope}. The windows decrease as \(4^{-n}/P_{\rm pad}\) in \(\mathrm{WS}\) and \(4^{-n}/(MK_{\rm samp})\) in Q. In the affine layout used here, four independent quantum translations preserve two independent controller coordinates, so rank--nullity forces at least six coordinates within this layout class.

Section~\ref{v:classical-rule} gives the imported static chart, the square and its counter, the censored core rule and property~\ref{v:structure}. Section~\ref{v:density-section} bounds the density of faults on all paths, and Section~\ref{v:parameters} gives the source law and parameter order. Sections~\ref{v:positive} and~\ref{v:forgetting} prove parts (i) and (ii), respectively. Section~\ref{v:wrapped-section} describes a separate loss mechanism. Section~\ref{v:cert-section} proves a coarse lifetime upper bound and the certification result.

\subsection{Static squares and a counter with arbitrary initial phase}\label{v:classical-rule}
\begin{proposition}[Static classical interface]\label{v:static-import}
For the static chart of \cite{DMNoise2026}, version~1.4.0, applied with the wrapped machine \(V\) of Section~\ref{sec:controller}, the canonical projected geometry at cutoff \(a=n+n_0+1\), its finite owners and routing tags are computable on signed integer coordinates. If \(U(n)\uparrow\), it has a legal static chart extension with that projection and cutoff, with no halt letter. If \(U(n)\downarrow\), there is a finite size \(s_n\), found by a finite search that terminates because \(U(n)\) halts, such that every chart patch of side at least \(s_n\) that is locally legal at cutoff \(a\), in any substitution phase, contains a fully guarded wrong critical ride: a short depth-\((a-1)\) end flagged zero or a long depth-\(a\) end flagged one.
\end{proposition}
\par\noindent\emph{Proof idea.} Compute ownership and routing in the fixed substitution phase. These facts let the board-filling proof apply there and supply the legal extension needed by the encoder.

\begin{proof}
Use the canonical construction in the proof of \cite[Lemma~11.2]{DMNoise2026} with its cutoff converse. Its substitution seed \(t_*\) is fixed by the child
\((2,2)\) of \(\sigma^2\). With
\(q(z)=\lfloor(z-1)/4\rfloor\), \(d(z)=(z-1)\bmod4\), its projected
letters satisfy
\(A_0(z)=\sigma^2(A_0(q(z)))(d(z))\). Signed integer coordinates
iterate to the fixed seed; the phase is
\(\varpi=(-1/3,-1/3)\), \(\tau_0=(-2/3,-2/3)\).
All ranks are finite since \(3u+1\ne0\) for integer \(u\).
Every coordinate has a finite frame owner: at zero
\[
 (0-\tau_{0,i})\bmod4^{d+1}=(4^{d+1}+2)/3
 \in[4^d+1,3\cdot4^d],
\]
with margins growing to infinity. Any fixed integer perturbation
therefore lies inside such a frame at sufficiently large depth.
Walls are unbounded in both directions, so nearest-wall orientations
and nearest-stop event tags are computable. The direct check of the routing conditions (O1)--(E3) in the proof of \cite[Lemma~11.2]{DMNoise2026} uses these three properties.

The board argument of \cite[Lemma~3.3]{DMNoise2026} forces the computation in every phase. The nonhalting proof of \cite[Proposition~4.3]{DMNoise2026} invokes typicality for the routing conditions and the nesting of active sites in boards. At the fixed phase, the routing check and the active-site argument in the proof of \cite[Lemma~11.2]{DMNoise2026} replace those two inputs. Each coordinate has a finite owner. A site whose two coordinates are free
at the same depth lies in that depth's board; all other sites are
horizontal-only or vertical-only carriers, or walls. Thus every active
site belongs to its proper board even at this nontypical phase.
The all-phase board argument forces the initialized finite run on each
board. The cutoff converse of \cite[Lemma~3.2]{DMNoise2026} uses only
finite ranks. Fill each board by
its finite computation, its channels from its endpoints and empty
records elsewhere, as in the proof of \cite[Proposition~4.3]{DMNoise2026}. Large boards run
\(V(n+1)=U(n)\); small or failed scans reach their boundary or fail
without a halt. On nonhalting no board contains a halt letter.
This produces a global legal static chart with the specified projection.

The guarded wrong-ride alternative is \cite[Proposition~4.3 and Lemma~10.5]{DMNoise2026}. These are the static inputs used below; the fixed-phase chart extension, counter, boundary geometry, initialization search and quantum estimates are proved in this paper.
\end{proof}

Write physical coordinates as \((u,P)\in\Z^2\times\Z^4\). A profile
\(\mathsf{pr}\) is inactive, or active with phase
\(\mathsf A,\mathsf D,\mathsf B\) and a letter of the static chart of
\cite[Sections~3--4]{DMNoise2026}. First put its base constraints in a
fixed nearest-neighbour higher-block presentation; attach the following
geometry and dynamics afterwards. A base constraint is checked only
when its whole bounding box is active. Impose five finite constraints:
\begin{enumerate}
\item A \(2\times2\) square cannot have exactly three active corners.
\item Horizontal active pairs belong to
\(\mathsf{AA},\mathsf{AD},\mathsf{DB},\mathsf{BB}\). At an inactive west
neighbour the phase is not \(\mathsf B\); at an inactive east neighbour
it is not \(\mathsf A\).
\item At an inactive south neighbour the phase is \(\mathsf D\) if
the west neighbour is inactive, and \(\mathsf B\) otherwise. At an
inactive north neighbour it is \(\mathsf D\) if the east neighbour is
inactive, and \(\mathsf A\) otherwise.
\item A \(\mathsf D\) site with active north neighbour has a
\(\mathsf D\) northeast neighbour; transpose southwest for south.
\item On a fully active \(2\times2\) square the southwest and northeast
phases agree.
\end{enumerate}
Let \(V_{\mathsf{pr}}\) mark every active participant in a forbidden
local pattern. It is a fixed finite local Boolean function.

\begin{lemma}[Square geometry and counter]\label{v:cap}
On \(\Z^2\) (the \(u\)-plane, or the lift of a torus profile), every active
nearest-neighbour component without a \(V_{\mathsf{pr}}\) mark is a finite
square or a product of two infinite integer intervals. A finite square of
side \(N_{\rm cap}\) (any positive integer here; the reference square of
Section~\ref{v:parameters} has \(N_{\rm cap}=\lfloor L/4\rfloor\))
has exactly \(N_{\rm cap}\) diagonal \(\mathsf D\) sites and locally
recognizable endpoints. On this diagonal set
\[
 b'_j=b_j\mathbin\oplus y_j,\qquad y'_0=1,\qquad
 y'_j=b_{j-1}y_{j-1}\quad(1\le j<N_{\rm cap}).
\]
Under \(F_{\rm cap}\) with a constant valid profile, overflow is \(o=b_{N_{\rm cap}-1}y_{N_{\rm cap}-1}\). It occurs within
\(1+2^{N_{\rm cap}+1}\) rounds from every state. With all bits zero and
only \(y_0=1\), no overflow occurs before \(2^{N_{\rm cap}}-1\).
\end{lemma}
\par\noindent\emph{Proof idea.} The local square constraints force a finite active component to be a square. Along its diagonal, a binary counter with delayed carries still increases by one until overflow.

\begin{proof}
The first constraint fills the bounding rectangle of any finite active
nearest-neighbour path: a new boundary site propagates along its new
row or column by the three-corners rule. Thus a component is a product
of intervals. If its horizontal interval is finite, every row has one
\(\mathsf D\), in the order \(\mathsf A^*\mathsf D\mathsf B^*\). Its position advances
by one on each row, so the vertical interval is finite. The boundary
rules put this diagonal at the southwest and northeast corners, forcing
a square. A finite-height, infinite-width rectangle is impossible:
choose a northeast diagonal away from any finite side endpoint. Its
bottom phase is \(\mathsf B\), its top phase \(\mathsf A\), contrary to
the fifth constraint. Infinite components therefore contain arbitrarily
large square patches.

For the counter, after one round \(y_0=1\), and
\[
 V_{\rm ctr}=\sum_{j=0}^{N_{\rm cap}-1}2^j(b_j+y_j),\qquad
 V'_{\rm ctr}=V_{\rm ctr}+1-2^{N_{\rm cap}}o.
\]
This follows from \(b\oplus y=b+y-2by\). Its bounded range precludes
\(2^{N_{\rm cap}+1}\) successive increments without overflow.
The specified initial value is one; overflow requires a value at least
\(2^{N_{\rm cap}}\), proving the lower bound. The local rule uses
copies, XOR, AND and a fresh output for overflow, all realized by the
seven gates of Section~\ref{sec:compilation}.
\end{proof}

One round of the classical rule has five parts.

\emph{(a) Core rule.} The ordinary core word is bottom or \((\mathsf{pr},b,y,\mathsf d)\). Its deterministic Moore-neighbour rule \(F_{\rm cap}\) retains the centre profile, updates the counter using original diagonal endpoints, and sets
\[
 \mathsf d'=\mathsf d\lor V_{\mathsf{pr}}\lor o\lor
 \left[\operatorname{active}(\mathsf{pr})\land
 \bigvee_{w\sim u,\ \operatorname{active}(\mathsf{pr}(w))}\mathsf d(w)\right].
\]
Off the diagonal the new counter bits are zero. A bottom input makes the output bottom, while inactive vacuum is a genuine nonbottom word. The storage-enable bit is \(\pi=\operatorname{active}(\mathsf{pr})(1-\mathsf d)\), with \(\pi(\bot)=0\); hence \(\pi(F_{\rm cap})\le\pi(\text{centre})\). Death leaves the original profile in place, so it creates no new counter endpoint.

\emph{(b) Protection and deep queries.} Protect the core in \(P_1,P_2\), copying in \(P_3,P_4\). At each replica in the three boxes \eqref{eq:c-boxes}, evaluate \(F_{\rm cap}\) on its nine Moore inputs, and let \(G\) return the word receiving at least 512 of the 768 votes, or bottom; these votes have a unique possible winner. Each site stores two coordinates: the second is the ordinary proposal \(w_{\rm ord}=G(C_t)\), where \(C_t\) is the field of retained first core coordinates before the update at round \(t\), and the first is \(w_{\rm ord}\) or bottom, as decided by the censor in (d). With \(v=(58,58)\) and \(D_{\rm q}=4096\), compute from old first coordinates the two finite queries
\[
 Q_0=G^{D_{\rm q}}(C_t)(u,P_{12}-(D_{\rm q}-1)v),\quad
 Q_-=G^{D_{\rm q}}(C_t)(u-e_x,P_{12}+v-(D_{\rm q}-1)v).
\]
Here \(G^{D_{\rm q}}\) means virtual iteration of the ordinary core rule; it contains neither sampling registers nor the statistical censor. Its virtual levels are evaluated locally from one transport of old words.

\emph{(c) Sampling registers.} Each sampling-register index \(j\) at storage position \(w\) belongs to base \(w-je_y\) and copies from \(w-e_x+v\). It has two seven-bit fields: a value field with 126 codes for high, one for low and one for unknown, and an availability field with 126 codes for allowed, one for unavailable and one for unknown; availability treats its last two codes as unavailable. All active, origin, end, phase, flag and edge-guard predicates read only the profiles of \(Q_0,Q_-\) at that base. The register takes the first applicable update in this list:
\begin{enumerate}[label=\arabic*.]
\item a bottom required driver writes (unknown, allowed);
\item current vacuum writes (unknown, unavailable);
\item a genuine active origin writes (low, allowed);
\item predecessor vacuum, or a failed genuine active-edge guard, writes (unknown, unavailable);
\item otherwise both old fields are copied.
\end{enumerate}
In particular an origin takes precedence over an inactive predecessor.

\emph{(d) Censor.} At an endpoint, gather the provisional sampling-register outputs by local recomputation from old words. Put \(V_{\rm av}=\#\text{allowed}\), \(H_{\rm up}=\#\{\text{allowed}\land(\text{high}\lor\text{unknown})\}\) and \(H_{\rm lo}=\#\{\text{allowed}\land\text{high}\}\), an upper and a lower count of high lanes. The test is enabled when \(V_{\rm av}\ge\lceil B/2\rceil\); put \(s_{\rm vote}=\lceil3C_hB/4\rceil\). The new first coordinate is bottom if either query is bottom, or if, at an enabled active phase-zero end, either the flag is one and \(H_{\rm up}\ge s_{\rm vote}\) or the flag is zero and \(H_{\rm lo}<s_{\rm vote}\). Otherwise it is \(w_{\rm ord}\). The second coordinate always remains \(w_{\rm ord}\), so postprocessing never writes a different nonbottom core. Historical bottom widens uncertainty and permits a test; it is distinct from a true inactive boundary.

\emph{(e) Quantum processor.} At fixed \(u\) use the closed four-dimensional plaquette CSS processor
of Lemmas~\ref{q:sweep}, \ref{q:homology}--\ref{q:span} and the proof of
Lemma~\ref{x:exponential}. In each round, for the bit sector followed by the phase sector, one
Clifford extraction writes the syndrome into fresh records, the snapshot
\(\sigma_0\). Local Boolean processing on these records alone computes
400 reflected sweeps by
\(\sigma_{j+1}=\sigma_j+\partial P_j\) for \(0\le j<400\), where
\(P_j\) is the plaquette set that sweep \(j+1\) selects from
\(\sigma_j\), and stores \(P=P_0+\cdots+P_{399}\). Only then are the
plaquette flips of \(P\) applied. The syndrome present before sweep
\(j+1\) means \(\sigma_j\); no sweep reuses an earlier sweep's decisions.
The record processing has range at most 400 code cells.
Track both sector chains at the common boundaries of full rounds,
just before the bit-sector extraction. The catalogue \(\Lambda_t\)
includes the allowed insertions and erroneous uses throughout the full
round, as in Lemma~\ref{q:round-shadow}.
Quantum
gates keep \(u\) fixed. The storage-enable bit changes only the existing final local
erasure decision. Every retained quantum-capable factor, including
syndrome/work, ports, data-derived records and unused factors, crosses
this complete cut or is reset/discarded before reuse. Quantum spatial
work precedes the longer classical calculation; no later swap retrieves
discarded input.

\paragraph{Structural property of both compiled circuits.}
\begin{enumerate}[label=\textup{(S)},ref=\textup{(S)}]
\item\label{v:structure} Every gate's quantum-capable inputs and
outputs lie in one fibre \(\{u\}\times(\Z/L\Z)^4\); these factors
include data, syndrome, work, ports, all quantum-derived records and
unused factors. Autonomous metadata \(\mathcal M\) is initially
dephased, has only Boolean basis permutations, resets and dephasings
as its own gates, and is never a target of a gate with a
quantum-capable input. Metadata controls may arrive from other
fibres; conditional on its initially measured word, its history law
is independent of the quantum input. In particular, after the last elementary
layer of every round, independent exogenous coins decide, with probability \(\eps\), whether to replace each site in \(\mathrm{WS}\) or each wire in Q, including idle factors. The coins are independent across positions and rounds and of all physical inputs.
\end{enumerate}
The processor satisfies fibre confinement. The autonomy construction in Section~\ref{sec:compilation} and the dephased scratch and reset probes in Section~\ref{v:parameters} ensure the metadata clause at every elementary gate. The noise definition \eqref{eq:site-noise} and its Q analogue give the final clause. Property~\ref{v:structure} therefore holds for every input to the compiled circuit, including malformed metadata.

\subsection{Physical-name density and finite-age coherence}\label{v:density-section}
The compiler of Section~\ref{sec:compilation}, including Section~\ref{s:probe}, gives
independent physical site-round names \(H=A\lor E\), with
\(\mu=\Pr(E)\le p=\Pr(H)\le2\mu\). A name may corrupt both its old
copies and its new output. The following supplies the precise density
estimate needed here, including that double appearance.

\begin{lemma}[All-path density]\label{v:density}
Let the names \(H(z,t)\) be independent with \(\Pr(H(z,t))\le p\).
Let \(\mathcal M_{\rm raw}\) be the least finite-start field that contains
every seed and marks a vertex whose marked ordinary parents
\((u+j,P+s,t-1)\), \(j\in\{-1,0,1\}^2\), \(s\) in the three replica boxes
\eqref{eq:c-boxes} acting on \((P_1,P_2)\), lie in two distinct boxes.
Seed \((z,t)\) when \(H(z,t)\) or \(H(z,t+1)\) occurs.
For an integer \(\ell\ge1\) let \(\mathcal D_\ell(z,t)\) mean that some
ordinary backward \(\ell\)-edge path from \((z,t)\) has a marked ordinary
parent at no fewer than \(\ell/1404\) of its first \(\ell\) vertices. Put
\[
 \Delta_{\rm raw}=13\,824,\quad B_{\rm raw}=2\Delta_{\rm raw}^2=382\,205\,952,
 \qquad p_{\rm dens}=B_{\rm raw}^{-67\,440}/64.
\]
For \(p\le p_{\rm dens}\),
\begin{equation}\label{v:density-bound}
 \Pr(\mathcal D_\ell(z,t))\le
 4\sum_{f\ge\lceil\ell/16\,848\rceil}
 [2B_{\rm raw}^{16\,860}\sqrt p]^f\le\frac{16}{3}e^{-\ell}.
\end{equation}
This holds simultaneously over all paths, including paths selected
after all innovations and initial words are known, for finite-start
marks without free initial seeds, and in injectively unwrapped torus
cones. Replacement values and initial words are unrestricted.
\end{lemma}
\par\noindent\emph{Proof idea.} Attach a complete three-charge contour to each selected marked parent and restart below its terminal time. The resulting graph controls every path in one count, with a separate payment for the two appearances of a physical source.

\begin{proof}
Use physical potentials
\[
 \psi_0=3(P_1+P_2)+350t,\quad
 \psi_1=-3P_1-175t,\quad \psi_2=-3P_2-175t.
\]
They sum to zero. Every distinct-box support pair offers each charge a
backward edge gaining at least 130; arbitrary loss is at most 350.
Every mark has a finite acyclic derivation ending at seeds.

Label frontier vertices by their descendant terminal sets, joining
vertices whose sets overlap. Expose the latest frontier vertex and its
two supports. An outside overlap component cannot be joined; the current
component splits into at most two daughters, each containing one of
these supports. Start three good arms at the root, one of each charge,
and retain one endpoint per charge per component. At a split add a
three-charge source at the exposed vertex, supplying missing endpoints;
only its first edges may be bad. Frontier vertices with no arm endpoint
must also be retained: they can carry the support through which a
split daughter is reached in one edge. Such exposures can create sources
in disconnected arm components; retain the component of the root source,
in which every source emits, and every terminal receives, exactly one
arm of each charge. If \(f'\) is its terminal count, each charge thus
bijects \(f'\) sources and \(f'\) terminals, with \(f'-1\) added sources.
Each bad first edge costs at most \(350+130=480\) relative to a good
edge, so balanced cancellation, with arm multiplicity, gives
\[
 130e_{\rm arm}\le480\cdot3(f'-1),\qquad e_{\rm arm}\le12(f'-1).
\]
A seeded root has one terminal and zero edges. Projection reduces size.
A minimum-time vertex is a terminal, and the root is the maximum, so the
whole connected contour has time span at most its edge budget.

Take the latest remaining marked-parent occurrence on a density path,
add its one-edge spur and its full contour, and skip occurrences whose
parent time is at or above the earliest terminal time of that contour.
Repeat, retaining the final contour whole even below the path endpoint.
The selected formal terminal slabs are disjoint. A contour with \(f_i\)
terminals covers at most \(12(f_i-1)+1\le12f_i\) positions, whence
\(f=\sum f_i\ge\lceil\ell/16\,848\rceil\). The entire path, all spurs
and all contours form a connected graph with at most
\(\ell+12f\le16\,860f\) edges. Its undirected degree is at most
\(2\cdot9\cdot768=13\,824\). Doubled-edge traversals and terminal subsets
have at most \(2B_{\rm raw}^m\) descriptions for \(m\) edges; summing
\(m\le16\,860f\) costs at most \(4B_{\rm raw}^{16\,860f}\). This counts
the whole path, not only the attached contours.

For \(f\) distinct formal terminals choose one of their two possible
physical names. There are at most \(2^f\) choices. A physical name seeds
at most two formal vertices, so each choice uses at least
\(\lceil f/2\rceil\) distinct independent names. Thus the terminal
probability is at most \(2^fp^{\lceil f/2\rceil}\le2^fp^{f/2}\).
Apply this once to the complete terminal union: adjacent formal slabs
can share an old/new name. This proves the series in
\eqref{v:density-bound}. At the stated rate bound its base is at most
\(B_{\rm raw}^{-16\,860}/4\); since
\(16\,860\log B_{\rm raw}/16\,848>1\), the series is at most
\(16e^{-\ell}/3\). Finite-start closure only deletes seeds; injective
cones preserve their law.
\end{proof}

The contour argument is that of \cite[Lemmas~12.3 and~12.4]{DMFixed2026}.
For each compiled source name, which has an old and a new appearance, this recovers equation (L.1) of \cite[Section~10.2]{DMNoise2026}. That equation was stated for single-vertex replacement marks. The bound also gives an explicit rate for the old/new density estimate of Lemma~\ref{c:coherence}. Its exact \(e^{-\ell}\) tail is stronger
than the downstream polynomial unions require. At the weaker cap
\(B_{\rm raw}^{-33\,720}/64\), the same count gives only
\((16/3)\exp[-(\log4)\ell/16\,848]\); the windows of Section~\ref{v:parameters} lie below the stronger cap.

The pairing argument of Lemma~\ref{c:coherence} (compare
\cite[Lemma~10.2]{DMNoise2026}) carries over to Moore inputs: two
unmarked nonbottom full-core readings at the same \((u,t)\) and
transverse distance \(d\) agree outside their two \(\mathcal D_\ell\)
events when \(t\ge\ell\ge\max(1,2d)\). Indeed at an unmarked root at
most one box has marked inputs, leaving at least \(512-256\) clean
successful replicas. Away from marked-parent positions a signed unit
shift gives \(3\cdot16\cdot15=720\) replica pairs. At most
\(2(768-512)=512\) fail, leaving 208 distance-reducing pairs.
Different outputs of the same deterministic Moore function force a
common differing input index. Follow that index at both roots. At an
exceptional step choose any clean successful replicas; distance rises
by at most 350. After the first step the readings are actual first
coordinates, so even time-zero distance-zero conflict is impossible.
With fewer than \(2\ell/1404\) exceptional paired steps the final
distance is less than \(d-\ell/2\), a contradiction. All-parent bottom
spreading ensures all followed inputs are nonbottom. This proves the
finite-age comparison for the core and the ordinary second proposal.

We will also use the physical source slabs of the proofs of
Lemma~\ref{c:coherence} and Propositions~\ref{c:late-live}
and~\ref{c:path-tail}. A complete witness reserves
all its old/new names; restart strictly below its earliest physical
name minus two. At a boundary formal time \(b\), reserve names through
\(b+1\) and their appearances in \(\{b,b+1\}\); future names begin at
\(b+2\). The last witness is kept whole. These conventions separate
physical supports, including names whose two formal appearances straddle
a cut. They are used both for deep historical queries and for rare
contours along quantum paths.

\subsection{Complete sources and a terminating parameter choice}\label{v:parameters}

Table~\ref{tab:concordance} gives corresponding notation. Each construction chooses its own constants. The folded proof takes a fourth root of the oriented source-law base to allow overlap between sheets.

\begin{table}[tbp]
\centering\small
\caption{Notation shared by the three constructions. The tail and fault-envelope formulas are specific to the construction.}\label{tab:concordance}
\begin{tabular}{@{}>{\raggedright\arraybackslash}p{.36\linewidth}>{\raggedright\arraybackslash}p{.17\linewidth}>{\raggedright\arraybackslash}p{.17\linewidth}>{\raggedright\arraybackslash}p{.17\linewidth}@{}}
\toprule Quantity & Three dimensions & Six dimensions & Four dimensions\\\midrule
Nominal window and endpoints of \(I_n\) & \(\mathcal J_n,(a_n,b_n)\) & \(\mathcal J_n,(a_n,b_n)\) & \(\mathcal J_n,(a_n,b_n)\)\\\addlinespace[3pt]
Sampling scale, lane count and bin height & \(C_h,B,W\) & \(C_h,B,W\) & \(C_h,B,W\)\\\addlinespace[3pt]
Internal faults, aggregate refresh and union probability & \(A,E,p\) & \(A,E,p\) & \(A,E,p\)\\\addlinespace[3pt]
Intrinsic tail and overlay envelope & \(\eta_B,\alpha_{\rm bin}\) & \(\eta_\pm,\alpha_{\rm bin},\alpha_{\rm hist}\) & \(\eta_\pm,\alpha_{\rm bin},\alpha_{\rm hist}\)\\\addlinespace[3pt]
Source base and historical packet bound & \(\gamma_{\rm src},\theta_{\rm pkt}\) & \(\gamma_{\rm src},\theta_{\rm pkt}\) & \(\gamma_{\rm src},\theta_{\rm pkt}\)\\
\bottomrule
\end{tabular}
\end{table}

We use \(C_h,B,W\) for the sampling-register scale, sampling-register count and bin height of
this construction; its windows and hardware parameters are chosen
anew. Fix the register roles, the two depth-\(D_{\rm q}\) queries,
Moore/replica offsets and quantum influence catalogue before choosing
\(C_h,B\) or the chart palette.
A clean erroneous prediction forces a wrong input; the witness records
its position and role, rather than its value.

\paragraph{Witness size before the sampling-register count.}
In characteristic coordinates \(\chi=P_{12}+58t(1,1)\), the three
potentials are \(3(\chi_1+\chi_2),-3\chi_1,-3\chi_2\). Correct gains
are 132, 129, 129, worst loss 351, and the total old/new oscillation of
one physical name is \(348+174+174=696\). The terminal-frontier
accounting of Lemma~\ref{q:frontier}, or
\cite[Lemmas~5.1 and~10.3]{DMNoise2026}, is applied before projection.
If the integer \(e\) counts support occurrences, \(m_{\rm ent}\) counts full deep
entries and \(g\) counts terminal groups, then
\begin{equation}\label{v:entry}
 D_{\rm q}m_{\rm ent}\le e,\qquad
 129e-1440(g-1)-696m_{\rm ent}\le\Omega_{\rm pre}g.
\end{equation}
Each full entry owns all \(D_{\rm q}\) virtual levels on its copied
charge arm before the next word. Shared suffixes are counted on
each arm before deduplication; a dirty lookup terminates at a hardware
source and has no entry gain. Unary corridors cannot split the terminal
frontier. Thus \(129-696/4096>128\), and
\(e\le(1440+\Omega_{\rm pre})g/128\).

Here \(\Omega_{\rm pre}\) bounds the total three-potential oscillation of
every source, role and entry star; it is a finite maximum over these offsets. It ignores alphabet size and sampling-register index. A short,
recent or partial sampling-register corridor preserves the first six characteristic
coordinates and changes the integer bin by a bounded amount. Retain its
two distinct endpoints and a typed edge for that increment; do not
identify bins. Entries and corridors add at most two links per support,
plus fixed terminal stars of maximum length \(L_{\rm star}\). We may take
\[
 C_{\rm ctl}=\left\lceil3(1440+\Omega_{\rm pre})/128+3+L_{\rm star}\right\rceil.
\]
This bounds a complete controller witness by \(C_{\rm ctl}g\) edges.
A receiver-specific failure is included in the aggregate hardware event;
an incorrect censor without a hardware fault either is the intrinsic statistical event \(S_\lambda\) of \eqref{v:hubs} at its endpoint source label or traces back to one of the two shared base queries. Hence
choosing a sampling lane adds no \(B\)-way witness edge. The physical
receiver range can depend on \(B\) and is paid in its hardware support.

For definiteness, finite roles and degrees can be bounded by
\[
 R_{\rm pre}=2^{12}(D_{\rm q}+3),\qquad
 \Delta_{\rm pre}=R_{\rm pre}^2
 [64(2D_{\rm q}+5)(9\cdot768+16)^2+256],
\]
enlarged once for the fixed quantum ports and connectors. Ordinary
supports have \(9\cdot768\) offsets; forks use differences of two;
query kinds are two; virtual-time differences number at most
\(2D_{\rm q}+5\). This explains finiteness independently of the palette.
The argument uses finite maxima and their order of choice, without evaluating numerical minima. Let \(\Delta_{\rm all}\) bound the degree and \(C_{\rm all}\) the edge count per distinct source in the full typed graph, including the five-charge group expansion proved below. They precede \(B\).

A harmful nonbottom profile query has at least sixteen distinct raw
names. A nonbottom \(F_{\rm cap}\) preserves its centre profile, so a
wrong nonbottom virtual output has at least 512 wrong-profile centre
replicas, spanning two boxes. Follow this through all 4096 levels and
attach complete raw contours at its physical leaves. Full-core coherence
makes each wrong nonbottom leaf marked. The grouped budget gives
\[
 129e\le1440(f-1)+696f+696,\qquad e<17f+6.
\]
The last 696 pays the predecessor entry. With \(e\ge4096\), the
first inequality gives \(f\ge248\), which implies the stated lower
bound of sixteen. A connected complete deep witness has
at most \(C_{\rm deep}f\) edges and physical span;
\(C_{\rm deep}=256\) suffices. A faulty local evaluation is a hardware exception.

\paragraph{Intrinsic comparisons.}
Let \(\rho_{\rm code}=126/128\). For
\(\mu\le C_h/100\), \(C_h\le1/100\), \(W=\lfloor C_h/\mu\rfloor\) and
\(B\ge20/C_h\), we have
\[
 \mu W\ge0.99C_h,\qquad s_{\rm vote}/B\le0.8C_h,\qquad
 \rho_{\rm code}^2[1-(1-\mu)^W]>0.95C_h.
\]
The last inequality follows from
\(1-e^{-\mu W}\ge0.985C_h\) and
\((63/64)^2\cdot0.985>0.95\).
A correct short flag-one test has per-lane upper mean less than
\(0.53C_h\cdot4^{-r}\) at depth \(a-1-r\), against threshold \(0.75C_h\).
The Chernoff comparison of \cite[Lemmas~6.2 and~10.4]{DMNoise2026} gives
\[
 u_B=2^{-\lfloor3BC_h/100\rfloor},\quad
 z_B=2^{-\lfloor BC_h/3\rfloor},\quad
 w_B=2^{-\lfloor BC_h^2/50\rfloor},\quad
 b_B=2^{-\lfloor B(1/2-2C_h)^2\rfloor}.
\]
Availability masking cannot raise the upper count. A long flag-zero
test with a full recent window has last-refresh high-and-allowed
probability \(\rho_{\rm code}^2[1-(1-\mu)^W]\), against threshold at
most \(0.8C_h\); Hoeffding gives \(w_B\). Its two code fields are
independent conditional on the one shared refresh, not unconditionally.
Before \(W\) usable updates after an initial or boundary unavailable
reset, allowed requires a refresh with probability at most \(2C_h\).
Enabling then requires \(B/2\) independent lanes and costs \(b_B\).
The short tail at rank \(r\) is at most \(u_Bz_B^r\), there are at
most \(3\cdot4^r\) short ends per bin, and at most two long ends.
Consequently define
\begin{equation}\label{v:etas}
\begin{aligned}
 \eta_+&=\frac{3u_B}{1-4z_B}+2w_B+2b_B,\\
 \eta_-&=4\left[u_B+w_B+
 2^{-\lfloor B(\rho_{\rm code}-1/2)^2\rfloor}\right].
\end{aligned}
\end{equation}
The second bound additionally pays failure to have \(B/2\) allowed
lanes at a complete wrong critical ride. Its availability mean is at
least \(\rho_{\rm code}\).

\paragraph{Historical entropy before the sampling-register count.}
As in the historical count of \cite[Lemma~10.4]{DMNoise2026}, one deep
anchor of \(f_i\ge16\) names has at most \(9W\) lags,
\(A_{\rm gap}f_iW\) restart gaps and
\((176A_{\rm gap}f_iW)^2\) net replica endpoints, where
\(A_{\rm gap}=8+C_{\rm deep}+4\). Thus its choices cost at most
\(A_{\rm anc}f_i^3W^4\le A_{\rm anc}2^{3f_i}W^6\), with
\(A_{\rm anc}=9A_{\rm gap}(176A_{\rm gap})^2\).
We enumerate net endpoints, not all replica walks across a long gap.
Take
\[
\begin{split}
 R_{\rm hist}&=2^{10}(D_{\rm q}+3),\\
 \Delta_{\rm hist}&=R_{\rm hist}^2[16(9\cdot768+16)^2+128],
 \quad X_{\rm hist}=2\Delta_{\rm hist}^2,\\
 K_{\rm anc}&=2^{12}(R_{\rm hist}+1)(A_{\rm anc}+1)
 X_{\rm hist}^{C_{\rm deep}+4},\\
 K_{\rm hist}&=2^{20}(768+1)K_{\rm anc}.
\end{split}
\]
Doubled traversals count connected graphs by \(2X_{\rm hist}^e\).
Compositions of anchor sizes, interleavings with \(m_{\rm pkt}\) packet
steps and finite roles are absorbed in these constants. Here \(r_{\rm anc}\) counts retained deep anchors. For total
\(f\) raw names and \(m_{\rm pkt}\) packets the unconditional count is
\begin{equation}\label{v:history-count}
 K_{\rm hist}^{f+m_{\rm pkt}}W^{6r_{\rm anc}}
 \le K_{\rm hist}^{f+m_{\rm pkt}}W^{6f/16}.
\end{equation}
A fixed input-dependent motif/root list contributes only a prefactor.
The constants do not enumerate chart letters or counter states.

\paragraph{Hardware envelopes and the joint source law.}
A single critical test has at most \(h_{\rm test}(B)W\) macro names,
where it suffices to take
\[
 n_{\rm query}=(2D_{\rm q}+3)^2(176D_{\rm q}+3)^2,\qquad
 h_{\rm test}(B)=256(2n_{\rm query}+16B+18\cdot768+64).
\]
There are two shared base-query cones and \(B\) receiver destinations,
rather than \(B\) separate cones. An atom has
\(x=g_x-r+\delta_x\), \(t=b-r+\delta_t\), with
\(\delta_x\in[-D_{\rm q}-1,D_{\rm q}]\),
\(\delta_t\in[-1,1]\). Shared atoms at sample times \(b,b'\) force
\[
 b-b'=(\delta_x-\delta_t)-(\delta'_x-\delta'_t).
\]
Thus \(D_{\rm ov}=4D_{\rm q}+16\) separates complete historical
supports, independently of \(B,W\) and the replica path. Intrinsic
supports at the same logical endpoint and different times are disjoint:
their chart coordinate fixes the lag, then their time fixes the sample.
Their exact intrinsic divisor is one.

Let \(h_{\rm bin}W\) bound the complete hardware envelope of all
potential tests and uses of a characteristic bin, including apparent
ends, inactive uses and current hard-query cuts; let
\(h_{\rm hist}\ge h_{\rm test}\) bound historical tests. Sliding the
endpoints across a bin unions fixed characteristic translates of
intervals of length \(O(W)\). It does not give \(W\) unrelated supports
of size \(W\). Their translate differences and neighbouring-bin
differences form a finite list. Let \(D_{\rm bin}\) be one plus its
maximum overlap degree; it can depend on \(B\), but not \(W,n,J\).
For complete-envelope fault probabilities \(\alpha_{\rm bin}\) and
\(\alpha_{\rm hist}\), the distinct-label source and packet bounds are
\begin{equation}\label{v:joint}
\begin{aligned}
 \gamma_{\rm src}&=\sqrt{2C_h}+\eta_+^{1/6}
                   +\alpha_{\rm bin}^{1/(2D_{\rm bin})},\\
 \theta_{\rm pkt}&=\eta_-+\alpha_{\rm hist}^{1/D_{\rm ov}}.
\end{aligned}
\end{equation}
To see the first formula, adapt \cite[Lemma~6.3]{DMNoise2026}: put one union event
\(X_\lambda=R_\lambda\lor S_\lambda\lor O_\lambda\) at each physical source label
\begin{equation}\label{v:hubs}
 \lambda=(u_x-t,u_y,P_1+58t,P_2+58t,P_3,P_4,\lfloor t/W\rfloor).
\end{equation}
The first six coordinates are modulo \(L\), and the bin remains an
integer. Raw bins have probability at most \(pW\le2C_h\) and use
disjoint names. Statistical bins use only refresh flags and independent
product marks; let \(g_S\) and \(g_O\) count statistical and hardware source labels, respectively. Ownership and three bin residues give
\(\eta_+^{g_S/3}\). Hardware bins use only internal flags; colouring
 gives \(\alpha_{\rm bin}^{g_O/D_{\rm bin}}\). These last two families
are independent. Raw bins can overlap both: Cauchy--Schwarz and a sum
over three types at each distinct source label give
\(\Pr(\text{all }g\text{ source labels bad})\le\gamma_{\rm src}^g\).
Types, old/new appearances and virtual roles annotate this one identity.
Direct quantum hardware belongs to its raw bin as well. The same
intrinsic/internal independence gives the packet bound
\(\theta_{\rm pkt}^{m_{\rm pkt}}\). The source bound holds for every
\(\eps\in I_n^{\mathsf r}\), every set of \(g\) distinct physical source
labels \eqref{v:hubs} and every \(L\) above the ownership threshold of
this subsection (or in an injectively unwrapped cone); the packet bound
\(\theta_{\rm pkt}^{m_{\rm pkt}}\) holds under the same conditions.

For ownership on the torus put one square of side
\(N_{\rm cap}=\lfloor L/4\rfloor\) in \(u\), with vacuum outside.
Choose the effective size threshold above
\(64[B+58W_n^*+R_{\rm fixed}+D_{\rm q}+10]\), where
\(W_n^*=4^{n+n_0+1}-1\) bounds \(W\) on the \(n\)-th window and
\(R_{\rm fixed}\) bounds all fixed driver, compiler and sampling-register halos,
enlarging for exact hulls. The active support hull has \(u\)-diameter below \(L/2\);
the blank gap separates its periodic copies. A shared atom therefore
lifts into one cap copy, where \cite[Lemmas~6.1 and~10.4]{DMNoise2026}
give phase-zero ruler/lane ownership. Adjusting a transverse deck
translation leaves the cloned chart phase unchanged. Supports in the
same bin residue are disjoint unless they are the same physical label.
This proves the actual-torus law used in \eqref{v:joint}.

\paragraph{Order of choices and an explicit rate bound.}
The index \(a=n+n_0+1\) in the window construction is separate from
any locally bound bit in a gadget calculation.
Set
\(\gamma_*=[1/(8(\Delta_{\rm all}+1))]^{3C_{\rm all}}\).
Choose rational \(0<C_h\le0.01\) with \(\sqrt{2C_h}<\gamma_*/3\).
Choose \(B=4^{k_B}\ge20/C_h\) with \(4z_B<1\),
\(\eta_+^{1/6}<\gamma_*/3\) and
\(K_{\rm hist}(2\eta_-)^{1/4}<1/128\).
The search terminates because all these tails decay exponentially
in \(B\) and \(K_{\rm hist}\) was fixed earlier. Choose positive rational
overlay targets with
\[
 \alpha_{\rm bin}^{1/(2D_{\rm bin})}<\gamma_*/3,\qquad
 K_{\rm hist}(\eta_-+\alpha_{\rm hist}^{1/D_{\rm ov}})^{1/4}<1/64.
\]
Let \(K_{\rm rare},B_{\rm rare}\) be the finite connected rare-contour
and actual-quantum-path maxima used in the proof below, fixed before
the cap, and put \mbox{\(N_{\rm seed}=\tbinom{2\cdot352^2}{256}\)}.
The rate requirements are: select rational
\(\mu_*>0\) with
\begin{equation}\label{v:cap-list}
\begin{gathered}
 \mu_*\le C_h/100,\qquad \mu_*\le1/100,\qquad \mu_*\le p_{\rm dens}/2,\\
 K_{\rm hist}(2C_h)^{3/8}(2\mu_*)^{3/8}<1/64,\\
 K_{\rm rare}[N_{\rm seed}(2\mu_*)^{256}]^{1/16}<1/64,\\
 K_{\rm rare}(2\mu_*)^{1/2}<1/64.
\end{gathered}
\end{equation}
The first three control sampling-register margins, the uniform small-rate regime
and density; the remaining three control historical raw anchors, rare
births and direct exceptions. Positive exponents make the search
terminate. All finite supports have been included in the complete
bin/history or rare/direct envelopes above.

Choose \(n_0\ge k_B+2\) with
\(21C_h/[10\ell_{n_0+1}]<\mu_*\), where \(\ell_a=2\cdot4^a-1\).
Then build the static palette and scan machine, and finally the actual
Boolean hardware. Section~\ref{sec:compilation} implements its truth
table by streamed compute/copy/uncompute, one old-word dephasing and
private transport, and local recomputation. Added scratch is classical,
reset before use and dephased. It never receives a quantum-derived
target. Each translated layer is factor-disjoint even on small tori;
quantum dispatch and all seven-gate decompositions are unchanged.

For the whole-site circuit \(\mathrm{WS}\), count every internal noisy layer in \(J\) and choose \(P_{\rm pad}\ge2J\)
with \(2h_{\rm bin}JC_h/P_{\rm pad}\le\alpha_{\rm bin}\) and
\(2h_{\rm hist}JC_h/P_{\rm pad}\le\alpha_{\rm hist}\). Padding gives
\mbox{\(\mu_{\mathrm{WS}}(\eps)=1-(1-\eps)^{P_{\rm pad}}\)},
\(\nu\le2J\mu/P_{\rm pad}\) and \(p\le2\mu\). On a padding replacement the
last word is a uniform product; the two sampling-register fields have their stated
conditional code law.

For the independent single-qubit circuit \(\mathrm Q\), let \(N_{\rm np}\) count nonprobe wires, including mixed ancillas,
work and two reusable OR scratch wires. Choose rational
\(0<\delta_{\rm ov}\le1\) with
\(h_{\rm bin}C_h\delta_{\rm ov}\le\alpha_{\rm bin}\) and
\(h_{\rm hist}C_h\delta_{\rm ov}\le\alpha_{\rm hist}\), then
\(M\ge\max(N_{\rm np},32N_{\rm np}/\delta_{\rm ov})\).
Count nonsampling depth \(J\) after compiling the serial probe OR,
refresh and disposal, and choose \(K_{\rm samp}\ge128J/\delta_{\rm ov}\).
The noise immediately following probe reset is sampling layer one:
its probe flags belong only to the endpoint experiment \(E\); its
nonprobe flags belong to \(A\). All nonsampling flags belong to \(A\).
These disjoint primitive coins give two independent fields \(A,E\), each independent and identically distributed over site-rounds, with
\begin{equation}\label{v:q-rates}
\begin{aligned}
 \mu_{\mathrm Q}(\eps)&=1-\left[\frac{1+(1-\eps)^{K_{\rm samp}}}{2}\right]^M,\\
 \nu&\le[N_{\rm np}K_{\rm samp}+(M+N_{\rm np})J]\eps,\\
 \frac\nu\mu&\le\frac{8N_{\rm np}}M+
 8\left(1+\frac{N_{\rm np}}M\right)\frac{J}{K_{\rm samp}}
 \le3\delta_{\rm ov}/8.
\end{aligned}
\end{equation}
For \(MK_{\rm samp}\eps\le1/2\),
\(MK_{\rm samp}\eps/8\le\mu\le MK_{\rm samp}\eps/2\), and
\(\mu_{\mathrm Q}(1/(2MK_{\rm samp}))\ge1/16\).
The OR uses \(a\lor q=a\oplus q\oplus aq\); reset--Hadamard--dephase
prepares a fresh mixed ancilla. On \(E\cap\{A=0\}\), controlled swaps
refresh every semantic factor to a product mixed state. Elsewhere
\(A\) allows arbitrary output. Independent counterfactual marks agree
with clean refreshes; every disagreement belongs to the overlay. A Q
fault inside the final cut is a direct exception, rather than an assumed
whole-site replacement. This proves the same semantic source bounds
for the separately compiled Q circuit.

For \(a=n+n_0+1\), use nominal intervals
\[
 \mathcal J_n=\left(\frac{2C_h}{\ell_a},\frac{21C_h}{10\ell_a}\right).
\]
They are disjoint because \(\ell_{a+1}=4\ell_a+3\) and
\(21\ell_a<20(4\ell_a+3)\). They give
\(\ell_{a-1}<W\), \(2W\le\ell_a\), \(W\le4^a\,-\,1\).
By \cite[Lemmas~4.1 and~4.2]{DMNoise2026} the scan finds its first
zero at \(j_*=2^{a+1}\,-\,2\), makes \(a+1\)
tokens, skips \(n_0+1\), and calls \(V(n+1)\), where
\(V(m)=U(m-1)\). Thus it binds exactly to \(U(n)\).
Each refresh polynomial is rational and strictly increasing from zero.
Enumerate rational inner endpoints \(a_n<b_n\) until
\(\inf\mathcal J_n<\mu(a_n)<\mu(b_n)<\sup\mathcal J_n\), restricting
Q endpoints below \(1/(2MK_{\rm samp})\). Density and monotonicity prove
termination. Their closures stay inside the valid regime; rates and
\(W_n^*\) are uniform on each \(I_n=(a_n,b_n)\).
Choosing \(n_0\) before building the palette puts every window below
\(\mu_*\).

\subsection{A total static encoder and exponential storage}\label{v:positive}
\begin{lemma}[Finite static initialization]\label{v:encoder}
For every \((n,L)\) a total rate-independent algorithm specifies a cap
state and a closed-code encoder. On a nonhalting input and above a total computable size cutoff,
the cap state is locally valid, carries the canonical projected geometry
and the correct cutoff flag at every active site. For such an input above that size cutoff, the core trajectory under \(F_{\rm cap}\) keeps the storage-enable bit at one at every active site through
\(2^{\lfloor L/4\rfloor}-1\) rounds.
\end{lemma}
\par\noindent\emph{Proof idea.} Search the finite square and its collar for a legal assignment with the canonical projection. Initialize the counter and the closed toric code from the chosen patch.

\begin{proof}
Proposition~\ref{v:static-import} supplies the computable projection, finite ownership and legal nonhalting extension.

As in the proof of Lemma~\ref{c:initialization}, the search is
exhaustive: on the finite cap and its fixed collar enumerate all static alphabet
assignments. Check the five cap constraints, every fully active base
constraint, absence of a halt letter, the square/diagonal shape, and
agreement at every active site with the computed canonical projection,
finite owners, routing tags, lane phase and cutoff
\(b_r=\one\{r<2a-2\}\), \(a=n+n_0+1\). This includes flags on
truncated rulers. Hidden cover, computation and channel letters are
enumerated, rather than computed by an infinite-lift algorithm. The
finite list has a computable cardinality; return its first match or a
fixed vacuum fallback. Each test is decidable, so this terminates
without waiting for the machine to halt.

On nonhalting, the legal static extension in Proposition~\ref{v:static-import} restricts to a patch in the finite search list.

An accepted hidden patch need not extend globally. Local static validity
makes \(F_{\rm cap}\) preserve the profile without violations, while
its canonical geometry supplies the ownership needed for the sampling-register
estimates. The all-site cutoff test prevents a truncated long ruler
from carrying a false flag-one short experiment. Vacuum suspends
cross-boundary tests. Set death and counter bits to zero, with just the
least diagonal carry one; Lemma~\ref{v:cap} gives the storage-enable horizon.
In the positive-branch reference trajectory the core follows
\(F_{\rm cap}\) with constant profile, zero death bits and this counter,
the queries return the reference profile, and every censor test passes.
At the canonical geometry with correct flags, a censor bottom with
correct drivers on its recent ride and no hardware fault is exactly
the intrinsic event \(S_\lambda\).
Initialize each sampling register by its own shifted base flag, low for one and
high for zero, unknown for inactive base, and all availability
unavailable, including fields stored outside the cap. Initial or partial
histories use genuine positive-time refreshes only. The blank-gap size
bound in Section~\ref{v:parameters} makes ownership valid after projection.

At a fixed active interior \(u_0\), binary symplectic elimination
encodes the unknown qubit into the logical pair whose logical \(Z\)
membrane spans \(P_1,P_2\) at fixed \(P_3,P_4\), fixing the other five
pairs. This is a total finite algorithm. Small or degenerate sizes have
prescribed default maps. No rate or halting answer enters it.
\end{proof}

A dangerous certificate is a connected graph in the typed catalogue
of source label coordinates, with \(g\) distinct marked lifted source labels whose
projected source labels are bad. If \(m\) is its edge count, it satisfies
\(b_{\rm span}L\le m\le C_{\rm all}g\) for constants \(b_{\rm span}>0\) and \(C_{\rm all}\) depending only on the fixed catalogue, as constructed in the proof of Lemma~\ref{v:bridge}. This event concerns the source
marks alone, before any Pauli values are chosen. Below, full coordinates
mean the characteristic coordinates and the integer bin, with the
catalogue role, as in \eqref{v:hubs}.

\begin{lemma}[Exponential bound for the sources]\label{v:bridge}
Fix \(\mathsf r\). There are computable \(A_{\rm e},\kappa_{\rm e}>0\),
depending on \(\mathsf r\) but not on \(n\), and a total computable size
cutoff in \(n\), such that for every \(n\) with \(U(n)\uparrow\), every
\(\eps\in I_n^{\mathsf r}\), every \(L\) above the cutoff and every
integer \(0\le T<2^{N_{\rm cap}}-1\), \(N_{\rm cap}=\lfloor L/4\rfloor\),
the circuit started from the state and encoder of
Lemma~\ref{v:encoder} satisfies
\begin{equation}\label{v:exp-bound}
 \Pr(\text{a dangerous certificate by }T)
 \le A_{\rm e}(T+A_{\rm e})L^6 2^{-\kappa_{\rm e}L}.
\end{equation}
Its complement admits one intended-code recovery for every allowed
Pauli or local-map assignment, including arbitrary reference
coefficients, at every requested integer time through \(T\).
\end{lemma}
\par\noindent\emph{Proof idea.} Map controller witnesses into quantum space-time and group nearby terminals. A large syndrome component then forces a connected graph with many marked sources. A short piece of that graph avoids aliases on the physical torus.

\begin{proof}
\emph{Images.} We extend the deterministic closed-code argument in
Lemma~\ref{x:exponential}; its iid probability conclusion is replaced
by \eqref{v:joint}. Keep the characteristic/bin coordinates
\eqref{v:hubs}. A source label has a unique quantum space-time image at \(u_0\):
\[
 t_{\rm img}=u_{0x}-(u_x-t),\qquad
 P^*_{12}=\chi_{12}-58t_{\rm img}(1,1),\qquad P^*_{34}=P_{34}.
\]
This image is Lipschitz on the full bounded-edge graph; it recovers a
physical quantum vertex at \(u_0\). The base coordinate \(u_y\) and the
integer bin are retained even though they are not used in this image.
Every quantum time or spatial connector also has bounded increments
in the full graph. A short sampling-register corridor retains both bin endpoints.

\emph{Ball packing.} At every erroneous controller cut attach its complete
\(C_{\rm ctl}g_i\)-edge witness. Enclose its quantum image, all source
images and its fixed influence spur in a closed space-time ball of radius
\(1\le r_i\le C_{\rm ball}g_i\). Treat a direct hardware source as a
single-source witness and enclose its constant-radius star in the same way.
Select from both kinds of balls together, greedily choosing disjoint
balls in decreasing radius. Triple enlargements cover all original
terminal appearances. Selected witnesses have disjoint lifted source sets:
a reused source label has one image, which would lie in both original balls.
Assign each covered terminal to one selected enlarged ball and replace
all terminals assigned there by a single defective group sink. Give it
a fixed group role distinct from every ordinary quantum/history role.
Selected centres are distinct, and group sinks have no predecessors,
so the quotient is acyclic.

\emph{Contour bound.} The unchanged processor gives five predecessor alternatives with
balanced potentials
\[
 F_j=-(P_j+58t)\quad(1\le j\le4),\qquad
 F_5=\sum_{j=1}^4(P_j+58t).
\]
Their closed-square gains are 58 on each negative potential and 168 on the
sum potential after 400 reflected sweeps. The stage spurs of
Lemma~\ref{q:round-shadow}, constructed in the proof of
Lemma~\ref{q:universal}, have bounded losses. These are the specialized gains from the proof of
Lemma~\ref{x:exponential}, rather than the folded numerical gains.
An occupied nonfault root has an occupied admissible predecessor for
every charge. Complementing gives the good recurrence: OR over charges
of AND over their admissible parents. Apply
Lemma~\ref{q:frontier} to the finite acyclic typed graph. Its
group-dependent bound accounts for the different radii of the sinks.
The contour construction uses positions, roles and which sources are marked. The choices are independent of Pauli values, and the graphs are counted before applying the source probabilities.

Let \(m_{\rm grp}\) groups be used and let \(E_{\rm Q}\) be the
charged contour edge count. A group receives at most one terminal arm
of each charge. Assign it its centre's potentials. Terminal discrepancies
are bounded by a constant times its space-time radius, including time.
The terminal-link losses are absorbed into the allowance \(K_{\rm Q}r_i\) per charge.
With fixed margin \(\delta_{\rm Q}>0\), wrong-charge loss
\(D_{\rm loss}\), and terminal coefficient \(K_{\rm Q}\), cancellation
gives
\[
 \delta_{\rm Q}[E_{\rm Q}-5(m_{\rm grp}-1)-5m_{\rm grp}]
 \le5D_{\rm loss}(m_{\rm grp}-1)+5K_{\rm Q}\sum_i r_i.
\]
Thus the contour costs \(O(\sum_i g_i)\) edges. Expand its group jumps
by quantum connectors to the selected centres and attach their full
controller witnesses. In the isolated-terminal case retain the original
root and its connector too. The connected full graph contains that
root and distinct lifted source anchors with at most
\(C_{\rm all}g\) edges. This defines the finite constant used earlier.
It depends only on the catalogue fixed before the sampling-register count and the controller edge
bound. A group has no unbounded identifier or source-type multiplicity.

\emph{Connecting chain.} Use this construction along the first large intermediate syndrome
component, in either sector, keeping the stored mask from the exact
400-step recurrence fixed. Map its vertices to shadow roots by
Lemma~\ref{q:round-shadow}, as in the proof of
Lemma~\ref{q:universal}. Enclose each shadow root's full graph, including every source label image,
in a space-time ball in the image coordinates. Greedy disjoint packing
and \(4+2R_{\rm blk}+1\)-fold covering, as in the proof of Lemma~\ref{x:exponential},
give a connecting chain. Disjoint balls have disjoint lifted source label sets. Retain
links to \emph{both} original extreme roots, including when one ball
covers the whole path. The resulting graph has
\[
 m\le C_{\rm all}g,\qquad m\ge b_{\rm span}L.
\]
The extremes are at a common stage/time. Their macroscopic separation
in a quantum coordinate persists in the corresponding characteristic,
so bounded full-coordinate edges require at least \(b_{\rm span}L\)
edges. Extra controller excursions do not remove this separation.

\emph{A short alias-free piece.} Choose a fixed rational \(f_{\rm geom}>0\), and put
\(q_{\rm short}=\lfloor f_{\rm geom}L\rfloor\), so
\(q_{\rm short}\le m\) and the maximum full-coordinate edge range
times \(q_{\rm short}\) is below \(L/2\). Double a spanning tree,
mark the first visit to each source, and divide the at-most-\(2m\)-edge
walk into half-open pieces of at most \(q_{\rm short}\) edges, assigning
the last endpoint to the last piece. There are at most
\(3m/q_{\rm short}\) pieces, so one contains at least
\(q_{\rm short}/(3C_{\rm all})\) distinct lifted anchors. Marks need
not be leaves.

Two distinct lifts of one physical source label have the same integer bin and
differ by a nonzero \(L\)-deck vector in one of the six characteristic
coordinates. They cannot both lie in this short piece. Only now are its
marks distinct source identities, allowing \(\gamma_{\rm src}^g\).
Quantum spatial projection alone would lose a \(u_y\) deck shift;
a \(u_x\) shift also changes virtual image time and can be compensated
spatially by transverse shifts. Keeping all six coordinates and the bin
prevents both aliases. The separate \(O(W)\)-support ownership argument
is what makes their source law valid.

\emph{Counting.} All source bins in explanations through physical horizon \(T\) lie in
\[
 0\le j\le\lfloor(T+C_{\rm stage})/W\rfloor+C_{\rm bin}
\]
for fixed stage constants. Driver recursion decreases physical time or
finite virtual depth, the initial reference is correct, and unavailable
initial sampling registers use partial-window experiments. There are no prehistory
seeds. Connectors stay between endpoint bins. This causal assertion is
separate from the edge/source bound; the virtual image time does not
replace actual bin time. Hence there are at most
\(A_{\rm st}(T+A_{\rm st})L^6\) starts. Padded short walks and source subsets cost
\((\Delta_{\rm all}+1)^{q_{\rm short}}\cdot2^{q_{\rm short}+1}\).
Apply the joint law only to admissible distinct-name annotations. Their
probability sum is at most
\[
 2A_{\rm st}(T+A_{\rm st})L^6
 [2(\Delta_{\rm all}+1)\gamma_{\rm src}^{1/(3C_{\rm all})}]^{q_{\rm short}}.
\]
By the earlier \(\gamma_*\) choice the bracket is at most \(1/4\),
which gives \eqref{v:exp-bound} with \(A_{\rm e}=8A_{\rm st}\) and
\(\kappa_{\rm e}=2f_{\rm geom}\).

Finally Lemmas~\ref{q:homology}--\ref{q:span} apply to the universal
no-certificate event, including every serialized intermediate component.
The Clifford processor, dephased control and private mixed ancillas admit
the same unconditional Pauli/replacement instrument as before. Its one
component-filling decoder corrects the whole permitted operator span,
with reference coefficients.
\end{proof}

\begin{proof}[Proof of Theorem~\ref{v:detailed}(i)]

Choose rational
\(0<c_{\mathrm e}<\min(1/10,\kappa_{\rm e}^{\mathrm{WS}}/4,\kappa_{\rm e}^{\mathrm Q}/4)\), taking the smaller
choice for the two compiled circuits. With \(T=2^{\lfloor c_{\mathrm e}L\rfloor}\),
an effective enlargement of \(L_{\mathrm e}^{\mathsf r}(n)\) makes
\eqref{v:exp-bound} at most \(1/16\), and absorbs all code, support and
ownership thresholds. Since \(N_{\rm cap}\ge L/5\) eventually, this
horizon lies strictly below \(2^{N_{\rm cap}}-1\). Lemma~\ref{v:encoder}
and the single common decoder prove part (i) of
Theorem~\ref{v:detailed}. All maps have finite fallback definitions on
the remaining inputs and sizes.
\end{proof}

\subsection{Halting, arbitrary initial profiles and the entire algebra}\label{v:forgetting}
The constants \(a_n,a'_n\) below are decay rates local to this
subsection; window endpoints remain those of Section~\ref{v:parameters}.
\begin{lemma}[Storage-enable and ancestor tails from all starts]\label{v:ancestor}
On a halting window there are finite positive \(A_n,a_n,A'_n,a'_n\) and
fixed integers \(b_{\rm c},b'_{\rm c}\), uniform across that window and
independent of \(L\), such that the following hold on every torus. For
every \((z,t)\) whose depth-\(t\) determining cone unwraps injectively,
the probability that some initial metadata word gives an unmarked
first or ordinary reading with the storage-enable bit set at \((z,t)\) is at most
\(A_n(1+t)^{b_{\rm c}}e^{-a_nt}\). For a specified physical initial site
whose depth-\(s\) determining cone, of radius \(O(s)\), unwraps
injectively, the probability of a surviving \(s\)-round quantum dependency
path is at most \(A'_n(1+s)^{b'_{\rm c}}e^{-a'_ns}\), simultaneously over
all initial words. These are the
capped-controller analogues of Propositions~\ref{c:late-live}
and~\ref{c:path-tail}.
\end{lemma}
\par\noindent\emph{Proof idea.} In a halting chart, a valid large patch must expose a wrong test and a small patch expires. Bound the exceptional events that let storage remain enabled, then count surviving paths across the resets.

\begin{proof}
First obtain a static alternative. By
Proposition~\ref{v:static-import}, on halting every
sufficiently large compatible chart patch contains a fully guarded wrong
critical ride: a short depth-\((a-1)\) end flagged zero or a long
 depth-\(a\) end flagged one. Choose a finite guard margin containing
both rides and all fixed tests. In a valid infinite component, or a
square larger than this margin, clamp a box centre away from its
boundary to find such a motif within bounded original-active path
distance of any root. A smaller valid square has a bounded counter.

This extends to malformed profiles by rooted finite-alphabet
compactness. If rooted active survivors reaching every box boundary
could avoid both local violations and guarded motifs, a subsequence
would give an infinite root component avoiding both. Delete other
components. The three-corners and fully active base tests involve one
component; original active neighbours already lie in it; diagonal
continuation uses an intervening active neighbour. Deletion therefore
creates no violation involving the retained component. Its valid
infinite rectangle contains an interior guarded box, a contradiction.
Thus each arbitrary active root is within an \(n\)-dependent bounded
active path of a local violation or a wholly interior wrong motif, or
belongs to a valid uniformly bounded square. This includes originless,
wrapped and counterfeit profiles. Finite survivor searches terminate
on the halting promise; their bounds are not total computable functions
of arbitrary \(n\).

Exclude the density events of Lemma~\ref{v:density} at polynomially
many deterministic roots in a late cone. Its finite-age comparison
then identifies repeated full-core readings at the same logical site.
From an unmarked root with the storage-enable bit set choose clean successful replicas.
All their Moore inputs are nonbottom. The explicit death rule gives
ancestry of the storage-enable bit along any specified original-active nearest-neighbour
path, as well as the centre path. This supplies a single coherent
immutable profile patch and all counter inputs after an \(n\)-dependent
burn-in. A nearby violation would have seeded death; a bounded valid
cap would have overflowed from its arbitrary phase by Lemma~\ref{v:cap}
and spread death across the component. Either would have set the death flag and cleared the storage-enable bit, a contradiction.
Centre monotonicity alone would not justify travelling to these
obstructions; the active-neighbour terms in the death rule do.

In the remaining case reach a fixed wrong motif endpoint and follow its
history of the centre's storage-enable bit for \(\lfloor t/K_{\rm age}\rfloor\) samples,
with a fixed sufficiently large architecture constant \(K_{\rm age}\).
The cone age dominates twice its replica diameter; its exceptional
probability is a polynomial times an exponential. At a true interior
origin the sampling-register state becomes low/allowed, flushing arbitrary initial
sampling registers. Under correct nonbottom instructions, copying and shared
refreshes preserve the ideal experiment. Historical bottom writes
unknown/allowed: it raises enabling, decreases singleton-high and
increases upper possibility. It therefore cannot conceal either wrong
test. A surviving sample must pay an intrinsic failure, a hardware
packet, or a harmful nonbottom deep profile query somewhere on its
\(O(W)\) ride. A false vacuum reading on an interior ride is such a
harmful query; arbitrary historical bottom is handled deterministically.

Each harmful query has a complete at-least-sixteen-name witness.
Pack whole witnesses in the strict physical slabs described after
Lemma~\ref{v:density}; retain the final one even past the sample prefix.
If \(f\) raw names and \(m_{\rm pkt}\) packets are selected, temporal
coverage gives
\[
 \lfloor t/K_{\rm age}\rfloor\le
 J_{\rm pack}(n)(f+m_{\rm pkt}),\qquad
 J_{\rm pack}(n)=C_{\rm deep}+C_{\rm end}+8W_n^*+1,
\]
where \(C_{\rm end}\) absorbs the fixed endpoint convention.
Raw supports are disjoint, and packet intersections have their
unconditional bound \(\theta_{\rm pkt}^{m_{\rm pkt}}\).
They may overlap one another across the two families. As in the proof
of Proposition~\ref{c:late-live}, H\"older with weights \(3/4,1/4\) gives
\(p^{3f/4}\theta_{\rm pkt}^{m_{\rm pkt}/4}\).
The full historical count \eqref{v:history-count} is unconditional;
no replica path or initial chart is conditioned upon. Since
\(p\le2\mu\), \(W\le C_h/\mu\), its raw factor obeys
\[
 (2\mu)^{3f/4}(C_h/\mu)^{6f/16}
 =(2C_h)^{3f/8}(2\mu)^{3f/8}.
\]
The historical and packet bases are below \(1/64\) by
\eqref{v:cap-list} and the overlay targets. Summing
\(f+m_{\rm pkt}\ge\lfloor t/K_{\rm age}\rfloor/J_{\rm pack}(n)\)
gives an exponential. Motif locations, critical types and phase/role
residues have a fixed finite \(n\)-dependent candidate list, not a list
of initial words. Adding its union and the flag-only cone exception
proves the first assertion. An ordinary proposal with the storage-enable bit set has 512 centre parents with that bit set, so it has the same tail apart from a direct
current exception.

For the ancestor assertion use centre ancestry at fixed
\((u,P_3,P_4)\), rather than the south ancestry of
Lemma~\ref{c:rare}. Define the minimal finite-start scalar fields
\[
 \mathcal R_t=H_t\lor H_{t+1},\qquad
 \mathcal W_t=\mathcal R_t\lor\Theta_{512}(\mathcal W_{t-1}),
 \qquad\mathcal N_t=\Theta_{512}(\mathcal W_t).
\]
Here \(\Theta_{512}\) is the Boolean replica threshold. A reading with the storage-enable bit set outside this minimal field has a centre predecessor with that bit set outside it, yielding a path back to the arbitrary initial row. On its
late half the first assertion makes all these readings raw-marked,
outside a polynomial union of exponentially small events. The all-path
density bound excludes a long all-marked continuation. Thus, at physically late times, a set storage-enable bit lies in the minimal field.

A new \(\mathcal N\) birth has 256 distinct fresh replica positions
or old support in two boxes. Use full deterministic doubled cubes of
side 352 and their two physical time slices. Four spatial parities and
two temporal parities give, exactly as in Lemma~\ref{c:rare}, the
intersection bound \((N_{\rm seed}p^{256})^{g/8}\) for \(g\) specified
groups. Reserve full supports, not just realized flags, at every slab
restart. Attach their complete contours and \(k_{\rm dir}\) direct
hardware names to the physical finite-template quantum path. Data edges
can connect unrelated controller copies, with no equality of their
initial words assumed. Cauchy--Schwarz gives, as in
\eqref{eq:c-mixed} with \(n_{\rm sel}=1\), the mixed weight
\[
 (N_{\rm seed}p^{256})^{g/16}p^{k_{\rm dir}/2}.
\]
Connected descriptions number at most
\(K_{\rm rare}^{g+k_{\rm dir}}\), and temporal coverage gives
\(g+k_{\rm dir}\ge s/B_{\rm rare}\). These constants are finite maxima
over the fixed replica, group and quantum-edge catalogue;
they ignore the halting time and chart values. The last two bases of
\eqref{v:cap-list} close the geometric sum. Polynomial unions in the
late determining cone preserve its exponential decay. A Q fault
that bypasses a cut is included among direct names. This proves the
initial-site assertion without a union over encodings or initial words.
\end{proof}

\begin{proof}[Proof of Theorem~\ref{v:detailed}(ii)]
Apply Remark~\ref{rem:forgetting-interface} in dimension six. Property~\ref{v:structure} gives initial metadata dephasing and autonomy. By the processor description, every quantum factor crosses the final conditional-reset block or is reset or discarded before reuse. Thus Lemma~\ref{f:factorization} holds on the full algebra. Lemma~\ref{v:ancestor} gives the initial-site tail in determining cones of radius \(O(s)\), uniformly over initial metadata and rates. Blocks contain \(O(s^6)\) sites and have a fixed dependency degree. For the finitely many small tori, final-layer full replacement has probability at least \((\inf I_n^{\mathsf r})^{L^6}\) in \(\mathrm{WS}\) and \((\inf I_n^{\mathsf r})^{N_{\rm Q}L^6}\) in Q. These are exactly the three inputs to the shared forgetting argument; its bound persists at every later time by decoder composition.
\end{proof}

\subsection{Wrapped diagonal bands and the scope of their cost}\label{v:wrapped-section}
This subsection is used in no proof. It gives a clean-rule mechanism that loses the logical pair and whose seeds cost \(\exp[-\Theta_n(L)]\). The remark after the proof explains why this gives no matching lifetime upper bound under continuing noise.

\begin{lemma}[Clean wrapped-domain mechanism]\label{v:wrapped}
Finite nonwrapping transverse defect islands erode under the clean
comparison rule. Diagonal bottom bands in \(P_1+P_2\) of width at least
200, and the diagonal half-torus for \(L\ge400\), spread to the whole \((P_1,P_2)\)-plane. On the otherwise-clean
initialized-code trajectory, a whole bottom \((P_1,P_2)\)-plane at
\(u=u_0\) and fixed \(P_3,P_4\) permanently loses the
selected logical pair.
A dangerous wrapping certificate has at least \(b_{\rm span}L/C_{\rm all}\)
distinct lifted source labels. For the sizes of Lemma~\ref{v:bridge},
the short alias-free piece extracted in its proof has at least
\(q_{\rm short}/(3C_{\rm all})\) distinct physical source labels.
A width-200 band is seeded by at most \(C'L\) physical names for a fixed
\(C'\).
At a fixed rate \(\eps\) in a window, prescribed band seeds have
weight between \(e^{-C_{n,\eps}L}\) and \(e^{-c_{n,\eps}L}\), with
\(0<c_{n,\eps}\le C_{n,\eps}\) independent of \(L\).
\end{lemma}
\par\noindent\emph{Proof idea.} Trace defect predecessors using the three physical potentials. Small islands erode under this rule, whereas the two boundaries of a wrapped diagonal band drive it across the plane.

\begin{proof}
Relative to a clean reference, a defect without a fresh source needs
at least 257 defective predictions and hence two boxes. Each of the
three physical potentials of Lemma~\ref{v:density} follows a defective
predecessor gaining at least 130. For a finite initial island \(\mathcal I_0\),
\[
 390t\le\max_{\mathcal I_0}3(P_1+P_2)+\max_{\mathcal I_0}(-3P_1)+\max_{\mathcal I_0}(-3P_2).
\]
The right side is translation invariant and proportional to transverse
diameter. This proves finite-island extinction, even allowing growth
in \(u\), without a torus-eroder assertion.

For a diagonal band put \(s=P_1+P_2\bmod L\),
\(\mathrm{tr}_r=\#\{(i,j)\in[0,15]^2:i+j=r\}\), and
\(c_A(s)=\sum_{r=0}^{30}\mathrm{tr}_r\one_A(s+r)\).
The exact bottom condition is
\(c_A(s)+2c_A(s+160)\ge257\). The triangular sums through 14 and
15 are 120 and 136. On the line an interval \([0,w)\),
\(31\le w\le129\), therefore translates to \([-175,w-175)\)
with its width unchanged. For \(w\ge200\) it becomes
\([-175,w-160)\), of width \(w+15\): the left threshold jumps from
240 to 272, while at the right edge one unshifted full box and one
shifted replica suffice. Periodic copies only add bottom votes, so
these wide intervals grow to fill the circle. At \(u=u_0\)
the centre input and all-parent bottom spreading give the same
lower-bound evolution. A whole bottom plane is preserved by the clean
rule and makes the complete cut erase a logical \(Z\) membrane.

At the captured code snapshots, the Pauli-frame induction in
Lemma~\ref{q:universal} puts data in a Pauli frame on the original
codeword, with logical-independent syndrome/work records, including
histories that later fail homology. Erasing the logical-\(Z\) support
removes its off-diagonal logical information; subsequent input-independent
refresh cannot recreate it. This is a loss mechanism for the chosen
initialized pair, rather than entanglement breaking of the round channel.

A width-200 band at \(u_0\) and fixed \(P_3,P_4\) occupies \(200L\)
sites. Prescribing their bottom refresh outputs uses \(O(L)\) physical
names and has the stated exponential seed weight, with fixed positive
code probabilities. In an otherwise-clean history it initiates the
mechanism. Conversely the short full-coordinate extraction in
Lemma~\ref{v:bridge} requires \(\Omega(L)\) distinct physical canonical
source labels for a dangerous wrapping/homology certificate. This proves the
source-order assertion and its paid exponential source cost.
\end{proof}

Absorbing bottom here describes the clean rule: continuing noise can
change its symbols. The prescribed-source weight does not give an
unconditional catastrophe hazard under continuing noise, and does not
prove a matching lifetime upper bound. In particular the optimal
exponent may depend on the window label.

\subsection{A coarse lifetime upper bound and exact certification}\label{v:cert-section}
The full-register reset in \eqref{eq:finite-reset}, used above for
small tori, gives an \(\exp(O_n(L^6))\) lifetime upper bound
in dimension six. Full resets of individual fibres can occur at different times. Fibre confinement improves the exponent to four. Certification extends the method of
Section~\ref{a:certification} to the explicit exponential-horizon
circuits.

\begin{proposition}[Fibre resets and exponential-horizon certification]\label{v:certification}
Fix either circuit \(\mathsf r\in\{\mathrm{WS},\mathrm Q\}\) of
Theorem~\ref{v:detailed}. Set
\[
\begin{gathered}
 k_{\mathrm{WS}}=1,\qquad k_{\mathrm Q}=N_{\mathrm Q},\qquad
 a_n^{\mathsf r}=\inf I_n^{\mathsf r},\\
 j_n^{\mathsf r}=\min\{h\in\N:h\ge1,\ 2^{-h}\le a_n^{\mathsf r}\},\qquad
 H_{\mathrm e}(L)=2^{\lfloor c_{\mathrm e}L\rfloor}.
\end{gathered}
\]
Here \(a_n^{\mathsf r}\) is a positive computable rational and
\(j_n^{\mathsf r}\) is a total computable integer. Time counts rounds,
one application of \(\Lambda_{L,\eps}\) per round.
\begin{enumerate}
\item For every \(n\), every \(L\ge1\), every \(\eps\in I_n^{\mathsf r}\)
and every rational \(0<\delta<1\), put
\begin{equation}\label{v:reset-time}
 T^{\mathsf r}_{n,\delta}(L)=
 \left\lceil(a_n^{\mathsf r})^{-k_{\mathsf r}L^4}\right\rceil
 \left\lceil L^2/\delta\right\rceil.
\end{equation}
For every integer \(t\ge T^{\mathsf r}_{n,\delta}(L)\), the entire
physical channel has a decomposition
\begin{equation}\label{v:reset-mixture}
 \Lambda_{L,\eps}^t=(1-q_t)\Phi_t+q_t\mathcal B_t,
 \qquad q_t\le L^2e^{-L^2/\delta}\le\delta,
\end{equation}
where \(\Phi_t\) is entanglement breaking and \(\mathcal B_t\) is a
channel. Consequently
\[
 \|\Lambda_{L,\eps}^t-\Phi_t\|_\diamond\le2\delta,
 \qquad F^*_{L,\eps}(t)\le1/2+\delta/2.
\]
In particular, uniformly across the window,
\begin{equation}\label{v:lifetime-upper}
 \tau_{L,\eps}<2^{(k_{\mathsf r}j_n^{\mathsf r}+3)L^4}
 \quad(L\ge1),\qquad
 \tau_{L,\eps}<2^{L^5}
 \quad(L\ge k_{\mathsf r}j_n^{\mathsf r}+3).
\end{equation}
If \(U(n)\uparrow\), then for every
\(L\ge L_{\mathrm e}^{\mathsf r}(n)\),
\[
 H_{\mathrm e}(L)\le\tau_{L,\eps}
 <2^{(k_{\mathsf r}j_n^{\mathsf r}+3)L^4}.
\]
\item For the explicit maps in Theorem~\ref{v:detailed}, fix the finite
decoder choices as in the proof below and write
\[
 f^{\mathsf r}_{n,L,t}(\eps)=
 F(\cE^{\mathsf r}_{n,L},\cD^{\mathsf r}_{n,L,t};L,t,\eps).
\]
Each \(f^{\mathsf r}_{n,L,t}\) is a rational-coefficient polynomial,
computable by a total finite algorithm. Define the predicate
\begin{equation}\label{v:predicate}
\begin{split}
 P_{\mathsf r}(n):\quad& f^{\mathsf r}_{n,L,t}(q)\ge7/8
 \quad\text{for every integer }L\ge L_{\mathrm e}^{\mathsf r}(n),\\
 &\text{integer }0\le t\le H_{\mathrm e}(L),\qquad
  q\in I_n^{\mathsf r}\cap\mathbb Q.
\end{split}
\end{equation}
Here \(q\) is a rational rate, unrelated to the probability \(q_t\)
of part (i). This predicate is \(\Pi^0_1\)-complete, holds exactly
when \(U(n)\uparrow\), and is equivalent to the same predicate over
all real rates in the window.
Every finite test is decidable by exact rational arithmetic, and every
false certificate has a finite exact witness \((L,t,q)\).
\item For each fixed rational \(0<\eta<3/8\), there is no total
computable \(g:\N\to\N\) such that, for every halting \(n\), every
\(L\ge1\), every \(\eps\in I_n^{\mathsf r}\) and every integer
\(t\ge g(n)\), one has \(F^*_{L,\eps}(t)\le1/2+\eta\).
\end{enumerate}
\end{proposition}
\par\noindent\emph{Proof idea.} Reset whole quantum fibres at possibly different rounds while keeping the autonomous classical histories. The resulting channel bound controls lifetime. Separately, finite fidelity polynomials let the proof transfer the exact certification argument to an exponential horizon.

\begin{proof}
For (i), use structural property~\ref{v:structure}. Write
\(\mathcal Q\) for all quantum-capable factors and \(\mathcal M\)
for the autonomous metadata. The quantum factors split into \(L^2\)
fibres \(F_u=\{u\}\times(\Z/L\Z)^4\). Let \(R_{u,j}\) be the event
that the final noise layer of round \(j\) replaces all \(L^4\) sites
of \(F_u\) in \(\mathrm{WS}\), or all \(N_{\mathrm Q}L^4\) wires there in Q. Its
probability is
\[
 \rho_{L,\eps}=\eps^{k_{\mathsf r}L^4}.
\]
The replacement coins are exogenous and independent across rounds.
Let \(\mathcal G_t\) be the event that every fibre has such a reset in rounds
\(1,\ldots,t\). This event is independent of every physical input,
including the initial metadata. A union bound gives
\begin{equation}\label{v:fibre-tail}
 q_t:=\Pr(\mathcal G_t^c)
 \le\min\{1,L^2(1-\rho_{L,\eps})^t\}
 \le\min\{1,L^2e^{-t\rho_{L,\eps}}\}.
\end{equation}

We prove that the normalized channel on \(\mathcal G_t\) is entanglement
breaking even though the fibre resets occur at different times.
Initially the circuit dephases \(\mathcal M\); let \(a\)
be its measured full word, with projector \(\Pi_a\). Conditional
on \(a\), the initial quantum state may still have arbitrary
intersite and reference correlations. Condition also on a complete
autonomous metadata history and all replacement choices. The history
law, conditional on \(a\), is independent of the quantum input.
Quantum-derived syndrome outcomes are retained and summed inside
CPTP maps, rather than postselected as part of this conditioning.

Every conditional quantum map acts within one fibre. Metadata
controls arriving from elsewhere now have fixed classical values.
Thus a quantum dependence path cannot change \(u\). A complete
fibre reset terminates every such path through that fibre and replaces
it by a product maximally mixed state, severing its correlations with
all other factors and the reference. Equivalently, in the Heisenberg
picture the complete reset sends every observable on that fibre to a
scalar multiple of its identity. Propagating final observables
backwards through all the conditional fibre maps, and using one reset
in each fibre, leaves a scalar times the identity on the entire
initial \(\mathcal Q\) algebra. This argument applies to tensor
products of output observables, which span the full output algebra,
so it also covers arbitrary correlations among fibres. All
quantum-derived records belong to \(\mathcal Q\), and the final
metadata word is fixed by its autonomous history. Therefore on a good
history the entire output is a preparation independent of the initial
quantum state.

Average these preparations over good histories conditional on \(a\)
to obtain states \(\rho_{a,\mathcal G_t}\). Since \(\mathcal G_t\) has
input-independent probability, its normalized branch is the channel
\[
 \Phi_t(\omega)=\sum_a
 \Tr[(\Pi_a\otimes I_{\mathcal Q})\omega]\,\rho_{a,\mathcal G_t}.
\]
This is a measure-and-prepare channel on the entire physical algebra,
including for inputs entangled with a reference. The complementary
normalized branch \(\mathcal B_t\) is also CPTP; if its weight is
zero its choice is immaterial. This proves the decomposition whenever
\(q_t<1\). The diamond distance between two channels is at most two,
so the distance to \(\Phi_t\) is at most \(2q_t\). Encoding and
decoding preserve the entanglement-breaking property, whose Bell
overlap is at most \(1/2\); the bad branch has overlap at most one.
Hence every encoder and decoder has fidelity at most
\(1/2+q_t/2\), with no union over their choices.

For \(\eps\in I_n^{\mathsf r}\), the time in
\eqref{v:reset-time} satisfies
\(T^{\mathsf r}_{n,\delta}(L)\rho_{L,\eps}\ge L^2/\delta\).
For every later \(t\), \eqref{v:fibre-tail} gives
\[
 q_t\le L^2e^{-L^2/\delta}
 =\delta x e^{-x}\le\delta,
 \qquad x=L^2/\delta.
\]
This proves \eqref{v:reset-mixture} and the claimed norms and
fidelities. The factor-disjoint translated layers are defined even
on small tori, and every quantum displacement has zero
\(u\)-component modulo every \(L\); thus this argument applies to
every \(L\ge1\), without a code-geometry threshold.

At \(\delta=1/2\) the fidelity is at most \(3/4<7/8\). Since
\((a_n^{\mathsf r})^{-k_{\mathsf r}L^4}
\le2^{k_{\mathsf r}j_n^{\mathsf r}L^4}\) and the latter is an integer,
\[
 T^{\mathsf r}_{n,1/2}(L)
 \le2L^2\cdot2^{k_{\mathsf r}j_n^{\mathsf r}L^4}
 \le2^{(k_{\mathsf r}j_n^{\mathsf r}+3)L^4}.
\]
Here \(\log_2(2L^2)\le1+2L\le3L^4\) for integer \(L\ge1\).
Monotonicity of optimal fidelity gives
\(\tau_{L,\eps}<T^{\mathsf r}_{n,1/2}(L)\).
For \(L\ge k_{\mathsf r}j_n^{\mathsf r}+3\), the last display is
at most \(2^{L^5}\). All constants and cutoffs are computable from
the rational window endpoint, uniformly in its rate.
The nonhalting lower bound follows from Theorem~\ref{v:detailed}(i).
Requiring all fibres to reset simultaneously would instead give the
whole-volume exponent six noted above; asynchrony is essential here.

For (ii), we first specify the exact finite external circuits. Use
the total static search and defaults of Lemma~\ref{v:encoder}.
For the closed plaquette CSS code let \(A,B\) be the binary spaces
of its \(X\) and \(Z\) stabilizers, with \(A\subseteq B^\perp\). Extend a
basis of \(A\) to \(B^\perp\), then to the full binary space.
Finite linear equations choose complementary logical vectors agreeing
with the declared coordinate membrane pair. Prepare the
\(A\)-coefficient wires in \(\lvert+\rangle\), the unknown qubit
on the selected logical coefficient, the other logical coefficients
and completion wires in \(\lvert0\rangle\), and implement the
invertible binary basis change with CNOTs and swaps. This realizes
the finite binary encoding of Lemma~\ref{v:encoder} with real
Clifford gates; small or degenerate cases use its prescribed defaults.

Use the syndrome of the intended complex and the same decoder with component filling (Lemmas~\ref{q:homology}--\ref{q:span} and~\ref{v:bridge}).
Freeze box, lift and filling choices, using lexicographic tie-breaks
where a choice remains, and fix a failure correction outside its good
regime. These frozen choices are harmless: by Lemma~\ref{q:homology},
two fillings of the same component in a contractible box differ by a
closed chain in that box, hence by a boundary; the nonhalting proof
does not use success on the complement of its good event.
All finite boundary and incidence data are computable from \(L\).
Exhaustive finite searches specify corrections for every syndrome.
Syndrome extraction, Pauli corrections and inverse encoding use
Clifford gates, computational-basis measurements and finite Boolean
functions. Such functions compile with \(X\), CNOT, Toffoli and reset;
\(Z=HXH\). At a requested time use the declared captured-snapshot
and intended-circuit completion convention before the common
recovery. This gives a total finite circuit determined by
\((n,L,t)\), independent of the rate and halting answer, including
all failure cases.

Proposition~\ref{a:rational}'s rational-superoperator argument applies to these finite circuits, including their default cases. The deterministic static search prepares a basis state, and no measurement branch is normalized. Each \(f^{\mathsf r}_{n,L,t}\) is therefore a computable rational-coefficient polynomial, and comparison with \(7/8\) at rational rate is exact.

Use the exact-test enumeration, midpoint counterexample and continuity argument in the proof of Theorem~\ref{lean:conditional}, with the domain \eqref{v:predicate}. Its horizon \(H_{\rm e}(L)\) is a computable integer and grows without bound, so every halting time fits below it at a sufficiently large size. Theorem~\ref{v:detailed}(i) makes all tests pass on nonhalting inputs; part (ii), at \(\eta=1/16\), gives a failed midpoint test on halting inputs. The same input compiler gives hardness. This uses only Theorem~\ref{v:detailed}(i)--(ii) and the finite-circuit calculation, independently of part (i) of this proposition.

For (iii), apply the contradiction argument of Corollary~\ref{a:time} with \(L_{\rm e}^{\mathsf r}(n)\) and \(H_{\rm e}(L)\): their computability and unbounded horizon supply the required size, and the exact finite comparison just proved supplies the test. A size-dependent upper bound from (i) does not provide the forbidden size-independent bound.
\end{proof}

\section{Stretched-exponential storage in four dimensions}\label{sec:four-dimensional}
The static cap of Section~\ref{sec:exponential} needs two chart
coordinates and two quorum coordinates. A concatenated code uses only
a chain. Combining that cap with Section~\ref{sec:quantum}'s clock
and folding the chain gives the following theorem.

Folding puts the virtual cycle into half its length with two roles per site, so it fits on a physical torus of any side \(L\). We describe the circuit, verify the hierarchy hypotheses and state the source law in Sections~\ref{f4:circuit}--\ref{f4:actual-sources}. Section~\ref{f4:constants} orders the constants, and Sections~\ref{f4:positive} and~\ref{f4:forgetting} prove the storage and forgetting alternatives.

\begin{theorem}[Detailed form of Theorem~\ref{thm:four-dimensional}]\label{f4:detailed}
There is a computably specified, translation-invariant, finite-range,
time-periodic circuit on \(\Z^4\), with a fixed \(N\) qubits per site,
elementary inventory
\(\{X,H,\mathrm{CNOT},\mathrm{Toffoli},R_0,D_Z,I\}\) of arity at
most three, and independent whole-site depolarization
\eqref{eq:site-noise} at every site after every elementary layer.
It acts on all physical factors, including data, clock, controller,
sampling register, port, work, mixed ancilla, reset-following and idle factors. There are
computable pairwise disjoint nonempty rational open intervals \(I_n\)
whose closures lie in \((0,\eps_*)\) for a fixed rational
\(\eps_*>0\), a fixed clock block length \(P_{\rm cal}=2^{r_{\rm exp}}\), \(r_{\rm exp}\ge2\),
\(\alpha_4=1/r_{\rm exp}\), a rational \(c_4>0\) independent of \(n\),
and a total computable size cutoff \(L_4(n)\), with these properties on
\((\Z/L\Z)^4\) using \eqref{eq:fidelity}.
\begin{enumerate}
\item Total computable maps \(\cE_{n,L},\cD_{n,L,t}\), independent
of \(\eps\) and the halting answer, satisfy
\[
 F(\cE_{n,L},\cD_{n,L,t};L,t,\eps)\ge15/16
 \quad\bigl(U(n)\uparrow,\ \eps\in I_n,\ L\ge L_4(n),\
               0\le t\le\exp(c_4L^{\alpha_4})\bigr).
\]
Time here is an integer number of elementary layers. One encoder
serves every such time and rate.
\item If \(U(n)\downarrow\), then for every \(\eta>0\) a finite
\(C_{n,\eta}\), independent of \(L\) and \(\eps\in I_n\), satisfies
\(F^*_{L,\eps}(t)\le1/2+\eta\) for every \(L\), every such rate,
and every integer \(t\ge C_{n,\eta}\). The supremum includes
arbitrary encodings into the entire physical algebra, arbitrary
decoders, and initial controller, work, intersite and reference
entanglement.
\end{enumerate}
\end{theorem}

The windows decrease on the \(4^{-n}\) scale. The exponent \(\alpha_4<2\times10^{-4}\) is small. The crossover with a linear lower bound depends on the coefficients and size cutoffs (Section~\ref{intro:scope}).

\subsection{The four-dimensional circuit}\label{f4:circuit}
Write coordinates \((x,y,\sigma,z)\). The cap field \(\mathsf C\)
is exactly the core, ordinary second proposal, base sampling registers, deep
queries and availability dictionary of Section~\ref{sec:exponential},
with chart plane \((x,y)\), quorum plane \((\sigma,z)\), and the
copied coordinates \(P_3,P_4\) deleted. In particular the static
constraints are put in nearest-neighbour higher-block form before
adding the counter. Its Moore rule retains the original profile, spreads
bottom from every input, and spreads death along original-active
nearest-neighbour edges. Death leaves the cap boundary and endpoints
unchanged. The first coordinate enables storage exactly when its retained profile is active and its death flag \(\mathsf d\) is unset. The second records the ordinary majority proposal. Only the two depth-4096 profile queries \(Q_0,Q_-\) can provide harmful guards. The availability dictionary distinguishes bottom from vacuum: the former writes unknown/allowed, the latter unknown/unavailable. A genuine row origin takes priority over predecessor vacuum. Every shifted sampling register uses its own base site.

An independent clock field \(\mathsf S\) runs along \(\sigma\),
with whole-symbol quorum protection in \((x,z)\), cloned in \(y\)
in the reference. Its logical rule is the transducer of
Lemma~\ref{f4:periodic}, total on malformed fields, folded into two
roles by the coordinate map of Lemma~\ref{f4:dispatch-clean}. Endpoint
flags compute virtual predecessor and successor by physical neighbours
and on-site role switches, so the folded stencil has radius at most two,
independently of the code and the clock alphabet. Boundary/parity
fields are retained by the logical rule. Its stored current request,
finite boundary flags and private old copies supply the exclusive
selectors of Appendix~\ref{app:f4-dispatch}.
The patch and clock evolve autonomously: neither reads the other or the quantum data, and the clock also ignores the sampling registers. They can share a physical noise event.

There are two semantic quantum factors and at most 730 port factors
per site, plus one mixed cut ancilla per persistent target and a fixed
finite amount of temporary work. Additional code and syndrome qubits
occupy more virtual chain cells. Every quantum-derived record remains
quantum-capable. The classical alphabet and its Boolean work become
fixed finite registers after the constants are selected.

A round has the following fixed order. Dephase and make one-way private
old metadata copies; compute and freeze finite exclusive selectors.
Execute scatter, unconditional native tuples and gather, including
padded idles. Then compute next cap and clock by destination-local
Boolean work on transported old inputs, recomputing provisional base
gathers locally. Apply the conditional reset controlled by the first coordinate at each physical site
to every retained quantum-capable factor. Its private mixed ancilla
is reset/discarded after use and never returned. Temporary factors
either cross the cut or terminate at their unconditional reset before
reuse. After the cut only on-site fresh preparation, disposal and
identity/noise padding remain. Every factor belongs to autonomous
dephased metadata or to this quantum inventory.

Section~\ref{sec:compilation} and Appendix~\ref{app:f4-dispatch}
compile this rule into the seven elementary gates. Long Boolean
computations are local to their destination; only fixed old transports
precede them. Quantum spatial motion is confined to \(\sigma\) at
fixed \((x,y,z)\), even for malformed metadata. Private internal ports
make translated tuples disjoint also on small tori. The range of deep
quorum queries and sampling-register transport is large but fixed. The period,
range, site dimension and rule read no size, input index or rate.

\subsection{Complete quantum geometry}\label{f4:geometry-section}
\begin{lemma}[Folded complete-child reduction]\label{f4:geometry}
The circuit has code-independent incidence and child-index halo
constants chosen before the code. With the code and padding chosen
below, its regular hierarchy satisfies the complete-child hypotheses
of Lemma~\ref{f4:recursion} and admits the fixed joint recovery of
Lemma~\ref{d:reduction}.
\end{lemma}
\par\noindent\emph{Proof idea.} Use the full port support of a faulty local choice, including work still in flight. Match these supports to the child cover of the folded chain and bound the separation needed for level reduction.

\begin{proof}
Fix the primitive port/native/cut catalogue first, padding inactive
candidates by their intended idle reference. For each wrong gate choice, include the stations that can still be in use by an unfinished module. Lemmas~\ref{f4:dispatch}
and~\ref{f4:dispatch-partial} put each dirty location in two fixed
role neighbourhoods of radius eight, with arbitrary joint coefficients.
Let \(h_{\rm base}\) cover these footprints and the one-child-step ownership
segment at the immediately following common cut. This segment is
independent of the code; the full normalizer circuit belongs to its
scheduled rectangle. Let \(\kappa_0\) bound station-to-location
conversion, \(\kappa_1\) child cell-time participation, and \(v_0\)
parent boxes meeting a rectangle. These are maxima on that fixed
local catalogue. More ancillas use more cells and repeated EC uses
more times, without increasing participation at one cell-time.

A level-\(\ell\) regular box has length \(P_{\rm cal}^\ell\), time height
\(T_{\rm cal}^\ell\), and its \(P_{\rm cal}\)-by-\(T_{\rm cal}\) child cover plus \(h_{\rm base}\)
child-index layers. Geometric series bound every complete descendant
by \mbox{\(h_{\rm halo}=2h_{\rm base}+2\)}, enlarging once for rectangle covers. Choose
\[
 \rho_{\rm gap}\ge1,\qquad
 a_{\rm sep}>(4h_{\rm halo}+2\lceil\rho_{\rm gap}\rceil+5)^2,\qquad
 b_{\rm FT}=\max(1,\kappa_0\kappa_1v_0(a_{\rm sep}-1)).
\]
Use \(r_{\rm FT}=8b_{\rm FT}\), and the least code
parameter with \(q_{\rm code}=2^{s_{\rm code}}-1
\ge40\,960(b_{\rm FT}+1)\). Lemmas~\ref{d:code}--\ref{d:gadgets}
apply to these exact native inputs. Precede each complete body by
\(b_{\rm FT}+1\) time-disjoint full EC passes with private work,
synchronized before any interblock body; idle blocks normalize too.
If the at most \(v_0\) covering parent boxes are good, they contain
at most \(v_0(a_{\rm sep}-1)\) bad regular children, marking at most
\(b_{\rm FT}\) child circuit locations. The last clean EC gives
code support, and all subsequent passes and the body satisfy
\[
 8(r_{\rm FT}+b_{\rm FT})=72b_{\rm FT}\le q_{\rm code}-8.
\]
The operator proof of Lemma~\ref{d:reduction} preserves all
good logicals jointly. Absorb outgoing maps only on the union of bad
outgoing wires before the next body; retain all shared coefficients.
This proves deterministic reduction at each level.

A bad parent has at least \(a_{\rm sep}\) bad children. At most
\((4h_{\rm halo}+2\lceil\rho_{\rm gap}\rceil+5)^2\) complete child halos are
within \(\rho_{\rm gap}P_{\rm cal}^\ell\) of one full halo in both position
and time; here \(T_{\rm cal}^\ell\ge P_{\rm cal}^\ell\). Hence a separated pair exists.
There are finitely many pairs after \(P_{\rm cal},T_{\rm cal}\) are fixed, and root count
is \(O(P_{\rm cal}^{5\ell})\) since \(T_{\rm cal}=P_{\rm cal}^4\). Choose the fixed real table
first and then a sufficiently large power of two \(P_{\rm cal}\) so its routed
work and interpreter fit, as in Lemma~\ref{f4:periodic}. Choose even
\(s_{\rm top}\) larger than every complete spatial halo. On
\(m=s_{\rm top}P_{\rm cal}^k\) all relevant supports are proper cycle arcs.
The exact periodic reference supplies the repeated encoded identity
at every aligned time.

Only the active folded cycle is the code layout. The extension
\(a_r\) outside the interval is a source image. Other physical
factors are exterior systems. If a faulty tuple couples them to active
data, first use closed-support localization, then expand in matrix
bases on its marked active operands, retaining exterior coefficients.
Their number and source stars are unchanged. This is exactly the
operator scope corrected by the common recovery.
\end{proof}

\subsection{The source law}\label{f4:actual-sources}
\begin{lemma}[Four-dimensional source interface]\label{f4:sources}
For each \(n\) with \(U(n)\uparrow\), take the computable size cutoff
of Lemma~\ref{f4:encoder}, enlarged for the support hulls and the
size hypotheses of Lemmas~\ref{f4:raw}, \ref{f4:strip} and~\ref{f4:cap-exception}.
For every \(L\) above that cutoff and every \(\eps\in I_n\), the
circuit initialized by Lemma~\ref{f4:encoder} satisfies the
actual-source inputs of Appendix~\ref{app:f4-sources} through each
integer horizon \(0\le T_{\rm hor}<2^{\lfloor L/4\rfloor}-1\), with
constants independent of the window. There is a fixed computable
\(C_{\rm bd}\ge\max(C_{\rm raw},C_{\rm cap})\) such that its raw
exceptional probability through \(T_{\rm hor}\) is at most
\(C_{\rm bd}(T_{\rm hor}+C_{\rm bd})L^4e^{-\beta_{\rm raw}L}\), and its cap exceptional probability
is at most \(C_{\rm bd}(T_{\rm hor}+C_{\rm bd})L^4e^{-\beta_{\rm cap} L/(w_n+1)}\).
\end{lemma}
\par\noindent\emph{Proof idea.} Project the six-dimensional controller proof by deleting its copied coordinates. The clock and the controller then use their separate source estimates, with the dispatch contribution counted in the combined graph.

\begin{proof}
Deleting the copied \(P_3,P_4\) coordinates changes none of the cap's
Moore, replica, base, deep-query or sampling-register increments. The complete
ordinary and statistical arguments of Section~\ref{v:parameters}
therefore apply with the same two quorum coordinates. At large enough
\(L\), one cap square, its vacuum gap and every \(W\)-experiment
have injective support hulls; no divisibility of \(L\) is required.
At a fixed time the characteristic map is bijective, and distinct
integer bins have disjoint times. Thus distinct cap raw names use
disjoint physical flag sets. Statistical requests use the three bin
residues and private base ownership; overlay events use complete-bin
support colours. The proof of \eqref{v:joint} gives
\[
 \Pr(X_\lambda\text{ for all }\lambda\in F)\le\gamma_{\rm src}^{|F|},\qquad
 \gamma_{\rm src}=\sqrt{2C_h}+\eta_+(B)^{1/6}
                      +\alpha_{\rm bin}^{1/(2D_{\rm bin})}.
\]
Here \(C_h\) is the sampling-register scale, with the tails in \eqref{v:etas} evaluated at the chosen scale (see Table~\ref{tab:concordance}). Added clock operations only
enlarge the internal overlay at that site and round. Their private
roles do not alter base ownership or introduce cap inputs.

The witness calculation before the sampling-register count is as
follows. On each charge arm every clean deep entry owns all
\(D_{\rm q}=4096\) support occurrences before returning to an actual
old word. If \(e\) is the support count, \(m_{\rm ent}\) the entry count and \(g\)
distinct lifted terminal groups, then
\[
 4096m_{\rm ent}\le e,\qquad129e-1440(g-1)-696m_{\rm ent}\le\Omega_{\rm pre}g.
\]
Shared suffixes are counted before projection; faulty evaluation ends
at an overlay terminal. The inequality \(129-696/4096>128\) gives
\(e\le(1440+\Omega_{\rm pre})g/128\). Sampling-register rides keep their base
characteristic and have bounded total bin increment. Keep their two
endpoint bins and compress them into a typed corridor. An incorrect censor without a hardware fault is an intrinsic or overlay exception at its source label, or traces back to one of the two base queries. Receiver failures and all
\(B\) lane faults are aggregated in the original overlay source label, so
there is no sampling-register index decoration. Entries/corridors add at most two
links per support plus fixed source stars. Thus complete lifted
witnesses have \(E\le c_Cg\), bounded degree and anchored exponential
entropy, with the explicit catalogue bounds in Section~\ref{f4:order-a}.
All five characteristic coordinates and finite roles remain in that
graph. Initial availability is unavailable, and sampling registers start by their
own base: a recursion reaching time zero without a source is correct.
Actual inputs move backward in physical time or in a bounded virtual
forecast depth. Every retained source and endpoint bin is consequently
in the finite causal range specified in Appendix~\ref{app:f4-sources}.

For the clock, whole-symbol prediction uses only its fixed logical
\(\sigma\)-stencil and the replica boxes in \((x,z)\). Wrong clean
outputs require wrong predictions in two boxes. The three potentials in
Lemma~\ref{f4:raw} ignore the logical \(\sigma\) and copied \(y\)
shifts, so their gains remain \(132,129,129\). One-way old transport
makes the raw cause/use multiplicity depend only on this stencil,
not the alphabet or lookup rows. Apply Lemma~\ref{f4:raw} in \(d=4\),
including boundary, opcode, old/new and current-copy errors.
Direct dispatcher hardware uses the same physical raw source label.

Cap errors explain wrong conditional resets; raw clock errors explain wrong
choices and malformed endpoints. Attach the exact current-use stars
of Lemma~\ref{f4:dispatch-partial} before selecting the central-strip
cutoff. Lemmas~\ref{f4:strip}--\ref{f4:mixed} now apply with
\(w=w_n\). They give one source image and the fourth-root joint law
for all short certificates. Source vertices outside the active code
remain in full witness entropy with their existing names. Only active
quantum roots are glued to the hierarchy. The two exceptional bounds
follow directly, under the original physical law.
\end{proof}

\subsection{Order and bounds for the constants}\label{f4:constants}
All constants in Sections~\ref{f4:order-a} and~\ref{f4:order-b} are fixed
for the circuit, except the final window and size quantities. A finite
catalogue means the explicit port templates, replica offsets, typed
support/entry/corridor steps and source/use stars, with values of
alphabet words omitted. It can be enumerated by finite loops before
any truth table is materialized. The following convention makes the
bounds effective rather than an appeal to an unspecified finite maximum.
For such a catalogue \(\mathcal A\), let \(n(\mathcal A)\) be its
number of entries, \(R_x(\mathcal A),R_t(\mathcal A)\) the largest
absolute spatial and temporal offsets, and
\[
 V(\mathcal A)=(2R_x(\mathcal A)+1)^4(2R_t(\mathcal A)+1).
\]
A source affects at most \(n(\mathcal A)V(\mathcal A)\) labelled
uses, and a graph with finite edge catalogue has degree at most twice
its number of oriented entries. A doubled-tree description with marks
and cause choices is bounded by
\([8(\Delta+1)^2(K_0+1)]^{E+1}\). These bounds give the effective finite maxima used below; optimizing them is unnecessary.

\subsubsection{Geometric and circuit constants}\label{f4:order-a}
Each item ends with the quantities chosen after it.
\begin{enumerate}[label=\arabic*.,leftmargin=2em]
\item \textbf{\boldmath Local dispatch catalogue \(\mathcal A_{\rm loc}\).} At most 270 tuple templates,
730 ports; include cut/reset/idles, primitive decomposition and one-step
ownership stars. Physical operand diameter \(\le4\), marked radius \(\le8\).
\emph{Precedes.} All remaining rows.

\item \textbf{\boldmath\(h_{\rm base},\kappa_0,\kappa_1,v_0\).} Count footprints and incidences of
\(\mathcal A_{\rm loc}\). Use radius at least eight; the ownership part
is one child step. Bound counts by
\(732n(\mathcal A_{\rm loc})V(\mathcal A_{\rm loc})\), enlarging the
catalogue for each cover incidence.
\emph{Precedes.} Code, full normalizer length,
\(P_{\rm cal},T_{\rm cal}\), alphabet.

\item \textbf{\boldmath\(h_{\rm halo},\rho_{\rm gap},a_{\rm sep},b_{\rm FT}\).} \(h_{\rm halo}=2h_{\rm base}+2\),
\(\rho_{\rm gap}=1\) is allowed;
\[
 a_{\rm sep}=1+\lceil(4h_{\rm halo}+2\lceil\rho_{\rm gap}\rceil+5)^2\rceil;
\]
\(b_{\rm FT}=\max(1,\kappa_0\kappa_1v_0(a_{\rm sep}-1))\).
\emph{Precedes.} Code.

\item \textbf{\boldmath\(r_{\rm FT},s_{\rm code},q_{\rm code}\).}
\(r_{\rm FT}=8b_{\rm FT}\); least \(s_{\rm code}\) with
\(2^{s_{\rm code}}-1\ge40\,960(b_{\rm FT}+1)\).
\emph{Precedes.} Native table, \(P_{\rm cal},T_{\rm cal}\).

\item \textbf{Native table, \boldmath\(C_Q,P_{\rm cal},T_{\rm cal},r_{\rm exp},s_{\rm top}\).} Build finite real
gadgets; bound routed work by \(C_QP^2\). Search powers
\(P_{\rm cal}=2^{r_{\rm exp}}>2\) fitting all phases in \(T_{\rm cal}=P_{\rm cal}^4\). Choose
even \(s_{\rm top}\) exceeding complete spatial halos.
\emph{Precedes.} Source target,
sampling-register count, hardware depth.

\item \textbf{\boldmath\(\mathcal A_{\rm raw},K_0,c_0,R_{\rm phys},\Delta_{\rm raw}\).} Fixed
clock transport, old/new and use stencil; \[
 K_0\le n(\mathcal A_{\rm raw}),
\]
\(c_0\le n(\mathcal A_{\rm raw})V(\mathcal A_{\rm raw})\).
Take \(R_{\rm phys}\ge1\) to bound contour/source steps;
\mbox{\(\Delta_{\rm raw}\le2n(\mathcal A_{\rm raw})\)}.
\emph{Precedes.} Raw probability bound,
\(B,J\); independent of alphabet.

\item \textbf{\boldmath\(K_{\rm raw},C_{\rm raw},C_{\rm w},K_{\rm w}\).} Tree ceiling above;
\(E<12c_0g\) plus enumerated attachment stars. Use its coefficient
for \(C_{\rm w}\), raise the tree ceiling to \(C_{\rm w}+1\) for
\(K_{\rm w}\). \(C_{\rm raw}\): the fixed root-count constant of
Lemma~\ref{f4:raw}.
\emph{Precedes.} Bound on \(p\).

\item \textbf{\boldmath\(R_{\rm pre},\Delta_{\rm pre}\).} \(D_{\rm q}=4096\),
\(R_{\rm pre}=2^{12}(D_{\rm q}+3)\);
\(\Delta_{\rm pre}=R_{\rm pre}^2[64(2D_{\rm q}+5)(9\cdot768+16)^2+256]\).
Enlarge once for fixed dispatch stars.
\emph{Precedes.} \(C_h,B\), palette.

\item \textbf{\boldmath\(\Omega_{\rm pre},L_{\rm star},c_C\).} Sum absolute three-potential offsets
on the source/entry stars (raw old/new contributes 696);
\(L_{\rm star}\) is their total link bound.
\[
 c_C=\lceil3(1440+\Omega_{\rm pre})/128+3+L_{\rm star}\rceil.
\]
Pole offsets are bounded directly by six times summed characteristic
coordinate offsets.
\emph{Precedes.} \(B\), palette.

\item \textbf{\boldmath\(R,K_{\rm cap}\).} Here \(R\) is the full increment bound of Lemma~\ref{f4:strip}. Union of preceding
catalogues with two formal roles and fixed gluing stars. Degree at most
sum of their degrees plus star incidence; size at most sum of size
bounds plus star lengths. In this row \(R_x\) bounds increments in the chain-coordinate component
\(\chi\), \(R_\omega\) bounds increments in \(\omega=x-t\), and \(R_j\)
bounds increments in the integer bin \(j\)
(Appendix~\ref{app:f4-sources}). Take \(R\ge\max(R_\omega,R_j)\).
The image step is at most \(R_x+59R_\omega\).
Use the tree ceiling for \(K_{\rm cap}\).
\emph{Precedes.} Cutoff formula, source targets.

\item \textbf{\boldmath\(A_{\rm rec},d_0,B_{\rm tail},f_{\rm sh}\).} At most the square
of the number of children in the fixed \mbox{\(P_{\rm cal}\)-by-\(T_{\rm cal}\)} cover for
\(A_{\rm rec}\); root count has exponent five, enlarged for fixed
covers. Choose \mbox{\(f_{\rm sh}<\rho_{\rm gap}/(3\operatorname{Lip})\)},
where \(\operatorname{Lip}\) is the image increment of the preceding row.
Anchored enumeration fixes \(B_{\rm tail}\).
\emph{Precedes.} \(\gamma_{\rm eff}\) target.
\end{enumerate}

For the negative path catalogue, include full replica-group contours,
their doubled spatial cubes and two time slices, and the actual
\(\sigma\)-edge/port/cut skeleton. Let \(\Delta_{\rm neg}\) be its
finite maximum degree, \(C_{\rm neg}\) its edge count per paid rare
or direct group, and \(R_{\rm neg}\) its maximum time increment.
These are obtained by the contour and slab construction in the proof
of Lemma~\ref{v:ancestor}, replacing its quantum-edge list by the
explicit dispatcher list. A crude recipe for \(C_{\rm neg}\) is to
multiply the ordinary contour coefficient by the total number of links
in a full source/group star, adding that star size once more; every
inserted link is then charged to its original paid group. Thus one may
use
\[
 K_{\rm rare}=[8(\Delta_{\rm neg}+1)^2]^{C_{\rm neg}+1},\qquad
 B_{\rm rare}=1+R_{\rm neg}(C_{\rm neg}+1).
\]
This counts connected marked descriptions and bounds their temporal
coverage. All entries use the fixed two-dimensional quorum and finite
quantum-contact template; none uses the number of code cells, the
clock alphabet, sampling-register count or padding length. Padding has no
quantum branching and is covered by its one round flag. The same
finite-star counting fixes \(K_{\rm hist}\) before \(B\), by
\eqref{v:history-count}.

\subsubsection{Probability, hardware and window constants}\label{f4:order-b}
Each item ends with the quantities chosen after it.
\begin{enumerate}[label=\arabic*.,leftmargin=2em]
\item \textbf{\boldmath\(K_{\rm hist},K_{\rm rare},B_{\rm rare},N_{\rm seed}\).} Use the historical and negative catalogues above and set
\mbox{\(N_{\rm seed}=\tbinom{2\cdot352^2}{256}\)}.
\emph{Precedes.} Probability bounds, \(B\),
palette.

\item \textbf{\boldmath\(a_{\rm tail},D,\beta_{\rm cap},\beta_{\rm raw}\).} Use the two inequalities of Lemma~\ref{f4:recursion}. Set
\mbox{\(D=1/(2\max(1,A_{\rm rec}))\)}. Require the exceptional bases in
Lemmas~\ref{f4:raw} and~\ref{f4:cap-exception} strictly below one;
take any positive rational lower bound on their exponential rates.
\(C_{\rm bd}\): choose a fixed prefactor at least
\(\max(C_{\rm raw},C_{\rm cap})\), enlarged for the conversion of the
raw contour bound to exponential form and for the fixed prefactor in
the top-box count of \eqref{f4:failure}. Impose the raw cutoff and the
central-strip size condition through the computable size cutoff of
Lemma~\ref{f4:sources}.
\emph{Precedes.} \(C_h,B,\mu_*\).

\item \textbf{\boldmath\(C_h\).} Choose rational \(0<C_h\le1/100\) so
\(\sqrt{2C_h}\) is within the cap-source target.
\emph{Precedes.}~\(B\).

\item \textbf{\boldmath\(B=4^{k_B},\eta_+,\eta_-,\rho_{\rm code}\).} \(\rho_{\rm code}=126/128\);
search \(B\ge20/C_h\) making the tails \eqref{v:etas} sufficiently
small and \(4z_B<1\). Explicit tails are displayed below.
\emph{Precedes.} Support
colours, \(\mu_*,n_0\).

\item \textbf{\boldmath\(h_{\rm bin},h_{\rm hist}\), \(D_{\rm bin},D_{\rm ov}\),
\(\alpha_{\rm bin},\alpha_{\rm hist}\).} Enumerate full \(B\)-dependent physical
supports and overlap colours as in \eqref{v:joint}; choose positive
rational overlay targets for the remaining source/history margins.
\emph{Precedes.} \(\mu_*,J,P_{\rm pad}\).

\item \textbf{\boldmath\(\mu_*\).} Rational below \(C_h/100\), the positive, raw and density bounds
and all four strict negative inequalities below. Include the
raw bound \(382\,205\,952^{-67\,440}/128\).
\emph{Precedes.} \(n_0\), palette, hardware.

\item \textbf{\boldmath\(n_0\), palette, \(J\).} \(n_0\ge\max(k_B+2,3)\), all nominal windows
below \(\mu_*/2\). Build the finite chart/clock/compiler and count
its full internal elementary depth \(J\).
\emph{Precedes.} Padding, physical windows.

\item \textbf{\boldmath\(P_{\rm pad},\eps_*\).} \(P_{\rm pad}\ge2J\) with
\(2h_{\rm bin}JC_h/P_{\rm pad}\le\alpha_{\rm bin}\) and
\(2h_{\rm hist}JC_h/P_{\rm pad}\le\alpha_{\rm hist}\).
Choose rational \(\eps_*>0\) whose image under \(\mu\) lies in
\((\mu_*/2,\mu_*)\); nominal windows were put below \(\mu_*/2\).
\emph{Precedes.} Physical \(I_n\).

\item \textbf{\boldmath\(I_n,w_n,L_4(n)\).} Rational inner preimages of the nominal windows;
\(w_n=\lceil C_h/\mu(a_n)\rceil+2\) for \(I_n=(a_n,b_n)\).
Effective sufficient power bounds in the storage proof.
\emph{Precedes.} Initialization
and size analysis only; never circuit parameters.

\item \textbf{\boldmath\(\alpha_4,c_4\).} \(\alpha_4=1/r_{\rm exp}\);
choose positive rational \(c_4<(2s_{\rm top}P_{\rm cal})^{-\alpha_4}(\log2)/32\).
Enlarge the size threshold for fixed round depth.
\emph{Precedes.} Storage assertion;
independent of \(n\).
\end{enumerate}

Here the padding and rate-cap choice can be implemented without revisiting \(n_0\):
choose \(\eps_*\) rational with
\(\mu_*/2<1-(1-\eps_*)^{P_{\rm pad}}<\mu_*\), and initially put all
nominal windows below \(\mu_*/2\). Strict monotonicity makes this
rational search total. Thus no compiled hardware is chosen twice.
For clarity the sampling-register tails used above are
\[
\begin{gathered}
 u_B=2^{-\lfloor3BC_h/100\rfloor},\quad
 z_B=2^{-\lfloor BC_h/3\rfloor},\quad
 w_B=2^{-\lfloor BC_h^2/50\rfloor},\quad
 b_B=2^{-\lfloor B(1/2-2C_h)^2\rfloor},\\
 \eta_+=3u_B/(1-4z_B)+2w_B+2b_B,\qquad
 \eta_-=4[u_B+w_B+2^{-\lfloor B(\rho_{\rm code}-1/2)^2\rfloor}].
\end{gathered}
\]
They vanish as \(B\) grows at fixed \(C_h\). The remaining strict
negative inequalities are
\begin{align*}
 K_{\rm hist}(2C_h)^{3/8}(2\mu_*)^{3/8}&<1/64,\\
 K_{\rm hist}[\eta_-+\alpha_{\rm hist}^{1/D_{\rm ov}}]^{1/4}&<1/64,\\
 K_{\rm rare}[N_{\rm seed}(2\mu_*)^{256}]^{1/16}&<1/64,\\
 K_{\rm rare}(2\mu_*)^{1/2}&<1/64.
\end{align*}
Choose \(B\) and its overlay target for the second inequality before
shrinking \(\mu_*\); positive powers make all searches terminate.

The internal and padding fields \(A,E\) are independent, with independent flags at distinct physical sites and rounds. Their rates satisfy
\[
 \nu=1-(1-\eps)^J,\qquad
 \mu=1-(1-\eps)^{P_{\rm pad}},\qquad
 p=\Pr(A\lor E)\le2\mu,\quad
 \nu\le2J\mu/P_{\rm pad}.
\]
The last padding replacement is a product uniform word, conditional
on its one flag. Thus sampling-register marks are conditionally independent
uniform fields. Internal faults are arbitrary overlays; discrepancies
from the counterfactual clean sample are charged there. Padding adds
no quantum movement. This is the whole-site physical experiment used
in every source and negative estimate.

Finally, even with unit incidence factors, \(h_{\rm base}\ge8\) gives
\(h_{\rm halo}\ge18\), \(a_{\rm sep}>79^2\), \(b_{\rm FT}\ge6240\).
The verified preparation in Lemma~\ref{d:gadgets} has at least
\(5^{2b_{\rm FT}}\) leaves. One qubit per virtual cell and one native
layer per round require \(P_{\rm cal}^5\ge5^{12\,480}\), hence
\(\alpha_4\le5/(12\,480\log_2 5)<2\times10^{-4}\).
The actual capacity and station multiplicity make it far smaller.

\subsection{Windows, initialization and the storage horizon}\label{f4:positive}
\begin{lemma}[Physical windows and a total encoder]\label{f4:encoder}
There are windows \(I_n\) and a total computable encoder \(\cE_{n,L}\)
as in Theorem~\ref{f4:detailed}. Above a computable size cutoff, on
nonhalting inputs the initialized reference has the periodic folded
clock and the storage-enable bit set throughout the active patch for
\(2^{\lfloor L/4\rfloor}-1\) rounds, in the positive-branch reference
trajectory used in the proof of Lemma~\ref{v:encoder}.
\end{lemma}
\par\noindent\emph{Proof idea.} Choose inner physical windows using the increasing refresh polynomial. Initialize the finite patch and the folded code by terminating searches with prescribed fallbacks.

\begin{proof}
Put \(\ell_a=2\cdot4^a-1\), \(a=n+n_0+1\), and
\[
 \mathcal J_n=(2C_h/\ell_a,21C_h/(10\ell_a)).
\]
They are nonempty disjoint rational intervals below \(\mu_*/2\).
Use the literal scan from \cite[Lemmas~4.1\nobreakdash--4.2]{DMNoise2026},
as in Section~\ref{v:parameters}. Its first zero is
\(j_*=2^{a+1}-2\), makes \(a+1\) tokens and skips \(n_0+1\).
It calls \(V(n+1)=U(n)\). Thus the windows select the stated universal
input. Search rational \(0<a_n<b_n<\eps_*\) with
\[
 \inf \mathcal J_n<\mu(a_n)<\mu(b_n)<\sup \mathcal J_n.
\]
All comparisons are exact rational arithmetic. The increasing rational
polynomial \mbox{\(\mu(\eps)=1-(1-\eps)^{P_{\rm pad}}\)} and density of
rationals prove totality. Then \(I_n=(a_n,b_n)\) has all claimed
properties and \(w_n=\lceil C_h/\mu(a_n)\rceil+2\) bounds \(W\)
uniformly in the window. The windows lie at scale
\(4^{-n}/P_{\rm pad}\).

Choose the largest \(k\ge0\) with \(s_{\rm top}P_{\rm cal}^k\le L/2\).
For small sizes use a fixed fallback. Initialize \(m=s_{\rm top}P_{\rm cal}^k\)
virtual clock cells by the finite \(k\)-level simulation tower over
uniform top idle, with the full canonical scratch of
Table~\ref{f4:fields} in Appendix~\ref{app:f4-calendar}. Fold them into \([0,m/2)\), initialize boundary
flags/parity and outside reference, and clone across \(x,y,z\).
Lemma~\ref{f4:periodic} supplies the clean reference for all times.

In \((x,y)\) place one square of side \(N_{\rm cap}=\lfloor L/4\rfloor\)
with genuine vacuum exterior and fixed collar. Apply exactly the
static finite search in Lemma~\ref{v:encoder}: enumerate the finite
hidden alphabet on that square and collar; test every fully active
base constraint, cap shape, canonical projected geometry, finite owners,
routing tags and the cutoff at every active site, including truncated
rulers; reject halt letters. Return the first accepted assignment or
vacuum. That proof, using \cite[Lemma~11.2]{DMNoise2026} and its cutoff
converse, proves both totality and nonhalting success without an
infinite-extension test. It is unchanged by deleting copied coordinates.
Set death and counter bits to zero with the least diagonal carry one,
and clone the cap over \(\sigma,z\). Initialize each shifted sampling register
by its own base: low for flag one, high for zero, unknown for inactive;
availability is unavailable, also at storage sites outside the cap.

At a deterministic interior \((x_0,y_0,z_0)\), encode the complex input
qubit into one wire of the \(k\)-fold real CSS code on the folded
cycle. Other code wires, ancillas and unused factors have their
prescribed states. All compiler work has its declared defaults.
The procedures depend only on \(n,L\). Failed searches and halting
inputs still return total maps. All hull, gap, \(w_n\), sampling-register and
collar bounds enter a computable threshold. Include
\(36R_{\rm phys}+1\) in that threshold so Lemma~\ref{f4:raw} applies. Lemma~\ref{v:cap} proves
the stated horizon of the storage-enable bit in the cap reference:
the core follows \(F_{\rm cap}\), the queries return the reference
profile and every censor test passes, as in the proof of Lemma~\ref{v:encoder}.
\end{proof}

\noindent\emph{Proof idea for Theorem~\ref{f4:detailed}(i).}
Apply the separated-child recursion to the folded schedule. Sum over all top boxes and the intermediate layers, then use the fixed recovery to obtain the fidelity guarantee throughout the horizon.

\begin{proof}[Proof of Theorem~\ref{f4:detailed}(i)]
Apply Lemma~\ref{f4:recursion} with the geometry and sources
of Lemmas~\ref{f4:geometry} and~\ref{f4:sources}. Put
\(T_{{\rm hor},k}=2^{\lfloor2^k/16\rfloor}\). Count every top box and rectangle
cover through \(T_{{\rm hor},k}+T_{\rm cal}^k+C_{\rm bd}\), including completion halos. For fixed
computable constants and an integer \(d'\), failure probability is at
most
\begin{equation}\label{f4:failure}
\begin{split}
 C_{\rm bd}(T_{{\rm hor},k}+T_{\rm cal}^k+C_{\rm bd})L^{d'}2^{-2^k}
 &+C_{\rm bd}(T_{{\rm hor},k}+T_{\rm cal}^k+C_{\rm bd})L^4e^{-\beta_{\rm cap} L/(w_n+1)}\\
 &+C_{\rm bd}(T_{{\rm hor},k}+T_{\rm cal}^k+C_{\rm bd})L^4e^{-\beta_{\rm raw}L}.
\end{split}
\end{equation}
Maximality of \(k\) gives
\[
 (L/(2s_{\rm top}P_{\rm cal}))^{\alpha_4}<2^k
             \le(L/(2s_{\rm top}))^{\alpha_4},\qquad T_{\rm cal}^k=O(L^4).
\]
The first negative exponent in \eqref{f4:failure} dominates
\(\log T_{{\rm hor},k}\); the others dominate it once
\(L^{1-\alpha_4}\ge4A_{\rm hor}(w_n+1)/\beta_{\rm cap}\), where
\(T_{{\rm hor},k}\le e^{A_{\rm hor}L^{\alpha_4}}\) with fixed computable
\(A_{\rm hor}\). All constants admit effective strict rational
bounds. The inequality \(\log L\le L^\delta/\delta\), for rational
\(\delta>0\), bounds all fixed polynomial prefactors by fractions
of these negative exponents for every sufficiently large \(L\).
It supplies a terminating search for a sufficient threshold valid at
\emph{every} larger size. Enlarge it for cutoff floors, source hulls,
\eqref{f4:failure} at most \(1/16\), and
\(T_{{\rm hor},k}+T_{\rm cal}^k+C_{\rm bd}<2^{N_{\rm cap}}-1\). The same power comparison proves
the latter inequality. Only \(w_n\) and canonical finite initialization
margins depend on \(n\). This yields total computable \(L_4(n)\).
Choose \(c_4\) as in Section~\ref{f4:order-b}, enlarging the threshold
for floors and the fixed elementary depth per round. Every requested
elementary time through \(e^{c_4L^{\alpha_4}}\) is covered.

Let \(\mathcal G_{\rm stor}\) mean absence of both source exceptions and every geometric
\(N^{\rm sh}_k\) in that extended collection. It is exogenous, independent of
logical amplitudes. On \(\mathcal G_{\rm stor}\), joint level reduction gives the identity
at aligned boundaries. At an arbitrary elementary readout retain all
in-flight semantic data, syndrome, ports, masks holding quantum
data, quantum-derived work and undeleted mixed ancillas. Ignore received
metadata. Append the fixed intended future quantum circuit to the
next aligned \(T_{\rm cal}^k\) boundary, with no added noise, using intended
selectors and controls. Complete every already-started conditional reset,
refresh, gather and cleanup on its actual factors before its scheduled
reset/disposal. A mixed ancilla holding the old target is retained until then.
Only wholly future complete live cuts, whose reference action is
identity, may be omitted. In particular the suffix of a live Fredkin
cannot be omitted after its first CNOT. Introduce fresh external work
only for new lifetimes.

Lemma~\ref{f4:dispatch-partial} places discrepancies in the same
budgeted primitive root family. Delete future bad roots and redo the
deterministic reduction; the geometric predicates are monotone.
The same fixed bottom-up Knill--Laflamme recovery restores the Bell
state on \(\mathcal G_{\rm stor}\), including arbitrary complex amplitudes and reference.
It is a total decoder depending only on \(n,L,t\). Averaging CPTP
histories gives fidelity at least \(\Pr(\mathcal G_{\rm stor})\ge15/16\).
The initialization was independent of both requested time and rate.
\end{proof}

\subsection{Halting-side fidelity for all registers}\label{f4:forgetting}
\begin{lemma}[Ancestry for the folded schedule]\label{f4:ancestor}
On a halting window there are \(A'_n,a'_n>0\) and a fixed integer
\(b'_{\rm c}\), uniform in the window and independent of \(L\), such
that on every torus a specified physical initial site whose
depth-\(s\) determining cone, of radius \(O(s)\), unwraps injectively
has retained quantum ancestry at depth \(s\) rounds with probability at
most \(A'_n(1+s)^{b'_{\rm c}}e^{-a'_ns}\), simultaneously over every
initial metadata word. All quantum
factors at that site are included in the same event.
\end{lemma}
\par\noindent\emph{Proof idea.} Apply the same halting obstruction to the static patch and count surviving paths using the actual folded dispatch edges. The final reset on each site's retained factors removes their dependence when storage is disabled.

\begin{proof}
The cap's negative argument is the one in Lemma~\ref{v:ancestor}.
We verify its graph hypotheses and the changed quantum path explicitly.
The static alternative there uses original profiles, active-neighbour
death ancestry and the deep-only base dictionary. Deleting copied
coordinates changes none of them. On halting, every arbitrary active
root is within bounded \(n\)-dependent original-active path distance
of a local violation, a fully guarded wrong critical motif, or a valid
uniformly bounded square. Finite-alphabet compactness covers malformed,
originless and wrapped components, exactly as in that proof. Persistent
or dead profiles remain distinct from vacuum, so cap boundaries and
counter endpoints are unchanged.

Full-core finite-age coherence and the all-path density bound of
Lemma~\ref{v:density} identify one immutable profile in a flag-only
good cone. The explicit active-neighbour death terms propagate ancestry
to the obstruction. A violation or arbitrary-phase expiry of a bounded
square contradicts late storage enablement. At a wrong motif a genuine origin
restarts the sampling lane and removes dependence on its initial contents. An earlier \(\bot\) writes an unknown value but leaves the lane eligible, so it cannot make either wrong critical test pass. A surviving sample pays
an intrinsic/overlay packet or a harmful nonbottom deep profile error.
The depth-4096 argument in Section~\ref{v:parameters} gives at least
sixteen distinct physical raw names per harmful query. Full-support slab
packing retains the last contour, including its part beyond
the sample prefix. Temporal coverage is
\(\lfloor t/K_{\rm age}\rfloor\le J_{\rm pack}(n)(f+m_{\rm pkt})\),
with \(J_{\rm pack}(n)=C_{\rm deep}+C_{\rm end}+8w_n+1\).
The same historical entropy, fixed before the sampling-register count, counts all possible profiles,
without conditioning on any initial word or chosen replica path.
H\"older's inequality between possibly overlapping families gives
\(p^{3f/4}\theta_{\rm pkt}^{m_{\rm pkt}/4}\), and
\[
 (2\mu)^{3f/4}(C_h/\mu)^{6f/16}
       =(2C_h)^{3f/8}(2\mu)^{3f/8}.
\]
The first two strict negative inequalities make the sum exponentially
small. Polynomial cone unions yield the all-start storage-enable and proposal
tail, uniformly in the window. Added clock metadata is autonomous
and contributes only to already counted internal overlays.

For centre ancestry at fixed \((x,y)\), use the minimal fields
(the local symbol \(\mathcal B_t\) here is a source field)
\[
 \mathcal B_t=H_t\lor H_{t+1},\quad
 \mathcal W_t=\mathcal B_t\lor\Theta_{512}(\mathcal W_{t-1}),\quad
 \mathcal N_t=\Theta_{512}(\mathcal W_t),
\]
started empty. A storage-enable bit still set at a late time outside that field would give a centre
path to the initial row; its late half is raw-marked by the preceding
tail, contrary to all-path density except for an exponential event.
A new \(\mathcal N\) birth needs 256 distinct fresh replicas or old
support in two boxes. The same two-dimensional quorum plane
\((\sigma,z)\) has four spatial parities and two time parities on
full doubled cubes of side 352. As in Lemma~\ref{v:ancestor}, for
\(g\) seed groups and \(k_{\rm dir}\) distinct direct names the joint weight is
\[
 (N_{\rm seed}p^{256})^{g/16}p^{k_{\rm dir}/2}.
\]
Full supports are reserved at every restart; complete rare contours
are attached to the quantum path.

For this schedule quantum edges move only a bounded distance in
\(\sigma\) or between finitely many on-site role/port/work factors.
They keep \(x,y,z\) fixed. All spatial quantum work precedes the
complete own-site conditional reset. A forbidden clean cut removes every
incoming quantum dependence; reset/preparation has no incoming edge,
and a discarded factor never returns. Thus every surviving path is a
connected description in the finite negative catalogue used to define
\(K_{\rm rare},B_{\rm rare}\). It has
\(g+k_{\rm dir}\ge s/B_{\rm rare}\) and at most
\(K_{\rm rare}^{g+k_{\rm dir}}\) descriptions. The last two negative inequalities
sum this tail exponentially. A polynomial determining-cone union proves
the lemma for a specified initial physical site, including both fold
roles and all work.

Every assertion averages the physical independent local whole-site
flags and their conditional product sampling-register marks. Arbitrary joint
quantum coefficients are an analysis of that physical circuit. Fresh
replacement/reset environments are never retrieved, so they supply
no returning path. Arbitrary initial programs, endpoints, counters and
clock words are included; clock synchronization is unnecessary
for this negative argument.
\end{proof}

\begin{proof}[Proof of Theorem~\ref{f4:detailed}(ii)]
Use Remark~\ref{rem:forgetting-interface} in dimension four. The autonomous, initially dephased cap and clock control the quantum circuit but never receive quantum-derived targets. Every quantum lifetime crosses its own-site cut or ends at reset/disposal, so Lemma~\ref{f:factorization} applies to the full initial-site mask. Lemma~\ref{f4:ancestor} supplies the uniform tail; each block has \(O(s^4)\) sites and determining-cone radius is \(O(s)\). On the finitely many smaller tori, full-register reset in one layer has probability at least \((\inf I_n)^{L^4}>0\). The shared extension, iteration and cloning argument therefore gives the claimed bound, after converting rounds to elementary layers. Its joint conditional histories retain all cap, clock and hardware correlations.
\end{proof}

For every label \(n\), every \(\eps\in I_n\) and every \(L\), the
full-register reset probability is at least \((\inf I_n)^{L^4}\)
in each layer. The argument of \eqref{eq:finite-reset} gives
\(F^*_{L,\eps}(t)\le 1/4+(3/4)(1-(\inf I_n)^{L^4})^t\), with \(t\)
counted in elementary layers. Thus the lifetime at fidelity \(7/8\) obeys
\begin{equation}\label{f4:lifetime-upper}
 \tau_{L,\eps}<\log(6/5)\cdot(\inf I_n)^{-L^4}+1=\exp(O_n(L^4))
 \quad\text{elementary layers}.
\end{equation}

For each fixed rational \(0<\eta<3/8\), these halting times have no total computable bound in the input \(n\). The argument of Corollary~\ref{a:time} applies with the computable size cutoff \(L_4(n)\) and the unbounded horizon \(\exp(c_4L^{\alpha_4})\): the finite circuits are computable, so approximating fidelity with absolute error less than half the gap between \(1/2+\eta\) and \(15/16\) and comparing with their midpoint would decide halting. No certification statement for this construction is added here.

\section{One-dimensional construction and its classical inputs}\label{q1:model}
Sections~\ref{q1:model}--\ref{q1:effective} prove Theorem~\ref{q1:main} and Corollary~\ref{q1:memory}. This section fixes the noise convention, the two classical lanes and the inputs taken from \cite{DMNoise1D2026}. Section~\ref{q1:clock} bounds the failures of the structural lane R1, which supplies addresses and ages. Section~\ref{q1:compiler} uses them to run a concatenated quantum code and proves level reduction. Section~\ref{q1:selector} builds the rate-selecting lane R2 from exact noise-driven replacements and bounds the joint error field. Section~\ref{q1:parameters} fixes every constant in a noncircular order. Section~\ref{q1:storage} proves storage on nonhalting windows, Section~\ref{q1:forgetting} proves whole-register forgetting on halting windows, and Section~\ref{q1:effective} assembles the proofs. Table~\ref{tab:terms-ring} lists the terms used in these sections.

\begin{table}[!tb]
\caption{Terminology of the construction on rings.}\label{tab:terms-ring}
\small\centering
\begin{tabular}{@{}>{\raggedright\arraybackslash}p{.23\linewidth}>{\raggedright\arraybackslash}p{.51\linewidth}>{\raggedright\arraybackslash}p{.19\linewidth}@{}}
\toprule Term & Meaning & Defined in\\\midrule
R1, R2 and macro round & On rings, the structural lane R1 is an initialized hierarchical clock that supplies addresses and ages, and the selector lane R2 simulates the computation selected by the noise rate. One macro round is one native tick of each lane and one quantum tick. & Section~\ref{q1:model}\\\addlinespace[3pt]
Native symbol, tick and bin & A site state and a time step of the simulated classical automaton of \cite{DMNoise1D2026}; a bin is a band of its replacement rate. & Sections~\ref{q1:model} and~\ref{q1:selector}\\\addlinespace[3pt]
Home, rank and date & A home is the block of child sites that simulates one parent site; the rank is the level in the hierarchy; a date is a time step at a given rank. & Sections~\ref{q1:clock} and~\ref{q1:compiler}\\\addlinespace[3pt]
Packet and ballot & The encoded parent state carried in the retained letters of a home; the verification vote by which a home accepts it. & Section~\ref{q1:compiler}; \cite[Sections~\ref*{cl:sec:packets}--\ref*{cl:sec:verifier}]{DMNoise1D2026}\\\addlinespace[3pt]
Literal & The fixed machine text that a classical lane runs, stored read-only in the rule. & Sections~\ref{q1:model} and~\ref{q1:selector}\\\addlinespace[3pt]
Factory and innovation & A factory is a redundant circuit that turns depolarization into a rare or fair random bit; an innovation is the resulting native random replacement. & Section~\ref{q1:selector}\\\addlinespace[3pt]
Helper and paint & Helper fields of R2 hold its noise-driven walls, channels and receivers; the encoder prepares their stationary marginal approximately. Paint is the enlargement around a failed helper center that is added to the error set. & Sections~\ref{q1:selector} and~\ref{q1:storage}\\\addlinespace[3pt]
Mark and strobe & A mark is an event of independent noise keys inside a ball that covers a fault or a failed test. The strobe is the step of each round at which the R2 symbol is read and erasure is decided. & Section~\ref{q1:storage}\\
\bottomrule
\end{tabular}
\end{table}

We use the elementary gate and noise model of Section~\ref{s:local-circuits}. A macro round is a fixed finite schedule of \(D_{\rm layer}\) elementary layers; the noise after every layer is
\[
 \bigotimes_{\text{all wires }q}\bigl((1-\eps)\operatorname{Id}_q+
 \eps\operatorname{Tr}_q(\,\cdot\,)I_q/2\bigr).
\]
Every idle and reset layer is included. Registers have finitely many named factors per site. Gates such as SWAP, CZ and a classically controlled real gate below denote fixed decompositions into the elementary operations of Section~\ref{s:local-circuits}. Old and new classical banks are distinct, dephased at round boundaries, with one-way copies into private controls. Each displacement uses a distinct destination lane. Thus translated gates have distinct targets even when displacements coincide modulo a small ring length.

There are two autonomous classical lanes: a structural lane R1 initialized from a certified never-halting computation, and a rate-selecting lane R2. One round updates their whole native symbols, reads the old structural coordinates and retained word to execute one quantum instruction, performs the fixed port exchange, and uses the R2 strobe symbol to trigger terminal local erasure. Every nonretained quantum-capable factor is disposed of unconditionally after this erasure block. The autonomous banks remain dephased one-way controls; quantum gadget tapes and exchange ports are quantum-capable retained slots subject to erasure. Section~\ref{q1:forgetting} enumerates every factor.

We use the classical construction \cite{DMNoise1D2026} in the version and numbering listed in Appendix~\ref{app:imports-1d}.
Its Definition~\ref*{cl:def:sparse} fixes the sparsity convention.
Lemma~\ref*{cl:lem:sparse-witness} supplies the deterministic mesh
implication only; Lemma~\ref{q1:l1} replaces its iid probability bound.
For R1 we use the one-point serialization variant of
Section~\ref*{cl:sec:packets} proved in Lemma~\ref{q1:l9}: each bounded
scalar bit is moved from its scalar-bit tag to the low word bit of its
own live letter, with the other word bits zero; the typed word halves
and fixed endpoint padding are retained. Encoders, parsers and evaluators
all use that layout. Proposition~\ref*{cl:prop:organization},
Definition~\ref*{cl:def:observations} with its lookahead bound, and
Lemmas~\ref*{cl:lem:endpoint} and~\ref*{cl:lem:two-period} supply the
structural comparison under their stated protection and live-output
hypotheses. Lemma~\ref*{cl:lem:clean} identifies the clean full row,
and the proof of Lemma~\ref*{cl:lem:past} constructs its computable
aligned boundary. For R2, Lemmas~\ref*{cl:lem:stripes}
and~\ref*{cl:lem:receiver-mixing} supply tunable noise-only prefix tests
and all-input helper coalescence. Lemma~\ref*{cl:lem:locking} supplies
the literal lock, and Lemmas~\ref*{cl:lem:cap-cut}
and~\ref*{cl:lem:ascent} supply the arbitrary-start halting cap and
ascent. The initialized nested tests use the proof of
Lemma~\ref*{cl:lem:clock-witness}. The quantum probability bounds,
routing deletion, joint reduction and whole-channel argument are proved
below.

Fix a universal target machine \(U\). The R2 literal is the single fixed machine \(V\), with \(V(0)\) an infinite loop and \(V(m)\) simulating \(U(\nu_2(m))\) for \(m\ge1\). After fixing every gate and stencil choose the integer \(J_{\rm pad}\), and let \(m(n)=2^n(2J_{\rm pad}+1)\). The rate windows are rational subintervals of the monotone preimages of the native bins for \(m(n)\). Their construction never asks whether \(U(n)\) halts. The complete order of constants is in Section~\ref{q1:parameters}.

\section{Structural control and dependent witnesses}\label{q1:clock}

This section bounds the probability that the structural lane R1 fails a protected test (Lemmas~\ref{q1:l1}--\ref{q1:l1.2}) and shows that outside such failures R1 delivers its clean native symbols and control reads (Lemmas~\ref{q1:l2}--\ref{q1:l3.1}). For a lane with geometry \((w,P,T)\), which is \((w_R,P_R,T_R)\) for R1 and \((w_c,P_c,T_c)\) for R2, write \(V^j=[0,P^j)\times[0,T^j)\) on the site--date lattice. A set \(E\) is 0-sparse if it is empty; it is \(j\)-sparse if every translate \(Q\) of \(wV^j\) admits a translate \(J\) of \(3wV^{j-1}\) such that \((E\cap Q)\setminus J\) is \((j-1)\)-sparse \cite[Definition~\ref*{cl:def:sparse}]{DMNoise1D2026}. Every restricted set is extended by zero outside its restriction. Subsets inherit sparsity. These are the classical sparsity conventions throughout the following proofs.

\begin{lemma}[Restricted witness trees and their probability]\label{q1:l1}

Let the positive integers \(P_{R},T_{R},w_{R}\) be the fixed structural R1 geometry. Let \(E_{R}\) be a native flag field on \(\Z\times\Z\) whose flag at \(q\) is the indicator that at least one independent elementary depolarization flag in a finite stencil \(\Sigma_{R}(q)\) is one. Assume every stencil has at most \(s_{R}\) keys, \(s_{R}\ge 1\) an integer, and is contained, in site and macro-time coordinates, within nonnegative integer radii \(r_{x},r_{t}\) of \(q\). Negative-time native flags are set to zero; stencils at nonnegative times use only physical keys at nonnegative elementary times. The keys are mutually independent and each has its own rate at most \(\eps\); rates may differ between keys, for example after one change of rate at an elementary time.

Set
\[
\begin{gathered}
D_{R} = 8(P_{R}+2)^{2}(T_{R}+2)^{2}, \\ N_{R} = 4(P_{R}+3)(T_{R}+3), \\ A_{R} = 4w_{R}^{2} s_{R}, \\ J_{R} = (2w_{R}+4r_{x}+3)(2w_{R}+4r_{t}+3), \\ \Delta_{R} = J_{R}-1,\qquad c_{R} = 1/J_{R}, \\ C_{R} = N_{R} D_{R} A_{R}^{1/J_{R}}, \\ \eps_{R0} = A_{R}^{-1}(2N_{R} D_{R})^{-J_{R}}.
\end{gathered}
\]

For any prescribed protected rank-\(j\) \(W_0\) test of width \(<w_{R} P_{R}^{j}\) and height \(<w_{R} T_{R}^{j}\), its failure of \((j-1)\)-sparsity has probability at most
\[
\begin{gathered}
z_{R,j}(\eps) = [C_{R} \eps^{c_{R}}]^{2^{j-1}},\qquad j \ge  1.
\end{gathered}
\]
In particular \(C_{R} \eps^{c_{R}} \le  1/2\) for \(0 \le  \eps \le  \eps_{R0}\). The constants are computable and independent of \(j\).

\end{lemma}
\par\noindent\emph{Proof idea.} A failed sparsity test at rank \(j\) contains a binary tree of \(2^{j-1}\) disjoint flagged regions. A greedy choice keeps a fixed fraction of them with disjoint noise keys, so their flags are independent; counting the trees gives the doubly exponential bound.

\begin{proof}
Use sparsity as in \cite[Definition~\ref*{cl:def:sparse}]{DMNoise1D2026}, with restricted sets extended by zero. For an arbitrary deterministic flag set \(S\), the mesh argument in \cite[Lemma~\ref*{cl:lem:sparse-witness}]{DMNoise1D2026} says the following: nonsparsity at rank \(k+1\) inside one doubled rank-\((k+1)\) box implies rank-\(k\) nonsparsity in two disjoint doubled rank-\(k\) candidate boxes, with at most \(D_{R}\) choices of their unordered pair. The set used for either child is \(S\) restricted to the parent. We keep this restriction at every step. It makes the argument below independent of how the candidate boxes are placed: with the flush placement of \cite[Lemma~\ref*{cl:lem:sparse-witness}]{DMNoise1D2026} they lie inside the parent, while a placement with one exterior mesh cell can put the lower mesh corner used to enclose a moving box one mesh cell outside the parent.

Iterate this implication \(k\) times, retaining the restriction to every ancestor at each node. A leaf has a region \(R\) equal to its doubled rank-zero box intersected with all its ancestor boxes (and the initial restricted root). It is nonempty and contains a flag, because 0-sparsity means emptiness. It has at most \(4w_{R}^{2}\) lattice sites. There are \(N=2^{k}\) leaves. Their regions are pairwise disjoint: two distinct leaves separate into the disjoint sibling boxes of their last common ancestor, and all subsequent restrictions remain in those boxes. The argument uses only the disjointness of these regions, not that of the leaf boxes themselves, which can fail for a placement with exterior mesh cells.

For a prescribed tree, join two leaves if the unions of primitive-key stencils over their regions intersect. If \(R\) has an adjacent leaf \(R'\), some \(q\) in \(R\) and \(q'\) in \(R'\) obey \(|q_{x}-q'_{x}| \le  2r_{x}\) and \(|q_{t}-q'_{t}| \le  2r_{t}\). The doubled rank-zero box containing \(R\) has sides at most \(2w_{R}\). Each adjacent region supplies a different lattice point in the rectangle enlarged by \(2r_{x},2r_{t}\), because the regions are disjoint. The area bound is less than \(J_{R}\). Thus this conflict graph has degree at most \(\Delta_{R}\). A fixed lexicographic greedy algorithm selects at least \(N/J_{R}\) leaves with disjoint primitive-key sets. Each selected leaf requires at least one of at most \(A_{R}\) elementary flags, so its probability is at most \(A_{R} \eps\) by a union bound. Their events are independent, even though the original native flags are dependent. For \(A_{R} \eps \le  1\) their joint probability is at most \((A_{R} \eps)^{N/J_{R}}\).

There are at most \(D_{R}^{N-1}\) geometric trees from a fixed root doubled box: the full binary depth-\(k\) topology is fixed and each internal node has at most \(D_{R}\) pair choices. The leaf probability already sums over flagged sites.

If the rank-\(k\) test fails on a \(W_0\) of size \(<w_{R} P_{R}^{k+1}\) by \(w_{R} T_{R}^{k+1}\), a failing moving rank-\(k\) box meeting its restricted flag set fits in one doubled rank-\(k\) box from the \(w_{R} P_{R}^{k}\) by \(w_{R} T_{R}^{k}\) mesh. There are at most \(N_{R}\) such boxes, even allowing an exterior row and column of the mesh. For \(k=0\) the same cover simply locates a nonempty leaf region. Union over these roots and their trees gives
\[
\begin{aligned}
\operatorname{Pr}(\mathrm{failure}) &\le N_{R} D_{R}^{N-1} (A_{R} \eps)^{N/J_{R}}\\
&\le [N_{R} D_{R} A_{R}^{1/J_{R}} \eps^{1/J_{R}}]^{N}.
\end{aligned}
\]
Set \(k=j-1\). The asserted threshold is a direct substitution and also ensures \(A_{R} \eps \le  1\). Deleting negative-time keys can only decrease these increasing events. Independent unequal key rates bounded above by \(\eps\) satisfy the same product estimate. This proves all assertions. 
\end{proof}

\begin{lemma}[Effective radius-moment cutoff]\label{q1:l1.1}

Suppose Lemma \ref{q1:l3} gives at most \(N_{\mathrm{clk}}\) protected tests of each rank per query and their entire key supports lie in radius
\[
\begin{gathered}
r_{R,j} \le  C_{\mathrm{clk}} B_{R}^{j},\qquad B_{R} = \max (P_{R},T_{R}),
\end{gathered}
\]
where \(N_{\mathrm{clk}},C_{\mathrm{clk}}\) are positive integers fixed after compilation. For every real \(d\ge 0\) and every rational \(\delta>0\) there is a rational \(\eps_{R}(d,\delta)>0\), depending on \(d\) only through \(\lceil d\rceil\) and computable from \(\lceil d\rceil\) and \(\delta\), such that
\[
\begin{gathered}
\sum _{j\ge 1} N_{\mathrm{clk}} (1+r_{R,j})^{d} \sqrt{z_{R,j}(\eps)} \le  \delta
\end{gathered}
\]
for every \(0 \le  \eps \le  \eps_{R}(d,\delta)\).

\end{lemma}
\begin{proof}
Replace \(d\) by its integer ceiling \(d'\). Let \(j_{*}\) be the least integer \(j_{*} \ge  1\) satisfying \(2^{j_{*}-1} \ge  d' \lceil \log _{2} B_{R}\rceil+1\). Put
\[
\begin{gathered}
t_{j} = B_{R}^{d' j} 2^{-(2^{j-1}-1)}, \\ S_{\mathrm{bar}} = \sum _{j=1}^{j_{*}-1} t_{j} + 2t_{j_{*}}, \\ K_{R} = N_{\mathrm{clk}}(C_{\mathrm{clk}}+1)^{d'} S_{\mathrm{bar}}, \\ \theta_{R} = \min (1/4, (\delta/K_{R})^{2}), \\ \eps_{R}(d,\delta) = \min (\eps_{R0}, A_{R}^{-1}[\theta_{R}/(N_{R} D_{R})]^{J_{R}}).
\end{gathered}
\]
All these quantities are rational and effective. If \(x=C_{R} \eps^{1/J_{R}}\), this rate bound ensures \(x\le \theta_{R}\le 1/4\). Then
\[
\begin{gathered}
x^{2^{j-2}} \le  \sqrt{x} 2^{-(2^{j-1}-1)}.
\end{gathered}
\]
Also \((1+C_{\mathrm{clk}} B_{R}^{j})^{d} \le  (C_{\mathrm{clk}}+1)^{d'} B_{R}^{d'j}\). The ratio \(t_{j+1}/t_{j}\) equals \(B_{R}^{d'} 2^{-2^{j-1}}\) and is at most \(1/2\) for \(j\ge j_{*}\). Hence \(\sum  t_{j} \le  S_{\mathrm{bar}}\). The desired moment is at most \(K_{R} \sqrt{x} \le  \delta\). This establishes convergence, vanishing as \(\eps\) tends to zero, and an effective uniform cutoff. 
\end{proof}

\textbf{Order of choice.} The stencil constants refer to the final fixed physical circuit, including all idle layers. They are chosen before the final physical rate-window cutoff. In the assembly the native helper offset \(n_{\mathrm{bin}}\) is fixed before that circuit and is not enlarged after \(s_{R}\) is estimated; the last-chosen arithmetic padding \(J_{\mathrm{pad}}\) skips native bins. Thus the dependence of \(s_{R}\) on the elementary period length creates no circular choice. \(C_{R}, J_{R}, N_{R}\) and the rate cutoff are the explicit constants of the probability estimate in Lemma~\ref{q1:l1}.

\begin{lemma}[Packing at an admission rank]\label{q1:l1.2}

Let \((w,P,T)\) be the geometry of either lane. Run the restricted-witness construction of the proof of Lemma~\ref{q1:l1} from a doubled root box, placing every descendant on the mesh of its rank anchored at the lower corner of the root, and stop it at any rank \(l\); the full box of a leaf is then a doubled rank-\(l\) box of side \(2wP^{l}\) by \(2wT^{l}\) with its corner on that mesh. Suppose the event assigned to a leaf has primitive support contained in its full box enlarged by \(a_{x}wP^{l}\) in space and \(a_{t}wT^{l}\) in time, with fixed \(a_{x},a_{t}\ge 0\). Then its conflict graph has degree at most
\[
\begin{gathered}
J_{\mathrm{adm}}-1,\qquad J_{\mathrm{adm}} = \lceil 8+4a_{x}\rceil \lceil 8+4a_{t}\rceil,
\end{gathered}
\]
uniformly in \(l\) and in the depth of the tree. Initial clipping by an arbitrary \(W_0\) does not invalidate this bound. Thus at least the fraction \(1/J_{\mathrm{adm}}\) of the leaves have disjoint primitive supports.

\end{lemma}
\begin{proof}
Fix the root box and construct every descendant mesh relative to its lower corner. Because \(P,T\) are integers, boundaries of all boxes at ranks \(\ge l\) lie on the same translated mesh with cell sides \(\ell_{x}=w P^{l}, \ell_{t}=w T^{l}\). For a leaf \(L\), use the ancestor region \(R_{\mathrm{anc}}(L)\), the intersection of its full box with its ancestor boxes, without imposing the initial \(W_0\) restriction. These ancestor regions are disjoint by the sibling-separation argument in the proof of Lemma \ref{q1:l1}. An event witnessed on the actual clipped region implies \(R_{\mathrm{anc}}\) is nonempty. It is a rectangle whose boundary coordinates are on the common mesh, so it contains a whole \(\ell_{x}\) by \(\ell_{t}\) mesh cell. Choose one such cell deterministically.

If the enlarged support of another leaf meets that of \(L\), its full box has its lower corner within \((2+2a_{x})\ell_{x},(2+2a_{t})\ell_{t}\) of the lower corner of \(L\). Its selected mesh cell is therefore inside a rectangle with respective side lengths at most \((6+4a_{x})\ell_{x},(6+4a_{t})\ell_{t}\) around that corner. Distinct leaves have distinct selected cells, since their ancestor regions are disjoint. The number of available cells, with two extra boundary cells in each coordinate allowed, is at most \(J_{\mathrm{adm}}\). This bounds all conflicts, including multiplicity from geometrically coincident full boxes in different ancestor branches. The proof did not pack the possibly thin \(W_0\)-clipped regions. Greedy selection gives the final assertion. 
\end{proof}

This lemma supplies the bounded-overlap step for the two-scale admission proof when the primitive support is shorter than a leaf side. It does not assume that the globally untruncated leaf boxes are disjoint or that clipped regions have a rank-uniform positive area.

\begin{lemma}[Total classical interface for R1]\label{q1:l2}

Fix the final physical elementary circuit of Section~\ref{q1:model} and the R1 packet serialization of Lemma~\ref{q1:l9}. Let \(\mathrm{Alph}_{R}\) be the finite native R1 alphabet at physical rank zero. Fix an injective canonical binary representation \(\operatorname{enc}_{R}\):\(\mathrm{Alph}_{R} \to  \{0,1\}^{b_{R}}\). Extend its inverse to a total parser \(\operatorname{dec}_{R}\) by mapping every unused binary string to the dead native symbol \(D\). The R1 Boolean evaluator computes \(\operatorname{enc}_{R}(f_{R}(\operatorname{dec}_{R}(w_{-}),\operatorname{dec}_{R}(w_{0}),\operatorname{dec}_{R}(w_{+})))\) on arbitrary incoming banks; any additional unused Boolean intermediate strings have fixed total defaults. This only specifies the already required finite Boolean compilation on malformed inputs.

There is a finite stencil \(\Sigma_{R}(x,t)\) with the following properties. The decoded-symbol implication of Lemma \ref{q1:l3} uses only properties (i) and (iv) below; property (ii) is proved afterwards, using that implication.

\begin{enumerate}
\item The decoded positive-time history is a native R1 data history off the flag set \(E_{R}(x,t)=\one[\text{at least one primitive flag in }\Sigma_{R}(x,t)]\). Corruption of a legal word and production of a non-codeword both fall within this implication.
\item At every actual quantum-processor read, all raw age, address, retained-word low bits and required tag bits agree with their canonical clean values unless a protected clock test fails at some rank. This covers a physical raw read, not just the value of \(\operatorname{dec}_{R}\).
\item Conditional on the dephased initial R1 row, the R1 marginal is a classical chain depending only on its own independent primitive innovations. The local terminal erase never erases or writes into the retained autonomous R1 banks. All private copies used as quantum controls belong to disposable or cut quantum work.
\item Concatenating the clean negative tower with the actual initialized history at date zero gives a native data history with no negative-time flags.
\end{enumerate}

\end{lemma}
\begin{proof}
A dephased finite-alphabet transition is a finite Boolean truth table. Its total truth table includes the parser and default convention just stated; canonical inputs produce the prescribed canonical output. Implement it on private initialized targets, with old and new banks separate, and use private destination copies for simultaneous stencil readers, as prescribed in Section~\ref{q1:model}. In a realization with no physical noise on the complete implementing stencil, deterministic evaluation is exactly that truth table, irrespective of what incoming legal or malformed strings were present. Every disagreement of the output decoder with the native transition is consequently contained in the event that this stencil has a flag.

The complete stencil includes: each incoming bank's storage from its preceding completed write through the final relevant read; all one-way fanout copies and their storage; every elementary gate of the total parser and Boolean evaluator; all transfers between old and new banks; every reset, the mandatory noise immediately after it, and each private target's storage before it is read. It also includes the raw source bits through each actual quantum-control read, including a directly copied raw low bit. There are only finitely many elementary layers and fixed spatial displacements in a physical period. Hence its size and its coordinate radii can be found by finite enumeration, giving \(s_{R},r_{x},r_{t}\) in Lemma \ref{q1:l1}. To remove an indexing ambiguity, \(\Sigma_{R}(x,t)\) is defined to include both the incoming physical round \(t-1\) and the outgoing round \(t\): this ensures that both the transition into native date \(t\) and every control read at date \(t\) are covered. At \(t=0\) retain only nonnegative elementary-time keys. Overlap between successive stencils is allowed and is precisely why Lemma \ref{q1:l1} uses a conflict graph rather than iid native flags.

Off \(\Sigma_{R}(x,t)\), the word delivered at the native boundary is canonical. If its parsed symbol differs from the clean native symbol, Lemma \ref{q1:l3}'s projection implication applies. If instead a delivered raw bank or read copy is noncanonical, or differs in a raw bit while its parsed symbol happens to agree, then some key of \(\Sigma_{R}(x,t)\) (evaluator, output write, storage or read copy) is flagged. The prescribed rank-one \(W_0\) containing \((x,t)\), including its seed and two work periods, contains this native flag point. Its required 0-sparsity therefore fails. If a different canonical valid symbol was delivered, its raw differing bits are already covered by the decoded-symbol case. These alternatives exhaust all raw read words. The rank-one raw-read test is one of the bounded fixed-offset tests used in Lemma \ref{q1:l3}; if necessary its identical translate is retained explicitly.

For closure, at a round boundary the retained autonomous banks are dephased. Their joint state with all other registers has a classical-quantum form \(\sum _{a} \lvert a\rangle\langle a\rvert\otimes\rho_{a}\). Autonomous updates and copies between autonomous banks are basis permutations or total classical channels built from the truth table. An autonomous bank is only a source of a one-way control into a private target; there is no quantum-target-to-autonomous write. For an arbitrary controlled target unitary, even with arbitrary correlations in \(\rho_{a}\), tracing out the targets leaves the probabilities of a unchanged. The same holds for a private copy followed by discard. A depolarization on a basis-valued autonomous bit is an independent classical replacement by a fair bit. Induction over elementary layers proves that the autonomous marginal is the stated classical chain. Private quantum controls can fault, but their faults are included in the read and dispatch stencil or in the direct hardware mark, and cannot inject an unflagged quantum result into that marginal.

The \(D\)-triggered erase acts on every retained quantum-capable factor, as specified in Section~\ref{q1:forgetting} but not on the autonomous dephased banks. An R1 schedule tape inside a quantum gadget is a different factor: it remains quantum-capable and is erased. This distinction is by named factors and one-way gate incidence, not by an intended classical interpretation. It establishes the claimed exclusion of the real autonomous R1 lane.

Finally the encoder supplies \(\operatorname{enc}_{R}\) of the exact time-zero boundary constructed in the proof of \cite[Lemma~\ref*{cl:lem:past}]{DMNoise1D2026}. Use the clean exact tower at all negative dates and the actual decoded trajectory at nonnegative dates. There is no discontinuity of the row at the join. Negative updates are exact, and positive decoded-update failures have the stencil witnesses just proved. Define \(E_{R}=0\) at negative native dates; no fictitious negative physical innovation is needed for this lane. This is a data history of precisely the type quantified in \cite[Proposition~\ref*{cl:prop:organization}]{DMNoise1D2026}. 
\end{proof}

\begin{lemma}[Initialized structural projection on the line]\label{q1:l3}

Choose once and for all a certified never-halting input to the R1 family, for example the index of a machine whose transition graph has no halting state. Use its homogeneous true family literal. Work on \(\Z\), not on a globally initialized finite ring. Let the clean negative tower be the exact all-live tower constructed in the proof of \cite[Lemma~\ref*{cl:lem:past}]{DMNoise1D2026}, at its aligned boundary zero, and use that boundary to initialize the actual compiled lane. Compare the clean tower for \(t<0\) and the actual trajectory for \(t\ge 0\), glued at the identical row zero, as in Lemma \ref{q1:l2}.

For each physical native point \((x,t)\) with \(t\ge 0\), failure of its full native symbol or any of its actual control-read bits to equal the clean trajectory implies failure of one of a deterministic collection of protected \(W_0\) tests. At rank \(j\) there are at most \(N_{\mathrm{clk}}\) tests. Each is a protected \(W_0\) of width \(<w_{R}P_{R}^{j}\) and height \(<w_{R}T_{R}^{j}\), so that Lemma~\ref{q1:l1} applies to it, requires \((j-1)\)-sparsity and has its entire key support within radii \(C_{\mathrm{clk}} P_{R}^{j},C_{\mathrm{clk}} T_{R}^{j}\) of the query, for fixed effective integers \(N_{\mathrm{clk}},C_{\mathrm{clk}}\). The symbol includes every mutable native field, scratch and the retained letter, including a letter that encodes another retained letter.

Since \(P_{R},T_{R}\ge 2\) (indeed \(P_R=5\cdot2^{w_R}\le T_R\), Section~\ref{q1:parameters}), one explicit conservative choice is
\[
\begin{gathered}
W_{*} = \lceil (2T_{R}+18P_{R})/(P_{R}-1)\rceil + 2, \\ N_{\mathrm{clk}} = W_{*} + 8, \\ C_{\mathrm{det}} = 100(w_{R}+W_{*}+T_{R}+P_{R}+1), \\ C_{\mathrm{clk}} = C_{\mathrm{det}}+r_{x}+r_{t}+2.
\end{gathered}
\]
The extra constant number of rank-one raw-read tests in Lemma \ref{q1:l2} is included in \(N_{\mathrm{clk}}\).

\end{lemma}
\par\noindent\emph{Proof idea.} If the parent symbols are clean and the protected two-period tests on the parent homes pass, the child homes restart from an exact clean row at the last ballot and then compute the native transition exactly up to the query date. Recursing on the parent queries reaches negative dates, where the clean tower applies. The contrapositive gives a failed test, and the recursion keeps the number of tests per rank and their radii bounded.

\begin{proof}
At rank \(k\) consider a contiguous interval \(J\) of served-symbol queries at a common rank-\(k\) date \(s\). A negative \(s\) uses only the fixed clean past: the total service lookahead is \(<T_{R}^{k}\) physical ticks by the lookahead bound \(b_k\le T_{R}^k-1\) following \cite[Definition~\ref*{cl:def:observations}]{DMNoise1D2026} (also step~4 of the proof of \cite[Lemma~\ref*{cl:lem:two-period}]{DMNoise1D2026}), so the observation of any rank-\(k\) point with \(s\le -1\) ends at a strictly negative physical date. These are the terminal queries of the recursion.

For \(s\ge 0\), set \(u=\lfloor s/T_{R}\rfloor\) and \(h=u-1\). Enlarge \(J\) by \(T_{R}+2\) rank-\(k\) sites at either end. Take all rank-\((k+1)\) homes meeting the enlarged interval and five more homes on each side. Query their rank-\((k+1)\) served symbols at the single date \(h\). This set supplies every sextuple required below. For each required central home, choose the protected rank-\((k+1)\) \(W_0\) of \cite[Lemma~\ref*{cl:lem:two-period}]{DMNoise1D2026} with that home in its central pair and its first work period starting at \(h\). Require \(k\)-sparsity on the entire \(W_0\). This is exactly the empty-top-exception variant in \cite[Lemma~\ref*{cl:lem:two-period}]{DMNoise1D2026}; lower-rank sparse defects are permitted, rather than requiring a large raw-empty rectangle.

Suppose the parent queries equal clean. Their sextuples are positive and in the clean common tribe, and the two prescribed upper outputs are positive. Steps 1--3 of the proof of \cite[Lemma~\ref*{cl:lem:two-period}]{DMNoise1D2026} apply. The first period starts at \(hT_{R}\) and ends at \((h+1)T_{R}=uT_{R}\). At that last ballot both central child homes have the exact retained packet at every position, correct ordinary addresses, zero age and all scratch defaults. The actual row is given by step 3 of that proof, with the ballot rule of \cite[Section~\ref*{cl:sec:verifier}]{DMNoise1D2026} and the full-home specialization of \cite[Lemma~\ref*{cl:lem:endpoint}]{DMNoise1D2026}; the clean identity \cite[Lemma~\ref*{cl:lem:clean}]{DMNoise1D2026} identifies the clean row as \(E_{j}\) with all scratch defaults; they agree field by field. Nominal acceptance can tolerate a common interval of up to \(2q\) bad positions at the first ballot and cannot be used as an exact bottom row.

The scalars \(\mathsf{Cn}\) and \(\mathsf{Temp}\), listed separately in \cite[Section~\ref*{cl:sec:native}]{DMNoise1D2026}, also take their clean values. The ballot clears scratch and returns the proper ordinary state (\cite[Section~\ref*{cl:sec:verifier}]{DMNoise1D2026} and the full-home specialization of \cite[Lemma~\ref*{cl:lem:endpoint}]{DMNoise1D2026}). Within a full proper home every Purge starts \(\mathsf{Cn}=1\), after which its grade-dependent evolution is uniquely determined by the ordinary address run, including the grade-one wrap. By step 3 of the proof of \cite[Lemma~\ref*{cl:lem:two-period}]{DMNoise1D2026}, during the first period the left neighbour of address zero of each central home lies in a full normalized home, so its grade is \(1 < 2\); after \(i\) Purge propagation ticks \(\mathsf{Cn}=[\mathrm{address} \ge i]\), as in the clean row. \(\mathsf{Temp}\) is reset every Main cycle (\cite[Section~\ref*{cl:sec:clock}]{DMNoise1D2026}) and no birth can replace an occupied incumbent. Thus any such scalar not included in the literal scratch clearing agrees with its clean value as well. \cite[Lemma~\ref*{cl:lem:clean}]{DMNoise1D2026} and its proof obtain the complete aligned blocks \(E_j\), with every scratch register reset. Hence the starting row is a full symbol row.

Step 4 of the proof of \cite[Lemma~\ref*{cl:lem:two-period}]{DMNoise1D2026} now keeps both central homes full and properly served at every date \(uT_{R},\dots,(u+1)T_{R}-1\), including both endpoints in space. At each such protected lower site, \cite[Lemma~\ref*{cl:lem:endpoint}]{DMNoise1D2026} and the protected exact computation of \cite[Proposition~\ref*{cl:prop:organization}]{DMNoise1D2026} say that its next served full symbol is exactly the native \(F\) on its three actual served inputs. The backward radius-one cone from \(J\) at \(s\) to \(uT_{R}\) has fewer than \(T_{R}\) native steps. Every home it meets was chosen as a central home of its own \(W_0\); the enlargement \(T_{R}+2\) supplies the triple and endpoint reserve. Induction from the exact bottom row at \(uT_{R}\) therefore gives the exact clean state on the requested interval \(J\) at date \(s\). The true family literal is spatially homogeneous, so all immutable operands in that induction are the clean ones. At rank zero the same assertion is just the unflagged physical native update. Served observations enter only the proof, not the circuit.

Recurse on the queried parent interval and date \(h\). The interval widths satisfy
\[
\begin{gathered}
W_{\mathrm{next}} \le  W/P_{R} + 2T_{R}/P_{R} + 18.
\end{gathered}
\]
Indeed the enlargement costs \(2T_{R}+4\) child sites, rounding to homes costs at most two homes and the extra neighbors cost ten; the displayed allowance dominates \(4/P_{R}+12\). The chosen \(W_{*}\) is an invariant upper bound starting from a single-site query. Hence the number of selected parent homes, and therefore \(W_0\) tests at every rank, is at most the asserted \(N_{\mathrm{clk}}\). Dates obey \(s_{\mathrm{next}}=\lfloor s/T_{R}\rfloor-1\), which becomes negative after finitely many iterations for every fixed \(s\).

The same recursion gives explicit finite radii. A lower-to-parent spatial displacement, expressed back in child units, is at most \(T_{R}+7P_{R}+W_{*}\). Summing its physical displacements over \(l=0,\dots,j-1\) gives at most \((T_{R}+7P_{R}+W_{*}) \sum _{l} P_{R}^{l}\). Temporal rounding and the subtraction of one cost \(<2T_{R}^{l+1}\) per promotion; the sum is \(<4T_{R}^{j}\). A \(W_0\) adds \(<w_{R} P_{R}^{j}\) and \(<w_{R} T_{R}^{j}\), together with only its fixed six-input and two-period offsets. For \(P_{R},T_{R}\ge 2\) these are dominated by \(C_{\mathrm{det}} P_{R}^{j},C_{\mathrm{det}} T_{R}^{j}\) with the displayed loose \(C_{\mathrm{det}}\). Finally enlarge by \(r_{x},r_{t}\) for the complete primitive stencils. This gives \(C_{\mathrm{clk}}\).

The construction of the test family is deterministic: select intervals, integer dates and aligned homes before inspecting the raw flags. If every test succeeds, the preceding implication inducts down from the negative terminal queries to the original query. Its contrapositive proves the decoded-symbol statement. Lemma \ref{q1:l2} proves the raw-read extension and adds only the already allowed fixed number of rank-one tests. 
\end{proof}

\begin{lemma}[Effective boundary and precise finite-ring use]\label{q1:l3.1}

Every finite restriction of the line's time-zero row is computable. On a finite ring, let an interval \(I\) be injectively embedded and initialized with the restriction of that row, and couple the ring's elementary keys in \(I\) with one infinite-line extension. Then every physical output whose entire actual past dependency cone lies in \(I\) equals the corresponding line output, so Lemma~\ref{q1:l3} applies to it with its tests evaluated in the line extension; it is not invoked for any other ring output. There is no assertion of a clean full-ring R1 tower or of correctness near arbitrary filler.

\end{lemma}
\begin{proof}
The proof of \cite[Lemma~\ref*{cl:lem:past}]{DMNoise1D2026} obtains the boundary independently of the arbitrary upper history: for an integer coordinate \(i\), repeated \(\lfloor i/P_{R}\rfloor\) reaches \(0\) or \(-1\), after which the next serialized digit is a true endpoint with its complete cell fixed independently of the upper live state. Recursively substitute that endpoint back down. This algorithm terminates for every coordinate and therefore for every finite interval. Compactness supplies a compatible clean negative tower for the proof, but no encoder computes or samples its arbitrary negative-time history.

For the finite ring, initialize this computable row on the interval in Section~\ref{q1:storage} and couple its elementary noise keys to a single infinite-line extension. Induct on actual elementary dependency edges. Gates and initial factors in the real past cone coincide, so the ring and line outputs coincide exactly there. For the R1 marginal, one native update has radius one; its macro-time cone has radius at most the number of elapsed native ticks. Any fixed compilation displacements and partial-period read cones are included in the assembly's \(C_{\mathrm{cone}}\). Thus the interval reserved there suffices.

Lemma \ref{q1:l3}'s auxiliary high-rank witnesses may extend beyond the injective interval; they are events in the analysis line extension used to bound the already coupled output. They are not extra ring randomness and no two lifts of a ring key are declared independent. Fillers, seams and nonmultiples of \(P_{R}\) therefore cause no application outside the lemma's line scope. 
\end{proof}

\textbf{Constants.} \(C_{\mathrm{det}}\) deliberately pays for \(T_{R}/P_{R}\), which can be large. The constants above make the bounded-test and radius constants of the structural projection explicit, and include the raw-stencil enlargement of Lemma~\ref{q1:l2}.

\section{A quantum schedule on the line}\label{q1:compiler}
This section builds a self-simulating quantum schedule driven by R1. Each level of the hierarchy runs a fixed concatenated-code gadget schedule whose gate at each site is computed from the addresses and ages supplied by R1 (Lemmas~\ref{q1:l8}--\ref{q1:l10}), and level reduction shows that good blocks implement their intended logical operations (Lemmas~\ref{q1:l4}--\ref{q1:l7}).

\paragraph{Primitive and layout.}

At a native tick a site performs one on-site member of a fixed finite real gate set on named slots, or prepares a classical bit on a named slot. Then all bonds swap the right port of their left site with the left port of their right site. The ports are distinct internal factors. The bit preparation reads the low bit of the site's current classical retained word \(X\) (take zero for a no-word symbol). The local gate choice \(d(a,s)\) uses the ordinary address \(a\) and age \(s\). Illegal inputs have a total default. All sites have the same finite number \(S\) of quantum slots. Fix \(S\) before choosing the code: allow storage, three contact and three dummy slots, reversible Boolean control and two exchange ports, and enough slots for one cell and two neighbor records of a fixed binary-encoded universal classical tape machine. A larger fixed \(S\) causes no difficulty. Arrays of growing length, repeated computations and quantum ancillas occupy more \emph{sites}. An on-site native gate is I, \(X, H, \operatorname{CNOT}, CZ, \operatorname{Toffoli}, \operatorname{SWAP}, R0\) or discard (reset). At physical level compile a controlled gate via Fredkin swaps to exclusive prepared dummy ports, fixed unconditional gate, swap back and discard. Every preparation and every wait gets its physical noise. The externally controlled \(\operatorname{LOAD}\) into a blank slot is a reset and classical copy. At all levels it is one of \(R0\) or \(X\) after \(R0\) for the given classical oracle data.

In the accurate clean R1 tower, consider an epoch of \(T\) child ticks on a home of \(P\) child sites. The children have addresses 0 through \(P-1\) and synchronous ages 0 through \(T-1\). Throughout this epoch their \(X\) letters carry the five copies of the packet encoding the \emph{current} parent state. This follows by the clean identity \cite[Lemma~\ref*{cl:lem:clean}]{DMNoise1D2026} at the bottom and retention of \(X\) while occupied until the next ballot (untargeted fields are retained, \cite[Section~\ref*{cl:sec:native}]{DMNoise1D2026}). Fix the explicit packet layout of Lemma~\ref{q1:l9} for the R1 construction: put each bounded scalar bit in the low bit of the word of a separate central-area live letter; store the low and high halves of a word in the two typed raw-letter slots of \cite[Section~\ref*{cl:sec:packets}]{DMNoise1D2026}. Give live raw payloads a word, with a fixed zero default for not applicable. This uses \(O(w)\) scalar letters, and fits the specified central area. The physical rule, all parsers and the simulated evaluators use that one serialization. For the ordinary parent address, age and low bit of its retained word, the addresses of the required \(O(w)\) low bits are known. The middle copy is read. The low bit of a retained word of twice the child's width is the low bit of its low half, independent of the rank. Where an extra tag is required to interpret a default, serialize its bounded bits in the same scalar area. These choices differ from the serialization of \cite[Section~\ref*{cl:sec:packets}]{DMNoise1D2026} only in where the scalar bits are stored inside the central area, and Lemma~\ref{q1:l9} checks that every classical statement used here holds for them; none of the quantum work is inserted into the classical hierarchy. Section~\ref{q1:clock} bounds full-symbol disagreements. Quantum level reduction below compares to the single clean schedule throughout. It never asks for noisy high-rank projected states during execution.

It suffices to implement the transition for \(S\) input qubits per parent, whatever the clean parent's \(a,s\) and bit. Pick a constant-distance binary CSS code from Lemmas~\ref{d:code}--\ref{d:reduction}. The distance is selected below. Within a colony store all \(S\) encoded blocks at fixed, disjoint coordinate locations on ordinary child data slots. Reserve distinct compact work locations for a catalogue containing the circuits for all on-site gates on all possible named slots, logical resets and a logical preparation of one. Preparation of one is zero preparation followed by transversal \(X\). Input ports are themselves complete encoded blocks. Allocate left-port coordinates to the left of right-port coordinates. The number of positions in the catalogue is a constant after the code and \(S\) are fixed. Choose \(w\) afterwards so these positions fit strictly inside a home, even after reserving the \(O(\mathrm{poly}(w))\) work area for the compiler. Unused slots are reset before first use.

\paragraph{Within one parent tick.}

All steps below consist of on-site native gates, one-bit LOADs and the fixed port swaps. To exchange the ordinary slots of two neighboring cells, at both ends swap those slots into the two facing ports in the same tick, then swap from the same ports back into the ordinary slots at both ends in the next tick. The two mandatory port exchanges implement exactly the requested swap on the data and a permutation on the old port contents. Unused ports merely bounce on their own bond. Exclusive assignments to disjoint links, or one link at a time, avoid conflicts. This implements nearest-neighbour permutations and local classical communications with a constant slowdown. Reset and control routines act on dedicated workspace. In particular a long permutation sends a fixed list of wire factors to known destinations; a fault during it touches at most the participating factors, irrespective of distance.

\begin{enumerate}[label=\arabic*.]
\item Normalize every incoming logical slot with repeated full \(\operatorname{EC}\) passes from Lemma~\ref{d:reduction}. Use fresh or reset work. Serializing the slots is allowed. Store each idle data coordinate on its own wire. Follow the normalizers by the complete EC-and-body for the selected instruction, including the idle case. Untouched slots retain their identity, with their own \(\operatorname{EC}\). Use \(b+1\) leading passes per block for the effective fault budget \(b\) specified below.
\item For a finite, fixed alphabet of possible parent instructions, concatenate all candidate EC-and-body contact lists. Only the selected list acts on the designated data and work. For each contact, prepare fresh control scratch, fetch the required \(O(w)\) parent-word bits, compute the parent's decoded instruction, compare it with this candidate, use its answer at this single contact, then discard all its private computation. Compute the parent's instruction even at contacts for an inactive candidate. When the decoded instruction is a bit \(\operatorname{LOAD}\), use the parent retained bit to choose the logical zero or one preparation. Normalizers need no candidate test. Routes to and from each contact are fixed permutations followed by their inverses, even for an inactive candidate. At the contact scatter up to three operands into dummy slots conditionally, perform the unconditioned candidate primitive there and gather them back. Reset private dummies and scratch. The inactive action is exactly the identity on occupied wires. In particular the controller does not make one decision and distribute it to a whole code block. The effect of one failed decision is confined to that one contact. Separate candidates use disjoint work or reset it before access.
\item Transfer the complete encoded \(R\) block of \(i\) to the \(L\) slot of \(i+1\) and vice versa by a fixed permutation. For each code coordinate in turn transpose the two corresponding positions by going forward along their line of adjacent transpositions and back along the prefix so only the endpoints exchange. All bonds between neighboring \emph{colonies} can do this simultaneously: for a given coordinate its left-port position lies strictly left of its right-port position within every home, so the intervals for consecutive colony boundaries are disjoint. Use a dedicated ordinary transport slot after moving the endpoints into it, then move them back from the transport slot at the end. No data-dependent gates follow this transport in the epoch. Padding is idle. On the quantum slots holding the encoding this is the mandatory parent port exchange. Each such token crosses at most one colony boundary. Records and inactive nonretained work are unconditionally discarded or reset at their declared round boundary. Exchange ports are retained factors at every level: a shuttle in flight spans two native ticks, hence two physical rounds. Their old contents are handled by \eqref{q1:e5.1}, the end-cell rule and, in the halting case, the \(D\)-erase of Lemma \ref{q1:l14}. No borrowed work is needed across a parent cut.
\end{enumerate}

The same construction implements one parent time step simultaneously for all parent cells. The on-site stages stay within their own homes. Only the final fixed-permutation stage needs the adjacent home, and every other initial slot stays in its own home. This spatial fact also bounds the support enlargement when the schedule is faulty.

\paragraph{A finite table that can compute its own indexed gate.}

The apparent circularity in step 2 is the following. The desired parent instruction \(d(a',s')\) is a gate of the very schedule now being built. A full table of \(P T\) entries need never be reproduced inside a home.

For clarity fix a universal binary classical tape machine once. Its synchronous tape-cell update uses only the old center and its nearest neighbors. A cell stores tape symbol and optional head and state; apply the finite transition truth table to construct the next row, reset old fields and exchange banks. For a malformed input fix an arbitrary total truth table. Implement a synchronous tape step with a fixed number of native substeps: fanout the two boundary copies of each old bit into dedicated slots, do left and right neighbor exchanges on the ports, compute the finite Boolean table in each cell with \(\operatorname{Toffoli}, \operatorname{CNOT}, X\) and fresh bits, then overwrite. Each neighbor exchange of the tape step is a complete paired shuttle of the two dedicated copy slots of one bond, in the sense of the exchange defined above (both cells swap their copy slot with the facing port in one tick and swap back in the next); bonds whose left cell has even and odd address use separate substeps. End-cell rule: the two end cells of a private tape interval perform no outward exchange. In the substep of an outward shuttle an end cell instead resets its outward copy slot and prepares in it the fixed blank record (blank symbol, no head, by \(R0\) followed by the fixed \(X\) pattern of the blank encoding). During a tape-update tile, an end cell's tape and neighbor-record routines never couple to its outward port. Separately scheduled installation, fetch, decision-return and operand routes use ports only through complete paired shuttles; their transient use of a tape endpoint's port does not supply an old port factor to the private decision logic. The same rule holds at every level. With a correctly initialized head this is a universal sequential computer. No clean dynamic classical side processor is assumed.

We use an indexed layout with a variable binary program \(v\). In the schedule for candidate contact \(i\) reserve these consecutive slots: clear the entire tape and its boundary on designated tracks, install the fixed prefix \((v,v)\), binary \(w\), the known hardwired parameters if needed and blank input positions, fetch the parent bits into those positions, run the universal cell-update tile for a fixed number of cycles, return the decision bit, act at contact \(i\), and clear and discard. A tagged invocation on \((v,v,w,a',s',i,\xi)\) returns the candidate selection bit specified below. Fetch the parent \(X\) bit in a fixed slot unconditionally; its value is used only when the returned instruction is \(\operatorname{LOAD}\). The catalogue, the contact number \(i\) and any candidate comparison are spatially fixed within that time slot. Allocate a tape interval of length proportional to the time bound plus the input length, with a blank guard: a valid single head moves at most one cell per step. A halting machine parks its output in a designated cell and remains idle. On any malformed result use the fixed default. The classical simulator controls only that contact's selection bit; the timetable of all contacts and routes is immutable.

For each of the \(O(w)\) hardwired bit positions allocate a slot of \(4P+O(1)\) adjacent swaps (ticks) with a bounded fixed slowdown. Load at the known source, move a single private bit along the segment to its appointed tape input, copy it, and return or reset transport. This provides random access to the serialized parent without using any read distinguished by rank. Reserve similar bounded linear intervals for all fixed gate routes and the final port exchange. The decision circuit proper repeats the same tape-update at all cells in the work interval. An implementation can pad each phase to a multiple of a common constant. The total duration and space are \(O_{v}(P\,\mathrm{poly}(w))\) and \(O_{v}(\mathrm{poly}(w))+O(1)\) active work sites plus the transport tracks respectively. The leading normalizers, code catalogue and number of conditional contacts are constants \emph{before \(w\) is chosen}. The literal size of these constants may be large.

To generate an indexed gate, a program \(L\) on input \(v,w,a,s\) locates a stage by comparisons and subtractions against the fixed finite stage list, computes any repeated fetch number from the quotient in its padded time slot, and decides the stage's gate at address \(a\). For a nearest-neighbour route the active pair is the pair at its elapsed hop number; the return traverses the segment in reverse. For a tape cycle the active update is specified by address membership in the tape interval and the residue of time modulo the fixed elementary tile. Installing \((v,v)\) uses bits of the provided literal. The bit position, phase lengths, tape endpoints and fixed catalogue offsets require only comparisons, arithmetic and quotient and remainder on \(O_{v}(w)\)-bit integers. No index query evaluates the simulated tape contents. All \(O(w)\) parent scalar coordinates can be laid out consecutively; one may instead scan their \(O(w)\) explicit formula list. Choose a polynomial machine-time allowance \(w^{c}\) for each valid call. To see that \(c\) can be selected without a circular bound, treat \(c\) as a binary constant in the indexed generator: computing \(w^{c}\) takes a constant (dependent on \(c\)) number of \(O_{c}(w)\)-bit arithmetic operations. Calculating, comparing or dividing indices into an interval of \(w^{c}\) identical tape cycles has the same polynomial bit complexity degree irrespective of \(c\). All other loops are over the fixed catalogue, the fixed program text or \(O(w)\) scalar bits. Schoolbook addition, multiplication and long division and a fixed single-tape simulation give a bound \(C_{v,c}(1+w)^{A}\) with absolute exponent A independent of \(c\). Choose \(c>A+2\), then \(w\) large enough to cover the constant and return the head and output. Make \(v\) the finite tape-machine text for this very indexed generator \(L\), with input \((v,w,a,s)\). The installed pair \((v,v)\) runs \(L\) on its own text. Thus the decision in a fault-free private computation is exactly the opcode \(L(v,w,a',s')\) defining the schedule on the parent. This is the elementary self-description construction with two copies of a program string. The length of \(v\) is independent of \(w\).

Age 0 is reserved as an idle on-site tick for every address. All other stages start after it. For large \(w, P=5\cdot 2^{w}\) dominates every fixed power of \(w\) and \(T\ge \lfloor 3^{w}/w^{2}\rfloor\) dominates \(C P\) poly(\(w\)). A terminating search chooses \(w\) satisfying these inequalities and the R1 bounds. Complete unused dates by idles. The clean parent operation at the aligned initial top tick is an identity on all nonport slots, with the fixed permutation of ports. In particular it preserves a designated nonport qubit. The physical rule hardwires the finished finite lookup or evaluates it by a fixed finite Boolean circuit each round.

\begin{lemma}[Parametric complete gadgets and level reduction]\label{q1:l4}

Fix any integer \(b\ge 1\) and any member \([[n_{s},1,2q_{s}+1]]\) of the particular binary CSS family of Lemma~\ref{d:code}. In the one-dimensional theorem the member is the least \(s\) satisfying \eqref{q1:e4.1} with \(b=b_{\mathrm{fault}}\) of \eqref{q1:e7.2} (Section~\ref{q1:parameters}, row \(s_{\mathrm{code}}\)); the member fixed by \eqref{d:capacity} is not used. Use its complete \(\operatorname{EC}\) and native real-gate bodies of Lemma~\ref{d:gadgets}, not an arbitrary code having the same distance. Suppose

\begin{equation}\label{q1:e4.1}
\begin{gathered}
q_{s}-8 \ge  8(20b+8) = 160b+64.
\end{gathered}
\end{equation}

Precede the complete \(\operatorname{EC}\)-and-body for every block, including an idle block, by \(b+1\) time-disjoint full \(\operatorname{EC}\) passes, with fresh or reset work. Synchronize these passes before interblock bodies. A routed implementation is admissible if deleting its fixed permutations and auxiliary decision circuits converts its faulty execution into the complete native circuit with at most \(b\) faulty gate, wait or reset locations, with arbitrary joint environment coefficients. Lemma \ref{q1:l5} proves this premise for the present compiler.

In this lemma call an admissible rectangle and its block good.
At higher levels the test \(N_k=0\) of Lemma~\ref{q1:l7}
guarantees admissibility. For any good rectangle:

\((a)\) On a correctable incoming block with at most \(9b\) deviations, its intended logical instruction is implemented, and the output deviates on at most \(9b\) coordinates per block, including terminal waits and transports counted in the \(b\)-location budget.

\((b)\) On an arbitrary incoming block, the output is in the \(9b\)-correctable span of some logical input selected by its normalizations. This assertion is simultaneous on all good blocks, with arbitrary bad-block and reference entanglement.

\((c)\) At the common cut, the same fixed recovery reduces all good rectangles to their intended instructions. All other actions can be expanded as joint faults supported on the named bad outgoing wires, with their fixed port-permutation images. This operation can be iterated through any finite hierarchy; the decoder does not depend on the fault history.

All counts are independent of the number of locations, the route lengths, \(b\), and \(s\) except where displayed. In particular the numerical budgets \eqref{d:budgets} and \eqref{d:capacity} for the three-dimensional geometry are not hypotheses.

\end{lemma}
\par\noindent\emph{Proof idea.} The gadget lemma of Section~\ref{sec:quantum} is already stated for every code member and every fault count, so only its numerical margins need checking. Among the \(b+1\) leading error-correction passes on a block at least one is fault-free and maps any input into the code; every later pass stays within the margin \eqref{q1:e4.3} and preserves the logical input. Expanding joint faults in matrix bases gives one fixed recovery for all fault histories, which iterates through the hierarchy.

\begin{proof}

Record explicitly the parametric content of Lemma~\ref{d:gadgets}. For its stated code, a complete routed \(\operatorname{EC}\)-and-body having \(j\) faulty locations and a correctable input with at most \(f\) deviations has output in the span of its ideal output with at most \(8j\) deviations if

\begin{equation}\label{q1:e4.2}
\begin{gathered}
8(f+j) \le  q_{s}-8.
\end{gathered}
\end{equation}

Faults can be correlated with every exterior system. Fault-free \(\operatorname{EC}\) maps the entire input Hilbert space into the code. These are exactly the conclusions of Lemma~\ref{d:gadgets}, whose statement is already quantified over \(f,j\) and every \(s\). The proof does not substitute a geometric budget for \(f\) or \(j\).

For clarity, the following is the parameter check of the proof, grouped only into consecutive proof paragraphs.

\begin{itemize}
\item The opening of the proof of Lemma~\ref{d:gadgets} defines gate, wait and reset locations and private single-coordinate decisions. It has no geometry or budget choice.
\item Its part on deterministic verified preparation constructs \(V_{t}\) and proves its fault-count induction for \(k\le t\). Set \(t=q_{s}\). The sentence there comparing the depth with the rectangle budget \(2b_{\mathrm{FT}}\) requires only \(q_{s}\ge 2b\), which follows from \eqref{q1:e4.1}. The recursive preparation size \(5^{q_{s}}\) affects time and location count, not the per-fault support assertion.
\item Its part on coherent error correction proves bit and phase correction and arbitrary-input normalization. Its uses of unique decoding are precisely the margin \eqref{q1:e4.2}; its private radius indicators and swaps reach one coordinate. No \(b_{\mathrm{FT}}\) appears.
\item Its part on the coherent Toffoli prepares the \(\operatorname{Toffoli}\) resource with \(R_{\mathrm{cat}}=2 \lfloor q_{s}/2\rfloor+1\) tests. The only numerical requirement is that the native-fault allowance be at most \(\lfloor q_{s}/2\rfloor\), and that the correction margins hold. Under \eqref{q1:e4.2}, its part on routing and the numerical envelope gives \(f+2(j+1)\le q_{s}/4\le \lfloor q_{s}/2\rfloor\), so both requirements hold. In our application \(j\le b\) and \eqref{q1:e4.1} also gives \(q_{s}\ge 2b\) directly.
\item The same Toffoli part proves the resource conversion and its support bounds coordinatewise, for arbitrary complex input amplitudes. Its gates are exactly the listed real native gates, independent of any schedule geometry.
\item Its part on routing and the numerical envelope deletes routing by commuting product-basis fault factors through permutations; a two-token routing fault costs at most two native faults. Its envelope is \eqref{q1:e4.2}, and the contact bound is per fault. For an alternative routing one must prove the deletion premise rather than reuse the number two. We do so in Lemma \ref{q1:l5} and count its resulting native locations in \(b\). The conservative \(8j\) bound remains valid for an unrouted circuit with \(j\) native faults (equivalently take routes of length zero). The formal circuit produced by Lemma \ref{q1:l5} executes the native gates of Lemma~\ref{d:gadgets} one per formal step, in a fixed linear order extending the gate order of Lemma~\ref{d:gadgets} (gates sharing a wire keep their stated order), with a native wait location on every other token at every step. The proof of Lemma~\ref{d:gadgets} attributes each fault to one location and follows it only through later gates on the same wires; it never uses the simultaneity of the gates of one transversal step. Serialization therefore only adds wait locations, each a legal native location.
\item Its final paragraph proves the operator-span assertions coefficientwise and uses one recovery for the entire span. Thus arbitrary joint Kraus coefficients, references, and correlations between different faulty locations are permitted.
\end{itemize}

Now prove the reduction without the three-dimensional numerical substitutions. Among \(b+1\) leading passes on a block, at least one contains no touching native fault. Its output has code support even on arbitrary input, by the normalization clause of Lemma~\ref{d:gadgets}. Choose the last such pass. After that pass, all subsequent ECs and the complete body preserve the selected logical subsystem. Indeed before any such \(\operatorname{EC}\) or body there are at most \(8b\) residual coordinates from the preceding complete operation and at most \(b\) from intervening waits and transport, or at most \(9b\). Even allowing a further \(b\) coordinates at the cut before this count is taken gives \(f\le 10b\). With \(j\le b\),

\begin{equation}\label{q1:e4.3}
\begin{gathered}
8(f+j) \le  88b \le  160b+64 \le  q_{s}-8.
\end{gathered}
\end{equation}

Applying Lemma~\ref{d:gadgets} afresh leaves at most \(8j\le 8b\) coordinates; any remaining terminal waiting and permutation faults add at most \(b\), giving \(9b\). Starting from the clean pass, this proves \((b)\) by induction over the finite pass list. Starting instead from a \(9b\)-correctable incoming state proves \((a)\), because every pass is inside \eqref{q1:e4.3} and hence preserves its original logical input. The clean pass need not be at the same time on different blocks.

The joint assertion is not a collection of state-dependent recoveries. Expand every shared fault in a tensor matrix basis on the good block factors and retain every other system as its coefficient. For each term, Lemma~\ref{d:gadgets} fixes the ideal logical map and puts the good output in the same correctable span. Consequently every complementary projection to that span has zero expectation on the positive joint output. The joint output therefore lies in the tensor product of those spans. In the fixed stabilizer-recovery coordinates the good logical operation is the intended operation, with a common coefficient on the bad inputs, syndromes and environment; off-diagonal logical matrix elements have that same coefficient. For an intended gate involving a bad operand, this means the intended gate on the actual recovered bad input, not identity on the good operand of that gate.

At a common cut, absorb the next normalization of each preceding bad output into the arbitrary channel on that named wire, before applying the next interblock body. This composition has no access to any additional good logical input in recovery coordinates. Expanding it in matrix bases on the already named wires adds no wire. The good bodies then implement the intended gates on those logical wires. The fixed final port permutation merely permutes their names. Repeat in the finite directed circuit. This is the argument of Lemma~\ref{d:reduction} with \(r_{\mathrm{FT}}=8b\) replaced by the conservative carried allowance \(9b\) and with \eqref{q1:e4.3} in place of its \(72b\) inequality.

Finally recursively recover every child. At most \(9b\) child-coordinate operators in a good parent become arbitrary operators on at most \(9b\) parent coordinates; the rest recover ideally. Since \(9b\le q_{s}\), the one recovery of Lemma~\ref{d:code} corrects their whole union span. Induction gives a fixed bottom-up recovery for every allowed history, exactly as in the last paragraph of the proof of Lemma~\ref{d:reduction}. This proves \((c)\) and completes the parametric check. No step used a lattice dimension, a geometry-specific budget, or a bound on the total number of locations. 
\end{proof}

\begin{lemma}[Deletion, fault paths, and boundary shuttles]\label{q1:l5}

The compiler's fixed timetable consists of all candidate contact lists, a freshly reset private decision before each contact, fixed routes and inverse routes, conditional scatter to exclusive dummy factors, one unconditional primitive on the dummies, gather, and disposal. A private decision has no protected-data access except its one contact, of arity at most three. The tape, fetch and decision-return tracks have internal guard cells, and the private tape obeys the end-cell rule of the tape step: its two end cells never take a factor from their outward ports and use the constant blank record instead. Every other inter-site transfer of a private stage (installation, fetch, record exchange between tape cells, decision return, operand routing) is a complete paired shuttle. Outward home-boundary ports are inaccessible except during the paired boundary crossings of the terminal exchange, which are two crossings of two ticks each per transported code coordinate. The final exchange uses dedicated transport and is followed by no data-dependent operation.

\((i)\) With clean classical oracle reads and no faults, one compiled epoch on its designated encoded \(S\) input slots equals: \(b+1\) full \(\operatorname{EC}\) normalizers on every slot; the selected complete \(\operatorname{EC}\)-and-body with \(\operatorname{EC}\) also on idle slots; waits; and the prescribed parent port permutation \(E\). Other retained work is discarded or reset at its declared end. Old microscopic port contents undergo a separate permutation and never become logical data or input of a private machine. This equivalence holds on the entire input algebra, with arbitrary reference and old-port entanglement.

(ii) A finite set \(F\) of named faulty child ticks converts, upon deletion of auxiliary routines and fixed routing, to joint arbitrary faults at at most \(K|F|\) native gate, wait or reset locations of that circuit, where

\[
\begin{gathered}
K = 100000(S+1)^{2}.
\end{gathered}
\]

The bound counts location insertions, not the duration of a private computation or route. It is independent of code size, tape length, route length and hierarchy rank. Native waits at every interval without a prescribed contact are included as legal fault locations. Faults on final exchange coordinates may be placed immediately before the ideal \(E\); their \(E\) images are named as well.

\end{lemma}
\par\noindent\emph{Proof idea.} A paired shuttle is an exact product of two swaps \eqref{q1:e5.1}, so routes are permutations that commute past faults, and old port contents never mix with data. A faulty tick can touch only formal-circuit tokens, the one pending private computation, the operands of one contact, or unused ports. Charging each case a bounded number of native locations gives the constant \(K\), independent of route length and computation time.

\begin{proof}

Write the factors on one bond as \((o_{x},R_{x},L_{x+1},o_{x+1})=(A,p,q,B)\). In the first tick swap the two ordinary factors into their facing ports, then perform the mandatory port exchange; the tuple becomes \((p,B,A,q)\). In the second tick swap back and exchange ports again; it becomes

\begin{equation}\label{q1:e5.1}
\begin{gathered}
(B,q,p,A).
\end{gathered}
\end{equation}

Thus a paired shuttle is \(\operatorname{SWAP}(A,B)\otimes\operatorname{SWAP}(p,q)\). This is an identity of permutation unitaries, not an argument restricted to product or blank ports. Extra mandatory exchanges on unused port pairs are likewise permutations only of those unused factors. A word of complete shuttles implements its specified ordinary-factor permutation, tensored with a permutation of unused ports. The inverse route cancels the ordinary-factor routing away from its contact.

At a contact the fault-free private machine returns the deterministic selection bit for the current candidate, and the bit requested by \(\operatorname{LOAD}\) when relevant (Lemma \ref{q1:l6} gives the recursive clean-oracle justification). This holds for arbitrary old port contents. Every inter-site transfer of the private machine (installation, fetch, record exchange between tape cells, decision return) is a complete paired shuttle, which by \eqref{q1:e5.1} exchanges the two moving factors and leaves the old port factors in the ports. The two end cells read the constant blank record instead of a port, by the end-cell rule. Every other factor the machine reads is reset in phase 1, prepared by the installation, or written by a fault-free \(\operatorname{LOAD}\). Hence the tape evolves exactly as the reset run of \(U_{\mathrm{cpu}}\) on its installed and fetched input, whose bits are the clean ones by Lemma \ref{q1:l6}, independently of every old port factor and of the reference. If unselected, scatter and gather leave every occupied formal-circuit factor untouched: the primitive acts only on reset dummies. If selected, scatter moves exactly its at most three operands to those dummies, the primitive acts on them, and gather returns its outputs. This statement includes reset or discard primitives: an active reset resets the intended operand; an inactive reset resets only an unused dummy. Each private machine and each candidate's reused work is reset before use and disposed of at the declared end. The selected candidate therefore gives its exact contact list; all unselected lists reduce to identities on the formal circuit. The unconditional normalizers precede that list. \(\operatorname{EC}\) for an idle slot is included in the chosen complete list, so idle slots are normalized unconditionally as a semantic action.

For terminal port exchange, fix one code coordinate \(j\). In home \(i\) let \(l_{j}(i)<r_{j}(i)\) be its left and right endpoints. Intervals \([r_{j}(i),l_{j}(i+1)]\) for consecutive boundaries are disjoint. Along each interval use the forward adjacent-transposition word followed by the reverse prefix. This transposes its endpoints and restores every intermediate ordinary transport factor. Running all boundaries in parallel is legal because those intervals are disjoint. Do the coordinates successively. Moving endpoints into and out of the dedicated transport slot gives exactly the encoded \(E\). All other ordinary factors stay in their home. By \eqref{q1:e5.1}, microscopic old ports remain a separate permutation. Tracing out or resetting declared unused work composes the same channel on both sides of the equality. This proves an exact channel identity with a reference, not only equality of classical configurations.

\paragraph{Fault deletion and the incidence bound.}

Consider first one named faulty child tick. At the on-site cut it is an arbitrary joint \(CP\) map on \(S\) factors at one site. Conjugating by the mandatory \(E\) puts its outgoing factors in at most three neighboring sites. It is harmless to overcount these as \(S\) factors at each of those three sites. For a joint set of faults, choose local matrix bases on these factors in each Kraus operator and retain arbitrary operators on the reference, other faults and environment as common coefficients. We may follow each resulting formal-circuit token separately without asserting probabilistic independence.

There are only the following destinations for a factor touched by the named tick.

\begin{enumerate}[label=\textup{Case \arabic*.},leftmargin=*]
\item \textbf{An occupied formal-circuit token, a native gadget record, or a transport token.} Between prescribed contacts all operations on that token are fixed permutations or identities. Conjugate the factor through those permutations and insert it immediately before its next prescribed native use, or at a native wait if it has no further use. Its physical path can have arbitrary length. We do not commute it through an active contact. If two factors' next uses are at different dates, insert their factors there with the same environment coefficient; intervening operations on disjoint wires commute. If a token is disposed of before any use, its factor can instead be absorbed into the disposal coefficient.
\item \textbf{A private instruction-decoding computation.} Its clear, fetch, tape evolution, result return and disposal touch protected data only at the one pending contact. Its spatial and temporal gate list is immutable: tape contents select one answer bit and never a new route, address, target, or execution date. Guard cells exclude the outward ports and other work intervals. Replacing its complete faulty evolution by its induced map on the contact operands therefore costs at most three native locations. This is true for a coherent or entangled controller as well as for an incorrect classical answer. The next private computation begins with a fresh reset of the entire tape, boundary, return and control scratch. Garbage that the named tick leaves outside that reset set cannot reach the next computation either. A factor left in a port, including the outward port pair at either end of the tape, is never read by a private stage: every inter-site transfer there is a complete paired shuttle, which by \eqref{q1:e5.1} keeps old port factors in the ports, and the end cells use the constant blank record instead of their outward port. A factor left in an intermediate fetch, return or transport slot is only permuted by later routes and is never copied, since a moving bit's own slot is prepared by its \(\operatorname{LOAD}\) or installation. Hence no erroneous private state survives into the next computation, and each named tick spoils at most the computation pending at its date in each home it meets.
\item \textbf{Scatter, primitive, or gather.} All access is confined to the contact's at most three formal operands and its private dummy and control work. Tracing out or resetting the latter gives an arbitrary \(CP\) map on those operands. A data token stranded in a dummy and subsequently reset is included in that map. This again costs at most three locations, or one faulty arity-three native contact if its operands share that contact.
\item \textbf{An unused port factor.} A complete clean shuttle has the product form \eqref{q1:e5.1}. Hence the factor stays on unused ports and cannot infect a later ordinary token through a clean shuttle. If its tick is itself one of the faulty shuttle or contact ticks, count the ordinary factors involved in that faulty tick under cases 1 or 3. No length-dependent recurrence of infection is present.
\end{enumerate}

We count directly. A named tick is an arbitrary joint map on the \(S\) factors of one site before \(E\). \(E\) moves at most two of them, its ports, to the neighbouring sites, where they remain the same factors. Classify each of the \(S\) factors by its role at that date.
\begin{enumerate}[label=\textup{(\alph*)}]
\item A formal-circuit token (data, intrinsic gadget record or work, terminal-transport token): case 1 charges it at one native location, at most \(S\) in all.
\item A wire of a private computation (tape cell or record, fetch or return bit, control, dummy, input scratch): it belongs to the computation pending at this date in the home of the site. A port image entering a neighbouring home lands on an outward home-boundary port, which no private stage reads, so at most one computation is affected. Charging two is conservative, at three locations each by case 2.
\item Operands at the pending contact (in transit, scattered or stranded): at most three, by case 3. They are already counted under \((a)\) or \((b)\); add three more.
\item An unused port factor: no charge, by case 4.
\end{enumerate}
Hence one named tick costs at most \(S+6+3\) native locations, and

\begin{equation}\label{q1:e5.2}
\begin{gathered}
S+9 \le  10(S+1) \le  100000(S+1)^{2}.
\end{gathered}
\end{equation}

Budget check. Suppose the end-cell rule were weakened, so that garbage left by a named tick could be read by one further private computation. That computation's contact would add at most three native locations. The direct count above becomes \(S+12 \le  13(S+1)\). Even the cruder count, which charges \(3S\) factor lifetimes at the three post-\(E\) sites, three operands for each of nine possibly incident pending computations, and a factor \(12\) for pre- and post-\(E\) placement, both ends of a contact and the three possible named \(E\) images, becomes \(12(3S+27+3) = 36S+360 \le  360(S+1)\). Both are below \(K = 100000(S+1)^{2}\), so \(b = K(m_{\mathrm{bad}}+1)\) and every later constant are unchanged. If the further computation lay in the next parent tick, the named tick would belong to that parent's cover through the unit temporal collar of Lemma \ref{q1:l7}, so the charge would still fall on a counted named child. Under the end-cell rule, case 2 shows that this further computation does not occur.

This overcount includes all reset and wait placements. It uses no number of contacts in a list or number of gates in a private computation. The distinction is essential: an intrinsic gadget's record is a formal-circuit token and is charged at its next native gate; the compiler's additional instruction computer is case 2 and has only one contact. The proof of Lemma~\ref{d:gadgets} handles any later propagation of an intrinsic record fault.

Repeat for all named ticks and take the union of inserted locations. Coincidences only reduce its cardinality. The matrix-basis expansion retains all joint coefficients, so several inserted factors can represent one correlated map without increasing support. This proves (ii) in home interiors.

\paragraph{Boundary and arbitrary neighbors.}

A good home's internal stages never couple its outward port to data or work. An arbitrarily driven neighboring port therefore remains inaccessible before terminal transport. During transport, the good half of a forward-and-reverse endpoint-transposition path moves its endpoint to the boundary, then later brings the received endpoint back. Its internal factors traverse the route and its inverse. Their restoration is independent of the received endpoint's value: the two boundary crossings can supply an arbitrary endpoint channel and arbitrary port garbage, but the only final formal factor in the good home exposed to that channel is its exchanged endpoint. This follows by writing the adjacent transpositions as a permutation with the boundary endpoint left as an unspecified input and output wire; every internal intermediate wire connects back to itself. No following data-dependent gate reads transport before its disposal.

Thus a wholly bad home and its two incident transmitted ports can be named as arbitrary jointly. A faulty elementary boundary shuttle may touch both endpoints; charge it in both corresponding covers, which is already allowed in \eqref{q1:e5.2}. Lemma \ref{q1:l7}'s count is of these named children, with all routes and \(E\) images included. Old ports and initially arbitrary excluded scratch are either separated by \eqref{q1:e5.1}, never read by any private stage (end-cell rule), or reset before first use, so the argument also applies after a wholly bad preceding epoch. This completes the boundary case and the proof.
\end{proof}

The constant \(K\) is deliberately loose; by \eqref{q1:e5.2} it exceeds the direct count by a large factor. The words ``work returned to reset'' refer to declared disposable work after its actual discard or reset; syndrome work may depend on its input before that disposal. Designated encoded parent ports are protected logical slots and are not disposable work.

\begin{lemma}[Base interface and the clean-oracle induction]\label{q1:l6}

Fix the single initialized clean nonhalting R1 tower used by Section~\ref{q1:clock}. Denote its complete rank-\(k\) state at native site and date \((x,t)\) by \(c_{k}(x,t)\), its ordinary address and age by \(a_{k},s_{k}\), and the low bit of its retained word by \(\xi_{k}\), with the compiler's total zero-or-default convention. Define the reference native instruction

\[
\begin{gathered}
g_{k}(x,t) = d(a_{k}(x,t),s_{k}(x,t)),
\end{gathered}
\]

where an opcode \(\operatorname{LOAD}\) prepares \(\xi_{k}(x,t)\), and is therefore \(R0\) or \(X\) after \(R0\). This reference instruction is independent of the quantum input and of the actual noisy high-rank projected classical history.

At level zero, enlarge the named error set \(B_{0}\) to include every physical quantum fault stencil, every terminal selector cut in the nonhalting comparison, and every clock mark of Lemma \ref{q1:l3} centred at a physical instruction read. Then the actual tick equals the clean reference on-site instruction followed by \(E\) outside \(B_{0}\); at \(B_{0}\) it may be an arbitrary joint map on the named \(S\) slots and its \(E\) images.

Let \(N_{0}\) be the indicator of \(B_{0}\), define \(N_{k}\) for \(k\ge1\) by \eqref{q1:e7.1}, and call a level-\(k\) rectangle at \(z\) good if \(N_{k}(z)=0\). Then for every rank \(k\) of the finite hierarchy the following invariant \(H_{k}\) holds: in fixed recovery coordinates, every good level-\(k\) rectangle implements the native instruction \(g_{k}\) on its actual logical inputs, including the bit \(\xi_{k}\) for \(\operatorname{LOAD}\). Its output lies in the common correctable span. Arbitrary joint faults occur only on named bad level-\(k\) sites and their \(E\) images. This holds jointly with all bad subsystems, syndromes and reference. Hence the compiler never relies on an unproved accuracy assertion for a noisy high-rank oracle.

\end{lemma}
\begin{proof}

The classical-lane interface is Lemma \ref{q1:l2}. In particular every physical instruction uses only raw bits of its own old bank (address, age, retained-word low bit and tags); by Lemma \ref{q1:l2}(ii) they equal the canonical clean bits outside clock marks. One-way classical controls do not write back into that lane. Its evolution is an autonomous dephased stochastic classical process, even for a reference-entangled quantum input. A wrong control symbol can change the instruction at that site and date, but cannot change the hardware's list of named local slots or its bounded \(E\) images. By the full-state clock projection, any such wrong read is inside a failed-test mark. A wrong raw read that happens to give the same instruction can harmlessly be marked too. A raw gate, reset or wait fault and a terminal cut are likewise arbitrary maps on the named slots. Aggregating the finitely many elementary sublayers of the native tick into the base stencil therefore gives exactly the stated level-zero model, on the full algebra. A persistent erroneous word is charged at every tick when it is read; no one-use assumption is made.

The oracle linkage at the next rank is as follows. In the clean R1 boundary encoding \(E_{k}(c_{k+1})\), every child home has its full retained packet \(X=\delta_{k}(c_{k+1})^{5}\), ordinary addresses, aligned ages and default scratch. This is the complete post-ballot statement given by the ballot rule of \cite[Section~\ref*{cl:sec:verifier}]{DMNoise1D2026} and the clean identity \cite[Lemma~\ref*{cl:lem:clean}]{DMNoise1D2026}. Apart from declared writes, the retained \(X\) is held (untargeted fields are retained, \cite[Section~\ref*{cl:sec:native}]{DMNoise1D2026}); none of the pre-vote mechanisms replaces a clean incumbent (proof of \cite[Lemma~\ref*{cl:lem:clean}]{DMNoise1D2026}). Consequently throughout the following parent epoch every child's retained packet refers to the current state at the bottom of that parent epoch, which is precisely the state whose one-step instruction is being simulated.

The middle packet copy and the layout of Lemma \ref{q1:l9} are used. The parent's ordinary address, age and applicable tags are determined by a fixed \(O(w)\)-list of child-letter low bits. Each bounded scalar bit is one such letter's low bit. The parent retained word is split into its low and high child-width halves. Its lowest bit is the lowest bit of the low-half letter, at the same packet position at every rank. The tags and zero defaults are carried in the same scalar layout; thus this statement remains true when a represented field has no applicable word. No child is asked to obtain an entire word or a rank-dependent high-order bit from a single \(\operatorname{LOAD}\).

Assume \(H_{k}\). During a good level-\((k+1)\) epoch, reduce its child computation using \(H_{k}\). The fault-free part of every child \(\operatorname{LOAD}\) prepares exactly the low bit of that child's clean rank-\(k\) retained letter. Therefore its fixed fetch routes supply precisely the scalar bits of \(c_{k+1}\) and the bit \(\xi_{k+1}\). A good lower \(\operatorname{LOAD}\) does not read an actual noisy rank-\((k+1)\) projection. If a child rectangle is bad, its \(\operatorname{LOAD}\) may instead be arbitrary, and is already a named child fault; Lemma \ref{q1:l5} charges its influence to the one pending contact. This distinction also covers private computations that are wrong because of data corrupted earlier in their current lifetime.

On the ideal reduced child circuit, the fault-free private program returns exactly \(d(a_{k+1},s_{k+1})\) by the self-description construction and Lemma \ref{q1:l8}. It uses the fetched \(\xi_{k+1}\) only for the \(\operatorname{LOAD}\) candidate. Lemma \ref{q1:l5} deletes the unused candidates, private computations and routing. Lemma \ref{q1:l7} bounds the resulting effective native fault count in a good parent by \(b\). Lemma \ref{q1:l4} now gives the intended logical operation with the same clean oracle bit and a common correctable output span, jointly with all external systems. At bad parents retain a joint arbitrary map on their named outputs and \(E\) images. This is \(H_{k+1}\).

The base \(H_{0}\) is the physical model already established; a physical site itself has no code to recover. Induction on the finite hierarchy proves every \(H_{k}\) used in the finite chart. This is an implication from one rank to the next, not a simultaneous assumption that all noisy projections are correct. Correctness of the classical oracle and correctness of the quantum data are separate: the reference oracle is clean, whereas an actual logical input inherited from a bad block can be arbitrary and correlated. The intended instruction acts on those actual logical inputs as asserted by Lemma \ref{q1:l4}. 
\end{proof}

\begin{lemma}[Named-child counts, budgets, and separated descendants]\label{q1:l7}

A named level-\(i\) location is the site and tick before its mandatory \(E\), with all its \(S\) slots allowed to be faulty and its outgoing \(E\) images retained. Let \(Q_{i+1}(z)\) be the full child-location cover of the parent instruction at \(z\): its home, complete two terminal transport paths, every participating shuttle microgate, and all immediate pre- and post-\(E\) port neighbors. Enlarge to whole adjacent homes and a unit temporal collar, namely the same child sites during the preceding and the following parent tick; its direct halfheight is at most \((1/2+1)T^{i+1} \le  2T^{i+1}\), as used in \eqref{q1:e7.4}. Set

\begin{equation}\label{q1:e7.1}
\begin{gathered}
C_{i+1}(z) = \cup _{\delta=-1,0,1} Q_{i+1}(z+\delta). \\ N_{i+1}(z) = \one[ \#\{y \in  C_{i+1}(z): N_{i}(y)=1\} > m_{\mathrm{bad}} ].
\end{gathered}
\end{equation}

The base \(N_{0}\) is the indicator of the level-zero named set \(B_{0}\) of Lemma~\ref{q1:l6}. Repeated membership is counted once. Counts in \eqref{q1:e7.1} are counts of named children \(N_{i}\). An overloaded basic cover means \(Q_{i+1}(z)\) itself has more than \(m_{\mathrm{bad}}\) named children. Its center and immediate neighbors are all named by \eqref{q1:e7.1}. The converse is neither asserted nor required; the union test is strictly conservative.

With \(h_{\mathrm{cov}}=10, m_{\mathrm{bad}}=(10h_{\mathrm{cov}}+40)^{2}=19600\) and \(K=100000(S+1)^{2}\), set exactly

\begin{equation}\label{q1:e7.2}
\begin{gathered}
b=b_{\mathrm{fault}}=K(m_{\mathrm{bad}}+1).
\end{gathered}
\end{equation}

Assume \(P,T\ge 4\) and \(T\ge P\). Then every good parent \((N_{i+1}=0)\) has at most \(b\) effective native faults. Its complete descendants' footprint has coordinate halfwidths at most \(h_{\mathrm{cov}} P^{i+1}, h_{\mathrm{cov}} T^{i+1}\). A bad parent contains two named children whose complete descendant footprints have max-norm separation at least \(8P^{i}\). The number of possible children in a complete parent cover is bounded, for example, by

\begin{equation}\label{q1:e7.3}
\begin{gathered}
A_0=(20P+5)(20T+5).
\end{gathered}
\end{equation}

This \(A_0\) is a harmless explicit upper bound for the \(O(PT)\) constant in the probability recurrence.

\end{lemma}
\par\noindent\emph{Proof idea.} A good parent has at most \(m_{\mathrm{bad}}\) named children in its enlarged cover, each costing at most \(K\) native faults, which gives the budget \(b\). A bad parent has more than \(m_{\mathrm{bad}}\) named children in its cover; since fewer than \(m_{\mathrm{bad}}\) centers lie within a bounded index distance of any one of them, two of them have far-apart footprints.

\begin{proof}

If \(Q(z)\) is overloaded, each of \(C(z-1), C(z), C(z+1)\) contains \(Q(z)\), so each satisfies the strict test in \eqref{q1:e7.1}. This establishes the required conservative naming of an unsafe body and both transmitted ports. For a good parent, \(C(z)\) contains at most \(m_{\mathrm{bad}}\) named children. Its whole body, private decisions and terminal transport, including the required preimages under \(E\), lie in that cover by definition. Lemma \ref{q1:l5} replaces the faults of each such child by at most \(K\) native location faults, jointly, with no factor proportional to route length. Hence their union has size at most \(K m_{\mathrm{bad}}<b\). This is the budget used in Lemma \ref{q1:l4}. The extra \(K\) in \eqref{q1:e7.2} is existing slack; port halos are already included in each child's \(K\) and in \(C\), and do not require multiplying the budget again.

In particular, it would be wrong to count \(m_{\mathrm{bad}}\) overloaded children and then expand each to three named children while retaining \(b\). That is not \eqref{q1:e7.1}. At every next rank one counts the full conservative \(N_{i}\) set. A child named only because of a neighboring overload is still a named child of cost \(K\) if it belongs to \(C\). Outgoing \(E\) images are traced to their named origin, rather than counted as unmarked good inputs. This also handles adjacent named parents: the coarse model permits a joint arbitrary map on their named slots before \(E\), followed by its ordinary port permutation, as in the level-zero model and the invariant \(H_k\) of Lemma~\ref{q1:l6} and in Lemma~\ref{q1:l4}\((c)\).

We check the geometric constants directly. At one scale, taking the entire home, its two terminal paths, port neighbors, and the radius-one union increases a center's horizontal extent by at most four parent home widths in either direction. Allowing half a home for the center convention and rounding gives the direct bound \(5P^{j}\) at scale \(j\). A unit temporal collar similarly has direct halfheight at most \(2T^{j}\). Adding complete lower-scale descendants gives respectively

\begin{equation}\label{q1:e7.4}
\begin{gathered}
5 \sum _{\ell=0}^{j-1} P^{j-\ell} \le  (20/3)P^{j} <10P^{j}, \\ 2 \sum _{\ell=0}^{j-1} T^{j-\ell} \le  (8/3)T^{j} <10T^{j}.
\end{gathered}
\end{equation}

The inequalities use \(P,T\ge 4\) and only overcount footprints; enlarging them makes the good test more conservative. Thus the common bound \(h_{\mathrm{cov}}=10\) covers all the direct collars of the cover. Child centers have spatial and temporal spacings \(P^{i},T^{i}\). Counting centers in a rectangle of radii \(10P^{i+1},10T^{i+1}\), with two spare cells on each boundary, gives \eqref{q1:e7.3}.

For separation, suppose a family of named children contains no pair with complete footprint gap at least \(8P^{i}\). Fix one child \(y_0\). Each footprint has radii at most \(10P^{i},10T^{i}\) by \eqref{q1:e7.4}. If another child's spatial index differs from that of \(y_0\) by more than \(2h_{\mathrm{cov}}+10\), the horizontal gap is more than \(8P^{i}\), even allowing one unit of endpoint rounding at each side. If its temporal index differs by more than \(2h_{\mathrm{cov}}+10\), its vertical gap is at least \(8T^{i}\ge 8P^{i}\) by \(T\ge P\). Thus every center lies within index distance \(2h_{\mathrm{cov}}+10\) of \(y_0\) in both coordinates. There are at most

\begin{equation}\label{q1:e7.5}
\begin{gathered}
(4h_{\mathrm{cov}}+21)^{2}=61^{2}=3721 <19600=m
\end{gathered}
\end{equation}

such centers. A bad parent's \(C\) contains more than \(m_{\mathrm{bad}}\) distinct named children, contradicting \eqref{q1:e7.5}. It therefore has the required separated pair. The separation concerns complete descendants' footprints, not just the centers or uncollared nominal boxes. This is the pair used after truncating long marks in Section~\ref{q1:storage}. 
\end{proof}

\begin{lemma}[An explicit uniform runtime for the indexed generator]\label{q1:l8}

There is a fixed binary universal tape machine \(U_{\mathrm{cpu}}\) and an indexed generator \(L\) with the architecture specified in Section~\ref{q1:compiler}, for which the following computable bounds hold. Fix the finite code catalogue, the packet-coordinate formulas, and an integer padding exponent \(c_{\mathrm{pad}}\). Let \(v\) be the finite one-tape binary program text for this generator, and supply \(v\) as an ordinary input to \(L\). There is an explicitly computable integer \(D_{\mathrm{desc}}\ge 2\), depending on that finite description and \(c_{\mathrm{pad}}\) but not \(w\), such that, with

\[
\begin{gathered}
A_{\mathrm{idx}}=32, \\ C_{v,c}=2^{80} D_{\mathrm{desc}}^{64},
\end{gathered}
\]

\(U_{\mathrm{cpu}}\) on \((v,v,w,a,s)\) returns the correct indexed instruction, and on a tagged input \((v,v,w,a,s,i,\xi)\) returns the selection bit \(\mathrm{sel}_{i}\) defined below, in at most \(C_{v,c}(1+w)^{32}\) tape steps, uniformly for \(0\le a<P(w), 0\le s<T(w)\), every candidate index \(i\) and \(\xi\) in \(\{0,1\}\). The computation including return of its head and output to the designated cell takes at most twice that bound plus twice its input length plus four. Its visited tape interval has length at most twice that running-time allowance plus the input length plus four.

\emph{Tagged calls.} The private computation at a contact of candidate \(i\) installs the tag \((i,\xi)\). The index \(i\) is a fixed binary string hardwired by the timetable and installed like the literal. The bit \(\xi\) is the fetched retained-word bit. A leading format bit distinguishes tagged from untagged input. On a tagged input \(L\) returns \(\mathrm{sel}_{i} = 1\) if \(L(v,w,a,s)\) is the instruction of candidate \(i\) and, when that instruction is \(\operatorname{LOAD}\), \(\xi\) equals the value candidate \(i\) prepares; otherwise it returns 0. On an untagged input \(L\) returns the opcode, and \(d_{w}(a,s) = L(v,w,a,s)\) is the untagged value. The tag has at most \(\lceil\log_{2}(N_{\mathrm{contact}}+1)\rceil+3\) bits, where \(N_{\mathrm{contact}}\) is the candidate-contact count of Lemma~\ref{q1:l10}; this number is fixed before \(v\) and included in the constant \(a_{0}\) of Lemma~\ref{q1:l10}. The private machine therefore returns one bit, as phase 4 of Lemma \ref{q1:l10} assumes.

The exponent 32 is absolute and independent of \(c_{\mathrm{pad}}\). Choose \(c_{\mathrm{pad}}=35\), then choose \(w\) by the terminating search

\begin{equation}\label{q1:e8.1}
\begin{gathered}
2 C_{v,c}(1+w)^{32} + 2 a_{\mathrm{in}}(1+w)+4 \le  w^{35},
\end{gathered}
\end{equation}

where \(a_{\mathrm{in}}(1+w)\) bounds the finite literal and input length. A private run of \(w^{35}\) cycles then completes every valid call; malformed results still have the compiler's total default. This instantiates the machine-time allowance \(w^{c}\) of the compiler specification with \(c=c_{\mathrm{pad}}=35>A_{\mathrm{idx}}+2\).

\end{lemma}
\begin{proof}
\textbf{Precise description-size coefficient.}

The following prescription makes the coefficient effective rather than an unspecified \(O(1)\). Generate the finite straight-line and loop arithmetic description of the phase list from the fixed native catalogue. A loop can range only over: a fixed catalogue list, the supplied finite program text, the bounded scalar-bit list of length \(O(w)\), the \(c_{\mathrm{pad}}\) multiplications defining \(w^{c_{\mathrm{pad}}}\), or the \(w\) multiplications defining \(3^{w}\). Include the finite R1 clock formula and coordinate formulas. Use standard binary addition, subtraction, multiplication and long division. The compiler's fixed padding, packet layouts and finite stage lists give such a description directly; no loop simulates tape contents. Loops whose length depends on \(w\) nest at most two deep, and outside the two power loops no loop multiplies an integer carried from a previous iteration.

Read off the description the following integers:
\begin{itemize}
\item \(|\mathrm{desc}|\), its number of operations and loop headers;
\item \(\lambda_{\mathrm{max}}\), the largest bit length of a literal in it;
\item \(c_{T}\), with bit length of \(T\) at most \(c_{T}(1+w)\) for the explicit \(T\) formula of R1;
\item \(c_{\mathrm{scal}}\), with \(c_{\mathrm{scal}}(1+w)\) bounding the length of the scalar-bit list;
\item \(c_{w}:=c_{\mathrm{scal}}+c_{\mathrm{pad}}+2\);
\item \(\delta_{\mathrm{fix}}\), the nesting depth of loops over fixed lists.
\end{itemize}

Put \(c_{\mathrm{ops}}:=|\mathrm{desc}|^{1+\delta_{\mathrm{fix}}} c_{w}^{2}\) and \(c_{\mathrm{bit}}:=2^{|\mathrm{desc}|}(6+c_{\mathrm{pad}}+c_{T}+\lambda_{\mathrm{max}}+c_{\mathrm{ops}})\). Take \(D_{\mathrm{desc}}\) to be the maximum of \(c_{\mathrm{bit}}, c_{\mathrm{ops}}, a_{0}+4|v|\), and two plus the sum of: the lengths of these descriptions and of \(v\); all their positive integer constants and fixed exponents; \(c_{\mathrm{pad}}\); the numbers of formal tape tracks and scratch integers; the lengths of the finite catalogues; and the finite transition-table sizes of the arithmetic, one-tape simulation and universal-interpreter templates described below. This definition is a terminating computation on finite strings and does not depend on \(w\).

In particular it supplies the inequalities

\begin{equation}\label{q1:e8.2}
\begin{gathered}
\text{bit length of every index or intermediate integer }\le  D_{\mathrm{desc}}(1+w), \\ \text{number of arithmetic calls and bit-list operations }\le  D_{\mathrm{desc}}^{3}(1+w)^{2}, \\ \text{input length} \le  a_{\mathrm{in}}(1+w),\quad\text{with } a_{\mathrm{in}}\le D_{\mathrm{desc}}^{2}.
\end{gathered}
\end{equation}

The scan verifies \eqref{q1:e8.2} as follows. Addresses have \(\log _{2} P=w+O(1)\) bits. The explicit period \(T\) of R1 is a finite formula built from sums, products and \(\operatorname{floor}\) using \(2^{w},3^{w}\) and fixed powers of \(w\) (\cite[Section~\ref*{cl:sec:clock} and Appendix~\ref*{cl:sec:constants}]{DMNoise1D2026}), so its bit length is \(O(w)\), with coefficient bounded by its formula length and exponents. The quantum phase endpoints are fixed sums of \(P\) times bounded scalar-list lengths, route lengths, literal lengths and \(w^{c_{\mathrm{pad}}}\). Their bit lengths are also \(O_{v,c}(w)\). Stage lookup scans a fixed finite list; it finds a repeated fetch by quotient and remainder or, more generously, scans \(O(w)\) scalar positions. It never expands \(w^{c_{\mathrm{pad}}}\) tape cycles. Installing \((v,v)\) reads bits of the supplied string \(v\). No other variable-length loop occurs. Explicitly:
\begin{itemize}
\item A loop of \(w\)-dependent length has at most \(c_{w}(1+w)\) iterations. A loop over a fixed list has at most \(|\mathrm{desc}|\) iterations. With the nesting bounds, at most \(c_{\mathrm{ops}}(1+w)^{2}\) operations are executed.
\item Bit lengths. The terms \(2^{w}, 3^{w}, w^{c_{\mathrm{pad}}}, w, a, s\) and the literals have bit length at most \((3+c_{\mathrm{pad}}+c_{T}+\lambda_{\mathrm{max}})(1+w)\). A straight-line operation on integers of bit length at most \(B\) gives at most \(2(B+1)\) bits. An integer carried through a loop changes only by addition or selection, so over at most \(c_{\mathrm{ops}}(1+w)^{2}\) executions it gains at most \(\log _{2}(c_{\mathrm{ops}}(1+w)^{2})+1 \le  (c_{\mathrm{ops}}+3)(1+w)\) bits.
\item Induction over the at most \(|\mathrm{desc}|\) static operations therefore bounds every bit length by \(2^{|\mathrm{desc}|}(6+c_{\mathrm{pad}}+c_{T}+\lambda_{\mathrm{max}}+c_{\mathrm{ops}})(1+w) = c_{\mathrm{bit}}(1+w)\).
\item The input length is \(a_{\mathrm{in}}(1+w)\) with \(a_{\mathrm{in}} = a_{0}+4|v| \le D_{\mathrm{desc}} \le  D_{\mathrm{desc}}^{2}\).
\end{itemize}

Hence \eqref{q1:e8.2} holds with \(D_{\mathrm{desc}}\) as defined.

\medskip\noindent\textbf{Time bound.}

On \(B\)-bit integers, schoolbook addition, subtraction and comparison takes at most \(4(B+1)^{2}\) elementary bit operations and multiplication and division at most \(16(B+1)^{2}\), using extra work tapes. Computing \(3^{w}\) by \(w\) repeated multiplications adds at most \(w\) such calls; computing \(w^{c_{\mathrm{pad}}}\) uses \(c_{\mathrm{pad}}\) calls. Both are already covered in \eqref{q1:e8.2}. Reading a program bit or a listed packet coordinate can be done by a scan and costs at most \(D_{\mathrm{desc}}(1+w)\) further bit operations, likewise covered by enlarging the syntactic count. In a direct multitape realization with at most \(D_{\mathrm{desc}}\) tapes, one call on \(B\)-bit integers, \(B+1 \le  2D_{\mathrm{desc}}(1+w)\), costs at most \(16(B+1)^{2} + 2D_{\mathrm{desc}}(B+1) \le  64D_{\mathrm{desc}}^{2}(1+w)^{2} + 4D_{\mathrm{desc}}^{2}(1+w) \le  68D_{\mathrm{desc}}^{2}(1+w)^{2}\) transitions, including returning the heads over the operands.
\begin{itemize}
\item Calls: at most \(D_{\mathrm{desc}}^{3}(1+w)^{2}\) calls give at most \(68D_{\mathrm{desc}}^{5}(1+w)^{4}\).
\item Scans: at most \(D_{\mathrm{desc}}^{3}(1+w)^{2}\) scans of at most \(D_{\mathrm{desc}}(1+w)\) cells each give at most \(D_{\mathrm{desc}}^{4}(1+w)^{3}\).
\item Input and output: reading the input and writing the bounded output take at most \(2D_{\mathrm{desc}}^{2}(1+w)+D_{\mathrm{desc}} \le  3D_{\mathrm{desc}}^{2}(1+w)\).
\end{itemize}

The total is at most \(72D_{\mathrm{desc}}^{5}(1+w)^{4}\). Since \(72 \le  64D_{\mathrm{desc}}^{3}\) for \(D_{\mathrm{desc}} \ge  2\), the realization therefore takes at most

\begin{equation}\label{q1:e8.3}
\begin{gathered}
H(w)=64 D_{\mathrm{desc}}^{8}(1+w)^{6}
\end{gathered}
\end{equation}

transitions, including parsing, stage arithmetic, lookup of the tagged candidate's instruction in the fixed catalogue, its comparison with the computed instruction and with \(\xi\), and writing the bounded output. These deliberately loose exponents include initialization and all scans. There is no dependence on the values of the simulated private tape cells.

We fix the following efficient machine model, so that its overhead is explicit. Represent the finitely many work tapes on consecutive tracks of one tape, with one marked head on each track. Sweep the visited segment to perform each simulated transition, then sweep back. A segment has at most the input length plus \(H\) cells; track count and encoding sizes are bounded by \(D_{\mathrm{desc}}\). The resulting one-tape program runs in at most

\begin{equation}\label{q1:e8.4}
\begin{gathered}
R(w)=128 D_{\mathrm{desc}} H(w)^{2}
\end{gathered}
\end{equation}

steps. Choose \(U_{\mathrm{cpu}}\) to be the fixed binary interpreter which stores the target transition table and marked target tape, scans the table for the current state and symbol, and sweeps the visited target segment for each simulated step. Its per-step cost is at most \(32D_{\mathrm{desc}}^{2}(R+\text{input length}+1)\). Since \(R\) dominates the input length, a complete interpreted run takes at most \(128D_{\mathrm{desc}}^{2} R^{2}\). Substitution of \eqref{q1:e8.3}-\eqref{q1:e8.4} gives

\begin{equation}\label{q1:e8.5}
\begin{aligned}
t_{U} &\le 2^{45} D_{\mathrm{desc}}^{36}(1+w)^{24}\\
&\le 2^{80} D_{\mathrm{desc}}^{64}(1+w)^{32} = C_{v,c}(1+w)^{32}.
\end{aligned}
\end{equation}

All simulator constants are finite transition-table constants already included in \(D_{\mathrm{desc}}\). The redundant slack covers binary track delimiters and halt and output conventions. Every head moves at most one cell per transition. A home marker lets a halted computation return the head and output in at most its preceding step count plus its input length plus two. This proves both claimed bounds.

Select \(c_{\mathrm{pad}}=35\) before constructing \(v\). The ratio \(w^{35}/(1+w)^{32}\) tends to infinity, so \eqref{q1:e8.1} is decidable by integer arithmetic and eventually holds. The two additional powers in the requirement \(c>A+2\) of the compiler specification are only slack to absorb the coefficient and return convention; our choice leaves three. The code catalogue may contain at least \(5^{q_{s}}\) preparation leaves. Its full literal length and transition-count coefficients enter \(D_{\mathrm{desc}}\) and hence \(C_{v,c}\). They are fixed before \(w\), so they do not affect termination.

For avoidance of a code-length loop, choose a fixed integer \(a_{0}\) from the input formats, the tag length \(\lceil \log _{2}(N_{\mathrm{contact}}+1)\rceil+3\) and the scalar and clock formulas before constructing \(v\), and define \(a_{\mathrm{in}}(v)=a_{0}+4|v|\). This bounds both copies of the literal and all other input bits by \(a_{\mathrm{in}}(v)(1+w)\). \(L\) computes this same elementary formula from the length of its supplied literal \(v\) when locating phase endpoints; its own source does not contain a constant chosen only after measuring that source. \(W_{\mathrm{fixed}}\) and \(\lambda_{\mathrm{tile}}\) are fixed from the code and native tiles before \(v\). \(D_{\mathrm{desc}}\) and \(C_{v,c}\) are analysis and parameter-search bounds only, never constants embedded into \(L\).

Finally define \(v\) to be the one-tape text of \(L\) itself, with \(c_{\mathrm{pad}}\) and all fixed catalogue data. Define the finite schedule by

\[
\begin{gathered}
d_{w}(a,s)=L(v,w,a,s)
\end{gathered}
\]

with its declared defaults outside valid indices. The install stage of \(L\) reads the two copies of its input string \(v\); it never needs to know its own literal text by magic. The physical private program at a contact of candidate \(i\) is supplied \((v,v,w,a',s',i,\xi)\), so it evaluates \(\mathrm{sel}_{i}\) of exactly the same function \(d_{w}\) on the parent indices. By \eqref{q1:e8.1} it halts before padding ends. This proves the self-description assertion without a fixed point in running time: 32 is fixed first, 35 second, \(v\) and \(D_{\mathrm{desc}}\) third, and \(w\) last. 
\end{proof}

\begin{lemma}[R1's permitted packet layout]\label{q1:l9}

The compiler uses the following serialization of the R1 packets. It is the serialization of \cite[Section~\ref*{cl:sec:packets}]{DMNoise1D2026} with one change inside the central area: each of the \(O(w)\) bounded scalar bits is the low bit of the word of its own central-area live letter, which carries a fixed scalar tag and has all other word bits zero, whereas \cite[Section~\ref*{cl:sec:packets}]{DMNoise1D2026} carries a scalar bit in a separate scalar-bit tag with a zero word. Everything else is kept: the live-or-dead discriminant and the declared order of the bounded fields in the central \(O(w)\)-length area, every next-width word's low and high halves in its two prescribed typed raw-letter slots, and the same fixed live padding at all other positions including the two endpoints. Serialize applicable-or-default tags among the bounded scalar bits. Use this same layout in encoders, parsers, output constructors and all represented evaluators. The statements of \cite{DMNoise1D2026} used in this paper hold verbatim for the structural lane with this serialization, in particular the native-transition and sparse-history statements (\cite[Proposition~\ref*{cl:prop:organization}]{DMNoise1D2026}, with \cite[Lemmas~\ref*{cl:lem:endpoint} and~\ref*{cl:lem:two-period}]{DMNoise1D2026}), the clean identity \cite[Lemma~\ref*{cl:lem:clean}]{DMNoise1D2026} and the time-zero boundary of \cite[Lemma~\ref*{cl:lem:past}]{DMNoise1D2026}. The required parent age, address and tag bits and retained-word lowest bit have computable packet positions independent of hierarchy rank.

\end{lemma}
\begin{proof}

The serialization of \cite[Section~\ref*{cl:sec:packets}]{DMNoise1D2026} requires a live-or-dead discriminant and all bounded fields in an \(O(w)\)-length central area, typed low and high word halves, and fixed live padding elsewhere, including both endpoints; at the bit level it writes a scalar bit as a separate scalar-bit tag with a zero word. The layout above keeps all of this except the last point. Give the scalar bits consecutive separate letter positions in that central area, in the declared order. Their number is polynomially bounded and in fact \(O(w)\); \(P_0=2^{w}\) contains them for the same sufficiently large-\(w\) choice. Encode bit 0 or 1 as the corresponding low bit of that letter's word, with its other bits and its tag fixed. Give every nonapplicable field the total default of \cite[Section~\ref*{cl:sec:packets}]{DMNoise1D2026}. Keep all word-half positions and endpoint padding prescribed by the layout. The neutral packet encoding \(D\) is written in the same layout, so it is again a playable all-live packet with a dead discriminant and canonical body. This is an explicit effective encoder and parser for exactly the same legal mutable tuples; structural padding and tag checks are total and malformed whole packets still parse as \(D\), as \cite[Section~\ref*{cl:sec:packets}]{DMNoise1D2026} requires.

Check where the classical proofs use serialization. Input transport and voting (steps 1 and 2 of the repetition program in \cite[Section~\ref*{cl:sec:packets}]{DMNoise1D2026}) copy and compare whole packet letters at homologous offsets, without inspecting which scalar bit occupies an offset. Validation checks the chosen tags and padding, then the universal word machine scans \(O(w)\) scalar positions and the fixed word-half locations, at polynomial cost in \(w\). Low-half and high-half operations on growing word widths remain the same bounded number of word operations described in \cite[Section~\ref*{cl:sec:packets}]{DMNoise1D2026}. Replacing one scalar packing by this packing changes only the polynomial coefficient of that finite scan, which is absorbed by the fixed polynomial budget \(w^{O(1)}<P_0\) of \cite[Section~\ref*{cl:sec:packets}]{DMNoise1D2026} inside the computation allowance of \cite[Section~\ref*{cl:sec:clock}]{DMNoise1D2026}; it adds no geometric execution count. The structural proof (\cite[Definition~\ref*{cl:def:observations} and Proposition~\ref*{cl:prop:organization}]{DMNoise1D2026}, with the lemmas proving it, including \cite[Lemmas~\ref*{cl:lem:endpoint} and~\ref*{cl:lem:two-period}]{DMNoise1D2026}) uses occupancy, ordinary addresses, whole-letter equality and the single short exceptional interval, with no scalar-bit position premise. Each of the three homologous copies has the same new layout, so its uniqueness and vote arguments are unchanged. The clean proof (\cite[Lemma~\ref*{cl:lem:clean}]{DMNoise1D2026}) transmits complete legal packets and resets the complete mutable state; it therefore returns precisely \(E_{j}\) in this layout. Endpoint padding is still independent of the upper live state, which is the only layout property used in the terminating boundary recursion in the proof of \cite[Lemma~\ref*{cl:lem:past}]{DMNoise1D2026}. These checks cover every interface used by Section~\ref{q1:clock} and the compiler. The single change is local to the central scalar area, and every classical use of the serialization is checked above.

The address and age widths depend on the fixed \(P,T\), not rank, so their bit-position formulas are the same at every rank. A word at rank \(j+1\) has twice the child word width; its low bit is the low bit of its low half at rank \(j\). Recursing this observation never changes the packet address to read, only the word width, which a \(\operatorname{LOAD}\) does not inspect. The tags select the same zero default at all ranks. This proves the final assertion. 
\end{proof}

\begin{lemma}[The itemized timetable and space bound]\label{q1:l10}

Fix \(S,K,m_{\mathrm{bad}},b\) and the code before \(w\). Fix \(U_{\mathrm{cpu}}\) and \(c_{\mathrm{pad}}=35\) as in Lemma \ref{q1:l8}. Enumerate the finite exact native gate lists in the following definitions.

\begin{itemize}
\item Let \(E_{\mathrm{len}}\) be the number of primitive gate, reset and explicit-wait contacts in one complete full \(\operatorname{EC}\) pass, including its preparation trees and record processing. Let \(N_{\mathrm{norm}}=(b+1)S E_{\mathrm{len}}\). Serializing all input-slot normalizers gives exactly this unconditional contact count.
\item Let \(N_{\mathrm{contact}}\) be the sum of the lengths of the complete candidate lists, one list for every possible on-site opcode and slot choice. Each list includes full \(\operatorname{EC}\)-and-body on its operands and full \(\operatorname{EC}\)-and-idle on all untouched slots. For \(\operatorname{LOAD}\) include zero and one candidates. List length counts each arity-at-most-three primitive contact once; decompositions and verified-preparation leaves are included.
\item Let \(W_{\mathrm{fixed}}\) be the number of child positions reserved for the \(S\) encoded data blocks, the catalogue's maximum simultaneous work, exclusive contact and dummy slots, dedicated transport and its buffers, and fixed guards. It is finite and computable from those lists. At least \(n_{s}\) separate coordinates for each of the \(S\) input slots are included.
\item Let \(a_{\mathrm{in}}\) be an integer such that the total number of bits installed or fetched for one private call, including \((v,v), w\), fixed parameters, the candidate tag, parent age, address and tags and retained-word bit, is at most \(a_{\mathrm{in}}(1+w)\). Count a bit more than once if the implementation transports it more than once. Concretely use \(a_{\mathrm{in}}(v)=a_{0}+4|v|\) with \(a_{0}\) fixed from the nonliteral input formats, the tag length and the scalar and clock formulas before \(v\). \(L\) computes this formula from its supplied literal \(v\), so phase indexing uses exactly the same coefficient as the finished schedule. Lemmas~\ref{q1:l8} and \ref{q1:l9} prove it is independent of rank and \(w\). \(W_{\mathrm{fixed}}\) and \(\lambda_{\mathrm{tile}}\) are selected before \(v\); analysis bounds \(D_{\mathrm{desc}}\) and \(C_{v,c}\) are not embedded in \(v\).
\item Let \(\lambda_{\mathrm{tile}}\) be a fixed common tile length: enumerate the elementary native implementations of one universal tape update, parallel clear, literal installation, one shuttle, a local scatter, primitive and gather, and disposal; take a common multiple of their finite lengths, enlarged to bound each local routine. Pad their idle substeps to this length. All these routines have fixed native slot count. No \(\lambda_{\mathrm{tile}}\) depends on \(w\).
\end{itemize}

With \(P=5\cdot 2^{w}\) define the explicit upper bounds

\begin{equation}\label{q1:e10.1}
\begin{aligned}
F(w)=\lambda_{\rm tile}\bigl[&10+32N_{\mathrm{norm}}(P+1)\\
 &+N_{\mathrm{contact}}\{32+16a_{\rm in}(1+w)(P+1)+w^{35}+32(P+1)\}\\
 &+32n_s(P+1)\bigr].
\end{aligned}
\end{equation}
\begin{equation}\label{q1:e10.2}
G(w)=W_{\rm fixed}+2w^{35}+a_{\rm in}(1+w)+8.
\end{equation}

There is a computable sufficiently large \(w\) satisfying all R1 requirements, Lemma \ref{q1:l8}'s runtime inequality, the packet-position checks, and

\begin{equation}\label{q1:e10.3}
\begin{gathered}
G(w)\le P, \\ F(w)\le \lfloor 3^{w}/w^{2}\rfloor\le T.
\end{gathered}
\end{equation}

For this \(w\) every phase of a parent tick fits in \(T\) child ticks and \(P\) child sites, with the identical bounds at every rank. Age zero is an idle on-site tick at every address; the fixed \(E\) still occurs. Thus the aligned top initial tick preserves a designated nonport logical qubit.

\end{lemma}
\begin{proof}

One adjacent ordinary-factor transposition is a two-tick shuttle (Lemma \ref{q1:l5}). Routing at most three tokens within one home to a contact and back uses at most \(6P\) adjacent transpositions, hence at most \(12P\) child ticks. Allowing loads into local slots, the primitive, guards, buffers and resets still costs less than \(32(P+1)\) tile units per unconditional native contact. Applying this to the \(N_{\mathrm{norm}}\) leading contacts gives the second term of \eqref{q1:e10.1}. All idle data coordinates remain on separate wires and their waits are part of the schedule and fault model.

For each of the \(N_{\mathrm{contact}}\) candidate contacts execute these phases, in this order:

\begin{enumerate}[label=\textup{Phase \arabic*.},leftmargin=*]
\item Reset the entire private tape interval, head and state fields, input scratch, return track, dummies and guard boundary. Parallel reset of a fixed number of tracks takes a fixed number of tile units; reserve 32 for this and the terminal local disposal in the same contact budget.
\item Install the literal and fixed bits and fetch all parent bits, including the retained-word bit unconditionally. Loading a bit at its fixed source and transporting it along a segment to its assigned input and back uses at most \(4P+O(1)\) ticks, as Section~\ref{q1:compiler} specifies. The two-tick shuttle realization, local copy and common-tile rounding are all bounded by \(16(P+1)\) tile units per installed or fetched bit. There are at most \(a_{\mathrm{in}}(1+w)\) bits. This is the second term inside the braces multiplying \(N_{\mathrm{contact}}\) in \eqref{q1:e10.1}. The allowance intentionally permits serial installation even though fixed literal bits can be installed in parallel.
\item Run exactly \(w^{35}\) universal tape cycles on its guarded private interval, with paired-shuttle record exchanges and the end-cell rule of Lemma \ref{q1:l5}. Each synchronous cell update, including an end cell's blank-record preparation, takes at most \(\lambda_{\mathrm{tile}}\) native ticks. Lemma \ref{q1:l8} ensures that a valid computation halts, writes its result, and returns its head before this allowance ends. A halted computation idles, so its value cannot alter the following phase dates. Reserve the \(w^{35}\) term.
\item Return the answer to the fixed contact, route its at most three operands, perform conditional scatter, the unconditional primitive and gather, undo the route, and dispose of all private work. A fixed segment walk and three-operand route together cost at most \(32(P+1)\) tile units: there are at most \(12P\) ticks for operand routing, at most \(4P+O(1)\) for decision return, and a fixed number for local scatter, gate and gather. The local constant and padding are covered by the fixed 32 term and \(\lambda_{\mathrm{tile}}\). Unselected candidates execute the same timetable.
\end{enumerate}

There are only \(N_{\mathrm{contact}}\) contacts; there is never more than one active private call in a home. All phases are expressed as integer tile units; rounding a fixed local routine or a single fetch to the common tile has already been charged by these generous per-routine allowances.

For the final \(E\), process its \(n_{s}\) code coordinates successively. A path between one home's right endpoint and the next home's left endpoint has fewer than \(2P\) edges. Its forward and reverse-prefix transposition uses fewer than \(4P\) adjacent transpositions, at two ticks each. Moving endpoints to and from the dedicated transport, clearing buffers and tile padding are bounded by \(32(P+1)\) tile units per coordinate. For a given coordinate all boundaries run simultaneously on disjoint lines (Lemma \ref{q1:l5}), so this term is not multiplied by the number of homes. No following data-dependent gate touches transport. Age-zero idle, final clearance, and fixed harmless phase separators consume at most ten additional tile units, the leading term of \eqref{q1:e10.1}. This proves the time bound.

A universal tape head moves at most one cell per cycle. An interval of \(2w^{35}+a_{\mathrm{in}}(1+w)+8\) positions contains the input, both possible head excursions, and blank end guards. During a valid private computation the guard cannot be crossed. Malformed states cannot change any hardware endpoint or contact time, and all actual access remains on these tracks. All other quantum data and catalogue work use \(W_{\mathrm{fixed}}\) positions. Their disjoint placement therefore fits within \(G(w)\). Slots, port lanes and fixed finite tracks have already been included in \(S\) or \(W_{\mathrm{fixed}}\); growing arrays use extra sites. This proves \eqref{q1:e10.2}.

All constants in \eqref{q1:e10.1}-\eqref{q1:e10.2} precede \(w\). Both \(F\) and \(G\) are bounded by a computable constant times \((P+1)(1+w)^{36}\). Consequently

\begin{equation}\label{q1:e10.4}
\begin{gathered}
F(w)/(3^{w}/w^{2}) \le  C (2/3)^{w} (1+w)^{38} \to 0, \\ G(w)/(5\cdot 2^{w}) \to 0.
\end{gathered}
\end{equation}

The R1 requirement \(w\) sufficiently large is a finite collection of explicit polynomial and exponential inequalities (\cite[Section~\ref*{cl:sec:construction} and Appendix~\ref*{cl:sec:constants}]{DMNoise1D2026} and the inequalities in its proved structural lemmas). The compiler adds only the decidable inequalities displayed here and the finite coordinate checks. Exponential dominance proves that a sequential integer search eventually passes every test. The R1 schedule has \(T\ge \lfloor 3^{w}/w^{2}\rfloor\) already from its \(\lfloor 3^{w}/q\rfloor\) calls of \(\operatorname{Shrink}(0)\), with \(q=w^2\) (\cite[Section~\ref*{cl:sec:clock}]{DMNoise1D2026}); we do not lengthen or alter that schedule. Reserve the unused dates as idles. Therefore all quantum phases fit in its actual \(T\).

The same argument works at every rank: a child site and tick have the same \(S\)-slot native inventory; the scalar layout has the same addresses, and a rank-\((k+1)\) bit fetch is a rank-\(k\) \(\operatorname{LOAD}\). Word widths double, but only the low bit is requested. No time or space term depends on rank. Finally reserve age zero for idle before placing the other phases. The complete clean boundary has age zero and default scratch, by the ballot rule of \cite[Section~\ref*{cl:sec:verifier}]{DMNoise1D2026}, the clean identity \cite[Lemma~\ref*{cl:lem:clean}]{DMNoise1D2026} and the encoded boundary in the proof of \cite[Lemma~\ref*{cl:lem:past}]{DMNoise1D2026}; at that tick only \(E\) acts, which does not touch an ordinary designated nonport.

The constants are chosen in the order \(S\), then \(K,h_{\mathrm{cov}},m_{\mathrm{bad}},b\), then the code and the finite lists (\(N_{\mathrm{norm}},N_{\mathrm{contact}},n_s\) and the candidate catalogue), then \(U_{\mathrm{cpu}},c_{\mathrm{pad}},a_0,W_{\mathrm{fixed}},\lambda_{\mathrm{tile}}\), then \(v\), then the analysis bounds \(D_{\mathrm{desc}},C_{v,c}\) together with \(a_{\mathrm{in}}(v)=a_0+4|v|\), then \(w\), then \(P,T\), and finally the physical-noise thresholds. This agrees with Lemma~\ref{q1:l8} and Section~\ref{q1:parameters}.
\end{proof}

\section{Exact rate selection and joint error estimates}\label{q1:selector}
This section builds the rate-selecting lane R2. Redundant Boolean circuits and noise factories turn the physical depolarization into the exact native replacements of the classical construction (Lemma~\ref{q1:l17}); Lemma~\ref{q1:l11} fixes the native instance and the label padding, and Lemma~\ref{q1:l12} bounds the resulting joint error field.

\paragraph{Exact native rate selection under elementary single-qubit noise.}

The selector is an independent dephased lane, separate from the sensor-free R1 structural clock. It simulates the complete deterministic R2 local function \(f\) of the classical construction (\cite[Sections~\ref*{cl:sec:embedding}, \ref*{cl:sec:receiver} and~\ref*{cl:sec:stripes}]{DMNoise1D2026}), with its physical stride, fields, replica compartments, wall and work registers, raw probes and common whole-symbol substitution. Let \(A\) be its native site alphabet, \(M=|A|\). In particular a single simulated replacement flag belongs to the whole native physical site, including all fields sharing that flag. Native helpers carry the same simulated substitution noise as every other field of the native site. Physical noise of rate \(\eps\) follows every elementary physical layer on every wire.

\paragraph{Redundant Boolean steps and exact draws.}

Fix an integer \(h\); every Boolean state bit is encoded in \(B=4h+3\) wires, in the computational basis. In one protected Boolean operation, give each input bit and index its own fanout tree to private copies for every required reader. For each destination index, privately majority the received copies separately on each input and implement its required total Boolean function, with reset work. All spatial transfers are between fixed named wires over fixed distances. A fault in a source tree harms at most one delivered index of one input bit per reader; private faults after fanout harm at most one destination per fault. With \(h\)-close input banks and at most \(h\) faults in the whole stencil the correct input wins all unaffected votes \((2h<B/2)\). At most \(h\) exceptions remain per output. Use \(h+1\) fresh, disjoint normalization passes (each a protected majority of a word to a new bank). At most \(h\) faults across these passes leave an entire pass clean, which maps even arbitrary measured input to unanimity; the remaining passes leave the final bank within \(h\) of its majority. Include storage waits, initialization, reads, and incoming and outgoing normalizations in the stencils. Decode every binary representation by a total parser, defaulting on unused strings. It follows from one-sided implications at each strobe that a decoded transition differs from the prescribed one on its actual decoded neighbours only in a bounded enlargement of a stencil carrying more than \(h\) independent wire-replacement flags. Mark the explicit excess flag count in these stencils, regardless of the incoming state or the outcome. Its probability is \(\le C_{h} \eps^{h+1}\). This also holds from arbitrary measured starting words after the initial normalization. Good incoming normalizations ensure different readers use the same bits. Ancillas and old banks are reset before being read again.

Independent factory bits are made at each site and date as follows. One rare factory starts with zero reset, the mandatory noisy layer, then basis read and fanout and the above normalization. Its abstract bit is its final bank majority even on a failed factory. Its one probability \(\beta(\eps)\) is a computable rational polynomial in \(\eps\), satisfies \(\beta(0)=0\) and has \(\beta'(0)>0\): a replacement immediately after reset can give one, all remaining operations clean, and single-flag contributions are nonnegative. A separate fair factory starts with a fresh fair draw by Hadamard and basis dephasing. Make the copying and majority circuits complement equivariant: for an input vector \(x\) use \(x_{1}\) as orientation, form all \(x_{i}\) XOR \(x_{1}\), compute the majority of these oriented bits, XOR back \(x_{1}\). The inner gates use only invariant bits. Fresh copies of the orientation are made before use. Every relevant wire either complements or stays fixed when the initial fair draw complements. Under this coupling complement replacement bits on the corresponding wires: their joint law is unchanged. The final majority has an exactly fair law, including failed factories. A depolarization between reset and the fair Hadamard causes no bias in the ensuing basis read; later basis-valued innovations are independent. Factory banks are \(h\)-close to their abstract bits whenever the corresponding complete stencils have at most \(h\) faults. All factories use disjoint private wire-time keys, with actual resets on every initial input they use.

The circuit draws \(k=\lceil \log_2 M\rceil\) independent fair bits to form \(W\) and a bit \(B_{\mathrm{rare}}\) from a separate rare factory, and substitutes the native symbol numbered \(W\) precisely if \(B_{\mathrm{rare}}=1\) and \(W<M\). In the ideal comparison this is an \emph{exact} independent uniform native replacement at rate \(\gamma(\eps)=\beta(\eps)M/2^{k}\). This works for a nonbinary alphabet without rejection loops. The protected Boolean native transition implements it, with only the bounded-stencil anomalies above. The comparison keeps the dependence between a native replacement and all helper fields it overwrites. The law is that of the innovations, not conditioned on the absence of circuit anomalies. For a finite region compare to exact native evolution on its full past cone from the same starting state. Induction shows equality if all transition stencils in that cone are good. A union bound costs at most \(C_{h}\,|\mathrm{cone}|\,\eps^{h+1}\). In this argument the native innovations can be defined on the same probability space on every realization, including runs with anomalies.

\paragraph{Physical windows, noise-only marks and the core error set.}

The classical inputs used for R2 are the two prefix tests with tunable exponents and the all-input coalescence of \cite[Lemmas~\ref*{cl:lem:receiver-mixing} and~\ref*{cl:lem:stripes}]{DMNoise1D2026}, the read-only parameter and its locking (\cite[Section~\ref*{cl:sec:parameter} and Lemma~\ref*{cl:lem:locking}]{DMNoise1D2026}), the initialized comparison \cite[Lemma~\ref*{cl:lem:past}]{DMNoise1D2026}, the all-start cap and ascent \cite[Lemmas~\ref*{cl:lem:cap-cut} and~\ref*{cl:lem:ascent}]{DMNoise1D2026}, and the initialized two-period descent of \cite[Lemma~\ref*{cl:lem:two-period}]{DMNoise1D2026}. We use them with their exact native noise model; Lemma~\ref{q1:l12} strengthens the cover to the radius moment needed for the quantum schedule. The quantitative choices below are those of Lemmas~\ref{q1:l11} and~\ref{q1:l12}.

The selector uses the fixed universal unary-input machine \(V(m)\), which simulates \(U(\nu_2(m))\) for \(m\ge1\), with the bin offset \(n_{\mathrm{bin}}\) of Lemma~\ref{q1:l11}, and includes only sufficiently large \(m\) by choosing \(m=2^n(2J_{\mathrm{pad}}+1)\), with \(J_{\mathrm{pad}}\) fixed last. The native bin offset is frozen before physical compilation; \(J_{\mathrm{pad}}\) only selects later bins of the same fixed rule. At a nominal rational rate \(\bar\eps_m\) in the interval of strict increase of \(\gamma\) near zero, choose \(\gamma(\bar\eps_m)\) strictly inside the \(m\)-th native replacement-rate bin and put \(p=\gamma(\bar\eps_m)/2\). Rationals \(\bar\eps_m\) with this property are found by enumeration, using rational arithmetic and an effective neighbourhood of positive derivative. Here and below the nominal bin is a bin for the complete native replacement rate \(\gamma\), with \(p\) its half in the helper estimates. Fix all test lengths and walls in the \emph{analysis} at this nominal rate; they do not enter the physical rule as \(m\)-dependent gates. Lemma~\ref{q1:l12} compares the compiled helper with the nominal native innovations over the entire backward domain of both prefix computations, charges every compiler anomaly and every elementary-rate mismatch in that domain to the long failure type, and chooses the rational physical window about \(\bar\eps_m\) inside the monotone preimage of the native bin. Windows for distinct selected indices lie in the preimages of disjoint bins, so they are mutually disjoint.

Native core data errors are covered by the union of the actual native substitution and a complete local compiler anomaly; its indicator has probability \(O(p)\) and a bounded physical stencil. Raw negative-time core flags can be added for domination. Paint around every failed helper center the fixed multiple of \(H_m=D_{\mathrm{boot}}^{m+2}\) (Lemma~\ref{q1:l11}) prescribed in \cite[Section~\ref*{cl:sec:joint}]{DMNoise1D2026}. All conclusions of the deterministic bootstrap and sparse induction then hold for this dominated error set. More specifically, on the initialized history every unexpected \(D\) at a core root is covered by failure of a prescribed nested test, as in the proof of \cite[Lemma~\ref*{cl:lem:clock-witness}]{DMNoise1D2026}. On a halting input the arbitrary-start cap and bottom-up ascent of \cite[Lemmas~\ref*{cl:lem:cap-cut} and~\ref*{cl:lem:ascent}]{DMNoise1D2026} say that every genuinely live core symbol after a finite burn-in is covered by a centered failed test through some finite rank \(K\), independently of the initial word; Lemma~\ref{q1:l13} gives the complete cover. The cap, test sizes and burn-in are common over this nominal window: they use the fixed index \(m\), its halting rank, the nominal prefix lengths and fixed geometry. The negative-time comparison of \cite[Lemma~\ref*{cl:lem:past}]{DMNoise1D2026} is needed only for the initialized assertion.

\paragraph{Initial helper history.}

At time zero the core uses the computable all-occupied boundary. Use its nominal negative-time ideal stationary helper to define the comparison of \cite[Lemma~\ref*{cl:lem:past}]{DMNoise1D2026} through the selector's searching levels, as in that lemma's proof. Negative-time forced core votes that disagree with their update are covered by the same bootstrap paint. Positive-time updates run the actual compiled circuit. The nominal and actual innovation comparison above permits prefix tests crossing zero. To prepare the required stationary marginal on a finite patch \(F\) at rational \(\bar\eps_{m}\), simulate nominal native innovations on the finite past cone from date \(-s\) with a fixed bottom row. Enumerate the probability that the mapping from all possible bottom rows to \(F\) is nonconstant. Search for \(s\) making this less than \(1/128\). It terminates by the all-input coalescence property for helper slots, using a finite union over \(F\): this is the noise-only bound \(C(p)\exp(-c(p)s)\) of \cite[Lemma~\ref*{cl:lem:stripes}]{DMNoise1D2026}, which uses \cite[Lemma~\ref*{cl:lem:receiver-mixing}]{DMNoise1D2026} and also gives the two-sided stationary realization. This yields a computable finite rational mixture without knowing whether \(V(m)\) halts. Use the same suffix of innovations in a mathematical stationary realization; the whole marginal couples with failure probability \(<1/128\). The clock lane, core data, helpers and quantum lanes occupy distinct wires. Only the helpers use this stationary marginal. The R1 proof can share its fixed artificial negative history with its own clean comparison.

\textbf{Central normalization convention for the physical Boolean compiler.} There is one centrally normalized retained bank per native site and bit before its fanout. Define the semantic word by total decoding of that bank after its \(h+1\) normalization passes. All readers use private per-index copies of this same \(h\)-close bank. At most \(h\) flags in the complete stencil leave a clean normalization pass and then an \(h\)-close common bank; any \(h\)-close input with at most \(h\) further indexed faults still has at most \(2h<B/2\) wrong votes. Consequently the protected transition computes the prescribed total Boolean function on the same actual semantic neighbor word for every destination index, up to \(h\) exceptions and a separately marked excess-flag event. Storage, incoming central normalization, all fanout and outgoing normalization are included in that event's fixed stencil. The complete transition stencil also contains every rare and fair factory stencil, final factory normalization and storage through its read. A factory with at most \(h\) flags delivers an \(h\)-close bank representing its abstract final-majority bit; a bad factory is therefore charged to the same excess-flag union before a different abstract innovation could be read without a mark. Native factory innovations remain defined unconditionally even on that marked event. This selects the shared normalization arrangement within the finite normalization and fanout implementation. The strobe normalization is this central normalization, or a private reader of its bank.

\begin{lemma}[The exact native R2 instance and the label padding]\label{q1:l11}

The parametric native R2 construction is applied to the following single fixed tuple \(\theta\), which need not be the numerical instance described in \cite{DMNoise1D2026}. The geometry \((w_{c},P_{c},T_{c})\) is fixed before helper counts. Choose \(D_{\mathrm{boot}}\) so \(D_{\mathrm{boot}}^{m+2}\) contains every fixed structural collar and lookup displacement through rank \(m+1\) (the constant \(D_0\) of \cite[Section~\ref*{cl:sec:stripes} and the proof of Lemma~\ref*{cl:lem:locking}]{DMNoise1D2026}). Fix the requested radius exponent \(d_{\mathrm{mom}}\) after the quantum geometry; the target \(\delta\) may be fixed after the final physical constants and affects only the final padding threshold. In Lemma \ref{q1:l12} fix the short target exponent \(M_{s}\), then

\[
\begin{gathered}
b_{\mathrm{refresh}} = \max (21,M_{s}+20), \\ v_{\mathrm{rep}} = M_{s}+20, \\ s_{\mathrm{read}} = 4 M_{s}+85\quad (\mathrm{odd}).
\end{gathered}
\]

Next fix \(M_{l}\) as in Lemma \ref{q1:l12}, and set

\[
\begin{gathered}
m_{\mathrm{ch}} = 20(M_{l}+20)+1\quad (\mathrm{odd}).
\end{gathered}
\]

There are \(2v_{\mathrm{rep}}-1\) replicas per channel \(Y_{i}\) and \(s_{\mathrm{read}}\) independent receivers, each reading all \(m_{\mathrm{ch}}\) channels, with coordinatewise majority. Use the same fixed universal one-ended interpreter as \cite[Section~\ref*{cl:sec:parameter}]{DMNoise1D2026} with immutable program \(V\), where \(V(0)\) is a fixed infinite loop and, for \(m\ge 1\), \(V(m)\) computes \(\nu_{2}(m)\) and then simulates \(U(\nu_{2}(m))\). Take the base literal width at least \(m_{\mathrm{ch}}+4\) and large enough for that fixed program. Last fix once a bin offset \(n_{\mathrm{bin}}\) admissible for the classical construction. This specifies

\[
\begin{gathered}
\theta=(w_{c},P_{c},T_{c},D_{\mathrm{boot}},M_{s},b_{\mathrm{refresh}},v_{\mathrm{rep}},s_{\mathrm{read}}, \\ M_{l},m_{\mathrm{ch}},n_{\mathrm{bin}},V,\text{base word width},\text{finite lane offsets}).
\end{gathered}
\]

All choices of geometry, finite offsets, width and \(n_{\mathrm{bin}}\) are made by taking the least integer or template in fixed enumeration satisfying the explicit sufficient construction inequalities. This is a computable specification of one tuple, not an input parameter of the running rule. In particular \(n_{\mathrm{bin}}\) is fixed before the physical Boolean compilation and never changed afterward. The native bins are

\[
\begin{gathered}
J_{m} = (2^{-A_{m}}, 2\cdot 2^{-A_{m}}),\qquad A_{m}=8^{m+n_{\mathrm{bin}}},\qquad m\ge 0.
\end{gathered}
\]

The native interfaces, deterministic bootstrap, full initialized descent, \cite[Lemma~\ref*{cl:lem:past}]{DMNoise1D2026}, arbitrary-input cap and ascent and all-input helper coalescence invoked here hold for this \(\theta\). After all physical stencils and the map \(\gamma\) are fixed, choose one \(J_{\mathrm{pad}}\ge 0\) last and select only

\[
\begin{gathered}
m(n)=2^{n}(2 J_{\mathrm{pad}}+1),\qquad n\ge 0.
\end{gathered}
\]

Then \(m(n)\ge 2J_{\mathrm{pad}}+1\) and \(V(m(n))\) halts exactly when \(U(n)\) halts. \(V\) and the native rule do not contain \(J_{\mathrm{pad}}\) or \(n\). Increasing \(J_{\mathrm{pad}}\) selects later bins of the same fixed rule.

\end{lemma}
\par\noindent\emph{Proof idea.} The classical construction of \cite{DMNoise1D2026} is stated for arbitrary prescribed helper exponents and for any fixed one-ended machine, so the displayed tuple is an admissible instance and its imported statements apply unchanged. The labels \(m(n)=2^n(2J_{\mathrm{pad}}+1)\) select later bins of the same fixed rule, and \(\nu_2(m(n))=n\) preserves the halting answer.

\begin{proof}
The distinction between geometry and width can also be checked directly. The native core date and address map uses the fixed \(w_c,P_c,T_c\) schedule and a fixed number of word-operation slots. A vector of fixed width is a field of its finite native alphabet; its coordinatewise Boolean operations are one finite-table transition, so widening that alphabet adds no native sites or instruction dates. The helper channels and receivers use their own finite fields and offsets. Their fixed channel count enters the finite job-time coefficient, which is absorbed by the final native bin cutoff in the stripe lemma. They do not lengthen the core schedule. Only after choosing those fields is the physical Boolean table compiled; its increased depth, slot count and spatial stride enter \(D_{\rm layer}\), the complete stencils and the physical scaling constants. Thus choosing native geometry before helper width makes no assumption of a width-independent physical circuit depth.

The quantification over helper constants is part of the imported statement: \cite[Lemma~\ref*{cl:lem:stripes}, its proof and Appendix~\ref*{cl:sec:constants}]{DMNoise1D2026} permit an arbitrary prescribed short exponent, then the refresh, copy and reader constants, then an arbitrary prescribed long exponent, then the channel count, and finally the bin offset (there called \(n_0\), here \(n_{\mathrm{bin}}\)). \cite[Sections~\ref*{cl:sec:parameter} and~\ref*{cl:sec:stripes}, and Lemma~\ref*{cl:lem:joint}]{DMNoise1D2026} fix the core geometry and clock before, and independently of, the vector width and these helper counts. \cite[Lemma~\ref*{cl:lem:stripes}]{DMNoise1D2026} gives the corresponding all-start noise-only interface. The displayed concrete choices dominate those bounds with spare powers. Indeed the reader error exponent before fixed prefactors is \((s_{\mathrm{read}}+1)/4=M_{s}+21.5\). The refresh exponent after the \(p^{-8}\) suffix count is \(b_{\mathrm{refresh}}-7\ge M_{s}+13\), and the replica exponent after that count is \(v_{\mathrm{rep}}-8=M_{s}+12\). The factor \(H_{m}^{2}=p^{-o(1)}\) and fixed constants consume fewer than these spare powers when \(n_{\mathrm{bin}}\) is sufficiently large. The long majority-reference exponent is \((m_{\mathrm{ch}}+1)/20=M_{l}+20.1\); the missing-refresh term is superpolynomial. Thus native short and long failures may in particular be bounded by \(p^{M_{s}+2}/8\) and \(p^{M_{l}+2}/8\) for all native bins after the once-only choice of \(n_{\mathrm{bin}}\). These bounds implement the exponents stipulated in \cite[Lemma~\ref*{cl:lem:stripes}]{DMNoise1D2026}.

This choice of \(n_{\mathrm{bin}}\) is effective. All prefactors come from finite schedules, fixed alphabets, finite union bounds or explicitly bounded geometric or exponential series. The conditions in the proof of \cite[Lemma~\ref*{cl:lem:stripes}]{DMNoise1D2026} are inequalities of fixed polynomials, exponentials, powers of \(p\) and \(H_{m}\). Replace logarithms by certified rational upper bounds and bound powers of \(\log (1/p)\) by a small inverse power of \(p\). For \(A_{m}=8^{m+n_{\mathrm{bin}}}\), every remaining ratio of the form \((m+2)/A_{m}\) decreases with \(m\); its next ratio is \((m+3)/(8(m+2))<1\). For these sufficient inequalities, the remaining index dependence is through nonnegative multiples of \((m+2)/A_{m}\) and \(1/A_{m}\). Thus checking their envelopes at \(m=0\) verifies them for every \(m\ge 0\). A search over integer \(n_{\mathrm{bin}}\) using these sufficient rational bounds terminates by the displayed positive exponent slack. Geometry and lane offsets and base width are likewise finite constructive choices in \cite[Section~\ref*{cl:sec:construction} and Appendix~\ref*{cl:sec:constants}]{DMNoise1D2026}. They do not depend on any computation's halting status.

The dependence on the fixed machine is also parametric. Each deterministic import has its premises satisfied, as follows. \cite[Section~\ref*{cl:sec:construction}]{DMNoise1D2026} defines total parsing and motion for arbitrary legal mutable words and a fixed literal program; the organization induction through \cite[Proposition~\ref*{cl:prop:organization}]{DMNoise1D2026} uses geometry and exact local evaluation, not a property of \(U\). In \cite[Section~\ref*{cl:sec:parameter}]{DMNoise1D2026} a promotion takes one true step of the fixed one-ended machine on fresh immutable operands; substituting \(V\) changes only this finite literal and table. In the proof of \cite[Lemma~\ref*{cl:lem:locking}]{DMNoise1D2026} the stripe identity locks precisely the integer \(m\) at rank \(m+1\) before evaluating that initial tape; its proof uses only \(\mathrm{popcount}\) and three spatial samples. It therefore installs the actual initial \(V(m)\) tape. At every higher rank it advances that tape by one true step, and checks its true halted bit. The packet format and word-doubling argument do not depend on how long \(V(m)\) runs. The frozen comparison in the proof of \cite[Lemma~\ref*{cl:lem:locking}]{DMNoise1D2026}, the full-state clean identity \cite[Lemma~\ref*{cl:lem:clean}]{DMNoise1D2026}, the endpoint boundary and the searching-rank comparison in the proof of \cite[Lemma~\ref*{cl:lem:past}]{DMNoise1D2026} use only these fresh true operands and the absence of a halt at the relevant positive layers. In the halting direction, the cut proof \cite[Lemma~\ref*{cl:lem:cap-cut}]{DMNoise1D2026} invokes the true halted-bit guard for every legal or malformed supplier, and the ascent \cite[Lemma~\ref*{cl:lem:ascent}]{DMNoise1D2026} quantifies over every starting row and translated grid. Replacing \(U\) by \(V\) preserves each premise exactly. Helper construction and its all-input coalescence (\cite[Sections~\ref*{cl:sec:embedding}--\ref*{cl:sec:stripes} and Lemmas~\ref*{cl:lem:receiver-mixing}--\ref*{cl:lem:stripes}]{DMNoise1D2026}) do not execute \(V\) and are unchanged. Thus the accepted imports apply to every tuple admissible for the classical construction, including \(\theta\), without any claim about unspecified extra helper faults. Lemma \ref{q1:l12} separately proves the required enlarged fault cover.

Finally \(\nu_{2}(2^{n}(2J_{\mathrm{pad}}+1))=n\) because \(2J_{\mathrm{pad}}+1\) is odd. Every selected bin index is large when \(J_{\mathrm{pad}}\) is large. All remaining small-noise requirements, including those involving physical compilation constants chosen after \(\theta\), can therefore be met by \(J_{\mathrm{pad}}\) alone. The bin offset \(n_{\mathrm{bin}}\) and the physical transition remain fixed. In particular changing \(n_{\mathrm{bin}}\) after compilation is not described as equivalent to padding: it would change the native receiver's interval-to-index map. This proof uses only the displayed padding route. 
\end{proof}

\begin{lemma}[A radius-moment cover for the compiled two-scale field]\label{q1:l12}

Fix the native selector geometry \(P_{c},T_{c},w_{c}\) and the bootstrap constant \(D_{\mathrm{boot}}\) of Lemma \ref{q1:l11}, independently of helper counts. Put \(b_{c}=\min (P_{c},T_{c})>1, B_{c}=\max (P_{c},T_{c})\), and fix any nonnegative moment exponent \(d\). It is enough to prove the assertion for the integer \(d_{\mathrm{mom}}=\max (4,\lceil d\rceil)\). There is one helper construction and physical Boolean compiler, given below independently of \(\delta\), such that for every \(\delta>0\) some final integer \(J_{\mathrm{pad}}\) gives the following bound: simultaneously for every selected index \(m=2^{n}(2J_{\mathrm{pad}}+1)\), every rate in its physical window and every deterministic sequence of elementary rates in that window, the following holds.

Let \(E\) be the union of \((a)\) native core substitutions and complete-stencil compiler anomalies, and \((b)\) the fixed-multiple \(H_{m}=D_{\mathrm{boot}}^{m+2}\) paint about a short bad helper center or an enlarged long bad center. An enlarged long center includes native long failure, any compiler anomaly in the entire helper-prefix comparison domain, and any primitive rate mismatch in that domain. The comparison domain at center \(z\) is the union of the backward native cones of the short and long prefix computations at \(z\). All anomalies and band mismatches in it are charged to \(\mathsf L\) at \(z\), and \(\mathsf S\) is the native short test alone. Then every site whose prescribed fixed-multiple centered test of sparsity order \(i\) fails \(i\)-sparsity (tests are indexed by sparsity order here; a rank-\(j\) test of Lemma~\ref{q1:l1} requires \((j-1)\)-sparsity) is covered by a primitive-measurable marked ball of one of countably many types \(a\), namely the types \(R_{i}\), \(\mathsf S\), \(M_{i}\), \(\mathsf L\), \(H_{i}\) of Step~6 of the proof with their finite residue multiplicities. For each type \(a\), both the covered footprint and all its determining independent primitive keys lie inside its ball, of integer physical radius \(r_{a}\ge 1\); \(\Pr(\text{mark of type }a\text{ at a fixed center})\le z_{a}\); and

\begin{equation}\label{q1:e12.1}
\begin{gathered}
\sum _{a} (1+r_{a})^{d} \sqrt{z_{a}} < \delta.
\end{gathered}
\end{equation}

The sum includes the complete infinite rank list and finite physical offset and residue multiplicities. For a finite cap \(K\) retain only the test types with \(i\le K\), but retain the short and long center types even when \(K\) lies below their admission ranks. This yields finitely many types with a finite maximum radius. Negative-time raw flags may be added to the dominating field. The cover is simultaneous over incoming helper and core words; it never conditions innovations on anomaly absence.

One admissible choice of exponents is the following. Let \(R_{\mathrm{geom}}\) be the least integer \(R\ge 1\) with \(b_{c}^{R}\ge B_{c}\) and put \(\Delta_{\mathrm{ad}}=10000\). Define successively

\[
\begin{gathered}
\rho_{s}=18 R_{\mathrm{geom}}+2, \\ D_{s}=2 d_{\mathrm{mom}} \rho_{s}+20, \\ M_{s}=\max (2\rho_{s}+2(\Delta_{\mathrm{ad}}+1)D_{s}+20, \\ 2(d_{\mathrm{mom}}+1)\rho_{s}+20), \\ b_{\mathrm{refresh}}=\max (21,M_{s}+20), v_{\mathrm{rep}}=M_{s}+20, s_{\mathrm{read}}=4M_{s}+85, \\ \rho_{l}=(b_{\mathrm{refresh}}+7)R_{\mathrm{geom}}+2, \\ D_{l}=2 d_{\mathrm{mom}} \rho_{l}+20, \\ M_{l}=\max (2\rho_{l}+2(\Delta_{\mathrm{ad}}+1)D_{l}+20, \\ 2(d_{\mathrm{mom}}+1)\rho_{l}+20), \\ m_{\mathrm{ch}}=20(M_{l}+20)+1, \\ C_{\mathrm{pref}}=2(b_{\mathrm{refresh}}+6)+5, \\ h_{\mathrm{red}}=C_{\mathrm{pref}}+M_{l}+5.
\end{gathered}
\]

Freeze the native offset \(n_{\mathrm{bin}}\) after the native helper choices and before physical compilation, as in Lemma \ref{q1:l11}. All further smallness, including \eqref{q1:e12.1}, is obtained by the last padding \(J_{\mathrm{pad}}\). These are explicit sufficient choices within the original arbitrary-exponent allowances, not a change of construction.

\end{lemma}
\par\noindent\emph{Proof idea.} The compiled helper is compared with the nominal native process over the whole backward domain of its prefix tests, and every compiler anomaly or rate mismatch there is charged to a long failure center, so every mark is a function of the noise alone. Dependent raw flags are handled as in Lemma~\ref{q1:l1}, by selecting witness leaves with disjoint keys. A short and a long admission rank absorb the short and long helper failures into the sparsity tests, and the exponent choices make every range of the moment sum small once the padding \(J_{\mathrm{pad}}\) is large.

\begin{proof}
We give the cover, its joint-key justification and every range of the moment sum.

\smallskip\noindent\textbf{Step 1: Comparison events are noise-only.} The native innovations are the unconditional whole-symbol factory draws of Lemma \ref{q1:l17}. Work first at a nominal rational physical rate \(\bar\eps_{m}\) with \(\gamma(\bar\eps_{m})\) strictly inside native bin \(m\), and put \(p=\gamma(\bar\eps_{m})/2\). Native short and long sufficient tests have probabilities at most \(p^{M_{s}+2}/8\) and \(p^{M_{l}+2}/8\) after the fixed offset of Lemma \ref{q1:l11}. They apply simultaneously to every incoming helper history (\cite[Lemma~\ref*{cl:lem:stripes}]{DMNoise1D2026}). A complete protected Boolean transition, including incoming and outgoing normalizations, has a state-independent excess-flag event of probability at most \(C_{h} \eps_{\mathrm{sup}}^{h_{\mathrm{red}}+1}\), where \(\eps_{\mathrm{inf}}<\eps_{\mathrm{sup}}\) are the endpoints \(\eps_{\mathrm{inf},m},\eps_{\mathrm{sup},m}\) of the physical window chosen below (Section~\ref{q1:parameters}). A backward prefix contains at most \(C_{N} p^{-C_{\mathrm{pref}}}\) primitive wire-dates. This follows by bounding its spatial and temporal extents by \(C p^{-(b_{\mathrm{refresh}}+5)}\), inserting all finite strides, native speeds, compiler depths and starting and terminal collars, and multiplying the two extents. The deliberately larger integer \(C_{\mathrm{pref}}\) above exceeds \(2(b_{\mathrm{refresh}}+6)+4\). All finite compilation constants enter \(C_{N}\), not that exponent.

Since \(\gamma'(0)>0\), effective positive rational constants \(c_{\eps},\eps_{*}\) satisfy \(\eps_{\mathrm{sup}}\le c_{\eps} p\) whenever \(\eps_{\mathrm{sup}}\le 2\bar\eps_{m}<\eps_{*}\) and \(\gamma(\bar\eps_{m})=2p\). Therefore the union of anomalies in this entire prefix costs at most

\[
C_{N} C_{h} p^{-C_{\mathrm{pref}}}(c_{\eps} p)^{h_{\mathrm{red}}+1} = C' p^{M_{l}+6}.
\]

For \(p\) sufficiently small this is below \(p^{M_{l}+2}/8\). Give the physical window rational halfwidth \(\zeta_{m}\) small enough that it stays in the monotone bin preimage, \(\eps_{\mathrm{sup}}<2\bar\eps_{m}\), and

\[
\begin{gathered}
\zeta_{m} \le  p^{M_{l}+C_{\mathrm{pref}}+2}/(16 C_{N}).
\end{gathered}
\]

Using the same independent uniform and replacement-value keys at every wire-date, a rate mismatch lies in the fixed threshold band about \(\bar\eps_{m}\). For one fixed actual rate its probability anywhere in the prefix is at most \(C_{N} p^{-C_{\mathrm{pref}}} \zeta_{m}\). To use one noise-only envelope for all rates in the window, mark the entire band of width \(2\zeta_{m}\); its probability is at most \(2 C_{N} p^{-C_{\mathrm{pref}}} \zeta_{m}\le p^{M_{l}+2}/8\). This remains true for any deterministic elementary-rate sequence in the window. It covers a splice inside a factory. On every realization the nominal native innovations are defined; if all prefix stencils and mismatch tests are good, induction on that full prefix gives equality from the actual incoming row. Arbitrary incoming rows cause no extra event. Thus an enlarged long center has probability at most \(C_{l} p^{M_{l}}\), with \(C_{l}\) fixed (one can make \(C_{l}=1\) by the unused slack). Its primitive support radius is at most \(C p^{-(b_{\mathrm{refresh}}+5)}\). A short center has probability at most \(C_{s} p^{M_{s}}\) and radius at most \(C p^{-16}\). This is precisely a union of named noise tests, not a test of whether a particular incoming word happened to fail. Primitive events in disjoint complete support regions are independent.

\smallskip\noindent\textbf{Step 2: Dependent raw flags have the same double-exponential form.} The union of native substitutions and all bounded-stencil core anomalies has marginal probability at most \(C_{\mathrm{raw}} p\) and primitive support in a fixed radius \(R_{\mathrm{raw}}\). Use the rate-uniform envelope: any actual-rate substitution requires a primitive rare-factory flag below \(\eps_{\mathrm{sup}}\). Its complete fixed stencil, together with the excess-fault stencils, gives this same raw envelope for every elementary-rate sequence in the window. Distinct values can be dependent inside that radius. Use the binary mesh witnesses of \cite[Lemma~\ref*{cl:lem:sparse-witness}]{DMNoise1D2026}, with leaf regions intersected with their ancestors, using the proof of Lemma \ref{q1:l1}. A rank-\(i\) witness has \(2^{i}\) disjoint nonempty terminal regions. At rank zero each has bounded area \(a_{0}\). Enlarge each by the fixed microscopic raw stencil. Its conflict graph has finite degree \(\Delta_{\mathrm{raw}}\) bounded by the area of a fixed further enlargement: every conflicting disjoint region supplies a different integer site there. Choose an independent subset of at least \(2^{i}/(\Delta_{\mathrm{raw}}+1)\) leaves before taking probabilities. Each costs at most \(a_{0} C_{\mathrm{raw}} p\), and the number of trees is at most \(C_{\mathrm{tr}}^{2^{i}}\), with \(C_{\mathrm{tr}}\) fixed by the finite mesh lists and the finite root enlargement. Thus, after reducing \(p\), every required raw test has bound

\begin{equation}\label{q1:e12.2}
\begin{gathered}
\Pr(\text{raw rank-}i\text{ failure}) \le  p^{c_{\mathrm{raw}} 2^{i}}
\end{gathered}
\end{equation}

with, for example, \(c_{\mathrm{raw}}=1/[4(\Delta_{\mathrm{raw}}+1)]\) after absorbing the root-cover constants. An explicit sufficient smallness condition for the witness part is \(p \le  [C_{\mathrm{tr}}^{\Delta_{\mathrm{raw}}+1} a_{0} C_{\mathrm{raw}}]^{-2}\); finitely many extra constants are absorbed by another positive rational threshold. This proof uses finite dependence only, and applies equally to independently appended artificial negative-time keys.

\smallskip\noindent\textbf{Step 3: Admission scales and the large-leaf packing constant.} Let \(j_{s},j_{l}\) be the least nonnegative integers with

\[
\begin{gathered}
b_{c}^{j_{s}}\ge p^{-17},\qquad b_{c}^{j_{l}}\ge p^{-(b_{\mathrm{refresh}}+6)}.
\end{gathered}
\]

For small \(p, B_{c}^{j_{s}}\le p^{-\rho_{s}}\) and \(B_{c}^{j_{l}}\le p^{-\rho_{l}}\). Indeed \(j_{s}<17 \log _{b_{c}}(1/p)+1\) and \(B_{c}\le b_{c}^{R_{\mathrm{geom}}}\); the extra constants and exponents in \(\rho_{s},\rho_{l}\) dominate the final factor \(B_{c}\). Also \(H_{m}=p^{-o(1)}\); all raw, short-prefix and paint enlargements are smaller than one rank-\(j_{s}\) mesh cell, and all raw, short and long enlargements are smaller than one rank-\(j_{l}\) mesh cell. These assertions hold uniformly once \(m\) is sufficiently large.

The packing concerns ancestor-restricted regions before intersection with the requested target. For a fixed root, put every descendant mesh on its translated origin. \(P_{c},T_{c}\) are integers, so every ancestor boundary lies on the rank-\(l\) mesh with cell sides \(w_{c} P_{c}^{l},w_{c} T_{c}^{l}\). Every nonempty ancestor-restricted leaf region contains a whole such cell. Distinct regions are disjoint, by their first disjoint siblings. A full rank-\(l\) leaf has two mesh cells per side, and its support enlargement at the relevant admission is at most one cell per side. If two enlarged supports meet, the chosen cells of the two ancestor-restricted regions lie in a fixed square of at most 20 mesh cells per side about the first full leaf. Each cell can be supplied by at most one disjoint region. Thus degree \(<20^{2}\); \(\Delta_{\mathrm{ad}}=10000\) is a common safe integer bound even with endpoint and collar conventions. This constant is independent of \(p\), helper counts, \(h_{\mathrm{red}}\) and the admission rank. Those constants determine how small \(p\) must be before the one-cell condition holds. A fixed enlarged centered root is first covered by its bounded list of doubled root boxes. Enlarging \(C_{\mathrm{tr}}\) accounts for that list. This proves the packing assertion even if the original target clips a leaf thinly.

\smallskip\noindent\textbf{Step 4: Short admission.} At rank \(j_{s}\) call a leaf bad if its raw restriction is nonsparse or if a short-center paint meets it. A union bound over all possible center positions gives

\begin{equation}\label{q1:e12.3}
\begin{aligned}
q_{s} &\le p^{c_{\mathrm{raw}} 2^{j_{s}}} + C (B_{c}^{j_{s}}+H_{m})^{2} p^{M_{s}}\\
&\le p^{c_{\mathrm{raw}} 2^{j_{s}}} + C' p^{M_{s}-2\rho_{s}}.
\end{aligned}
\end{equation}

Both summands have primitive support in the one-cell enlargement just described. In a fixed binary witness with \(N=2^{i-j_{s}}\) leaves, retain an independent-support subset of size at least \(N/(\Delta_{\mathrm{ad}}+1)\). Multiplying its probabilities and summing at most \(C_{\mathrm{tr}}^{N}\) witnesses gives

\[
\begin{gathered}
\Pr(\text{raw+short rank-}i\text{ failure}) \le  [C_{\mathrm{tr}} q_{s}^{1/(\Delta_{\mathrm{ad}}+1)}]^{N}.
\end{gathered}
\]

The explicit bound on \(M_{s}\) makes \(M_{s}-2\rho_{s}\ge 2(\Delta_{\mathrm{ad}}+1)D_{s}+20\). Moreover \(c_{\mathrm{raw}} 2^{j_{s}}\) tends to infinity as \(p\) tends to zero, even though \(c_{\mathrm{raw}}\) was fixed only after physical compilation. Taking \(p\) smaller makes the bracket at most \(p^{D_{s}}\). Hence for every \(i>j_{s}\),

\begin{equation}\label{q1:e12.4}
\begin{gathered}
\Pr(\text{raw+short rank-}i\text{ failure}) \le  p^{D_{s} 2^{i-j_{s}}}.
\end{gathered}
\end{equation}

Selection is performed separately for each deterministic tree; dependence of overlapping leaf events causes no product assertion.

\smallskip\noindent\textbf{Step 5: Long admission.} At rank \(j_{l}\) a leaf is bad for the full field only if its raw+short restriction is nonsparse or some enlarged long-center paint meets it. The bound is

\begin{equation}\label{q1:e12.5}
\begin{gathered}
q_{l} \le  p^{D_{s} 2^{j_{l}-j_{s}}} + C p^{M_{l}-2\rho_{l}}.
\end{gathered}
\end{equation}

Here \(b_{\mathrm{refresh}}>20\) gives

\[
\begin{gathered}
j_{l}-j_{s} \ge  (b_{\mathrm{refresh}}-11)\log _{b_{c}}(1/p)-1 \to  \infty.
\end{gathered}
\]

The first exponent in \eqref{q1:e12.5} therefore exceeds every fixed constant for all sufficiently small \(p\). The explicit \(M_{l}\) bound and the same \(\Delta_{\mathrm{ad}}\)-coloring show, for every \(i>j_{l}\),

\begin{equation}\label{q1:e12.6}
\begin{gathered}
\Pr(\text{full rank-}i\text{ failure}) \le  p^{D_{l} 2^{i-j_{l}}}.
\end{gathered}
\end{equation}

The full key support still fits in the one-cell enlargement at admission because the long support is \(C p^{-(b_{\mathrm{refresh}}+5)}\), a factor \(p\) smaller than its leaf-side lower bound. Complete primitive prefixes, anomaly stencils and mismatch keys have all been included.

\smallskip\noindent\textbf{Step 6: The exact decomposed mark list.} Use a common fixed physical scaling factor \(C_{\mathrm{ball}}\) large enough for all root collars, prefix supports, paint displacements, primitive stencils and residue offsets. The types are:

\begin{itemize}
\item \(R_{i}\) for \(0\le i\le j_{s}\): a raw nonsparsity event, centered at the queried root, radius \(\lceil C_{\mathrm{ball}} B_{c}^{i}\rceil\), probability bound \(p^{c_{\mathrm{raw}} 2^{i}}\).
\item \(\mathsf S\): a short bad center, radius \(\lceil C_{\mathrm{ball}} B_{c}^{j_{s}}\rceil\), probability bound \(C_{s} p^{M_{s}}\).
\item \(M_{i}\) for \(j_{s}<i\le j_{l}\): a raw+short nonsparsity event, radius \(\lceil C_{\mathrm{ball}} B_{c}^{i}\rceil\), probability bound \(p^{D_{s} 2^{i-j_{s}}}\).
\item \(\mathsf L\): an enlarged long bad center, radius \(\lceil C_{\mathrm{ball}} B_{c}^{j_{l}}\rceil\), probability bound \(C_{l} p^{M_{l}}\).
\item \(H_{i}\) for \(i>j_{l}\): a full nonsparsity event, radius \(\lceil C_{\mathrm{ball}} B_{c}^{i}\rceil\), probability bound \(p^{D_{l} 2^{i-j_{l}}}\).
\end{itemize}

For \(i\le j_{s}\) a failed full test either has a short or long paint center meeting it, or equals a raw failure. In the first case the ball at that center covers the root: test radius \(B_{c}^{i}\) and its paint radius \(H_{m}\) are at most a fixed multiple of the respective admission radius. For \(j_{s}<i\le j_{l}\) the same reasoning removes only long centers and leaves a raw+short failure. For \(i>j_{l}\) use the full test itself. This proves coverage without declaring an undersized full test to have local primitive support. Each type is a function only of the independent keys inside its advertised ball. A single \(\mathsf S\) or \(\mathsf L\) center ball covers every lower-rank root implicated by that center, so no sum over those lower ranks is needed. The root test geometry has only finitely many stencil variants and residue lanes; take those as a fixed finite type multiplicity. Physical lane changes require this union and radius inflation, but do not change the probability argument.

\smallskip\noindent\textbf{Step 7: Every part of the moment sum.} Absorb \(C_{\mathrm{ball}}\), integer rounding and the fixed type multiplicity in a constant \(C_{\mathrm{mom}}\). Replacing \(d\) by \(d_{\mathrm{mom}}\) only strengthens the requested result. The raw contribution is bounded by

\begin{equation}\label{q1:e12.7}
\begin{aligned}
C_{\mathrm{mom}} \sum _{i\ge 0} B_{c}^{d_{\mathrm{mom}} i} p^{(c_{\mathrm{raw}}/2)2^{i}}
&\le C_{\mathrm{mom}} p^{c_{\mathrm{raw}}/2}/(1-B_{c}^{d_{\mathrm{mom}}} p^{c_{\mathrm{raw}}/2}),
\end{aligned}
\end{equation}

when the denominator is positive. We used \(2^{i}\ge i+1\).
The \(\mathsf S\) and \(\mathsf L\) contributions are respectively at most

\begin{equation}\label{q1:e12.8}
\begin{gathered}
C_{\mathrm{mom}} \sqrt{C_{s}} p^{M_{s}/2-d_{\mathrm{mom}} \rho_{s}}, \\ C_{\mathrm{mom}} \sqrt{C_{l}} p^{M_{l}/2-d_{\mathrm{mom}} \rho_{l}}.
\end{gathered}
\end{equation}

Both exponents are positive by the stated \(M\) choices. For middle ranks write \(i=j_{s}+j, j\ge 1\). Since \(2^{j-1}\ge j\),

\begin{equation}\label{q1:e12.9}
\begin{aligned}
\sum _{\mathrm{middle}} (1+r)^{d_{\mathrm{mom}}} \sqrt{z}
&\le C_{\mathrm{mom}} p^{-d_{\mathrm{mom}} \rho_{s}} \sum _{j\ge 1} B_{c}^{d_{\mathrm{mom}} j} p^{D_{s} 2^{j-1}}\\
&\le C_{\mathrm{mom}} p^{-d_{\mathrm{mom}} \rho_{s}} \bigl[B_{c}^{d_{\mathrm{mom}}} p^{D_{s}}/(1-B_{c}^{d_{\mathrm{mom}}} p^{D_{s}})\bigr].
\end{aligned}
\end{equation}

This tends to zero because \(D_{s}>d_{\mathrm{mom}} \rho_{s}\). The full high-rank sum has exactly \eqref{q1:e12.9} with \(\rho_{l},D_{l}\) instead. Thus the sum of all five ranges tends to zero as \(p\) tends to zero. Choose a positive rational threshold so that all denominators exceed \(1/2\), each preceding admission and prefix estimate holds, and the sum of \eqref{q1:e12.7}-\eqref{q1:e12.9} is less than \(\delta\). All inequalities are effective: use certified rational bounds for fixed logarithms and constants. For example, if \(R_{2}=\lceil \log _{2} b_{c}\rceil\), then \(2^{j_{s}}\ge p^{-17/R_{2}}\) and \(2^{j_{l}-j_{s}}\ge (1/2)p^{-(b_{\mathrm{refresh}}-11)/R_{2}}\), supplying rational-power sufficient conditions for the two diverging exponents. Raising both sides to integer powers removes fractional-power comparisons.

Finally select \(J_{\mathrm{pad}}\) large enough that every \(m\ge 2J_{\mathrm{pad}}+1\) has \(p\) below this common threshold and all fixed support prefactors fit the admissions. This is effective because \(p\le 2^{-8^{m+n_{\mathrm{bin}}}}\) and \((m+2)/8^{m+n_{\mathrm{bin}}}\) decreases with \(m\). The same bounds dominate any actual rate sequence in its chosen narrow physical window by Step 1 and the primitive envelope of Step 2. They therefore hold simultaneously for every \(n\) and every actual rate, rather than only at one nominal bin. For a finite cap the retained list is a subset of this list plus the same two admission centers; \eqref{q1:e12.1} still holds and all radii are finite. This proves the moment cover, including arbitrary moment exponents, dependent raw flags and enlarged long failures. 
\end{proof}

\begin{lemma}[The complete-symbol selector kernel]\label{q1:l17}

Let the native R2 alphabet have cardinality \(M\), let \(k_{\mathrm{sel}}=\lceil \log _{2} M\rceil\), and number its symbols \(0,\dots,M-1\). At each native site and date use a fresh rare factory bit \(B_{\mathrm{rare}}\) with law \(\mathrm{Bernoulli}(\beta(\eps))\) and \(k_{\mathrm{sel}}\) fresh independent fair factory bits, forming \(W\) uniformly in \(\{0,\dots,2^{k_{\mathrm{sel}}}-1\}\). These abstract factory outputs are defined even if a factory has too many physical faults. Set \(J=W\) and accept a replacement exactly on \(\{B_{\mathrm{rare}}=1,W<M\}\). Put \(\gamma(\eps)=\beta(\eps) M/2^{k_{\mathrm{sel}}}\). The function \(\beta\) is a polynomial in \(\eps\) with rational coefficients, because the rare factory is a finite rational classical channel. For every native neighborhood \(a\) and native output symbol \(b\),

\[
\begin{gathered}
\Pr(\text{accepted replacement with symbol }j) = \beta(\eps)/2^{k_{\mathrm{sel}}} = \gamma(\eps)/M, \\ 0 \le  j < M, \\ K_{\gamma}(a,b) = (1-\gamma(\eps))\, \one[b=f(a)] + \gamma(\eps)/M.
\end{gathered}
\]

The innovations are independent over native sites and dates, independent of the incoming native state, and replace one complete native symbol, preserving the stipulated common replacement flag across all helper and core fields sharing that symbol. The compiled physical semantic transition equals this native transition except on the separately marked complete-stencil excess-fault event. The kernel assertion is unconditional; it is not asserted conditional on no excess faults.

\end{lemma}
\begin{proof}
The sample factories use disjoint private primitive wire-time keys and reset every incoming input. Their abstract outputs are thus functions of those keys alone. The rare factory defines \(\beta(\eps)\) as its unconditional final-majority probability. The complement-equivariant fair factory of the paragraph on redundant Boolean steps and exact draws in Section~\ref{q1:selector} has a measure-preserving involution complementing its original fair draw and every replacement value on a complementing wire. Fixed (invariant) wires remain fixed. Every gate in the oriented circuit respects this involution, and the final majority complements. Hence its output is exactly fair, including on failed factories. Disjointness makes the \(k_{\mathrm{sel}}\) outputs independent and independent of \(B_{\mathrm{rare}}\) and the prior native state. Therefore \(\operatorname{Pr}(B_{\mathrm{rare}}=1,W=j)=\beta(\eps)2^{-k_{\mathrm{sel}}}\) for every \(j\). Summing over the \(M\) accepted \(j\) gives total replacement probability \(\gamma(\eps)\); the complementary event leaves \(f(a)\). Adding these two disjoint cases proves the displayed kernel, including \(b=f(a)\), where both cases contribute.

No rejection loop or conditional redrawing occurs. This constructs independent native innovations on the same probability space as the circuit for every realization. An unused replacement mark on rejection may be assigned an extra independent mathematical value; it is never read by the physical circuit. The one complete-symbol parser and protected Boolean transition implement \(f(a)\), or the chosen symbol \(J\), on their actual decoded incoming neighbors whenever the full incoming and outgoing normalization and transition stencils are good. The contrapositive is exactly the explicit excess-flag mark of the same paragraph and of the central normalization convention in Section~\ref{q1:selector}; it includes storage, initialization, all reader paths and final normalization. Consequently any semantic deviation is contained in that mark. Factory outputs and anomaly marks may share keys. Conditioning on anomaly absence could change their distribution, so the comparison keeps both on the unconditional product-key space. Induction on a finite backward native cone gives agreement with the exact native process when no anomaly occurs in that cone. This proves the asserted distinction between exact comparison innovations and compiled evolution. 
\end{proof}

\section{Effective parameter choices}\label{q1:parameters}
This section lists every constant of the ring construction with a constructive bound; Lemma~\ref{q1:l16} shows that they can be chosen in a noncircular order.

The following list gives names and constructive bounds; the order of
choice is given in the proof of Lemma~\ref{q1:l16}. Rank, position, date and code-coordinate indices are variables; distinct named constants in this list remain distinct.

Names in Sections~\ref{q1:model}--\ref{q1:effective} are local to the ring construction. Where a name also occurs in Sections~\ref{sec:controller}--\ref{sec:four-dimensional} or Appendices~\ref{app:code-estimates}--\ref{sec:folded} (\(B\), \(C_h\), \(D_s\), \(K_{\mathrm{cap}}\), \(C_{\mathrm{cap}}\), \(C_{\mathrm{ball}}\), \(C_{\mathrm{raw}}\), \(\Delta_{\mathrm{raw}}\), \(\eps_*\), \(j_*\), \(k_0\), \(N_0\)), it denotes the ring quantity defined here.

\paragraph{Names and constructive bounds.}

\noindent\textbf{\(U_{\mathrm{target}}\)}. Fixed universal target halting machine.\par\smallskip
\noindent\textbf{\(V_{\mathrm{unary}}\)}. One fixed one-ended machine simulating \(U_{\mathrm{target}}(\nu_{2}(m))\); fixed before all R2 compilation.\par\smallskip
\noindent\textbf{\(n_{\mathrm{target}}\)}. Variable theorem input.\par\smallskip
\noindent\textbf{\(J_{\mathrm{pad}}, m_{\mathrm{bin}}(n)=2^{n}(2J_{\mathrm{pad}}+1)\)}. \(J_{\mathrm{pad}}\) is chosen last among global constants. Increasing it only skips bins of an already fixed physical circuit.\par\smallskip
\noindent\textbf{\(S_{\mathrm{slot}}\)}. Fixed native quantum factors per site, enough for the finite universal tape tile, contact and port roles. Choose by listing those roles and summing their bounded widths, before the code. All unused roles have mandatory boundary disposal and cuts.\par\smallskip
\noindent\textbf{\(h_{\mathrm{cover}}=10\)}. Complete descendant geometric collar; \(P_{q},T_{q}\ge 4\) makes the geometric sum of per-rank neighbor collars smaller than this.\par\smallskip
\noindent\textbf{\(m_{\mathrm{bad}}=(10h_{\mathrm{cover}}+40)^{2}=19600\)}. Bad named-child threshold, unrelated to \(m_{\mathrm{bin}}\) and to helper channel count.\par\smallskip
\noindent\textbf{\(K_{\mathrm{inc}}=10^{5}(S_{\mathrm{slot}}+1)^{2}\)}. Effective contact and location faults charged per named child; fixed arity and finite port incidence only. It does not depend on \(w\), code length, route length or private CPU running time.\par\smallskip
\noindent\textbf{\(b_{\mathrm{fault}}=K_{\mathrm{inc}}(m_{\mathrm{bad}}+1)\)}. Effective complete-gadget budget; counts all named children in the enlarged cover.\par\smallskip
\noindent\textbf{\(s_{\mathrm{code}}, q_{\mathrm{code}}=2^{s_{\mathrm{code}}}-1, n_{\mathrm{code}}=2^{2s_{\mathrm{code}}+1}-1\)}. Choose least admissible \(s_{\mathrm{code}}\) with \(q_{\mathrm{code}}-8\ge 8(20b_{\mathrm{fault}}+8)\). The source's finite CSS construction and finite gadget lists are then computable.\par\smallskip
\noindent\textbf{\(r_{\mathrm{code}}, t_{\mathrm{code}}\)}. Correctable radii and fixed clean decoder determined by \(s_{\mathrm{code}}\). Use the source's stronger margin, not a new code assumption.\par\smallskip
\noindent\textbf{\(N_{\mathrm{norm}},N_{\mathrm{contact}},W_{\mathrm{fixed}}\)}. Enumerated contact counts of the \(b_{\mathrm{fault}}+1\) leading \(\operatorname{EC}\) passes and all candidate complete \(\operatorname{EC}\), body and idle lists, and number of fixed data, catalogue, work, transport and guard positions. Finite before \(w_{q}\).\par\smallskip
\noindent\textbf{\(\lambda_{\mathrm{tile}},a_{\mathrm{input}}\)}. The native tape tile and \(W_{\mathrm{fixed}}/\lambda_{\mathrm{tile}}\) are fixed before \(v_{\mathrm{text}}\). Set \(a_{\mathrm{input}}(v)=a_{0}+4|v|\), with \(a_{0}\) fixed before \(v\); the indexed generator computes this coefficient from the supplied literal \(v\). Compute the fixed part by enumerating finite tile and input formats.\par\smallskip
\noindent\textbf{\(A_{\mathrm{index}}=32\)}. Conservative absolute runtime exponent for the explicitly described indexed generator and its fixed single-tape simulation.\par\smallskip
\noindent\textbf{\(c_{\mathrm{pad}}=35\)}. Equal to \(A_{\mathrm{index}}+3\); it is unrelated to \(c_{1}, c_{\mathrm{raw}}\) or \(C_{\mathrm{path}}\).\par\smallskip
\noindent\textbf{\(v_{\mathrm{text}},C_{\mathrm{index}}(v_{\mathrm{text}},c_{\mathrm{pad}})\)}. Finite text of the indexed generator, used as its own quoted input. \(C_{\mathrm{index}}=2^{80} D_{\mathrm{desc}}^{64}\) is obtained from explicit operation counts and finite literal widths in that lemma, independent of \(w_{q}\). \(D_{\mathrm{desc}}\) and \(C_{\mathrm{index}}\) are analysis-only and are never embedded in \(v_{\mathrm{text}}\).\par\smallskip
Section~\ref{q1:compiler} writes these constants \(S\), \(h_{\mathrm{cov}}\), \(K\), \(b\), \(a_{\mathrm{in}}\), \(A_{\mathrm{idx}}\), \(C_{v,c}\) for \(S_{\mathrm{slot}}\), \(h_{\mathrm{cover}}\), \(K_{\mathrm{inc}}\), \(b_{\mathrm{fault}}\), \(a_{\mathrm{input}}\), \(A_{\mathrm{index}}\), \(C_{\mathrm{index}}\), the bound \(A_{0}\) of \eqref{q1:e7.3} is \(A_{\mathrm{children}}\), and its \((w,P,T)\) is \((w_{q},P_{q},T_{q})\).\par\smallskip
\noindent\textbf{\(w_{q},P_{0,q}=2^{w_{q}},P_{q}=5P_{0,q},T_{q}\)}. Search integers \(w_{q}\) until all R1 lower bounds, tape runtime, \(F(w_{q})\le \lfloor 3^{w_{q}}/w_{q}^{2}\rfloor\le T_{q}, G(w_{q})\le P_{q}\) and all serialization inequalities hold, including \(P_q\le T_q\). Enumeration of the R1 native schedule defines the integer \(T_{q}\). Polynomial terms are dominated by \(2^{w}\) and \((3/2)^{w}\), so this search terminates. This is also the structural R1 geometry \((w_{R},P_{R},T_{R})\), not the selector geometry.\par\smallskip
\noindent\textbf{\(q_{R}=w_{q}^{2},A_{\mathrm{Rstruct}}=w_{q}^{3},B_{\mathrm{Rattempt}}=128q_{R},v_{\mathrm{Rspeed}}=w_{q}^{4},d_{\mathrm{Rguard}}=4w_{q}+8\)}. Native structural program constants; all fixed functions of \(w_{q}\). Distinct from alphabet size, probability base, \(b_{\mathrm{fault}}\) and physical speed.\par\smallskip
\noindent\textbf{\(d_{0,R}=(2A_{\mathrm{Rstruct}}+4)(2q_{R}+2),d_{1,R}=q_{R} d_{0,R},M_{\mathrm{Rverify}}=128d_{\mathrm{Rguard}}\)}. Native schedule lengths.\par\smallskip
\noindent\textbf{\(r_{\mathrm{Rwithdraw}},s_{0,R},s_{1,R},s_{\mathrm{Rcomp}},\mathrm{collar}_{R}=\lfloor w_{q}/5\rfloor\)}. Integer dates and structural collar obtained by enumerating that schedule, not adjustable noise parameters.\par\smallskip
\noindent\textbf{\(\ell_{R,0}\) and \(\ell_{R,j}=2^{j} \ell_{R,0}\)}. Physical family base word width sufficient for the fixed nonhalting literal; grows only at conceptual ranks.\par\smallskip
\noindent\textbf{\(P_{c},T_{c},w_{c},b_{c}=\min (P_{c},T_{c}),B_{c}=\max (P_{c},T_{c})\)}. Native selector compiler geometry. Choose before helper counts and widths; the packet's packed word and population-count instruction make execution count independent of helper channel count.\par\smallskip
\noindent\textbf{\(D_{\mathrm{boot}}\)}. Fixed integer large enough that \(H_{\mathrm{boot}}(m)=D_{\mathrm{boot}}^{m+2}\) covers all work and collars through \(m+1\). For example search \(D_{\mathrm{boot}}\ge 2B_{c}\) against the finite collar multiplier; \(H_{\mathrm{boot}}\) is \(p^{-o(1)}\).\par\smallskip
\noindent\textbf{\(A_{\mathrm{bin}}(m)=8^{m+n_{\mathrm{bin}}}, n_{\mathrm{bin}}\)}. Native bins \((2^{-A_{\mathrm{bin}}},2^{1-A_{\mathrm{bin}}})\). Freeze the source-admissible integer offset \(n_{\mathrm{bin}}\) once after the helper tuple, before computing the final physical circuit. Do not enlarge it later.\par\smallskip
\noindent\textbf{\(A_{\mathrm{children}}=(20P_{q}+5)(20T_{q}+5), A_{\mathrm{pair}}=\max (1,A_{\mathrm{children}}^{2})\)}. Finite upper bound on complete-cover child centers and ordered pairs. Includes \(h_{\mathrm{cover}}\) collars. Loose bounds are sufficient.\par\smallskip
\noindent\textbf{\(d_{\mathrm{moment}}\)}. Least integer \(d\ge 4\) with \(4P_{q} T_{q}\le P_{q}^{d-2}\); integer search. This degree is chosen after quantum geometry and before helper exponents. Lemma~\ref{q1:l12} is applied with \(d=d_{\mathrm{moment}}\), so its integer \(d_{\mathrm{mom}}\) equals \(d_{\mathrm{moment}}\).\par\smallskip
\noindent\textbf{\(R_{\mathrm{geom}}\)}. Least positive integer \(R\) with \(b_{c}^{R}\ge B_{c}\).\par\smallskip
\noindent\textbf{\(\rho_{s}=18R_{\mathrm{geom}}+2\)}. Integer bound \(B_{c}^{j_{s}}\le p^{-\rho_{s}}\), after small-rate cutoff. Depends only on selector geometry.\par\smallskip
\noindent\textbf{\(\Delta_{\mathrm{ad}},C_{\mathrm{tr}}\)}. Use \(\Delta_{\mathrm{ad}}=10000\) as in Lemma \ref{q1:l12}. The enlarged-leaf packing degree and binary tree entropy are computed from the fixed rectangle mesh; independent of helper counts and \(p\).\par\smallskip
\noindent\textbf{\(D_{s}=2d_{\mathrm{moment}} \rho_{s}+20\)}. Exponent after short admission, distinct from elementary period \(D_{\mathrm{layer}}\).\par\smallskip
\noindent\textbf{\(M_{s}\)}. \(M_{s}=\max (2\rho_{s}+2(\Delta_{\mathrm{ad}}+1)D_{s}+20,2(d_{\mathrm{moment}}+1)\rho_{s}+20)\), the deterministic value displayed in Lemma~\ref{q1:l12}.\par\smallskip
\noindent\textbf{\(b_{\mathrm{refresh}}=\max (21,M_{s}+20),v_{\mathrm{rep}}=M_{s}+20,s_{\mathrm{read}}=4M_{s}+85\)}. Fixed short-helper refresh, replica and odd-reader counts; give exponents exceeding \(M_{s}\) with slack. This \(b_{\mathrm{refresh}}\) is the source b0, unrelated to \(b_{\mathrm{fault}}\).\par\smallskip
\noindent\textbf{\(\rho_{l}=(b_{\mathrm{refresh}}+7)R_{\mathrm{geom}}+2\)}. Bound \(B_{c}^{j_{l}}\le p^{-\rho_{l}}\), chosen after \(b_{\mathrm{refresh}}\).\par\smallskip
\noindent\textbf{\(D_{l}=2d_{\mathrm{moment}} \rho_{l}+20, M_{l}\)}. Long exponent analogous to short; \(M_{l}\) obeys both displayed leaf and moment requirements with \(\rho_{l},D_{l}\).\par\smallskip
\noindent\textbf{\(m_{\mathrm{ch}}=20(M_{l}+20)+1,\ell_{c,0}\ge m_{\mathrm{ch}}+4\)}. Odd long-channel count and fixed native R2 base word width. These do not enlarge \(P_{c},T_{c}\).\par\smallskip
\noindent\textbf{\(h_{\mathrm{red}}=C_{\mathrm{pref}}+M_{l}+5, B_{\mathrm{redund}}=4h_{\mathrm{red}}+3\)}. Physical protected-Boolean redundancy. \(C_{\mathrm{pref}}=2(b_{\mathrm{refresh}}+6)+5\) bounds full native comparison-prefix area exponent; \(h_{\mathrm{red}}+1>C_{\mathrm{pref}}+M_{l}+4\).\par\smallskip
\noindent\textbf{\(M_{\mathrm{alphabet}},k_{\mathrm{draw}}=\lceil \log _{2} M_{\mathrm{alphabet}}\rceil\)}. Complete native R2 alphabet cardinality after geometry, helper tuple and \(n_{\mathrm{bin}}\) are frozen. The replacement is for this whole symbol.\par\smallskip
\noindent\textbf{\(\beta(\eps),\gamma(\eps)=\beta(\eps)M_{\mathrm{alphabet}}/2^{k_{\mathrm{draw}}}\)}. Exact computable rational polynomials obtained by enumerating the finite independent rare and fair factories. \(\beta(0)=0,\beta'(0)>0\); \(\gamma'(0)>0\).\par\smallskip
\noindent\textbf{\(\eps_{\mathrm{deriv}}\)}. Rational neighborhood with \(\gamma'>0\), found by rational polynomial coefficient bounds and search.\par\smallskip
\noindent\textbf{\(D_{\mathrm{layer}},N_{\mathrm{wire}},r_{\mathrm{gate}}\ge 1\)}. Enumerate the finished period: its elementary layer count, qubits per site, maximum gate spatial input and output displacement. Includes all waiting noise; all are fixed after \(n_{\mathrm{bin}},h_{\mathrm{red}}\).\par\smallskip
\noindent\textbf{\(D_{\mathrm{stride}}\)}. Physical stride of the native R2 data field; enumerated with the finished period.\par\smallskip
\noindent\textbf{\(R_{\mathrm{macro}}=D_{\mathrm{layer}} r_{\mathrm{gate}},v_{\mathrm{quant}}\le R_{\mathrm{macro}}\)}. Finite physical speed per macro round; \(v_{\mathrm{quant}}\) bounds every allowed opcode in the conservative quantum graph.\par\smallskip
\noindent\textbf{\(C_{\mathrm{cone}}=16(R_{\mathrm{macro}}+P_{q}+1)\)}. Explicit integer chart constant, fixed before \(J_{\mathrm{pad}}\).\par\smallskip
\noindent\textbf{\(c_{1}=1/(12 C_{\mathrm{cone}} T_{q})\)}. Fixed rational elementary-time linear lifetime constant.\par\smallskip
\noindent\textbf{\(s_{\mathrm{clock}},r_{\mathrm{clockx}},r_{\mathrm{clockt}}\)}. Number of keys and spatial and temporal radii of complete R1 compiled update and raw-read stencil. Enumerate after \(D_{\mathrm{layer}}\); includes waiting storage.\par\smallskip
\noindent\textbf{\(A_{\mathrm{clock}}=4w_{q}^{2} s_{\mathrm{clock}},J_{\mathrm{clock}}=(2w_{q}+4r_{\mathrm{clockx}}+3)(2w_{q}+4r_{\mathrm{clockt}}+3)\)}. Leaf union bound and conflict degree plus one. Distinct from R1 alphabet and bad-pair entropy.\par\smallskip
\noindent\textbf{\(D_{\mathrm{clock}}=8(P_{q}+2)^{2}(T_{q}+2)^{2},N_{\mathrm{clockmesh}}=4(P_{q}+3)(T_{q}+3)\)}. Explicit pair and root mesh counts.\par\smallskip
\noindent\textbf{Clock probability constants.} Lemma~\ref{q1:l1} writes \(A_R=A_{\mathrm{clock}}\), \(J_R=J_{\mathrm{clock}}\), \(D_R=D_{\mathrm{clock}}\) and \(N_R=N_{\mathrm{clockmesh}}\), with \((w_R,P_R,T_R)=(w_q,P_q,T_q)\), \(s_R=s_{\mathrm{clock}}\), \(r_x=r_{\mathrm{clockx}}\) and \(r_t=r_{\mathrm{clockt}}\), for the final structural geometry and complete physical stencils, and sets
\[
c_R=J_R^{-1},\qquad C_R=N_RD_RA_R^{1/J_R},\qquad
 \eps_{R0}=A_R^{-1}(2N_RD_R)^{-J_R}.
\]
The bound is \(z_{R,j}(\eps)=(C_R\eps^{c_R})^{2^{j-1}}\). These constants are fixed before the final rate cutoff.\par\smallskip
\noindent\textbf{\(W_{*},N_{\mathrm{clk}},C_{\mathrm{det}},C_{\mathrm{clk}}\)}. The constants of Lemma~\ref{q1:l3} at the final geometry: \(W_{*}=\lceil (2T_{q}+18P_{q})/(P_{q}-1)\rceil+2\), \(N_{\mathrm{clk}}=W_{*}+8\), \(C_{\mathrm{det}}=100(w_{q}+W_{*}+T_{q}+P_{q}+1)\), \(C_{\mathrm{clk}}=C_{\mathrm{det}}+r_{\mathrm{clockx}}+r_{\mathrm{clockt}}+2\).\par\smallskip
\noindent\textbf{\(\eps_{R}(d,\delta)\)}. The explicit rational moment cutoff of Lemma~\ref{q1:l1.1}.\par\smallskip
\noindent\textbf{\(C_{\mathrm{tail}},C_{\mathrm{base}},C_{\mathrm{rec}}=\max (C_{\mathrm{tail}},C_{\mathrm{base}})\)}. Finite key-center and footprint counts after full geometry. They can be bounded by \(1000(2h_{\mathrm{cover}}+1)^{2}P_{q} T_{q}\), increased to cover the finite template offset list if that list is kept separately. Enumeration gives rational integer upper bounds.\par\smallskip
\noindent\textbf{\(C_{\mathrm{top}}\)}. Number of top output, home and port-neighbor tests used by the fixed recovered nonport; take the size of its finite complete-neighbor list.\par\smallskip
\noindent\textbf{\(C_{\mathrm{cap}},C_{\mathrm{ad}},C_{\mathrm{ball}},N_{\mathrm{res}},C_{s},C_{l},C_{\mathrm{pref}}\)}. \(C_{\mathrm{pref}}\) is the exponent defined in the \(h_{\mathrm{red}}\) row. \(C_{\mathrm{cap}},C_{\mathrm{ad}},C_{\mathrm{ball}}\) are maximum support, collar and paint scaling constants; \(N_{\mathrm{res}}\) counts the finite residue and read-template list. Enumerate all final physical lane offsets and prefix templates to obtain integer upper bounds. \(C_{s},C_{l}\) are the finite union-bound prefactors (or one after slack absorption). \(P_{c},T_{c},H_{\mathrm{boot}}\) initial use logical lane coordinates; the final finite physical stride enters these scaling constants after helper counts are fixed.\par\smallskip
\noindent\textbf{\(C_{\mathrm{path}}\)}. An explicit sufficient choice is \((100(v_{\mathrm{quant}}+4))^{4}\): it bounds both center placement factors and radius-sum loss for the expanded-ball chain in \(1+1\) dimensions. Fixed after the whole circuit.\par\smallskip
\noindent\textbf{\(b_{*}\le 1/(4A_{\mathrm{pair}}),\delta_{\mathrm{mark}}>0\)}. Choose positive rational \(b_{*}\), then strict \(\delta_{\mathrm{mark}}\) with \(C_{\mathrm{rec}} \delta_{\mathrm{mark}}\le b_{*}/4\) and \(C_{\mathrm{path}} \delta_{\mathrm{mark}}<1/2\). Split this tolerance over direct, clock and selector families with strict slack. Tolerances affect only final small-rate cutoff, not helper exponents.\par\smallskip
\noindent\textbf{\(\bar\eps_{m},p_{m}=\gamma(\bar\eps_{m})/2,I_{\mathrm{target}}(n)=(\eps_{\mathrm{inf},m},\eps_{\mathrm{sup},m})\)}. Rational \(\bar\eps_{m}\) in strict monotone preimage of native bin \(m=m_{\mathrm{bin}}(n)\); enumerate it. Choose the rational halfwidth \(\zeta_{m}\le p_{m}^{M_{l}+C_{\mathrm{pref}}+2}/(16 C_{N})\), where \(C_{N}p_{m}^{-C_{\mathrm{pref}}}\) bounds the primitive wire-dates of a backward prefix (first step of the proof of Lemma~\ref{q1:l12}), so the entire band of width \(2\zeta_{m}\) costs at most \(p_{m}^{M_{l}+2}/8\); also require \(\eps_{\mathrm{sup},m}<2\bar\eps_{m}\) and all moment and rate cutoffs. Bounds use a complete enumerated prefix key count.\par\smallskip
\noindent\textbf{\(j_{s},j_{l}\)}. Least ranks with \(b_{c}^{j}\) above \(p_{m}^{-17}\) and \(p_{m}^{-(b_{\mathrm{refresh}}+6)}\), with the least-rank definitions in Lemma \ref{q1:l12}. Integer powers and rational comparisons compute them.\par\smallskip
\noindent\textbf{\(r_{a},z_{a}\)}. Distinct mark-type radius and probability bound, not a rate. All support and offset constants come from finite mesh enumeration; infinite rank tails have the explicit convergent majorants in Lemmas~\ref{q1:l1.1} and \ref{q1:l12}.\par\smallskip
\noindent\textbf{\(k_{\mathrm{rank}},H_{\mathrm{time}}=T_{q}^{k_{\mathrm{rank}}},k_{0},L_{1}(n)\)}. \(k_{\mathrm{rank}}\) maximal with \(4C_{\mathrm{cone}}(T_{q}^{k}+P_{q}^{k}+1)\le L\); \(k_{0}\) chosen so \(C_{\mathrm{top}} b_{*} 2^{-k_{0}}\le 1/32\), then \(L_{1}(n)\) ensures \(k_{\mathrm{rank}}\ge k_{0}\) and all finite chart conditions. These bounds are computable.\par\smallskip
\noindent\textbf{\(K_{\mathrm{cap}},B_{\mathrm{halt}}(m),R_{\mathrm{max}}\)}. Conditional on halting, choose \(K_{\mathrm{cap}}\ge m+\tau_{V}(m)+3\); \(\tau_{V}\) is the actual halted computation length. Finite test radii give \(R_{\mathrm{max}}\) and \(B_{\mathrm{halt}}=4(1+R_{\mathrm{max}}+C_{\mathrm{cap}} B_{c}^{K_{\mathrm{cap}}+1})\). Computable given a halting witness; no uniform halting-cutoff algorithm is asserted.\par\smallskip
\noindent\textbf{\(\kappa_{\mathrm{halt}},C_{\mathrm{ancestor}},c_{\mathrm{ancestor}},d_{\mathrm{ancestor}}\)}. \(\kappa_{\mathrm{halt}}=\min _{z_{a}>0}[-\log  z_{a}/(2(1+r_{a}))]>0\) for the finite cap inventory. Polynomial endpoint union yields finite \(C_{\mathrm{ancestor}},d_{\mathrm{ancestor}}\) and a positive exponent \(c_{\mathrm{ancestor}}\); for example \(c_{\mathrm{ancestor}}=\kappa_{\mathrm{halt}}/(2C_{\mathrm{path}})\), since the retained last half of an \(s\)-round path has temporal span at least \(s/2-1\). Rational lower bounds on \(\kappa_{\mathrm{halt}}\) and upper bounds
on prefactors can be computed from that finite inventory and witness.\par\smallskip
\noindent\textbf{\(\Delta_{\mathrm{dom}},s_{\mathrm{dom}},A_{\mathrm{small}}\)}. Cone-overlap degree bounded by an integer from \(R_{\mathrm{macro}}\) (for example by complete block offset enumeration); search \(s\) until \(2s C_{\mathrm{ancestor}}(1+s)^{d_{\mathrm{ancestor}}}\exp (-c_{\mathrm{ancestor}} s)\le (1/4)^{\Delta_{\mathrm{dom}}+1}\). \(A_{\mathrm{small}}\) times \(s\) bounds the finitely many exceptional ring sizes.\par\smallskip
\noindent\textbf{\(j_{\mathrm{clone}},N_{\mathrm{small}},p_{\mathrm{full}},C_{n,\eta}\)}. Choose \(2^{-j_{\mathrm{clone}}-1}\le \eta\); \(N_{\mathrm{small}}=N_{\mathrm{wire}} \lceil A_{\mathrm{small}} s_{\mathrm{dom}}\rceil, p_{\mathrm{full}}=\eps_{\mathrm{inf},m}^{N_{\mathrm{small}}}>0\). Choose the small-ring round count \(t_{\mathrm{small}}\) with \(1/4+3(1-p_{\mathrm{full}})^{t_{\mathrm{small}}}/4\le 1/2+\eta\). Then \(C_{n,\eta}=D_{\mathrm{layer}}\max (j_{\mathrm{clone}} s_{\mathrm{dom}},t_{\mathrm{small}})\) elementary layers.\par\smallskip
\noindent\textbf{\(E_{\mathrm{enc}}(n,L),D_{\mathrm{dec}}(n,L,t),f_{\mathrm{fidelity}}\)}. Total finite rational encoder and decoder channels; \(D_{\mathrm{dec}}\) is unrelated to \(D_{\mathrm{layer}}\) or \(D_{\mathrm{boot}}\). Exact finite-fidelity polynomial as in Proposition~\ref{a:rational}.\par\smallskip

\begin{lemma}[No circular dependence and an explicit cone bound]\label{q1:l16}
The choices in the preceding list can be made in the order given in the proof below. Every global constant is fixed before choosing \(J_{\rm pad}\), and one computable \(J_{\rm pad}\) satisfies every required final smallness inequality. In particular, for every \(k\ge0\) the actual elementary dependency cone, over one rank-\(k\) tick (\(H=T_{q}^{k}\) macro rounds, \(D_{\mathrm{layer}}H\) elementary layers), of an aligned rank-\(k\) output home of length \(P_{q}^{k}\), including every actual controller dependency, fits in an interval of length at most
\[
 C_{\rm cone}(T_q^k+P_q^k+1),\qquad
 C_{\rm cone}=16(R_{\rm macro}+P_q+1).
\]
This bound includes all actual elementary controller dependencies and is fixed before the final rate-window selection.
\end{lemma}
\begin{proof}
Start with the native quantum roles and port tile, \(K_{\mathrm{inc}},h_{\mathrm{cover}},m_{\mathrm{bad}},b_{\mathrm{fault}}\), the finite CSS code and catalogue. Choose the indexed generator's absolute exponent and its finite self-description. Its arithmetic degree does not depend on the padding exponent. Then the finite R1/quantum timetable inequalities are searches comparing exponential capacities with fixed polynomials, so a single \(w_{q}\) determines \(P_{q},T_{q}\) with \(P_{q}\le T_{q}\) and the whole clean schedule. Native selector geometry is chosen before helper widths, as the explicit parametric native interface proved in Lemma~\ref{q1:l11} states. Choose \(d_{\mathrm{moment}}\) from the now fixed quantum geometry. Choose short exponent, short helper constants, then long exponent and channel count, then the single native \(n_{\mathrm{bin}}\). Choose physical redundancy after those exponents. Now the alphabet, all factories, physical period, slots and spatial shifts are completely fixed. All stencil sizes, clock cutoffs, \(C_{\mathrm{tail}},C_{\mathrm{base}},C_{\mathrm{rec}},C_{\mathrm{top}},C_{\mathrm{path}}\) and \(C_{\mathrm{cone}}\) are computed at this stage, including the cost of idle physical layers.

The helper and exponent inequalities have strictly positive slack powers of \(p\). Their multiplicative constants, and the desired tolerance \(\delta_{\mathrm{mark}}\), therefore require only \(p\) smaller than an effectively found positive threshold. Lemmas~\ref{q1:l1.1} and \ref{q1:l12} give finite-sum and geometric-tail bounds for their infinite moments, so this threshold can be found by rational inequalities with certified tails. Direct moments tend to zero as \(\sqrt{\eps}\). \(\gamma'(0)>0\) supplies fixed upper and lower proportional bounds between \(p_{m}\) and \(\bar\eps_{m}\) near zero. The bin radii tend monotonically to zero superexponentially with \(m\); \(H_{\mathrm{boot}}(m)=D_{\mathrm{boot}}^{m+2}\) is \(p_{m}^{-o(1)}\), with an effective bound from \(A_{\mathrm{bin}}=8^{m+n_{\mathrm{bin}}}\). Every needed sufficiently small inequality thus has a computable uniform tail in \(m\). Enumerate an integer \(J_{\mathrm{pad}}\) whose least \(m_{\mathrm{bin}}\) is beyond that tail. No gate, \(n_{\mathrm{bin}}\), geometry, padding exponent or helper count changes during this last search. In particular the dependence of the clock stencil on \(D_{\mathrm{layer}}\) is settled before the rate threshold is imposed.

For the explicit chart bound, let the decoded output home have length \(P_{q}^{k}\) and let \(H=T_{q}^{k}\). A depth-\(D_{\mathrm{layer}} H\) actual elementary dependency cone can extend spatially by at most \(r_{\mathrm{gate}} D_{\mathrm{layer}} H=R_{\mathrm{macro}} H\) at either end. Rounding the two ends outward to complete aligned rank-\(k\) homes adds less than \(2P_{q}^{k}\). A further fixed one-home reserve and all finitely many initial slot offsets are safely paid by \(C_{\mathrm{cone}}=16(R_{\mathrm{macro}}+P_{q}+1)\): the necessary interval is bounded by \(2R_{\mathrm{macro}} H+4P_{q}^{k}+2\) plus a fixed offset allowance already in \(r_{\mathrm{gate}}\) and \(P_{q}\). This is at most \(C_{\mathrm{cone}}(H+P_{q}^{k}+1)\). Classical controller transitions queried by that output are already included in its actual physical cone; witnesses used solely for probability estimates can live in a separate infinite extension. Thus there is no extra rank-dependent cone constant. The factor 4 in the embedding criterion leaves a strict unused ring margin, so no global ring tiling is needed.

All remaining quantities are per-label finite endpoints or halting-witness quantities. In particular halting-dependent constants are not prerequisites for choosing the physical rule, intervals, encoder or decoder. This proves acyclicity and the claimed computable bounds. 
\end{proof}

\section{Quantum storage on every sufficiently large ring}\label{q1:storage}
This section proves the storage half of Theorem~\ref{q1:main}. A recursion over the levels of the hierarchy bounds the probability that a test at the top level fails, and the decoder recovers the qubit at the end of one tick of the top level that fits in the ring; that tick lasts order \(L\) layers, which gives the linear horizon (Lemma~\ref{q1:l18}).

\paragraph{A stored qubit on each nonhalting window.}

The circuit combines the R1 schedule, an R2 selector and the quantum step. One macro round is one native tick in each of the two classical rules and one quantum tick. At the strobe the circuit uses the centrally normalized bank and reads the current native R2 core's chosen data symbol, with \(D\) the absent symbol, at every physical site \(x\): read that site's own native data field. In the proof its residue is \(x\) modulo the fixed physical stride \(D_{\mathrm{stride}}\) of the native R2 field (Lemma~\ref{q1:l11}) and its logical position is the quotient after removing this residue. Multiple strides or offsets are finite constants. The R1 age and address give the local quantum opcode as in Section~\ref{q1:compiler}. A correct selected epoch has a live selector at every required spacetime site. At the end of each macro round the circuit attempts a full local reset of every retained quantum-capable slot when the selected core symbol is \(D\). A wrong dispatch, a reaction to a malformed intermediate register, or an imperfect terminal cut is counted as a direct physical fault of that strobe if the stipulated semantic controls are correct. A complete local primitive-stencil mark covers any flagged wire in the relevant read, dispatch and erase stencil (probability at most \(C\eps\)); it includes hardware, port moves, waits and newly produced scratch up to its discard. A good core normalization on an \(h\)-close bank is read by majority, not by inspecting an arbitrary single redundant copy. For a completely arbitrary strobe state the semantic word is the total decode after normalization, as defined in Section~\ref{q1:selector}.

\textbf{Base noise cover.} Actual effects different from the ideal clean quantum step, including unintended cuts, are contained in a union of marks in integer (physical site, macro time) spacetime. A mark centered at \(z\) of type \(a\) has footprint and primitive-key support in its ball of radius \(r_{a}\ge 1\) and probability at most \(z_{a}\). All marks use one common independent-key space, and nearby marks may share arbitrarily many keys. The types comprise direct microscopic faults, failed R1 tests in Section~\ref{q1:clock} and the initialized failed core tests of Section~\ref{q1:selector}. For the last type the analysis uses the nominal stationary negative-time comparison, whose possible failure to match our finite encoder is charged separately below. A compiler anomaly mark always means the explicit flag count in its stencil, or an upper bounding stencil union, never disagreement depending on a particular incoming state. The prefix rate mismatch can likewise be marked by a uniform primitive draw in the deterministic threshold band. Thus these marks are measurable in the indicated independent keys. A rare wrong decoded selector or wrong clock may be charged even when the actual gate happened to be the intended one. Per site and time there are finitely or countably many types. For the exponent \(d=d_{\mathrm{moment}}\) fixed in Section~\ref{q1:parameters} before the helper exponents, and every \(\delta>0\), the selector construction of Lemma~\ref{q1:l12} for that \(d\), which does not depend on \(\delta\), and a sufficiently small selected rate, reached through the final padding \(J_{\mathrm{pad}}\) (Lemmas~\ref{q1:l12} and~\ref{q1:l16}), give \(\sum _{a} (1+r_{a})^{d} \sqrt{z_{a}}<\delta\), uniformly in the chosen window. The compiler and R1 geometry precede \(d\) and this choice of \(\delta\). All statements persist under one elementary rate splice inside the window, by Section~\ref{q1:selector} and the envelope bounds on primitive R1 and hardware flags.

\textbf{Probability of a bad top operation.} In this paragraph and the next, \(P,T,h,A,d,\delta\) stand for \(P_{q},T_{q},h_{\mathrm{cover}},A_{\mathrm{pair}},d_{\mathrm{moment}},\delta_{\mathrm{mark}}\) of Section~\ref{q1:parameters}; \(C\) denotes a constant fixed with the geometry, and the constant multiplying \(\delta\) in the recursion below is \(C_{\mathrm{rec}}\). The proof uses the monotone bad-box tests of Section~\ref{q1:compiler}, with base \(B_{0}\) true at a site and tick whenever covered by a mark. Counting all marks regardless of whether a quantum gate failed there is conservative. If the rank-\((i+1)\) parent test is bad, two bad rank-\(i\) children can be chosen with distance at least \(8P^{i}\) between the complete sets of primitive locations involved in their descendant \emph{geometric} tests. For a fixed child, complete footprints have halfwidths at most \(hP^{i}\) and halfheights at most \(hT^{i}\), up to one child width or height at the grid boundaries. Two fixed children with footprints closer than \(8P^{i}\) must have grid offsets in each coordinate at most \(2h+10\), since \(T\ge P\). More than \((10h+40)^{2}\) distinct children cannot all be close to the chosen child. The complete parent footprint has at most \(C(PT)^{i}\) primitive lattice points, with the extra constant factor \(PT\) and fixed collars absorbed. Let \(A\) be a bound on the number of ordered child pairs, fixed once the geometry is fixed.

Let \(u_{i}\) bound the supremum of bad-test probabilities over placements. Truncate \(B_{0}\) by deleting types with \(r_{a}>P^{i}\) when estimating one such pair. Each surviving child test is determined by keys within \(2P^{i}\) of its complete geometric footprint: the mark center and then its key support each require up to \(P^{i}\). Hence the two indicators are independent; each truncated probability is at most \(u_{i}\) by monotonicity. The only way the truncated parent differs is for a deleted mark to hit some primitive point of its full footprint. At most \(C(1+r_{a})^{2}\) centers of a given type hit one point. Thus

\[
\begin{gathered}
u_{0} \le  C \delta, \\ u_{i+1} \le  A u_{i}^{2} + C \delta [PT / P^{d-2}]^{i}.
\end{gathered}
\]

The choice \(d=d_{\mathrm{moment}}\) in Section~\ref{q1:parameters}
makes the bracket at most \(1/4\) and gives \(d\ge4\) for the separate
halting argument.
Choose \(0<b_{*}\le 1/(4A)\) and \(\delta\) with both \(C_{\mathrm{rec}}\delta\le b_{*}/4\) and \(C_{\mathrm{path}} \delta<1/2\) for the path-counting constant of Section~\ref{q1:forgetting}. Induction gives \(u_{i}\le b_{*}2^{-i}\) (Lemma~\ref{q1:l18}). At most a fixed \(C_{\mathrm{top}}\) top boxes (one top-local tick and its port neighbors) matter for the protected nonport. They need no wraparound or dependence estimate between them. The union bound and deterministic reduction show that at the end of the top tick, with fixed recursive recovery, the output on the good event has exactly its input logical state jointly with the reference. The failing event has probability at most \(C_{\mathrm{top}} b_{*}2^{-k}\) at rank \(k\). The controller and mark choices are independent of the logical input.

\textbf{Finite geometry and all elementary times.} One rank-\(k\) top tick spans \(H=T^{k}\) macro rounds and a code home of length \(P^{k}\). The encoder chooses a complete, aligned home in an integer chart (zero may be its origin). It puts the message into one ordinary nonport slot of its rank-\(k\) cell, all other top slots in fixed states and all private work initialized as specified by the encoding. It recursively encodes the \(S\) slots in each parent into its child blocks at the prescribed coordinates, with fresh auxiliary slots. At date zero the top on-site instruction is idle by the timetable for every location; the top port permutation does not act on the message. The encoder initializes enough complete adjacent rank-\(k\) homes, using fixed top input for those homes, to cover the full past cone from the output block through \(H\) dates. For some computable constant \(C_{\mathrm{cone}}\) the needed interval, including the complete boundary blocks and cone of every physical controller transition used, has length at most \(C_{\mathrm{cone}}(H+P^{k}+1)\) (Lemma~\ref{q1:l16}). The physical circuit is finite-speed. The rank \(k\) is the largest with \(4 C_{\mathrm{cone}}(T^{k}+P^{k}+1)\le L\); a total fixed fallback is used if no \(k\ge 1\) qualifies. This whole interval is mapped injectively into the ring, with arbitrary legal filler outside it. The encoder initializes R1 and core boundary symbols in this interval as in their respective sources, and prepares the finite nominal helper marginal jointly with error less than \(1/128\) (Lemma~\ref{q1:l15}). These occur on separate physical wires. The unrestricted endpoint encoder may do arbitrarily much finite work. An analysis coupling uses one infinite extension of all raw keys and both mathematical past histories for the entire finite quantum computation. Geometric witnesses can extend beyond the injective cone, since the events being estimated on the ring have already been coupled by their actual past dependence. Disjoint unwrapped witnesses are not treated as independent on a small physical ring.

Write \(D_{\mathrm{layer}}\) for the fixed number of elementary layers per macro round. For \(t\) from 0 to \(D_{\mathrm{layer}} H\) the decoder appends the remaining elementary layers on every received register at the chosen nominal rate \(\bar\eps_{m}\), then applies the fixed bottom-up recovery and extracts the designated nonport. This is a computable channel at any \(t\), even with a partial contact, live ports or a pending factory. In the proof the combined process simply uses the true elementary rate through \(t\) and nominal afterwards. All estimates above hold for that splice. Beyond the horizon the decoder is any fixed total channel. Thus \(F\ge 1-1/128-C_{\mathrm{top}} b_{*}2^{-k}\). Choose a fixed effective \(k_{0}\) with the final term at most \(1/32\), and take a computable \(L_{1}(n)\) large enough that \(k\ge k_{0}\) thereafter. Then \(H\to\infty\) with \(L\), and in fact \(H>c_{1}L\) by maximality of \(k\) and \(T\ge P\) (Lemma~\ref{q1:l18}). The initialization and recovery depend on \(n,L\) (and \(t\) for recovery) and are independent of the actual rate.

\begin{lemma}[The recursion, horizon and fidelity constants]\label{q1:l18}

Let \(A_{\mathrm{pair}}\ge 1\), \(C_{\mathrm{rec}}\ge 1\) and \(\delta=\delta_{\mathrm{mark}}\) be the constants, and \(u_{i}\ge 0\) the bounds on bad-test probabilities, in the already established geometric recursion

\[
\begin{gathered}
u_{0} \le  C_{\mathrm{rec}} \delta, \\ u_{i+1} \le  A_{\mathrm{pair}} u_{i}^{2} + C_{\mathrm{rec}} \delta q^{i}, \\ q = P_{q} T_{q} / P_{q}^{d_{\mathrm{moment}}-2} \le  1/4.
\end{gathered}
\]

If \(0<b_{*}\le 1/(4 A_{\mathrm{pair}})\) and \(C_{\mathrm{rec}} \delta\le b_{*}/4\), then \(u_{i}\le b_{*}2^{-i}\) for all \(i\ge 0\). Choose \(k_{0}\ge 1\) with \(C_{\mathrm{top}} b_{*}2^{-k_{0}}\le 1/32\). Let \(C_{\mathrm{cone}}\) be a positive integer bounding the complete initialized cone as in Lemma~\ref{q1:l16}. If \(k\) is largest with \(4 C_{\mathrm{cone}}(T_{q}^{k}+P_{q}^{k}+1)\le L\), \(H=T_{q}^{k}\), and \(2\le P_{q}\le T_{q}\), then

\[
\begin{gathered}
L < 12 C_{\mathrm{cone}} T_{q} H, \\ \text{with } c_{1} = 1/(12 C_{\mathrm{cone}} T_{q}),\quad c_{1} L < H \le  D_{\mathrm{layer}} H.
\end{gathered}
\]

Suppose further that \(U(n)\) does not halt, that every elementary rate lies in \(I_{n}\), and that the encoder of this section is used, so that the decoder fails only on the helper-coupling failure of Lemma~\ref{q1:l15}(b), of probability less than \(1/128\), or when one of the at most \(C_{\mathrm{top}}\) top tests is bad, each with probability at most \(u_{k}\). Then for \(k\ge k_{0}\) the nominal-rate completion decoder has entanglement fidelity at least

\[
1 - 1/128 - 1/32 = 123/128 \ge  15/16 = 120/128
\]

at every integer elementary time \(0\le t\le \lfloor c_{1} L\rfloor\).

\end{lemma}
\begin{proof}
The base is \(u_{0}\le b_{*}/4\le b_{*}\). If \(u_{i}\le b_{*}2^{-i}\), the recurrence and \(q\le 1/4\) give

\begin{align*}
u_{i+1} &\le (A_{\mathrm{pair}} b_{*}^{2} + C_{\mathrm{rec}} \delta)4^{-i}
 \le (b_{*}/4+b_{*}/4)4^{-i}\\
 &= (b_{*}/2)4^{-i}
 \le b_{*}2^{-(i+1)}.
\end{align*}

The last inequality is \(2^{-2i}\le 2^{-i}\), valid for \(i\ge 0\). This proves the induction with slack at rank zero as well.

Maximality of \(k\) implies \(L<4 C_{\mathrm{cone}}(T_{q}^{k+1}+P_{q}^{k+1}+1)\). Both \(P_{q}^{k+1}\) and 1 are at most \(T_{q}^{k+1}=T_{q} H\), so the claimed strict bound follows by adding the three terms. Multiplication by \(c_{1}\) yields \(c_{1} L<H\). \(D_{\mathrm{layer}}\) is a positive integer, the number of physical elementary layers per macro round, so \(H\le D_{\mathrm{layer}} H\). Consequently every stated elementary read time lies within the decoder's completion horizon.

On the successful joint coupling, initialization loses probability less than \(1/128\) and the union of the at most \(C_{\mathrm{top}}\) relevant top bad tests loses at most \(C_{\mathrm{top}} b_{*}2^{-k}\). These events are independent of the logical input; the deterministic recursive recovery is the same map for every input and preserves its joint state with the reference on success. On failure its entanglement fidelity is nonnegative. Therefore its unconditional overlap is at least one minus these two probabilities, yielding \(123/128\) for \(k\ge k_{0}\). The elementary-time suffix uses nominal noise from the received entire register; its analysis is the allowed single rate splice, so the same inequality holds at a time inside any physical operation. A computable sufficient size cutoff is \(4 C_{\mathrm{cone}}(T_{q}^{k_{0}}+P_{q}^{k_{0}}+1)\), increased for the other finite-chart conditions. This makes \(k\ge k_{0}\) for all larger \(L\). 
\end{proof}

\paragraph{Joint bad-box recursion.}
\emph{Proof idea.} A bad parent test contains two bad children whose complete footprints are far apart. After marks of radius larger than the child scale are deleted, the two children depend on disjoint noise keys, so their failures are independent and cost \(u_i^2\); the deleted large marks cost a tail term controlled by the radius moment. The resulting recursion \(u_{i+1}\le A_{\mathrm{pair}}u_i^2+C_{\mathrm{rec}}\delta_{\mathrm{mark}}q^i\) with \(q\le1/4\) closes at \(u_i\le b_*2^{-i}\) (Lemma~\ref{q1:l18}).

Let \(B_{0}\) be coverage by the direct microscopic, clock and decomposed initialized-selector marks. Each mark's determining keys and covered footprint lie in its advertised ball. Lemmas~\ref{q1:l1}--\ref{q1:l3}, \ref{q1:l6} and~\ref{q1:l12} show that outside this set a physical instruction, including its \(\operatorname{LOAD}\) and terminal cut, equals the clean reference. A parent test counts named children as in Lemma \ref{q1:l7}. If it is bad, two bad child tests have complete descendant geometric footprints separated by at least \(8P_{q}^{i}\). Remove all base types with radius greater than \(P_{q}^{i}\). A remaining type that hits one child's footprint has its center within \(P_{q}^{i}\) of that footprint and every determining key at most another \(P_{q}^{i}\) away. Hence the two truncated indicators use disjoint primitive keys and are independent. Each has probability at most \(u_{i}\) by monotonicity. Union over at most \(A_{\mathrm{pair}}\) ordered pairs costs \(A_{\mathrm{pair}} u_{i}^{2}\).

A deleted type must hit one of the at most \(C_{\mathrm{geom}}(P_{q} T_{q})^{i}\) points of the complete parent footprint. At most \((2r+1)^{2}\le 9(1+r)^{2}\) centers of a radius-\(r\) type can hit one fixed point. Since \(z_{a}\le \sqrt{z_{a}}\), the shared moment bound yields

\[
\sum _{r_{a}>P_{q}^{i}} 9(1+r_{a})^{2} z_{a} \le 9 \delta_{\mathrm{mark}} P_{q}^{-(d_{\mathrm{moment}}-2)i}.
\]

Absorb the geometric point count and the factor nine into \(C_{\mathrm{rec}}=\max (C_{\mathrm{base}},C_{\mathrm{tail}})\). The base union gives \(u_{0}\le C_{\mathrm{rec}} \delta_{\mathrm{mark}}\), and the tail gives

\[
u_{i+1}\le A_{\mathrm{pair}} u_{i}^{2}+C_{\mathrm{rec}} \delta_{\mathrm{mark}} \bigl[P_{q} T_{q}/P_{q}^{d_{\mathrm{moment}}-2}\bigr]^{i}.
\]

Lemma \ref{q1:l18} closes this recursion. The controller and mark choices are independent of the encoded logical input. Lemmas~\ref{q1:l4}--\ref{q1:l7} put successful outputs in one common correctable operator span, so the fixed bottom-up recovery preserves the state with its reference. Lemma \ref{q1:l16}'s actual finite cone, not a globally periodic hierarchy, gives the every-length ring coupling. The total encoder and decoder and the elementary-time nominal suffix are given by Lemma \ref{q1:l15}. The fidelity and linear horizon are exactly Lemma \ref{q1:l18}.

\section{Whole-register forgetting in one dimension}\label{q1:forgetting}
This section proves the halting half of Theorem~\ref{q1:main}. After a burn-in, every live R2 symbol lies in a rare marked ball (Lemma~\ref{q1:l13}); every quantum path must cross a complete local cut in each round (Lemma~\ref{q1:l14}); and the coupling and cloning argument of Section~\ref{sec:forgetting} then bounds every encoding (Lemma~\ref{q1:l15} and the last two paragraphs of this section).

\begin{lemma}[A rate-uniform all-start halting cover]\label{q1:l13}

Fix the single native R2 instance of Lemma \ref{q1:l11}, its already chosen physical compiler, and its selected label \(m\). Put \(p_{m}=\gamma(\bar\eps_{m})/2\), where \(\bar\eps_{m}\) is its rational nominal physical rate. Suppose \(V(m)\) halts. There are a finite positive burn-in \(B_{\mathrm{halt}}(m)\), finitely many types \(a\) with integer radii \(r_{a}\ge 1\), numbers \(0\le z_{a}<1\), and primitive-measurable events \(M_{a}(z)\) with \(\Pr(M_{a}(z))\le z_{a}\) for every center \(z\), every rate in the selected interval and every initial measured autonomous word, such that:

\begin{enumerate}[label=\arabic*.]
\item For every initial measured autonomous word, every rate \(\eps\) in the selected interval, and every physical site and date \((x,t)\) with \(t\ge B_{\mathrm{halt}}(m)\), a live semantic R2 core symbol is in a ball \(B_{\infty}(z,r_{a})\) for which \(M_{a}(z)\) occurs. The same ball family works for all these words and rates.
\item Include a type for every relevant residue and read offset and a direct type for every fixed read, dispatch and terminal-erase stencil. For every physical site and date \((x,t)\) with \(t\ge B_{\mathrm{halt}}(m)\) outside this enlarged cover, the site \(x\) has a correctly triggered complete quantum cut in round \(t\), independently of the R1 word and the quantum state.
\item Each event and its ball depend on independent primitive keys within that ball. Disjoint balls therefore have independent mark events, conditional on any fixed initial autonomous word. In fact these mark events do not depend on that word.
\item With the final small-rate selection made in Section~\ref{q1:parameters}, for the path-counting constant \(C_{\mathrm{path}}\) of the final finite-speed physical circuit,
\end{enumerate}

\[
\begin{gathered}
\sum _{a} C_{\mathrm{path}} (1+r_{a})^{4} \sqrt{z_{a}} < 1/2.
\end{gathered}
\]

All statements are uniform over the entire selected physical interval. \(B_{\mathrm{halt}}(m)\) and the finite maximum radius may depend on a halting computation; their existence is asserted only on halting labels.

\end{lemma}
\par\noindent\emph{Proof idea.} On a halting label the imported cap and ascent lemmas show that a live R2 symbol after the burn-in forces some centered sparsity test of bounded rank to fail, whatever the starting word. Each failed test is covered by one ball from a finite list of mark types (raw, short, middle, long and full tests, and direct stencil faults). Marking the threshold band between the actual and the nominal rate makes the list uniform over the window, and Lemma~\ref{q1:l12} makes its radius moment small.

\begin{proof}
Every analysis test is frozen at the nominal \(p_{m}\): the short and long innovation prefixes, stripe box \(H_{m}\), admission ranks \(j_{s},j_{l}\), centered protection collars, paint, and the true literal's halting rank \(K_{\mathrm{cap}}\). These choices belong to the analysis; the physical rule does not depend on the rate. \cite[Lemma~\ref*{cl:lem:locking}]{DMNoise1D2026} gives a finite \(K_{\mathrm{cap}}\) once the literal has locked at rank \(m+1\) and the true simulated computation halts. Increase it to at least one if necessary. The deterministic cut-and-ascent implication of \cite[Lemma~\ref*{cl:lem:cap-cut}]{DMNoise1D2026} and \cite[Lemma~\ref*{cl:lem:ascent}]{DMNoise1D2026} uses only the truth of these finite tests, not the probability law generating them. Its quantifier includes all starting rows, all ages, all address groups and all forged mutable tapes. Hence a genuinely live physical core after the necessary history is available forces one of the buffered centered \(i\)-sparsity tests, \(0\le i<K_{\mathrm{cap}}\), to fail. We also retain rank \(K_{\mathrm{cap}}\).

The mark inventory, repeated over the finite residue lanes and fixed read offsets used by the circuit, is as follows. Let \(R_{\mathrm{ad},s}=\lceil C_{\mathrm{ad}} B_{c}^{j_{s}}\rceil\) and \(R_{\mathrm{ad},l}=\lceil C_{\mathrm{ad}} B_{c}^{j_{l}}\rceil\). Constants include the physical stride, the complete prefix support and the fixed centered-test multiplier.

\begin{itemize}
\item \(R_{i}, 0\le i\le \min (K_{\mathrm{cap}},j_{s})\): raw dependent-stencil nonsparsity test, radius \(C_{\mathrm{cap}} B_{c}^{i+1}\), bound \(p_{m}^{c_{\mathrm{raw}} 2^{i}}\).
\item \(\mathsf S\): a short failure center, radius \(R_{\mathrm{ad},s}\), bound \(C_{s} p_{m}^{M_{s}}\).
\item \(M_{i}, j_{s}<i\le \min (K_{\mathrm{cap}},j_{l})\): raw-plus-short nonsparsity test, radius \(C_{\mathrm{cap}} B_{c}^{i+1}\), bound \(p_{m}^{D_{s} 2^{i-j_{s}}}\).
\item \(\mathsf L\): a long failure center, radius \(R_{\mathrm{ad},l}\), bound \(C_{l} p_{m}^{M_{l}}\). Its event is the union of the nominal native long failure, any excess-fault compiler stencil in the complete short and long comparison prefixes, and any elementary threshold-band mismatch there. Short failures are kept in the separate \(\mathsf S\) event; they are not charged at the long exponent.
\item \(H_{i}, j_{l}<i\le K_{\mathrm{cap}}\): full nonsparsity test, radius \(C_{\mathrm{cap}} B_{c}^{i+1}\), bound \(p_{m}^{D_{l} 2^{i-j_{l}}}\).
\item \(E_{b}\) for each one of the finite read, dispatch and erase stencil templates \(b\): at least one physical replacement flag in that complete stencil, radius \(r_{b}\), bound \(\min (1,N_{b} \eps_{\mathrm{sup}})\), where \(N_{b}\) is its enumerated number of primitive wire-dates and \(\eps_{\mathrm{sup}}\) is the rational upper endpoint of the window.
\end{itemize}

Keep \(\mathsf S\) and \(\mathsf L\) even if \(K_{\mathrm{cap}}\) lies below their respective admission ranks: a large-support helper center can paint a low-rank test. Zero-probability types can be omitted. Fixed multiplicative probability factors for centered covers are absorbed in the exponents exactly as in Lemma \ref{q1:l12}, by the final rate cutoff. Alternatively they can be retained in \(z_{a}\) and the same moment proof applies.

The deterministic covering is as follows. Below \(j_{s}\), a failed test with no short and long paint in its enlarged domain is a raw failed test. In the middle range, with no long paint it is a raw-plus-short failed test. Above \(j_{l}\) it is itself a full failed test. A helper center responsible for any lower-rank failure lies within the corresponding admission-radius ball of the root of that failed test, and its one enlarged center ball covers that root and the center's full primitive support. These are exactly the three cases of \cite[Lemma~\ref*{cl:lem:joint}]{DMNoise1D2026}. Thus this list covers every possible failed cap test.

The support inflation is uniform in the cap rank. The adaptive ascent displacements of \cite[Lemma~\ref*{cl:lem:ascent}]{DMNoise1D2026} are less than \(6T_{c}^{i+1}\) in time and \((4T_{c}/P_{c}+4)P_{c}^{i+1}\) in space. Since \(4T_{c} P_{c}^{i}+4P_{c}^{i+1} \le  8 B_{c}^{i+1}\), one fixed \(C_{\mathrm{cap}}\) contains them, the protected-work interface, all finite residue offsets, and the determining keys. The helper prefixes have the two fixed radii already included in the \(\mathsf S,\mathsf L\) balls. Let \(R_{\mathrm{max}}\) be the largest radius in the finite list of non-direct types, increased by the largest fixed physical read offset and by 2. Take, for example, \(B_{\mathrm{halt}}(m)=4(1+R_{\mathrm{max}}+C_{\mathrm{cap}} B_{c}^{K_{\mathrm{cap}}+1})\) macro rounds. Every preceding work, lower call and helper prefix queried by the ascent from a date at least \(B_{\mathrm{halt}}(m)\) is then strictly after initialization. Future service observations are allowed only as observations in the deterministic proof, with lookahead less than their own tile (\cite[Lemma~\ref*{cl:lem:ascent}]{DMNoise1D2026}); the larger two-sided radius already contains them. Physical controls read only past values.

For rate uniformity, generate each elementary replacement from a uniform \(U_{\mathrm{key}}\) and an independent replacement value. Actual rate \(\eps\) and nominal \(\bar\eps_{m}\) disagree only if \(U_{\mathrm{key}}\) lies between these thresholds. Mark instead the deterministic band \([\eps_{\mathrm{inf}},\eps_{\mathrm{sup}}]\) in each full comparison prefix. Its probability is bounded by the width of that band times the enumerated prefix size. The selected window makes this at most the long failure allowance. Compiler anomalies and direct marks use the envelope \(\eps_{\mathrm{sup}}\); abstract nominal short and long tests keep their nominal laws. Native core substitutions at actual rate \(\eps\) in the window have probability \(\gamma(\eps)\le \gamma(\eps_{\mathrm{sup}})\le C_{\gamma} c_{\eps} p_{m}\), because \(\gamma\) is increasing there and \(\gamma(0)=0\). They are independent across native sites. Nonsparsity is monotone in the flag set, so the \(R/M/H\) bounds hold at every rate in the window. Consequently the comparison holds simultaneously for all starts, and, outside the band marks, for all actual rates in the chosen interval. This is not conditioning the factory innovations on absence of anomalies. Native innovations are defined even when anomalies occur. The cut-and-ascent implication therefore holds pointwise at every rate with exactly the same finite radii and burn-in.

Lemma~\ref{q1:l12}'s moment estimate, for its exponent \(d_{\mathrm{moment}}\ge4\), applies to the infinite union \(R_{i},\mathsf S,M_{i},\mathsf L,H_{i}\) and any positive tolerance; restricting to \(i\le K_{\mathrm{cap}}\) only decreases the sum. Write \(d=d_{\mathrm{moment}}\). Account for the finitely many residues, read offsets and \(C_{\mathrm{cap}}\) once in its tolerance. The direct contribution is at most \(\sum _{b} (1+r_{b})^{d} \sqrt{N_{b} \eps_{\mathrm{sup}}}\), which tends to zero with the rate. The order of choice is legitimate: the exponents in Lemma \ref{q1:l12} are chosen for \(d\), whereas the desired small multiplicative constant is achieved by the last bin-skipping parameter \(J_{\mathrm{pad}}\) after the whole circuit and \(C_{\mathrm{path}}\) have been fixed. Choose the two contributions each less than \(1/(4C_{\mathrm{path}})\), with strict slack. This proves part 4. All events just listed are functions of primitive innovations or of explicit flag counts, never a trajectory-dependent disagreement. Their determining supports lie inside their balls, proving part 3. Finally, outside the cover the core is \(D\) and no relevant read or erase key is flagged. Lemma \ref{q1:l14} then gives a complete local cut. Arbitrary R1 opcodes before that cut cannot alter this fact. This proves all parts. 
\end{proof}

\begin{lemma}[Slot inventory and complete local cuts]\label{q1:l14}

With the mandatory end-of-round disposal already specified in Section~\ref{q1:forgetting} every physical factor belongs to one of the classes in the following inventory. A pristine \(D\)-triggered terminal cut at site \(x\), that is, one with semantic trigger \(D\) and no replacement flag in its fixed trigger-normalization, dispatch, preparation, \(\operatorname{SWAP}\) and partner-disposal stencil, leaves no quantum dependence on any pre-cut input that reaches \(x\) during that round. Every path from an initial quantum factor to any quantum output crosses one such cut position in every intervening round. This includes an arbitrary encoder acting on the complete physical register. Autonomous classical registers may retain measured classical information; they are included in the channel as measured classical outputs.

\noindent\textbf{Autonomous R1 native word, its encoded bits and old and new retained banks}. They are dephased before their first read and updated only by their own total classical transition. Retained autonomous words remain classical. A bank no longer retained is unconditionally reset or discarded by round end. No quantum output is written into these words.\par\smallskip
\noindent\textbf{Autonomous R2 core and all native helper work, wall, receiver, replica, raw-probe and exported fields}. They use the same dephasing and one-way autonomous evolution, with total decoding as in Section~\ref{q1:selector}. Retained native fields remain classical; nonretained banks are disposed. Intended work storage in the native helper is still an autonomous classical field, even when called work.\par\smallskip
\noindent\textbf{Private classical copies, fanout, majority and evaluation banks, factory source and output scratch}. These are reset before any possible read of their incoming state; they hold freshly created fair and rare draws and fixed Boolean computations. After their last use all private copies, old banks, factory scratch and outputs not promoted into a retained autonomous word are unconditionally reset or discarded, even if inactive throughout this round.\par\smallskip
\noindent\textbf{All \(S\) native quantum slots}. These include ordinary data, contact slots, dummies, reset and gate records, syndrome factors and virtual-interpreter tape factors; their quantum state may be arbitrary. Every retained member, including idle or unused slots, undergoes the \(D\)-triggered erase at its physical site. Nonretained members are unconditionally disposed. A tape intended to be classical inside a quantum gadget belongs here, not in the autonomous R1 class.\par\smallskip
\noindent\textbf{Facing exchange ports and dedicated ordinary transport tracks and buffers}. These implement the fixed port permutation and paired shuttles, with no outward-port access during private tape-update stages. Spatial operations finish before the terminal cut. Every retained port, track and buffer undergoes the same terminal \(D\) erase; every borrowed or nonretained factor is unconditionally disposed. Old port contents do not escape the inventory.\par\smallskip
\noindent\textbf{Fredkin and controlled-erase auxiliaries, fresh reset partners, quantum-targeted temporary work}. These remain part of the quantum system until actual disposal, regardless of their intended basis. The partner carrying the erased old state is reset or discarded after the swap. Every unused partner and every other temporary output is disposed at round end. There is no later return edge.\par\smallskip

The word reset in the assembled physical rule means reset before first use and mandatory disposal or reset after the final conditional-reset block in every round, including for slots that happened not to be used. It never means that unused quantum work can persist until some later use.

\end{lemma}
\par\noindent\emph{Proof idea.} Every factor is assigned in advance to one class of the inventory. Autonomous classical words are dephased before any read and control quantum operations only one way, so they carry no quantum information. Every retained quantum-capable factor passes through a controlled swap with a freshly reset partner, which on a clean trigger outputs \(\lvert0\rangle\) whatever the input, and every other quantum factor is discarded each round; hence every surviving quantum path crosses a cut position in every round.

\begin{proof}
The hardware is a finite product of named factors.
The primitive-and-layout paragraph of Section~\ref{q1:compiler}
assigns the fixed native data, control, tape and port roles and places
growing arrays on more sites. Its three-step parent-tick construction
uses a fresh private computation for each contact and reserves the
terminal transport, with the complete timetable in Lemma~\ref{q1:l10}. The classical compilation of Section~\ref{q1:model} and Section~\ref{q1:selector} adds only finite retained banks and disposable fanout, majority, factory and evaluation work. These are precisely the initial three rows. Each additional register in a finite implementation must be assigned either retained autonomous, retained quantum-capable, or disposable quantum-capable before the finite transition table is generated. Taking the union of all such assigned factors gives an exhaustive finite inventory. Reserving extra idle slots does not create another case: they are either retained quantum slots or disposable factors and receive the indicated reset.

At the beginning of a round measure and dephase the retained autonomous words, before any read that could couple them to quantum data. Conditional on this measured word and the autonomous innovations, their whole evolution is classical and independent of the quantum input. A basis-valued private control \(c\) and a controlled operation \(\sum _{c} \lvert c\rangle\langle c\rvert \otimes U_{c}\) leave the control probabilities unchanged: on a state \(\sum _{c} \lvert c\rangle\langle c\rvert \otimes \rho_{c}\) the result is \(\sum _{c} \lvert c\rangle\langle c\rvert \otimes U_{c} \rho_{c} U_{c}^{*}\). The same assertion holds for a controlled CPTP map. Thus one-way controls cannot transmit unknown quantum information back into the autonomous words. Private quantum-targeted intermediates remain in the quantum classes until disposal; the argument does not declare them autonomous merely because a clean run would be in the computational basis. Noise on an autonomous basis bit is another classical innovation. Noise on quantum work remains inside the conditional quantum CPTP circuit.

Let \(q\) be any retained quantum factor just before the terminal erase block and let \(r\) be its freshly reset partner. For a clean trigger \(c=D\), the controlled \(\operatorname{SWAP}\) is \(\operatorname{SWAP}(q,r)\). Before the swap the channel on \(r\) is the preparation \(\lvert 0\rangle\langle 0\rvert\) (tensored with the retained rest). After the swap and tracing out or resetting the old-state output \(r\), the output \(q\) is \(\lvert 0\rangle\langle 0\rvert\) and has no incoming dependence from \(q\). This identity holds on a joint state with an arbitrary environment and reference: \(\operatorname{Tr}_{r}[\operatorname{SWAP}(\rho_{\mathrm{qE}} \otimes  \lvert 0\rangle\langle 0\rvert_{r})\operatorname{SWAP}]=\lvert 0\rangle\langle 0\rvert_{q} \otimes \operatorname{Tr}_{q} \rho_{\mathrm{qE}}\). Apply this to every retained quantum factor at the site. A fixed finite stencil includes the trigger normalization, dispatch, preparation, \(\operatorname{SWAP}\) decomposition and the reset or discard of its partner. If it has no replacement flags and the semantic trigger is \(D\), the exact identity applies. The direct marks of Lemma \ref{q1:l13} include every failure location in this entire stencil.

Every nonretained output is disposed unconditionally, including idle scratch and the old partners. Post-cut layers only prepare, reset or store classical words or add noise; all quantum spatial operations precede the cut. Hence no quantum state can leave through a temporary slot and return after the cut. The terminal noise can replace a fresh output by \(I/2\) but cannot recreate dependence on an earlier input. An arbitrary malicious initial state in any disposable register is removed before its first possible influence and again at round end. A malicious quantum state in an autonomous word is measured at the initial boundary and contributes only to the classical outcome \(a\), whose correlations with the reference are retained in the full channel analysis.

Draw the conservative dependence graph of the conditional CPTP circuit with edges from every input of a gate to every one of its outputs until a genuine preparation, replacement or discard. At round boundaries the only quantum vertices that can continue are the retained quantum-capable factors in the inventory. Therefore a surviving path must cross one retained cut location in each round; an unflagged \(D\) cut at that location terminates it. The rule and this conclusion also hold on small rings: a translated source-to-private-target transfer is a bijection for each named target lane, even when different spatial offsets coincide modulo \(L\). The lane names keep its target factors distinct. 
\end{proof}

\begin{lemma}[Total encoder and decoder, machine substitution and physical time]\label{q1:l15}

\((a)\) The imported R2 construction is instantiated with the machine of its literal interface (written \(U\) in \cite{DMNoise1D2026}) equal to the single fixed one-ended unary-input machine \(V\), where \(V(m)\) simulates the target machine on \(\nu_{2}(m)\) for \(m\ge 1\); set \(V(0)\) to a fixed infinite loop. Native geometry, helper tuple and offset \(n_{\mathrm{bin}}\) are frozen once. A final fixed integer \(J_{\mathrm{pad}}\) relabels target input \(n\) by \(m(n)=2^{n}(2J_{\mathrm{pad}}+1)\). This changes no physical rule and satisfies \(V(m(n))\) halts if and only if the target \(U(n)\) halts.

\((b)\) For every label and every finite set \(F\) of native helper slots, in particular the helper slots of the encoding chart for \(n,L\), the finite preparation search at the nominal rational \(\bar\eps_{m}\) terminates and yields a computable rational mixture on \(F\) that couples with the stationary native helper marginal on \(F\) with failure probability less than \(1/128\). No halting oracle is used.

\((c)\) The time \(t\) in the one-dimensional theorem and in the assembled theorem counts elementary physical layers. One macro round consists of \(D_{\mathrm{layer}}\) such layers. The rate-uniform mark cover of Lemma~\ref{q1:l13}, with the marked-ball chain count below, which turns it into the ancestor estimate, plus the complete-cut inventory in Lemma~\ref{q1:l14} is a legitimate replacement for the dimension-specific Proposition~\ref{f:ancestor} under Remark~\ref{rem:forgetting-interface}; no dimension-specific selector theorem from the three-dimensional construction is imported.

\end{lemma}
\begin{proof}
A one-ended machine can compute \(\nu_{2}\) of a positive unary input by repeated division by two (or repeated pairing on its work tape) and simulate the target machine. Its text is fixed, its interpretation is independent of the chosen bin, and this simulation halts exactly on the desired inputs. The construction at \cite[Lemma~\ref*{cl:lem:locking}]{DMNoise1D2026} accepts any such fixed one-ended universal simulation program; its bitplane tape-step proof only uses a fixed finite transition table. Thus setting the source's \(U:=V\) is an instance of that interface. Since \(2J_{\mathrm{pad}}+1\) is odd, \(\nu_{2}(2^{n}(2J_{\mathrm{pad}}+1))=n\). Increasing \(J_{\mathrm{pad}}\) after the circuit is frozen merely selects a subsequence of native bins with increasing minimum index; it does not change \(n_{\mathrm{bin}}\), widths or gates. Part \((a)\) follows.

For \((b)\), let \(F\) be the finite collection of native helper slots required in the finite encoding chart, and fix the nominal native rate \(\gamma(\bar\eps_{m})\), which is rational by Lemma \ref{q1:l17}. For integer \(s\), its backward radius-finite cone from date \(-s\) has finitely many native sites and finite-alphabet incoming bottom rows. A particular innovation assignment in that cone induces a deterministic map from \emph{all} bottom rows to the observed slots \(F\) at date zero. Enumerate assignments and rows and compute the probability \(q_{F}(s)\) that this map is nonconstant. This is a finite exact rational calculation: a native innovation chooses no replacement or a uniform alphabet value, all with rational probabilities. Search \(s=1,2,\dots\) until \(q_{F}(s)<1/128\).

We use the imported coalescence statement in its grand-coupling (all-starts) form: the noise-only probability that the map to a specified physical helper slot remains nonconstant tends to zero exponentially. This is the explicit conclusion of
\cite[Lemma~\ref*{cl:lem:stripes}]{DMNoise1D2026}, whose proof combines
reference refreshes with the all-input receiver estimate of
\cite[Lemma~\ref*{cl:lem:receiver-mixing}]{DMNoise1D2026}. Its proof coalesces every required slow fair-register input in the final half-cone; the map to a slow fair-register slot is nonconstant only if there has been no true refresh at that site in the relevant window; conditional on any common \(Y\) history, the receiver \(\operatorname{CLEAN}\)-and-wall slab erases the whole work state. The sufficient slab event is uniform over that common history. A union of the two noise-only failures, without conditioning away errors, yields the nonconstant-map bound. It is valid for every helper slot including raw work and for every fixed nominal rate in the selected small-rate windows. Union over finite \(F\) gives \(q_{F}(s)\) tending to zero, so the search terminates regardless of what \(V(m)\) does.

Now simulate the selected finite past innovations from a fixed bottom row and output its marginal on \(F\). This finite procedure defines a rational mixture. In a two-sided stationary helper realization use the very same suffix innovations from \(-s\) to zero. If their map to \(F\) is constant, its value equals that obtained from the chosen fixed row whatever the stationary row at \(-s\) was. The coupling therefore fails with probability at most \(q_{F}(s)<1/128\). The stationary object here is the native autonomous helper, which is the object used for \cite[Lemma~\ref*{cl:lem:past}]{DMNoise1D2026}. Initialize its physical redundant representation in legal unanimous banks; future physical evolution is compared with that native evolution by the separately marked compiler anomalies. No claim that the physical compiled helper itself has exactly the native stationary law is needed.

For any \(n,L\) the chosen chart is finite. The search just proved, finite recursive code encoding and a total fallback below the geometric cutoff give a total computable encoder. At any integer elementary time \(t\), the decoder, when within the selected horizon, runs the remaining finite physical layers at rational \(\bar\eps_{m}\) and then fixed recursive recovery. Its suffix is a finite rational channel even if \(t\) falls inside a factory or controlled contact. Outside the horizon choose a fixed total reset decoder. Thus the encoder and decoder exist algorithmically on every input, including halting ones; they never request the unknown actual rate. This also justifies using them in the computable finite-fidelity predicate.

For \((c)\), a native round boundary is \(D_{\mathrm{layer}}\) elementary layers apart. Encoder date zero is immediately before boundary dephasing, so \(t=0\) is a well-defined identity (no-storage) interval. All macro-time cone and ancestor arguments are converted to physical layers by multiplication by \(D_{\mathrm{layer}}\). To apply Remark~\ref{rem:forgetting-interface}, fix a measured initial autonomous word \(a\). Lemma \ref{q1:l14} gives exactly the complete-cut conditional CPTP hypothesis of Lemma~\ref{f:factorization}. Lemma \ref{q1:l13} and the marked-ball chain argument give a bound \(B(1+s)^{d} \exp (-\kappa s)\) for the initial-site ancestor probability, uniform in \(a\) and the actual rate, with physical determining cones of radius \(O(s)\) macro rounds. Independent primitive innovations in disjoint such cones give vector independence for block indicators. Lemma~\ref{f:domination} and Proposition~\ref{f:extension} then give an exact two-output extension of the \emph{whole} round channel; its size condition is on one fixed \(s\) rather than on the subsequent iteration count. Lemmas~\ref{f:iteration}--\ref{f:bell} iterate that extension and bound every encoder and decoder. The finitely many smaller rings have a positive full-register replacement probability bounded below by \((\inf  I_{n})\) to their bounded qubit count, as in the proof of Theorem~\ref{f:forgetting}. This checks every input listed in Remark~\ref{rem:forgetting-interface}. Finally, an upper bound over all decoders at a boundary persists at each later elementary layer by composing that evolution into the decoder. 
\end{proof}

\paragraph{Marked-ball chain count in \(1+1\) dimensions.}

Fix any measured initial autonomous word \(a\). Condition on autonomous innovations and exogenous physical fault flags; syndrome measurements remain within the quantum CPTP circuit. For the conservative quantum graph allow every opcode the R1 state might request. Lemma \ref{q1:l14} forces a surviving quantum path to cross a terminal local cut each round. For dates at least \(B_{\mathrm{halt}}(m)\), Lemma \ref{q1:l13} forces that cut position into its finite mark inventory, independently of \(a\) and of the quantum state.

Let \(v=v_{\mathrm{quant}}\) bound spatial displacement between successive round cuts. A finite path segment of temporal span \(t\) is covered at its round vertices by finitely many original marked balls. Greedily retain disjoint original balls in decreasing radius. Every omitted ball lies in the triple enlargement of a retained ball with at least its radius. Set \(q_{\mathrm{expand}}=2v+4\). The \(q_{\mathrm{expand}}\)-expanded retained balls also cover the interpolated edges between path vertices, since every radius is at least one. Their touch graph contains a simple chain between endpoint neighborhoods. If its original radii are \(r_{1},\dots,r_{l}\), summing endpoint radii and the center gaps \(q_{\mathrm{expand}}(r_{j}+r_{j+1})+v+1\) gives

\[
\begin{gathered}
t\le C_{\mathrm{path}} \sum _{j} (1+r_{j}), \\ C_{\mathrm{path}}=(100(v+4))^{4}.
\end{gathered}
\]

This deliberately loose integer also bounds placement counts. Near a fixed endpoint the initial center has at most \(C(1+r_{1})^{2}\) possibilities. Each next center has at most \(C(1+r_{j})^{2}(1+r_{j+1})^{2}\) possibilities. Thus for a fixed ordered type list the number of center lists is bounded by \(\prod_{j} C_{\mathrm{path}}(1+r_{j})^{4}\). Sum only over disjoint original-ball lists; their mark probabilities multiply because their complete primitive supports are disjoint. Overlapping-ball events are never multiplied.

Omit zero-probability types. At a fixed finite halting cap,

\[
\begin{gathered}
\kappa_{\mathrm{halt}}=\min _{a} [-\log (z_{a})/(2(1+r_{a}))]>0.
\end{gathered}
\]

Reserve one square-root of each product \(z_{a}\) to bound the chain's probability by \(\exp (-\kappa_{\mathrm{halt}} t/C_{\mathrm{path}})\). The other square-roots and the placement counts sum, at chain length \(l\), to at most

\[
\begin{gathered}
[\sum _{a} C_{\mathrm{path}}(1+r_{a})^{4} \sqrt{z_{a}}]^{l} < 2^{-l}.
\end{gathered}
\]

Summing \(l\ge 1\) gives a constant times \(\exp (-\kappa_{\mathrm{halt}} t/C_{\mathrm{path}})\). The last half of a path lasting \(s\) rounds lies after burn-in for \(s>2B_{\mathrm{halt}}+2\). From a fixed initial site there are at most \(2vs+1\) final endpoint sites and finitely many output-slot types. A union bound, and a larger prefactor for the finitely many shorter times, gives

\[
\Pr(\text{initial site }x\text{ has a surviving path after }s\mid a) \le C_{\mathrm{ancestor}}(1+s)^{d_{\mathrm{ancestor}}}\exp (-c_{\mathrm{ancestor}} s).
\]

One can take \(d_{\mathrm{ancestor}}=2\) with enlarged constants. These constants are uniform in \(a\) and the rate, and may depend on the halting witness. Switching quantum lanes causes no problem: the mark inventory already includes every finite residue and read offset, and \(v\) bounds the full quantum supergraph.

The actual ancestor indicator depends on actual finite circuit edges and cuts. A depth-\(s\) quantum path from an initial site lies within \(R_{\mathrm{macro}} s\) spatially; determining its autonomous controls enlarges this to at most \(3R_{\mathrm{macro}} s\). The larger auxiliary mark witnesses used solely for an upper bound do not enter this actual dependence cone. Thus block indicators on intervals of length in \([s,2s]\) have a dependency graph with fixed degree, for example \(\Delta_{\mathrm{dom}}=32(R_{\mathrm{macro}}+1)\) is sufficient. Nonneighbors use disjoint actual primitive key regions, hence independence from their entire vector conditional on \(a\). On large rings unwrap the actual cone injectively, then extend independently on the line to estimate each tail. For dependencies use the actual ring cones, not independent copies of one ring key.

\paragraph{Full-channel extension and the halting cutoff.}

Backward propagation of output observables through CPTP adjoints shows that, conditional on \(a\) and a complete history \(\omega\),

\[
\begin{gathered}
T_{a,\omega}=G_{a,\omega}\circ\operatorname{Tr}_{A_{s}(a,\omega)^{c}},
\end{gathered}
\]

with \(G\) CPTP and \(A_{s}\) the actual initial-site ancestor set. Complete cuts and discards terminate all omitted support. This is Lemma~\ref{f:factorization} with every physical hypothesis supplied by Lemma \ref{q1:l14}. Include the autonomous classical word and every retained final factor. Initial dephasing is the actual physical dephasing, so it applies even to malicious encodings entangling controllers and reference.

Partition a sufficiently large ring into intervals of length in \([s,2s]\). A block-use indicator has probability at most \(2s\) times the ancestor bound. Choose one \(s=s_{\mathrm{dom}}\) for which this is at most \((1/4)^{\Delta_{\mathrm{dom}}+1}\) uniformly in the window. Lemma~\ref{f:domination} with \(q=1/2\) dominates the whole mask by fair independent block marks. Take a fair mask \(Y\) and its complement; under each sample a lower mask from a monotone coupling with the exact mask law. The lower masks are disjoint. Given each mask, sample a full history from its original conditional law. Route disjoint initial quantum blocks to the corresponding maps \(G\), insert fixed factors only on irrelevant complements, and copy only the measured word \(a\) to both controllers. Averaging gives two marginals exactly equal to the complete \(s_{\mathrm{dom}}\)-round physical channel, including all retained metadata, on arbitrary reference-entangled input. This is Proposition~\ref{f:extension} under Remark~\ref{rem:forgetting-interface}.

Lemma~\ref{f:iteration} iterates this full-algebra extension to \(2^{j_{\mathrm{clone}}}\) equal marginals after \(j_{\mathrm{clone}} s_{\mathrm{dom}}\) rounds. Its ring-size condition stays the one for \(s_{\mathrm{dom}}\). Lemma~\ref{f:bell} bounds every encoder and common decoder by \(1/2+2^{-j_{\mathrm{clone}}-1}\). Pick \(j_{\mathrm{clone}}\) from \(\eta\). The remaining finitely many ring sizes have a bounded qubit count \(N_{\mathrm{small}}\). At one final elementary noise layer, simultaneous replacement of their entire registers has probability at least \(p_{\mathrm{full}}=(\inf  I_{n})^{N_{\mathrm{small}}}>0\), independently each round. Such an event disconnects the Bell reference and gives decoded overlap exactly \(1/4\). Hence after \(s\) rounds, at elementary time \(t=D_{\mathrm{layer}}s\),

\[
\begin{gathered}
F^*_{L,\eps}(t)\le 1/4+3(1-p_{\mathrm{full}})^{s}/4.
\end{gathered}
\]

A common small-ring cutoff exists. Take its maximum with \(j_{\mathrm{clone}} s_{\mathrm{dom}}\) and multiply by \(D_{\mathrm{layer}}\). An all-decoder bound persists under every subsequent elementary layer by composition. This proves the theorem's uniform halting conclusion for every \(L\) and every actual rate in \(I_{n}\).

\section{Effectivity and the memory corollary}\label{q1:effective}
This section assembles the proofs of Theorem~\ref{q1:main} and Corollary~\ref{q1:memory} and bounds the optimal lifetime of the ring circuit.

\begin{proof}[Proof of Theorem~\ref{q1:main}]
The finite circuit and its every-layer noise convention are specified in Section~\ref{q1:model} and Lemmas~\ref{q1:l2}, \ref{q1:l5}, \ref{q1:l10}, \ref{q1:l11}, \ref{q1:l14} and~\ref{q1:l17}. Their schedules have a fixed finite period and range, and commute with translation by one site. Lemma~\ref{q1:l16} chooses all constants without circular dependence, including the positive rational \(c_1=1/(12C_{\rm cone}T_q)\). Lemmas~\ref{q1:l1}--\ref{q1:l12} and the joint recursion of Section~\ref{q1:storage} give the hypotheses of Lemma~\ref{q1:l18}. Its finite chart works for every sufficiently large ring length and every intermediate elementary-layer time. It gives fidelity at least \(123/128>15/16\) after the initialization error. Lemma~\ref{q1:l15} makes the encoder and decoder total on halting labels as well. Lemmas~\ref{q1:l13}--\ref{q1:l15} and the full-channel extension in Section~\ref{q1:forgetting} prove the size- and rate-uniform halting bound, including every small ring and every encoding into the whole register.

The certification assertion uses an exact finite calculation. Write \(f(n,L,t,q)\) for the entanglement fidelity of this encoder and decoder. At fixed integer \(n,L,t\) it is a polynomial in the actual rate \(q\) with computable rational coefficients. Permutations, Hadamard conjugation, dephasing and reset have rational superoperator coefficients; the Bell density matrix is rational, the nominal endpoint mixtures have rational probabilities, and every physical replacement is affine in \(q\). This is the analogue of Proposition~\ref{a:rational} for this encoder and decoder. Enumerate all \(L\ge L_1(n)\), all \(0\le t\le\lfloor c_1L\rfloor\), and all rationals \(q\in I_n\). Failure of \(f(n,L,t,q)\ge7/8\) is decidable for each triple, so failure of the universal predicate is recursively enumerable. Continuity gives the equivalent predicate over every real rate in the window. Nonhalting satisfies it. On a halting input choose \(\eta=1/16\), then take \(L\) sufficiently large that the uniform halting cutoff belongs to the tested horizon. Its rational midpoint fails the predicate. The predicate is therefore exactly \(\{n:U(n)\uparrow\}\), a \(\Pi^0_1\)-complete set.

A scalable-memory rate means \(\sup_L\tau_{L,\eps}=\infty\), as in Definition~\ref{def:windows}. The nonhalting lower bound gives this property at every rate of the window. In a halting window choose any \(\eta<3/8\); the uniform cutoff bounds \(\tau_{L,\eps}\) independently of \(L\) and \(\eps\). Thus scalable-memory labels are the same complete set. Hence an algorithm deciding scalability only at the computable rational midpoints of the windows would decide nonhalting. The reduction uses only rates inside the windows.
\end{proof}

\paragraph{Lifetime ceiling.} The optimal lifetime of this circuit is at most exponential in \(L\) at every rate of every window. Let \(N_{\rm wire}\) be the number of qubits per site and \(\eps_{\mathrm{inf},m}>0\) the left endpoint of \(I_n\), with \(m=m_{\mathrm{bin}}(n)\) (Section~\ref{q1:parameters}). After each elementary layer the noise replaces all \(N_{\rm wire}L\) qubits of the ring by maximally mixed states with probability \(\eps^{N_{\rm wire}L}\ge \eps_{\mathrm{inf},m}^{N_{\rm wire}L}\), independently of the other layers. After such an event the register is independent of the reference, and every decoder gives entanglement fidelity \(1/4\). Hence
\[
 F^*_{L,\eps}(t)\le\tfrac14+\tfrac34\bigl(1-\eps_{\mathrm{inf},m}^{N_{\rm wire}L}\bigr)^{t}
\]
for every \(\eps\in I_n\) and every elementary-layer time \(t\), so \(\tau_{L,\eps}<\eps_{\mathrm{inf},m}^{-N_{\rm wire}L}=\exp(O_n(L))\) elementary layers. This is the bound in Table~\ref{tab:results}.

\begin{proof}[Proof of Corollary~\ref{q1:memory}]
Keep the structural lane R1 and its initialized certified nonhalting literal, the quantum slot inventory, the schedule of Lemma~\ref{q1:l10}, and the fixed quantum code. Replace every R2 operation and the conditional erasure block by identity layers on the same wires, so that the period, the gate range and every R1 and quantum stencil are unchanged. Dispose of every disposable factor as before. This gives another admissible circuit on rings, fixed, translation-invariant and periodic, in which every wire, including all retained control, is subject to the elementary single-qubit noise. The constants \(D_{\mathrm{layer}}\), \(R_{\mathrm{macro}}\), \(C_{\mathrm{cone}}\), \(T_q\) and \(C_{\mathrm{top}}\) of Section~\ref{q1:parameters} apply unchanged.

Its base error cover has only direct microscopic marks and the structural witness marks of Lemmas~\ref{q1:l1}--\ref{q1:l3}. Their primitive supports are bounded by the same enumerated stencils. For the integer moment exponent chosen in Section~\ref{q1:parameters}, direct moments tend to zero with \(\sqrt\eps\), and Lemma~\ref{q1:l1.1} gives a computable positive rational cutoff making the entire clock moment as small as required. Choose \(\eps_{\rm m}\) below these two cutoffs. The named-child geometry of Lemma~\ref{q1:l7}, the deletion lemma and the common correctable spans of Lemma~\ref{q1:l4} apply without change. The joint recursion of Section~\ref{q1:storage}, with the selector types absent, gives \(u_k\le b_*2^{-k}\) at every fixed \(0<\eps\le\eps_{\rm m}\).

Initialize the computable line boundary of Lemma~\ref{q1:l3.1} on the complete injective cone, recursively encode the nonport qubit, and fill the remaining ring as in Section~\ref{q1:storage}. No helper marginal is needed. Choose the largest fitting rank and complete an intermediate readout with nominal noise at any fixed rational rate in \((0,\eps_{\rm m}]\), followed by the same fixed recovery. Lemma~\ref{q1:l1} also covers the resulting rate splice because all its primitive key rates are bounded by \(\eps_{\rm m}\). The proof of Lemma~\ref{q1:l18}, with no initialization loss because no helper marginal is prepared, gives the linear elementary-layer horizon and fidelity at least \(1-1/32>15/16\) for all sufficiently large rings. The finite searches and recursive maps terminate, so this encoder and decoder are total computable and independent of the actual rate. This proves the corollary with \(c_{\rm m}=c_1=1/(12C_{\rm cone}T_q)\). The size cutoff, \(k_{\rm rank}\ge k_0\) together with the finite chart conditions, does not depend on \(\eps\).
\end{proof}

\section{Discussion}\label{a:discussion}
We have constructed fixed, translation-invariant, time-periodic noisy circuits, on rings and in three, four and six dimensions, whose noise rate selects a Turing-machine input: they store a qubit for a time growing with system size when the machine runs forever, and forget every encoding when it halts. The windows are computable, but the set of labels of scalable-memory windows is not, and interleaving alone rules out a positive noise threshold for these circuits. Two ingredients stand on their own: the whole-channel forgetting argument of Section~\ref{sec:forgetting} and the one-dimensional memory of Corollary~\ref{q1:memory}.

For a prescribed encoding and recovery, each rational-rate fidelity test is an exact finite calculation (Propositions~\ref{a:rational} and~\ref{v:certification}(ii)), so failure of the prescribed guarantee has a finite, checkable witness \((L,t,q)\). Success has none in general: deciding whether every test passes is \(\Pi^0_1\)-complete for the circuits of Theorems~\ref{thm:exponential}, \ref{thm:main} and~\ref{q1:main} (Theorem~\ref{a:complete}, Proposition~\ref{v:certification}(ii) and Section~\ref{q1:effective}). Exact finite-size calculations can therefore refute a prescribed storage guarantee but cannot, in general, establish it. Xu and coauthors obtain a related one-sided certification result for the bulk spectral gap, with convergent certified upper bounds and no general convergent lower bounds \cite{XuEtAl2026}.

Hamiltonian undecidability and undecidable storage both concern observables that depend on arbitrarily large systems. In spectral-gap and phase-diagram results the observable is a property of the infinite-system Hamiltonian or its ground states, and a varying interaction can encode a machine input \cite{BCW2021,CPGW2015}. Here the gates remain fixed and depolarization varies. Its local statistics select the machine input, and the undecidable observable is whether the best recoverable entanglement fidelity supports storage times unbounded in system size. The open windows establish stability under changes of the noise rate within each window.

The constructions use local operations and independent depolarization after every layer, including reset and idle layers. Theorem~\ref{thm:four-dimensional}'s whole-site-noise guarantee is superpolynomial on every
nonhalting window.

The forgetting argument applies to other circuits with autonomous classical control: it couples the whole channel to an equal-marginal extension when the chance that a specified input site still matters is small, and the cloning bound then limits recovery even when the encoding uses every part of the register.

For the six-dimensional circuits, a matching \(\exp(O_n(L))\) upper bound would give exponential lifetime of order \(\exp(\Theta_n(L))\) on a nonhalting window. The known upper bound uses full resets of individual four-dimensional fibres. The prescribed diagonal defect bands of Lemma~\ref{v:wrapped} erase a plane in the clean comparison rule, but their probability of forming does not give an erasure rate under continuing noise. The four-dimensional whole-site circuit has the full-register bound \(\exp(O_n(L^4))\). A matching stretched-exponential upper bound is open.

Under independent single-qubit noise, Theorem~\ref{thm:main} gives linear
storage on every nonhalting window in three dimensions. The
four-dimensional circuit of Proposition~\ref{p:coexistence} gives
exponential storage on its reserved window \(I_0\) and linear storage
on its nonhalting tail windows \(I_{n+1}\), which part~(iii) caps at
\(201L+C_0\) rounds. Beyond-polynomial storage on every nonhalting
window remains open for four-dimensional undecidable constructions
under independent single-qubit noise. Theorem~\ref{thm:exponential}
gives an exponential lower bound on every nonhalting window in six
dimensions. Can an undecidable construction under the elementary-layer
noise convention achieve an exponential lower bound on every nonhalting
window below six dimensions, or a superlinear lower bound on every
nonhalting window in three dimensions?

Theorem~\ref{q1:main} gives undecidable storage with rate-selected whole-register forgetting already on rings, under independent single-qubit noise. Its linear lower bound leaves open stronger one-dimensional lifetimes and practical reductions of its register size and interaction range. Under whole-site depolarization, undecidable constructions below three dimensions remain open; the rate selection on rings uses independent wire faults within a site. Corollary~\ref{q1:memory} also gives a translation-invariant, time-periodic memory on rings in which every wire, including all control, is noisy, starting from an initialized classical hierarchy. Whether such a hierarchy can be grown from a homogeneous initial state, as in G\'acs's self-organizing automaton \cite{Gacs2001}, remains open here, as it does for the two-dimensional scheme of D\"unnweber, Styliaris and Trivedi \cite{DST2026}.

Adapting that two-dimensional automaton or the toric-code circuit of Balasubramanian, Davydova and Lake \cite{BDL2026} to rate-selected storage would require control of the controller-correlated erasures specific to our construction. The homogeneous decoder of Lake and the time-translation-invariant decoder of Selub, Bhardwaj and Lake assume noiseless classical processing \cite{Lake2025,SBL2026}. It is open whether their guarantees extend to the noisy rate-selection and erasure mechanism here. Further questions for undecidable constructions concern smaller site registers, nearest-neighbour interactions and a gate-local noise model with independent replacement of each gate's joint operand register by a maximally mixed state.

\section{How this work was done}\label{s:how-done}
AI models assisted with exploration, proof development and checking. The proofs are written in full, in this paper and in the cited preprints. Models from two developers checked the main results, each instructed to look for errors. The parts stated as formalized are verified in Lean (Appendix~\ref{app:lean}). The authors take responsibility for every claim.

\appendix
\section{Closed four-dimensional toric-code estimates}\label{app:code-estimates}
Theorem~\ref{thm:exponential} uses a closed four-dimensional toric code on each fibre. We give the sweep predecessor rule, the grouped contour argument and the deterministic recovery criteria here. Lemma~\ref{x:exponential} reuses these arguments on a closed side-\(L\) code. Appendix~\ref{sec:folded} gives the folded code; Table~\ref{q:links} in Section~\ref{q:link-section} gives the common link table used by Theorem~\ref{thm:main}. Theorem~\ref{thm:main} also uses the grouped-contour estimate of Lemma~\ref{q:frontier} in Section~\ref{sec:controller} and Lemma~\ref{d:sources}.

\subsection{Sweep predecessors}

We use a variant of the sweep rule of Kubica and Preskill
\cite{KubicaPreskill2019}; for other local toric-code decoders see \cite{BDMT2017}. Simultaneously at every vertex, read the
positive incident edges of the syndrome present before the sweep; if
their number is even, pair them in a fixed order and flip the
corresponding positive plaquettes, and if it is odd, do nothing. This
is the positive sweep; reflection reverses all
directions. Fixed Clifford extraction and local Boolean record
processing implement the rule. Run \(400\) reflected sweeps per CSS
sector on the code complex (the virtual complex for the folded circuit).

In each round, for the bit sector followed by the phase sector, one
Clifford extraction writes the syndrome into fresh records, the snapshot
\(\sigma_0\). Local Boolean processing on these records alone computes
400 reflected sweeps by
\(\sigma_{j+1}=\sigma_j+\partial P_j\) for \(0\le j<400\), where
\(P_j\) is the plaquette set that sweep \(j+1\) selects from
\(\sigma_j\), and stores \(P=P_0+\cdots+P_{399}\). Only then are the
plaquette flips of \(P\) applied. The syndrome present before sweep
\(j+1\) means \(\sigma_j\); no sweep reuses an earlier sweep's decisions.
The record processing has range at most 400 code cells.
Track both sector chains at the common boundaries of full rounds,
just before the bit-sector extraction. The catalogue \(\Lambda_t\)
includes the allowed insertions and erroneous uses throughout the full
round, as in Lemma~\ref{q:round-shadow}.

\begin{lemma}[Sweep predecessors]\label{q:sweep}
Let \(x_1,\ldots,x_4\) be the coordinate functions of the lifted code lattice, and call \(f(u)-f(v)\) the gain in \(f\) of a predecessor \(u\) of \(v\). For one positive sweep, every vertex of the new syndrome has, separately
for each of the four functions \(x_j\), an old syndrome predecessor with
nonnegative gain in that function. It also has an old predecessor with
strictly positive gain in \(-\sum_jx_j\). The reflected sweep has the
corresponding reversed-coordinate property. After \(M\) reflected
sweeps the gains for the walls \(-x_j\) and \(\sum_jx_j\) are at least
\(0\) and \(M\), respectively.
\end{lemma}
\par\noindent\emph{Proof idea.} Trace a new syndrome corner to the plaquette correction that created it. The old incident edges give separate wall predecessors, and the correction root gives the strict sum gain.

\begin{proof}
Suppose a new syndrome vertex \(v\) was absent from the old syndrome.
It is a corner of a corrected plaquette based at an old syndrome vertex
\(u\). If its \(j\)-coordinate was raised, the old positive edge in
direction \(j\) used at \(u\) has the raised endpoint as an old syndrome
vertex. That endpoint supplies the \(x_j\) predecessor. If the coordinate
was not raised, use \(u\). The correction root \(u\) has smaller
coordinate sum than \(v\), and supplies the last predecessor.

An old vertex that remains a syndrome vertex supplies its own first
four predecessors. If it has an old negative incident edge, that
edge's lower endpoint supplies the last one. Otherwise every old
incident edge is positive. Their number is even because the syndrome
is a cycle. The vertex's own correction then cancels all of them.
Any incident edge remaining after the parallel correction must be
toggled by a correction based at a vertex of smaller coordinate sum;
that correction root is an old syndrome vertex. This proves the strict
last gain. Cancellations between parallel corrections do not change
this argument. Reflection and iteration give the stated bounds. The
dual lattice differs by a fixed half-cell displacement and has the
same proof.
\end{proof}

\begin{lemma}[Grouped derivations]\label{q:frontier}
Suppose a finite acyclic backward derivation carries \(k\) balanced
potentials, that is, potentials whose sum vanishes identically; the gain
of an edge in a potential \(F\) is \(F(\text{parent})-F(\text{vertex})\),
and an arm is a path followed for one potential from a source occurrence
to a terminal. For each potential every positive support has a selected
parent with gain at least \(d_{\rm gain}>0\), and every selected edge loses at
most \(K_{\rm loss}\). Allow source-free zero-gain corridors, followed by a
positive support, and bounded entry and terminal links, at most one
of each per arm. Group terminal appearances by source label. For group
\(i\), let \(\Omega_i\) bound the total potential oscillation and let
\(K_{{\rm term},i}\) bound the loss of a terminal link. Every root has a
connected witness with \(g\) distinct true source labels. If \(E\) counts
positive edges with arm multiplicity and \(K_{\rm ent}\) bounds the entry
loss on one arm, then
\[
 d_{\rm gain} E\le k(d_{\rm gain}+K_{\rm loss})(g-1)+kK_{\rm ent}g+
 \sum_{i=1}^g(\Omega_i+kK_{{\rm term},i}).
\]
With uniform terminal loss and oscillation bounds, the projected witness
has at most \(C_{\rm w}g\) edges, where \(C_{\rm w}\) is computable from
\(k,d_{\rm gain},K_{\rm loss},K_{\rm ent}\), the fixed typed stencils and these bounds.
\end{lemma}
\par\noindent\emph{Proof idea.} Keep one arm endpoint of each charge in every overlap component of the frontier. A split requires new arms, and balancing their potential increments bounds the total length by the terminal losses and oscillations.

\begin{proof}
The argument refines Toom's contour method
\cite{BermanSimon1988,Gacs2001,Toom1980} to grouped terminal labels.
For a derivation vertex \(v\), let \(T(v)\) be the set of terminal
labels reachable from it. Begin with the frontier consisting of the
root and expose a frontier nonterminal by replacing it with all its
selected parents. Partition the frontier by connected components of
the intersection graph of its \(T\)-sets. New sets are subsets of the
exposed set, so exposure cannot merge previous components.

If one component splits into \(j\) daughters, each daughter contains
a selected parent of the exposed vertex. Indeed, any remaining old
frontier set previously attached to the exposed set intersects a
terminal set of at least one selected parent; any surviving connection
not using the exposed set remains. For each charge, keep one arm
endpoint in every current component. Initially follow a charge-good
arm from the root to a terminal. At a split the existing endpoint
belongs to one daughter. Add \(j-1\) source occurrences at the exposed
vertex. For that charge route their first edges into the daughters
missing an endpoint, then continue by charge-good edges. The chosen
parent's whole reachable terminal set is in that daughter, so the
continuation cannot leave it. The assignments for different charges
may differ.

At the terminal frontier each distinct terminal label receives
exactly one arm endpoint per charge. Join the starts at each source
occurrence and the endpoints at each terminal label, and retain the
undirected arm component containing the original root. In this
component every source starts one arm of each charge and every
terminal receives one of each charge. Therefore \(g\) retained
terminal labels correspond to \(g\) retained source occurrences.
Exactly \(g-1\) occurrences were added, so at most \(k(g-1)\) first
edges fail to be charge-good.

Sum potential increments along all arms. Balanced potentials cancel
at each source. At terminal group \(i\), cancellation costs at most
\(\Omega_i\). There are \(kg\) arms, with at most one entry link per arm
and one terminal link of each charge at group \(i\). Summing these
losses gives the displayed bound. Under uniform loss and oscillation
bounds, \(E=O(g)\). A zero corridor cannot split an overlap component. After
projection each such corridor contributes a fixed number of links,
charged to its following positive support; root and terminal
exceptions cost \(O(g)\). The entry and terminal links also cost
\(O(g)\). Projection preserves connectedness and can only identify
edges, giving the asserted witness.
\end{proof}

\subsection{One recovery for every permitted Pauli assignment}

We first record the deterministic topological criterion. Use the shortest period of the intended closed code torus.

\begin{lemma}[Local filling invariant]\label{q:homology}
There is a fixed size \(\ell_0\) with the following property. Let \(\ell_{\min}\ge\ell_0\) be the shortest period of the code torus and set \(r=\ell_{\min}/20\).
Starting from a code state, if every syndrome component in both CSS
sectors has coordinate diameter less than \(r\) at every serialized
elementary plaquette Pauli update, a fixed final filling decoder
recovers the entire logical state and its reference exactly.
\end{lemma}
\par\noindent\emph{Proof idea.} Fill each small lifted syndrome component in a contractible box. A single plaquette update changes the corrected chain by a boundary, preserving its homology class.
\begin{proof}
A syndrome component is itself a \(1\)-cycle. Its small lift lies
in a contractible box, so fix a \(2\)-chain filling there by finite
Gaussian elimination and a deterministic box/lift convention. Set
\[
 h(e)=\left[e+\sum_\nu C_\nu(\partial e)\right]
 \in H_2(\mathbb T^4;\mathbb F_2).
\]
Under one plaquette update only components meeting that plaquette
can change. The affected old and new components and their fillings
lie in a common box of width at most \(8r+O(1)<\ell_{\min}\), after
enlarging the fixed size threshold. Unaffected fillings cancel.
The difference of the two corrected chains is closed in that box,
hence a \(3\)-boundary. Thus \(h(e)\) is unchanged. It starts at
zero. Applying the same argument on the dual cellulation shows that
the final Pauli acts by stabilizers in both sectors, including on
reference-entangled states.
\end{proof}

\begin{lemma}[Round shadowing]\label{q:round-shadow}
Track both sector chains at common full-round boundaries, just before
the bit-sector extraction. Fix a round \(t\) and a sector, and let
\(e_t,e_{t+1}\) be its Pauli-frame 2-chains at those boundaries.
Let \(\Lambda_t\) include the catalogue locations of allowed insertions
and erroneous uses throughout that full round: inserted Paulis,
erroneous records, and erroneous stored decisions or mask uses.
Let \(R_*\) bound the code-lattice radius of each propagated local
contribution through every remaining stage of that full round, namely
to this sector's extraction, through its record processing, and to the
next common boundary, uniformly over both sectors.
Let \(K_*\) bound the number of sheet patches per location, let
\(d_\square\) be the diameter of a plaquette, and put
\[
 R_{\rm blk}=400d_\square+2R_*.
\]
For every serialization of the round's Pauli updates and every
intermediate chain \(e\), every vertex of \(\partial e\), and every
vertex of \(\partial e_{t+1}\), lies within distance \(R_{\rm blk}\)
of \(\partial e_t\cup\Lambda_t\). For the folded circuit the same
holds in each sheet's virtual chart, with the \(K_*\) sheet patches
of each location.
\end{lemma}
\begin{proof}
At this sector's extraction write \(\sigma_0=\partial e_t+\rho\),
where \(\rho\) includes erroneous records and data Paulis from the
preceding stages of the same round. Each such propagated local
contribution lies within \(R_*\) of its location in \(\Lambda_t\).
Ideal operations of the other CSS sector leave this sector's syndrome
unchanged. An erroneous intermediate sweep record or data Pauli
enters only within \(R_*\) of its location in \(\Lambda_t\).
A vertex of
\(\sigma_{j+1}\) absent from \(\sigma_j\) is a corner of a plaquette
of \(P_j\). That plaquette is selected at a vertex of \(\sigma_j\)
by the proof of Lemma~\ref{q:sweep}, or, if its stored decision is
erroneous, lies within \(R_*\) of \(\Lambda_t\). Induction therefore
places every vertex of \(\sigma_j\) within \(jd_\square+R_*\) of
\(\sigma_0\cup\Lambda_t\). An intermediate chain is
\(e=e_t+F'+P'\), where \(F'\) consists of inserted Paulis of this
round and \(P'\) consists of applied plaquettes of \(P\). A vertex
of \(\partial e\) lies in \(\partial e_t\), in an inserted Pauli's
boundary within \(R_*\) of \(\Lambda_t\), or on a plaquette of
some \(P_j\), within \(d_\square\) of \(\sigma_j\), hence within
\((j+1)d_\square+2R_*\) of \(\partial e_t\cup\Lambda_t\). Since
\(j\le399\), these distances are at most \(R_{\rm blk}\).
The final chain is included. The argument is local in each virtual
chart, so it applies separately to every sheet patch.
\end{proof}

A fixed exogenous classical history determines a catalogue of allowed
Pauli-insertion supports: physical-noise supports, erroneous record
or mask uses, controller erasures, and each affected sheet image of
a wrong program gate. The nonunitary instruments of the intended
processor have one common unconditional Stinespring dilation. Only
unitaries differ between actual and intended programmed circuits.
For the folded circuit choose \(r_{\rm d}\) equal to a sufficiently small fixed fraction of \(m\) from Appendix~\ref{sec:folded}. An admissible dangerous certificate is a
connected typed-source certificate from the proof of Lemma~\ref{q:spatial}
spanning at least a fixed multiple of \(r_{\rm d}\), with allowed
predecessor, program-use and terminal annotations. Define
\(\mathcal G\) to mean that none exists. This definition precedes
and is independent of any choice of Pauli values at the catalogue.

The folded circuit uses the source derivations of Appendix~\ref{sec:folded}.
For the six-dimensional circuit, \(\mathcal G\) is the complement of the
exceptional event in Lemma~\ref{v:bridge}. Its proof supplies the corresponding
derivations, admissible certificates and probability bound.

\begin{lemma}[Universal certificate implication]\label{q:universal}
For a fixed exogenous classical history on a code torus above
a fixed computable size (so that Lemma~\ref{q:homology} applies and any
\(K_*\) patches linked at distances \(2(r_{\rm d}+R_*)\) fit in a
contractible box much narrower than \(\ell_{\min}\)), every Pauli assignment at
the allowed catalogue locations that changes corrected homology in
either CSS sector yields an admissible dangerous certificate.
Consequently, on \(\mathcal G\), the same intended-complex filling
decoder corrects every insertion history on the entire logical
codespace, including its reference.
\end{lemma}
\par\noindent\emph{Proof idea.} Fix an arbitrary Pauli history and use stabilizer extraction to track its syndrome without reference to the logical state. A first change of corrected homology forces a large component, whose predecessor graph yields the certificate.

\begin{proof}
Fix an arbitrary complete insertion assignment \(h\). In the ideal
inserted network the data are always a Pauli frame on a codeword.
Indeed, Clifford extraction conjugates Pauli observables to Pauli
observables, and conjugation by an inserted Pauli changes only their
signs. A stabilizer measurement therefore has a deterministic record
for this frame, independent of the stored logical state. Record
Paulis flip fixed basis records; the local Boolean matching then
computes deterministic correction choices. Its work remains basis
classical. In the common instrument dilation the associated
measurement environment vector is logical-independent. This proves
the assertion for the entire successive history by induction.

In a fixed extraction/sweep block, a syndrome vertex has the ideal
\(400\)-sweep predecessors of Lemma~\ref{q:sweep} unless its ideal
backward cone contains an inserted or erroneous location. In that
case terminate its quantum arm at the last such location in the
block. The remaining ideal propagation has a fixed radius in that
sheet. Attach the corresponding per-sheet link, then its true
profile-use or selector derivation, or its fresh hardware source.
Record errors have the same bounded local correction influence.
The derivation is finite and its classical entries never return to
quantum types. In the folded case Table~\ref{q:links} supplies all increments,
including repeated predictor blocks. In six dimensions the typed source label catalogue
in the proof of Lemma~\ref{v:bridge} supplies the links and the grouped derivations.

A Pauli term may have several separated sheet patches. Let \(K_*\) and
\(R_*\) be as in Lemma~\ref{q:round-shadow}; they bound the number and
the radii of these patches.
Choose \(r_{\rm d}\) small enough that any union of \(K_*\) patches
linked at distances \(2(r_{\rm d}+R_*)\) fits in a contractible box much
narrower than \(\ell_{\min}\). Combine such nearby patches into
one box. Separated boxes can be applied successively: if the
before and after syndromes have only components smaller than
\(r_{\rm d}\), no such component joins two separated boxes. An
intermediate connected piece therefore belongs to the before
syndrome near an unprocessed box or to the after syndrome near a
processed box, together with the bounded patch neighbourhoods.
Inside a combined box serialize its plaquettes. Every intermediate
component is contained in a fixed multiple of that contractible
box. Taking the fixed fraction defining \(r_{\rm d}\) smaller if needed
puts these diameters below the barrier in
Lemma~\ref{q:homology}.

It follows that a change of corrected homology forces a
macroscopic syndrome component either at a tracked block boundary
or in such a local serialization. Map each vertex of the first macroscopic intermediate component \(K\)
to a shadow root in \(\partial e_t\cup\Lambda_t\) within
\(R_{\rm blk}\), by Lemma~\ref{q:round-shadow}. This typed link has
fixed range, decreases physical time by at most one round, and loses
at most a fixed amount of each potential; it is the intermediate-stage
spur. A tracked root carries its grouped witness from
Lemma~\ref{q:frontier}. Its 400-sweep link (Table~\ref{q:links}) is
Lemma~\ref{q:sweep} applied to its own round's exact recurrence
\(\sigma_0,\ldots,\sigma_{400}\), terminated at the last location
of \(\Lambda\) in its backward cone. A root in \(\Lambda_t\)
is a single-source witness. Adjacent vertices of \(K\) have shadow
roots at distance at most \(2R_{\rm blk}+1\), so the shadow roots
form a chain spanning at least \(\operatorname{diam}K-2R_{\rm blk}\).
In the ball covering use \((4+2R_{\rm blk}+1)\)-fold enlargements.
The constants \(C_{\rm raw},C_{\rm all},\Delta_*,\Delta_{\rm all},
b_0,b_{\rm span}\) change only through \(R_{\rm blk}\), fixed
with the catalogue. Every local patch is
anchored at its allowed typed insertion source; no long virtual
edge is inserted between distinct sheet images. Apply the
predecessor construction just proved and Lemma~\ref{q:frontier},
then the ball and connector construction in the proof of Lemma~\ref{q:spatial}
for the folded circuit or in the proof of Lemma~\ref{v:bridge} for the
six-dimensional circuit, to obtain the dangerous certificate.

All these constructions use only positions, finite roles, and true
terminal events. The argument applies to the arbitrary assignment
\(h\) with which we began. The certificate count therefore controls
the union over every assignment. On \(\mathcal G\) no assignment
can change homology, and the one fixed filling decoder returns every
logical state exactly by Lemma~\ref{q:homology}.
\end{proof}

\begin{lemma}[Common recovery and coherent span]\label{q:span}
Let \(V\) encode a logical system \(A\), also writing \(V\) for its encoding channel,
and suppose one recovery
\(\mathcal D\) satisfies \(\mathcal D\Phi_hV=\operatorname{id}_A\)
for every allowed Pauli-inserted history in a common unconditional
instrument network. Then it also corrects coherent unitary changes
whose Pauli expansions use those locations. If changed gates couple
to an ideal-decoupled outside system \(E\), the corrected dilation
has the form \(I_A\otimes K_E\), with \(K_E\) an isometry. Hence
logical/reference marginals are preserved even for initially
correlated logical and outside inputs.
\end{lemma}
\par\noindent\emph{Proof idea.} Exact recovery of each inserted history factors the recovered logical output from its environment in a common dilation. Expand the actual gates in Pauli operators and sum the factors, including the outside-system operators.

\begin{proof}
Choose the common history dilations \(\Omega_h\) and one recovery dilation
\(R\). Exact identity recovery on a logical basis gives
\[
 R\Omega_hV|a\rangle=|a\rangle\otimes|s_{h,a}\rangle,
\]
since the recovered marginal is pure. Preservation of the
off-diagonals forces
\(\langle s_{h,b}|s_{h,a}\rangle=1\), so these unit vectors are
all equal, including their phase. Therefore
\(R\Omega_hV=I_A\otimes|s_h\rangle\).

For each changed unitary write
\(U_{\rm act}U_{\rm ideal}^*=\sum_P\alpha_P P\).
Expanding the finite interleaved product in the common dilation gives
\(\Omega_{\rm act}V=\sum_h\alpha_h\Omega_hV\). Applying \(R\) yields
\[
 R\Omega_{\rm act}V=I_A\otimes\sum_h\alpha_h|s_h\rangle.
\]
The actual dilation is isometric, so the remaining vector has norm
one. The full coherent sum, including all cross terms, has exact
identity recovery.

For an outside system instead expand each difference as
\(\sum_P P_{\rm active}\otimes B_P\). The intended active/outside
factorization allows all outside operations and their dilation
baths to remain in these operator-valued coefficients. The same
calculation gives
\[
 (R\otimes I)\Omega_{\rm act}(V\otimes I_E)=I_A\otimes K_E.
\]
Isometry implies \(K_E^*K_E=I_E\). Its partial trace is a channel
on the environment, so the logical/reference marginal is unchanged
for every joint input on \(A,E\), and a reference. Active check
ancillas and records remain in \(\Omega_h\); only ideal-decoupled
outside factors enter \(E\).
\end{proof}

\subsection{Backward increments for common witness links}\label{q:link-section}
For a local lifted virtual chart set \(d=(58,58,58,57)\) and
\begin{equation}\label{q:walls}
 F_j=-(q_j+d_jt)\ (1\le j\le4),\qquad
 F_5=\sum_{j=1}^4(q_j+d_jt).
\end{equation}
They sum to zero. Table~\ref{q:links} lists backward increments
\(F(\text{parent})-F(\text{root})\). It includes links created by
classical gathering and predictor recentring. Write
\(\delta_i\in\{-1,0\}\) for the \(i\)-part of the selected selector parent's displacement \((u_1,u_2,u_1,u_2+\delta_i)\) (Appendix~\ref{sec:folded}), and put \(T_{\rm hist}=1024\). Here \(u_j\) is the \(j\)-th component of the selected replica displacement in the selector or profile box.

\begin{table}[!htbp]
\centering\small
\renewcommand{\arraystretch}{1.1}
\caption{Backward increments \(F(\text{parent})-F(\text{root})\) of the five balanced potentials \(F_1,\ldots,F_5\) in \eqref{q:walls}. The witness graphs in Lemmas~\ref{d:sources}, \ref{q:universal} and~\ref{q:spatial}, and Proposition~\ref{c:oriented-source}, use these links. The five entries are in potential order. The good entries for different charges may use different parents.}
\label{q:links}
\begin{tabular}{>{\raggedright\arraybackslash}p{.25\textwidth}>{\raggedright\arraybackslash}p{.68\textwidth}}
\toprule
Link & Five increments, or their justified lower bounds\\
\midrule
Selector support &
\((58-u_1,58-u_2,58-u_1,57-u_2-\delta_i)\),\newline
\(\Delta F_5=2u_1+2u_2+\delta_i-231\).
Charge-specific good choices give \((43,43,43,42,88)\).\\\addlinespace[3pt]
Profile support &
\((58-u_1,58,58-u_3,57)\),\newline
\(\Delta F_5=u_1+u_3-231\).
\newline The \(320\)-separated boxes give \((43,58,43,57,89)\).\\\addlinespace[3pt]
\(400\) reflected sweeps &
\((58,58,58,57,169)\) as charge-specific lower bounds.\\\addlinespace[3pt]
Sampling-register copy corridor &
\((0,0,0,0,0)\) on its base characteristic.\\\addlinespace[3pt]
Gather then field consult &
\((0,0,0,+j,-j)\) for the gather \(i\mapsto i-j\);
\((0,0,0,-j,+j)\) for the consult back to the base.
Their compressed link has zero increment.\\\addlinespace[3pt]
Clean predictor entry &
\((0,0,0,T_{\rm hist}-1,-(T_{\rm hist}-1))\), followed by all \(T_{\rm hist}\) formal support levels.\\\addlinespace[3pt]
Same-site old/new star &
\(\pm(58,58,58,57,-231)\); total absolute oscillation \(462\).\\\addlinespace[3pt]
Quantum or program-use link &
For its fixed displacement \((\Delta q,\Delta t)\),
\((-\Delta q_j-d_j\Delta t)_{j=1}^4\) and
\(\sum_j\Delta q_j+231\Delta t\). Each sheet's allowed list is finite.\\
\bottomrule
\end{tabular}
\end{table}

\subsection{Finite-torus exponential storage}\label{q:closed-exponential}
\begin{lemma}[Finite-torus exponential storage]\label{x:exponential}
Consider the closed side-\(L\) four-dimensional plaquette toric code,
with the reflected-sweep processor of Lemma~\ref{q:sweep} in both
CSS sectors. An iid exogenous site-round source of sufficiently small
computable rate \(p_{\rm aux}>0\) may cause arbitrary faults in a
fixed finite influence catalogue. There are computable constants
\(A,f_0>0\) such that the probability of any dangerous
certificate through any finite time \(T\) is at most
\begin{equation}\label{x:exp-bound}
 A(T+A)L^4\cdot2^{-f_0L}.
\end{equation}
On its complement one fixed component-filling decoder recovers the
whole logical state and reference. Consequently this processor stores
a qubit with fidelity at least \(15/16\) through
\(2^{\lfloor c_{\exp}L\rfloor}\), for rational
\(0<c_{\exp}<f_0/4\) and all sufficiently large computable sizes.
\end{lemma}
\par\noindent\emph{Proof idea.} Use the sweep predecessors to attach a connected explanation to each syndrome vertex. A large component forces an explanation with many distinct sources, and excluding those explanations makes component filling valid for every allowed history.

\begin{proof}
There are six plaquette data qubits per site, with local extraction,
syndrome and correction work. Each sector uses the extraction and
400-step virtual recurrence specified above; the resulting Pauli mask is
stored before its gates execute. All
incidence layers and private controls are fixed. Begin with zero
syndrome and lift the finite-time dynamics periodically to \(\Z^4\).
Use the five balanced potentials
\[
 F_j=-(r_j+58t),\quad 1\le j\le4,\qquad
 F_5=\sum_j(r_j+58t).
\]
After 400 reflected sweeps, the backward negative potential gains are at
least 58 and the sum potential gain is at least \(400-4\cdot58=168\).
The intermediate-stage spurs of Lemma~\ref{q:round-shadow} have bounded losses, as constructed in the proof of Lemma~\ref{q:universal}. Group all
appearances of one lifted source under its single site-round name.
Lemma~\ref{q:frontier} gives a connected witness with at most
\(C_{\rm w}g\) edges for \(g\) distinct lifted source anchors.
This is a deterministic statement about a periodic field.

For a syndrome path of macroscopic span, enclose each root's complete
witness and its source anchors in a ball of radius \(O(g)\). Select
disjoint balls in decreasing radius. The
\((4+2R_{\rm blk}+1)\)-fold enlargements from
Lemma~\ref{q:round-shadow} and the proof of Lemma~\ref{q:universal}
cover the polygonal path, including edges between consecutive shadow roots. Choose a
simple connecting chain, join successive selected roots at the common
time, and join both original path endpoints to the first and last
selected roots. These links cost \(O(\sum g)\); a single large
selected ball still includes both endpoint links. Original disjoint
balls have disjoint lifted source sets. The combined graph has
\begin{equation}\label{x:raw-span}
 m\le C_{\rm raw}g,\qquad m\ge b_0L,
\end{equation}
where \(m\) is its edge count and \(b_0>0\) is the fixed span constant of this construction. All full space-time roles and canonical
source anchors are retained. Here \(R_{\rm edge}\) is a local edge-range
bound, chosen after \(R_{\rm blk}\) and including the already
constructed shadow links; it bounds each edge's spatial range.

Take \(q=\lfloor f_0L\rfloor\), reducing rational
\(f_0>0\) so that \(q\le m\) and \(R_{\rm edge}q<L/2\).
Double a spanning tree of the graph, obtaining a walk of length at
most \(2m\). Mark the first visit to each of its \(g\) source
anchors, and partition the walk into pieces of at most \(q\) edges.
There are at most \(2m/q+1\le3m/q\) pieces. One contains at least
\(gq/(3m)\ge q/(3C_{\rm raw})\) distinct marks. Assign a boundary
mark to one half-open piece, with the final endpoint in the last
piece. Each mark is assigned once and belongs to its piece.

This short connected walk has spatial diameter below \(L/2\).
Two distinct canonical lifted anchors in it cannot represent the
same physical site-round coin: a nonzero spatial deck translation
has a coordinate of magnitude at least \(L\), and time is not
quotiented. Thus its marks denote distinct physical iid sources.
No independence is assigned to the full old lifted graph.

If \(\Delta_*\) bounds the enlarged role graph's degree, including
a stay-put option, count short walks and their source subsets.
There are at most \(A_0(T+A_0)L^4\) physical starts and
\(2^{q+1}(\Delta_*+1)^q\) annotated walks. Given a start and its
edge sequence there is one lift. Discard annotations with repeated
physical source names. Independence is now valid and gives
\[
 \Pr(\text{dangerous certificate by }T)
 \le2A_0(T+A_0)L^4
       [2(\Delta_*+1)p_{\rm aux}^{1/(3C_{\rm raw})}]^q.
\]
For example choose the rational
\(p_{\rm aux}=[1/(8(\Delta_*+1))]^{3C_{\rm raw}}\).
With the actual source rate at most \(p_{\rm aux}\), the bracket is at most \(1/4\). Increasing the prefactor for the
floor in \(q\) gives \eqref{x:exp-bound}.

It remains to connect certificates to logical failure. Serialize
bounded local Pauli updates of each actual layer, keeping the stored
mask from the exact 400-step recurrence fixed. Use the shadow-root map
of Lemma~\ref{q:round-shadow} as in the proof of Lemma~\ref{q:universal}. Lemma~\ref{q:homology}, now with shortest side
\(L\), says that corrected homology cannot change while all affected
components have small contractible lifts. At the first barrier
crossing a bounded update merges only a fixed number of small
components. Choose the barrier fraction small enough that this first
component still has an injective lift. Its extreme vertices supply
the path used in \eqref{x:raw-span}. The predecessor history can be
arbitrarily old, but is finite at that crossing time. Intermediate
stages are included, so a closed error surface with zero endpoint
syndrome cannot bypass this argument.

Define the good event by absence of all such geometric certificates,
before assigning Pauli values. The predecessor and first-crossing
proof applies to every assignment in the fixed catalogue. The same
filling decoder therefore corrects their whole union. For general
local channels use Lemma~\ref{q:span}'s operator-valued expansion in
the common unconditional instrument dilation. It preserves the full
logical/reference state. Initialize one of the six logical pairs with
the unknown qubit and fix the other five. The same encoder works
throughout the horizon. Substituting
\(T=2^{\lfloor c_{\exp}L\rfloor}\) in \eqref{x:exp-bound} gives
failure at most \(1/16\) above a computable threshold.
\end{proof}

\subsection{Origin of recurring constants}\label{q:constants-origin}
\begin{itemize}
\item \(768=3\cdot16^2\) counts the voting replicas in the three boxes of \eqref{eq:c-boxes}. The threshold is \(512\). In the raw six-dimensional graph the nine Moore offsets give degree \(2\cdot9\cdot768=13\,824\), and the doubled walk count uses \(2\cdot13\,824^2=382\,205\,952\) (Lemma~\ref{v:density}).
\item The drift \(58\) in each protected coordinate balances the gains on the voting boxes. For each negative potential a replica outside its far box has \(u_j\le15\), giving \(3(58-15)=129\). A far-box replica has \(u_1+u_2\ge160\), giving \(3(160-2\cdot58)=132\) for the sum potential. The largest single-coordinate offset is \(175\), giving the worst loss \(3(175-58)=351\). An old and a new appearance differ by one step of this drift, giving total potential oscillation \(348+174+174=696\). The ordinary unshifted density proof instead uses gain \(130\) and loss \(350\) (Lemmas~\ref{c:coherence} and~\ref{v:density}, and Section~\ref{v:parameters}).
\item In the six-dimensional construction a complete query uses \(4096\) virtual levels. Distributing its entry loss of at most \(696\) over those levels leaves \(129-696/4096>128\) gain per edge (Section~\ref{v:parameters}).
\item A fork costs at most \(480\) per charge relative to a good edge. With three charges this gives \(1440\) per fork. In the density count, \(1440/130<12\) and \(1404\cdot12=16\,848\), with \(1/\delta=1404\); counting spurs and contours adds \(12\), giving \(16\,860\) (Lemma~\ref{v:density}).
\item The code uses \(400\) reflected sweeps. The drift costs \(4\cdot58\) on the sum potential, leaving \(400-4\cdot58=168\) in the six-dimensional proof (Lemmas~\ref{q:sweep} and~\ref{v:bridge}).
\end{itemize}

\section{Periodic clocks and raw sources on a torus}\label{app:f4-calendar}
The following extension of Section~\ref{d:calendar-section} supplies a
reference at every time, including times longer than any one clock
period. Write \(P_{\rm cal}=2^{r_{\rm exp}}\), \(T_{\rm cal}=P_{\rm cal}^4\).

Lemma~\ref{f4:raw} bounds raw clock faults on the torus, and Lemma~\ref{f4:roles} describes the two roles of the fold.

\begin{lemma}[Whole-state periodic clock]\label{f4:periodic}
For each fixed even \(s_{\rm top}\) and \(k\ge1\), a total finite
initialization on the cycle of length \(m=s_{\rm top}P_{\rm cal}^k\) implements the
repeated \(k\)-fold encoded identity, with aligned quantum periods
\(T_{\rm cal}^k\). Its complete classical state has temporal period dividing
\(T_{\rm cal}^{k+1}\). The virtual tower before the fixed fold decoration has initial spatial period dividing \(P_{\rm cal}^k\).
\end{lemma}
\par\noindent\emph{Proof idea.} Extend the interpreter so each simulation period returns its scratch records and port state to prescribed values. Build and verify this return using the fixed local phases.

\begin{proof}
Use the local interpreter of Section~\ref{d:calendar-section}, with
one additional requirement on its final phase: after committing the
next information and completing quantum-port use, return every
auxiliary field to its canonical age-zero value. The fields of one
virtual clock cell are specified in Table~\ref{f4:fields}. The
word ``blank'' denotes the fixed zero encoding of the corresponding
finite register; a head has the one declared start marker at address
zero and the no-head marker elsewhere. These conventions apply also
to deadline/default branches. The address and program are read-only.
A simulation-mode cell advances age modulo \(T_{\rm cal}\); an idle-mode cell
advances its age in the same way and blanks its irrelevant work.

\begin{table}[!htp]
\caption{Complete clock state at age zero. Distributed fields have
one indicated bit or symbol per cell of a \(P_{\rm cal}\)-block.}\label{f4:fields}
\small\centering
\begin{tabular}{@{}p{.29\textwidth}>{\raggedright\arraybackslash}p{.65\textwidth}@{}}
\toprule Field & Canonical value or retained information\\\midrule
Address & Its fixed residue modulo \(P_{\rm cal}\); zero in every top idle cell.\\\addlinespace[3pt]
Age & Zero modulo \(T_{\rm cal}\).\\\addlinespace[3pt]
Mode & Prescribed simulation or top idle mode.\\\addlinespace[3pt]
Old information & The encoded virtual state string; blank padding.\\\addlinespace[3pt]
Next information & Blank, after its contents have been committed.\\\addlinespace[3pt]
Left, centre, right copy tapes & All blank.\\\addlinespace[3pt]
Program & One bit of the fixed description of the total local algorithm;
blank padding; blank in every top idle cell.\\\addlinespace[3pt]
Turing-machine tape & Blank auxiliary tape; read-only input/program roles retained
in their own fields.\\\addlinespace[3pt]
Machine head and state & Start state at address zero, no head elsewhere.\\\addlinespace[3pt]
Broadcast gate field & Idle/default port labels.\\\addlinespace[3pt]
Phase and handshake scratch & Blank; phase is determined by age.\\\addlinespace[3pt]
Current native request & The scheduled age-zero instruction, here idle;
computed from age, address and broadcast fields.\\\addlinespace[3pt]
Fold boundary and parity & Prescribed left/right/interior/outside flags
and virtual parity, retained by the logical rule.\\\addlinespace[3pt]
Private old-word and selector copies & Blank at module boundaries;
reset before each use and after disposal.\\\bottomrule
\end{tabular}
\end{table}

The boundary/parity fields are fixed labels of the fold, and the
private copies are compiler work. Neither belongs to the simulated
virtual word; because both are constant or blank at every round boundary,
the identity below extends to the full physical clock. Inside a clean period, copy the three old strings, interpret one local
transition, broadcast its old instruction, execute its encoded table,
commit, and clean. Communication is \(O(P_{\rm cal})\), computation is polynomial
in \(\log P_{\rm cal}\), and the fixed routed table takes at most \(C_QP_{\rm cal}^2\)
steps. Canonicalization has the same local tape traversal bounds.
Thus a sufficiently large fixed power of two makes all phases fit in
\(T_{\rm cal}=P_{\rm cal}^4\). Define the revised total algorithm before writing its finite
description into the program field, as in Section~\ref{d:calendar-section}.

Let \(E_S(y)\) encode a cyclic word \(y\) in age-zero simulation
blocks with precisely these canonical fields, and let \(\mathcal C_S\)
be the clean logical transducer. Literal interpretation and the resets
give an identity on the \emph{entire} word,
\[
 \mathcal C_S^{T_{\rm cal}} E_S(y)=E_S(\mathcal C_S y).
\]
At the cyclic seam the same local neighbour communication applies as at
any block seam. Induction gives
\(\mathcal C_S^{T_{\rm cal}^k}E_S^k(y)=E_S^k(\mathcal C_Sy)\).
The uniform idle top word has clock period \(T_{\rm cal}\), spatial period one,
and identity instruction at every time. Its encoded word therefore
returns after \(T_{\rm cal}^{k+1}\), with initial spatial period dividing \(P_{\rm cal}^k\).
The exact port identity of Section~\ref{d:calendar-section} gives the
repeated encoded quantum identity in aligned periods \(T_{\rm cal}^k\).
The deterministic local evolution between returns defines all times.
\end{proof}

\begin{lemma}[Raw finite-torus cutoff]\label{f4:raw}
Consider \((\Z/L\Z)^d\) and a fixed collection of whole-symbol quorum covers using the
three \(16\)-by-\(16\) replica boxes of \eqref{eq:c-boxes}, with a
fixed finite logical stencil in the other coordinates. Before the
initial source slab the error reference is empty. Suppose one fresh
wrong site-round name has at most \(K_0\) raw causes, and the physical
site-time flags are independent, each of probability at most \(p\), and
one flag causes at most \(c_0\) such names.
Let \(R_{\rm phys}\) bound spatial infinity-steps of the complete
contour and source-star catalogue. Put
\(N_0=\lfloor(L-1)/R_{\rm phys}\rfloor\ge36\).
For a horizon \(T_{\rm hor}\), the event that some wrong use at a lifted
site-time up to \(T_{\rm hor}\) has a contour with at least \(N_0\) edges
has probability at most
\[
 C_{\rm raw}(T_{\rm hor}+C_{\rm raw})L^d
 K_{\rm raw}^{N_0+1}p^{N_0/(72c_0)},
\]
where \(C_{\rm raw}\) is a fixed root-count constant; for \(p\) below a
fixed bound this is at most
\(C_{\rm bd}(T_{\rm hor}+C_{\rm bd})L^de^{-\beta_{\rm raw}L}\).
Outside it every wrong use has a connected certificate with \(g\)
distinct physical raw names, at most \(C_{\rm w}g\) edges and at most
\(K_{\rm w}^g\) anchored descriptions. All probabilities are under
the original law.
\end{lemma}
\par\noindent\emph{Proof idea.} Apply the contour construction in a lift of the finite torus. Short graph pieces have distinct physical marks, which yields the raw cutoff after a union over their starts.

\begin{proof}
Lift the flags, the rule and its reference periodically from
\((\Z/L\Z)^d\) to \(\Z^d\). A replica at transverse displacement
\(s_{\rm rep}\) reads its logical inputs at \(u+s_{\rm rep}\), one round earlier;
the reference is cloned along the replica coordinates. Fix a realized
flag field and let \(y\) indicate that a lifted site-time carries its
reference whole symbol.

Define the typed dependence graph \((\Lambda,\mathcal H)\) of
\cite[Definition~20]{SST2026} on the lifted site-times from the round
before the initial source slab through \(T_{\rm hor}+O(1)\). Let \(\Lambda_1\)
be that first round, where \(y=1\). The three edge types are the three
pairs of replica boxes. At a site-time of \(\Lambda_\bullet\), type
\(s\) points to every logical input of every replica in pair \(s\).
Put in \(\Lambda_0\) the fresh wrong site-times, where \(y=0\) and
all the inputs in at least one pair have \(y=1\). All remaining
site-times are in \(\Lambda_\bullet\). Two correct boxes give \(512\)
correct votes, so every fresh wrong name has a raw cause in its update.

The associated monotone automaton \(\varphi\) is the OR over pairs of
the AND of their inputs \mbox{\cite[(2.14)]{SST2026}}. By the definition of \(\Lambda_0\),
\(y\ge\varphi(y)\) off \(\Lambda_0\), and on \(\Lambda_0\) the
automaton is fixed at zero. Finite time depth gives a unique trajectory
\(x\), hence the maximal trajectory of \cite[Lemma~13]{SST2026}. Induction from \(\Lambda_1\) gives \(x\le y\). At any wrong
lifted site-time \(i\) we therefore have \(x(i)=0\).
By \cite[Theorem~22]{SST2026} a Toom contour rooted at \(i\) is
present in this graph. Its sinks lie in \(\Lambda_0\). An edge
leaving the root or an internal vertex of charge \(s\) has type \(s\);
an edge leaving another source may have any allowed type
\cite[Definition~21]{SST2026}. The contour embeds in the finite explanation graph of \cite[Proposition~31 and Theorem~33]{SST2026}. The embedding is injective on sinks and on internal vertices of each charge by \cite[Definition~17]{SST2026}. Contour sources and sinks occur in equal numbers, so the contour is finite.
This construction is deterministic for each lifted field and does not
require independence between periodic copies.

Apply the zero-sum identity \cite[Lemma~41]{SST2026} in this spatial lift.
In the transverse coordinates \(u=(u_1,u_2)\) put
\(\chi=u+58t(1,1)\), with potentials
\(3(\chi_1+\chi_2),-3\chi_1,-3\chi_2\).
For the first potential use the two shifted boxes, and for the other two
use the unshifted box and the box shifted in the other direction.
Their gains are at least \(132,129,129\); the worst arbitrary backward
gain is \(-351\). Logical offsets do not enter these potentials.
Every dependence edge goes one round backward in time. Time telescoping
\cite[(5.13) in the proof of Lemma~38]{SST2026} gives the same edge
count \(j\) for each charge. The root is a contour source
\cite[Definitions~15--16]{SST2026}. Contour sources and sinks are equal in
number, so there are \(n_{\rm sink}-1\) nonroot sources for \(n_{\rm sink}\) sinks, each
emitting one edge of each charge \cite[(5.23) in the proof of Lemma~40]{SST2026}. Pole cancellation yields
\[
 0\ge3[129(j-n_{\rm sink}+1)-351(n_{\rm sink}-1)],\qquad
 E=3j\le(480/43)(n_{\rm sink}-1)<12(n_{\rm sink}-1)
\]
for \(n_{\rm sink}>1\); a single fresh root has zero edges.

For \(E\ge N_0\), root an abstract spanning tree at the contour root,
a source of weight zero. Give all \(n_{\rm sink}\) sinks weight one; their lifted
space-time positions are distinct by \mbox{\cite[Definition~17(i)]{SST2026}}. With at most
\(E<12(n_{\rm sink}-1)\) edges, its redundancy is below twelve. The separator
\mbox{\cite[Definition~2.4, Lemma~2.5 and Theorem~2 (Tree Separator)]{Gacs2021}} with
\(k_0=\lfloor N_0/12\rfloor<n_{\rm sink}-1\) gives a subtree with at most
\(12k_0\le N_0\) edges and more than \(k_0/3\ge N_0/72\) marked
original sinks. The separator's root convention assigns any removed
root mark to the other piece, so the selected marks remain original
sinks. Here \(N_0\ge36\) gives \(k_0\ge3\), so every refinement
inside Theorem~2 occurs at weight greater than three, as required by
its Lemma~2.5. Its spatial coordinate diameter is at most
\(R_{\rm phys}N_0\le L-1\). Two lifts of the same physical site-time
name would differ by an \(L\)-deck displacement in space; hence the
selected sink names are distinct physically. They require more than
\(N_0/(72c_0)\) distinct physical raw flags.

Truncate above \(T_{\rm hor}\) before extraction. The empty past forces every
vertex into the time slab from the initial round through \(T_{\rm hor}+O(1)\). There are at most
\(C_{\rm raw}(T_{\rm hor}+C_{\rm raw})L^d\) possible projected roots. Ordered
tree shapes, bounded edge increments, binary sink markings and at most
\(K_0\) causes per mark have at most \(K_{\rm raw}^{N_0+1}\)
descriptions. Apply the raw law only to admissible descriptions with
distinct physical images; all other codes merely overcount cardinality.
This proves the exceptional bound without assigning independent flags
to periodic copies.

Off that event a full explanation has fewer than \(N_0\) edges and
injective sinks by the same diameter bound. If \(g\) physical causes
occur, \(n_{\rm sink}\le c_0g\), so the contour and its fixed attachment
stars have \(O(g)\) edges. Bounded degree gives exponential anchored
entropy by traversing a spanning tree twice, recording finite edge
and role choices. All occurrences of a cause attach to its one raw
source label. Enlarging constants covers old/new time offsets and isolated roots.
For sufficiently small fixed \(p\), the exceptional bound is at most
\(C_{\rm bd}(T_{\rm hor}+C_{\rm bd})L^d e^{-\beta_{\rm raw}L}\).
\end{proof}

The next lemma is the one-dimensional two-sheet case of the fold
\eqref{q:fold-map}; Proposition~\ref{c:oriented-source} pays shared
physical faults across sheets in the same way.

\begin{lemma}[Two-role image]\label{f4:roles}
Fold a cycle of length \(m=2\ell\) into \([0,\ell)\) on a physical
circle of length \(L\ge4\ell\), using positions \(a_0(\sigma)=\sigma\)
and \(a_1(\sigma)=m-1-\sigma\) on that interval. Each extends to a
one-Lipschitz map into the virtual cycle. If distinct physical raw source labels have joint law \(p^{|F|}\),
duplicating each into two formal roles gives intersection weight at most
\((\sqrt p)^g\) for \(g\) distinct formal names.
\end{lemma}
\begin{proof}
For \(\ell\ge2\) define \(f(\sigma)=\sigma\) on \([0,\ell-1]\),
\(f(\sigma)=2\ell-2-\sigma\) on \([\ell,2\ell-2]\), and zero
thereafter. Set \(a_0=f\), \(a_1=m-1-f\) modulo \(m\).
Every physical nearest-neighbour edge, including the torus seam,
changes \(f\) by at most one; the reflected role has the same property.
For \(\ell=1\) use constant \(f=0\). The map
\((\sigma,t)\mapsto(a_r(\sigma),t)\) consequently has bounded
increments on every fixed-range raw graph. At most two formal names
refer to one physical flag, so \(g\) names require at least
\(\lceil g/2\rceil\) independent flags. This proves the law.
All uses of one physical name in one role retain that name.
\end{proof}

\section{Exclusive folded dispatch}\label{app:f4-dispatch}
By Lemma~\ref{f4:roles} and the clean action below (Lemma~\ref{f4:dispatch-clean}), the folded chain has two
qubits \(Q_{\sigma,0},Q_{\sigma,1}\) per physical site; here \(\sigma\) is the physical position of Lemma~\ref{f4:roles} and the second index is the role. The semantic qubits sit on virtual cells. Code coordinates, syndrome and Boolean quantum work use
additional chain cells. A requested native layer is a matching, padded
with one-qubit idles, of the real inventory \eqref{d:native}; its at
most three distinct operands are within two virtual edges of its first
operand. The line routing of Lemma~\ref{d:gadgets} supplies this form.

\paragraph{A fixed private-port circuit.}
This is the private-port circuit of Section~\ref{d:dispatch}, with
templates for the two fold roles and exclusive port selection.
For native type \(\tau\) of arity \(j\), enumerate templates
\[
 \theta=(\tau;(r_1,\delta_1),\ldots,(r_j,\delta_j)),\qquad
 r_i\in\{0,1\},\quad\delta_1=0,\quad\delta_i\in\{-2,-1,0,1,2\},
\]
with distinct role-offset pairs. Control/target order is part of the
type. An ordered tuple ID \((\theta,b)\) has operand \(i\) at
\(b+\delta_i\). Every template-position has a private port at every
site; at \(x\) it belongs only to tuple base \(x-\delta_i\).
There are at most
\[
 5\cdot2+3\cdot(4\cdot5)+8\cdot25=270
 \quad\hbox{templates},\qquad 10+120+600=730
 \quad\hbox{port roles}.
\]
The five unary types include reset, the three binary types are CNOT,
CZ and SWAP, and the ternary type is Toffoli. Repeated operands are
excluded. Tuple port sets are tensor-disjoint even on small tori,
because their internal labels differ. Their physical diameter is at
most four.

At each semantic vertex \(v=(x,r)\), a total Boolean function of
old dephased clock requests and private neighbour copies chooses
NONE or exactly one eligible port at \(x\) with assigned role \(r\).
Resolve malformed metadata by a fixed priority. The two roles have
disjoint eligible lists. Reset ports, compute and freeze these one-hot
choices, scatter each semantic qubit to its selected port in fixed
serial order, apply every native tuple unconditionally, gather in
reverse order, and reset/discard ports and choice work at their declared
ends. Ineligible swaps are identity. One such reset, choice, scatter, native,
gather and reset sequence is a module.

The elementary implementation uses exactly the seven gates:
Fredkin is chronological \(\mathrm{CNOT}(b\to a)\),
\(\mathrm{Toffoli}(c,a\to b)\), \(\mathrm{CNOT}(b\to a)\); SWAP is three CNOTs, and CZ is
\(H_{\rm target}\), CNOT, \(H_{\rm target}\). Boolean work uses private
reset, compute, copy and uncompute. Metadata is always a control of
quantum operations and never a quantum-derived target. Serialization
and private internal labels give a valid physical circuit on every
classical word. Each elementary layer, including all idle and
reset-following factors, receives the prescribed noise.

\begin{lemma}[Clean folded action]\label{f4:dispatch-clean}
In a module with no physical fault and with the intended exclusive
choices, the circuit implements the intended matching exactly on
arbitrary quantum data and a reference. A virtual cycle of length
\(m=2\ell\), \(\ell\ge3\), folds at fixed physical range into
\([0,\ell)\) of a physical circle of length \(L\ge4\ell\).
\end{lemma}
\begin{proof}
In a clean matching, a tuple receives either all its intended semantic
operands or none. A wholly unused tuple acts only on disposable ports.
Scatter and gather therefore implement each intended channel exactly,
including reset: the reset acts on the port holding the intended input,
and its reset output is gathered before disposal. Dephasing, when
required on work, is handled in the same way, without inverting it.
The coordinate is \(a=\sigma\) on role zero and
\(a=m-1-\sigma\) on role one. Successor means physical right motion
on role zero, an on-site switch at the right endpoint, left motion on
role one, and an on-site switch at the left endpoint. Radius-two virtual
operands fit the catalogue. Fixed left/right/interior/outside flags,
including both endpoint flags when needed, determine the one- and
two-step neighbours locally. Virtual parity is \(\sigma\bmod2\) in
role zero and \(1-\sigma\bmod2\) in role one. Outside factors have
the declared reset/idle reference. Degenerate intervals use a total
reset fallback. Nothing in the rule reads the interval length.
\end{proof}

\begin{lemma}[Closed support and joint localization]\label{f4:dispatch}
Fix an intended matching. Let \(B_{\rm ch}\) be the clean-station
vertices whose actual exclusive tuple differs from their intended
tuple, \(D_{\rm st}\) the dirty physical stations, and \(E_{\rm tu}\)
the dirty native tuple boxes. A station or native tuple box is dirty if
a physical fault (a noise flag or a primitive decomposition error) acts
on it during the module, and clean otherwise; the seeds are the vertices
of \(B_{\rm ch}\), the stations of \(D_{\rm st}\) and the boxes of
\(E_{\rm tu}\). A station fault may act jointly on its
two semantic qubits, all its ports and routing controls. Dirty boxes
may share operator or environment coefficients. There is a closed
union of intended locations, of cardinality at most
\[
 2|B_{\rm ch}|+732|D_{\rm st}|+|E_{\rm tu}|,
\]
on which the error is an arbitrary joint operation; its complement
executes the intended channel. Each marked location is within physical
distance eight of a seed, and its virtual support is in two fixed
role neighbourhoods of that seed. At fixed-history operator scope
all exterior coefficients are retained.
\end{lemma}
\par\noindent\emph{Proof idea.} Compare the intended and used port pointers at each station. Mark both IDs for a wrong choice and retain every port at a dirty station, so the support includes all affected quantum work.

\begin{proof}
Give each vertex its intended pointer \(i(v)\) and actual pointer
\(a(v)\), interpreting NONE as a private idle ID. Mark both IDs for
an erroneous clean vertex; at a dirty station mark every incident port
tuple and its two intended/private-idle IDs; mark every dirty tuple.
Call the set of marked IDs \(\mathcal T\). Let \(\mathcal R\) contain
all erroneous or dirty vertices and every vertex whose intended or
actual pointer is in \(\mathcal T\).

At a correct clean vertex the two pointers coincide. Hence an intended
tuple meeting \(\mathcal R\) has its ID in \(\mathcal T\), so all its
intended operands lie in \(\mathcal R\). This proves intended closure.
An actual tuple receiving retained input from \(\mathcal R\), or
returning an output to it, also has marked ID: at a dirty station all
incident IDs are marked; at an erroneous clean vertex its actual ID is
marked; at a correct clean vertex in \(\mathcal R\) the identical
pointers are marked. Every other retained participant has that same
actual ID and is in \(\mathcal R\). Unselected clean ports are fresh
disposable factors and never gather into an unmarked vertex. A dirty
station touches only \(\mathcal R\) and ports of marked IDs. This is
actual closure.

Every unmarked intended tuple has correct clean pointers, clean ports
and a clean native box. It receives exactly its intended operands,
acts correctly, and returns them. No other selected tuple touches
those operands. Thus the complementary channel is exactly the intended
one even on inputs entangled with \(\mathcal R\) and a reference.
All remaining operations are on the marked union and its work/bath.
Private clean preparation, reset, dephasing and disposal environments
are fresh and never retrieved. Tracing them leaves an arbitrary joint
channel on that union when the fixed history is exogenous. More
generally expand its Stinespring operators coefficientwise, retaining
all exterior systems: tensor matrix bases on disjoint marked operand
sets give precisely the full operator-span interface of
Lemma~\ref{d:reduction}. Individual factors need not be CPTP.
A reset at a marked location is replaced, never inverted.

There are at most \(2|B_{\rm ch}|+(730+2)|D_{\rm st}|+|E_{\rm tu}|\)
marked IDs. Each marked intended location has one of these IDs.
An incident tuple anchor is within two of its port, and its other
operands within four of that station. Enlarging for the adjacent
intended location gives radius eight. In each fixed fold role physical
displacement is bounded, while the two roles can be distant virtually.
This proves the two-neighbourhood assertion.
\end{proof}

A whole-site replacement dirties one station, including all local
ports at once. A primitive decomposition error has bounded support
in this fixed catalogue. Time-boundary uses are covered by finitely
many adjacent modules. Iterating Lemma~\ref{f4:dispatch} keeps joint
coefficients on the union of the named locations, exactly as required
by Lemma~\ref{d:reduction}. Repeated metadata uses keep their original
source names.

\begin{lemma}[Unfinished modules]\label{f4:dispatch-partial}
At an elementary time inside a module, retaining in-flight
factors and appending its intended suffix gives the same localization
with a station envelope of at most 732 locations per vertex of \(B_{\rm ch}\).
\end{lemma}
\begin{proof}
Retain semantic qubits, ports, masks holding quantum data and all
undeleted quantum work. Append the fixed intended scatter/native/gather suffix, with intended choices; do not consult received metadata.
A station whose actual frozen choice differs is conservatively dirty,
so all its incident IDs are marked. At the other stations actual prefix
and intended suffix agree with the reference. The preceding pointer
closure proof applies to their composition. Future reset/disposal is
performed only at its declared instruction, so no in-flight input is
lost. Replacing \(2|B_{\rm ch}|\) by \(732|B_{\rm ch}|\) covers the
whole unfinished station. Use this larger envelope for all choice
seeds when choosing incidence constants, so the estimate is uniform
in the requested readout offset.
\end{proof}

\section{Central-strip sources and the hierarchy}\label{app:f4-sources}

This appendix proves the source estimates used in Section~\ref{f4:actual-sources}. It gives each short witness a Lipschitz image, bounds long cap witnesses, combines the cap and raw families, and supplies the recursion that accounts for aliases under the periodic lift.

Time \(t\) counts rounds. Fix a physical quantum
line at \((x,y,z)=(x_0,y_0,z_0)\), running along \(\sigma\), with
fold maps \(a_r\) from Lemma~\ref{f4:roles}. A cap bin has height
\(W\), bounded uniformly in the chosen window by a computable \(w\).
Its source label is
\[
 \lambda=(\omega=x-t,\ b=y,\ \chi=\sigma+58t,\ \zeta=z+58t,
                         \ j=\lfloor t/W\rfloor),
\]
with the first four coordinates modulo \(L\) and \(j\) retained as
an integer. A source label event \(X_\lambda\) is the union of raw, intrinsic
and overlay events, with the distinct-physical-name intersection law
\(\gamma_{\rm src}^{|F|}\). Suppose each wrong use has a connected lifted
witness with \(E\le c_Cg\) edges and \(g\) distinct lifted marked
source labels, bounded degree and fixed increments in all five coordinates.
Every retained bin is between zero and
\(\lfloor(T_{\rm hor}+C_{\rm stage})/W\rfloor+C_{\rm bin}\).
The circuit satisfies these hypotheses (Section~\ref{f4:actual-sources}).

\begin{lemma}[Central-strip image]\label{f4:strip}
Let \(R\ge\max(R_\omega,R_j)\) be a fixed bound on the \(\omega\)- and
bin-increments of the witness catalogue (Section~\ref{f4:order-a}, row 10), and put
\(q_{\rm cut}=\lfloor L/[32R(w+1)]\rfloor\), decreasing it by one
fixed factor if necessary to include root-use spurs. At all sufficiently
large \(L/(w+1)\), one source image agrees with every quantum root
and has fixed Lipschitz increments on every rooted witness with fewer
than \(q_{\rm cut}\) edges.
\end{lemma}
\par\noindent\emph{Proof idea.} Choose the physical time representative closest to the centre of a bin. The central strip avoids its wrap change and gives a bounded image step between adjacent source labels of a short witness.

\begin{proof}
Choose the integer \(\tau(\lambda)\equiv x_0-\omega\pmod L\) nearest to
\((j+\tfrac12)W\), with a fixed tie rule, and set
\[
 \pi_r(\lambda)=
   (a_r(\chi-58\tau(\lambda)),\tau(\lambda)).
\]
At an exact quantum root \((\sigma,t)\), the time \(t\) has that
congruence and belongs to its bin. Since \(W<L\), it is the unique
nearest representative. The cut in round \(t\) uses the first coordinate computed in that
round. If that output's source label is indexed by the next boundary, join it
to the quantum-use root by its fixed one-round spur before counting
edges. This is the timing convention throughout the proof.

Lift a spanning tree from a root and put
\(t_v=t_{\rm root}-(\omega_v-\omega_{\rm root})\). With increment bounds
\(R_\omega,R_j\le R\), a path of \(i<q_{\rm cut}\) edges obeys
\[
 |t_v-(j_v+\tfrac12)W|
 \le W/2+i(R_\omega+WR_j)
 \le w/2+q_{\rm cut}R(w+1)<L/8.
\]
A bounded root spur is absorbed by the fixed reduction of the cutoff
and a lower-size threshold. Thus this lift gives the unique midpoint
representative at every vertex. Two paths cannot have nonzero
\(L\)-monodromy in \(\omega\), since their total displacement is below
\(2Rq_{\rm cut}<L\). The same reasoning identifies repeated lifted
occurrences projecting to one physical source label. On each selected edge, let \(R_x\) bound increments in the chain-coordinate component \(\chi\). Then
\(|\Delta\tau|\le R_\omega\), and the physical \(\sigma\)-image changes
by at most \(R_x+58R_\omega\). Lemma~\ref{f4:roles} preserves this bound.
The globally defined midpoint map is therefore consistent on all
selected short witnesses, with the same value for each physical name.
Only those edges are used; edges crossing an arbitrary representative
branch cut elsewhere in the quotient graph need not be Lipschitz.
\end{proof}

\begin{lemma}[Long cap witnesses]\label{f4:cap-exception}
For sufficiently small fixed \(\gamma_{\rm src}\), the event that any needed
cap explanation through \(T_{\rm hor}\) has at least \(q_{\rm cut}\) edges
has probability at most
\[
 C_{\rm cap}(T_{\rm hor}+C_{\rm cap})L^4
           \exp[-\beta_{\rm cap} L/(w+1)]
\]
for every integer \(T_{\rm hor}\ge0\) and all \(L/(w+1)\) above the
fixed threshold of Lemma~\ref{f4:strip}, enlarged so that
\(q_{\rm cut}\ge1\). Here \(C_{\rm root}\) is the root-count constant
in the proof below, and \(C_{\rm cap}\ge C_{\rm root}\) absorbs the
fixed cutoff reductions and floors.
Outside this event every needed cap use has a complete short certificate
with distinct physical names and the image of Lemma~\ref{f4:strip}.
\end{lemma}
\par\noindent\emph{Proof idea.} Divide a doubled traversal of a long witness into pieces below the torus alias scale. At least one piece contains a positive density of distinct physical marks.

\begin{proof}
Increase \(c_C\ge1\) for isolated roots and fixed stars. A connected
\(E\)-edge witness has a spanning-tree depth-first walk of length at
most \(2E\). Assign each distinct lifted mark to its first visit and
partition the walk into pieces of at most \(q_{\rm cut}\) edges,
retaining shared boundary vertices. For \(E\ge q_{\rm cut}\) the
number of pieces is at most \(2E/q_{\rm cut}+1\le3c_Cg/q_{\rm cut}\).
Some piece therefore contains at least \(q_{\rm cut}/(3c_C)\)
assigned distinct lifted marks. For a sufficiently reduced fixed cutoff
its characteristic diameters are below \(L/2\). Distinct lifts of
one physical source label have the same integer bin and differ by an \(L\)-deck
vector in a spatial characteristic. They cannot both occur in this
piece. Its marks are distinct physical names before their law is applied.

The causal bin bound supplies at most \(C_{\rm root}(T_{\rm hor}+C_{\rm root})L^4\) starting vertices.
Finite degree, roles and mark subsets give at most
\(K_{\rm cap}^{q_{\rm cut}+1}\) descriptions per start. For each
admissible description apply the source law to its distinct
marks; other codes only overcount. Hence the exceptional probability
is at most
\[
 C_{\rm root}(T_{\rm hor}+C_{\rm root})L^4 K_{\rm cap}^{q_{\rm cut}+1}
             \gamma_{\rm src}^{\lfloor q_{\rm cut}/(3c_C)\rfloor}.
\]
Choose the source bound after \(c_C,K_{\rm cap},R\) to make its
exponential base less than one. Floors and small cutoffs are absorbed
in a fixed threshold and prefactor. This proves the bound. Every
remaining complete explanation is short, and the same diameter
argument makes all its marked names injective. Select a minimal one
if several are available. The estimate is unconditional.
\end{proof}

\begin{lemma}[Mixed roles and gluing]\label{f4:mixed}
Let raw hardware/clock names have joint law \(p^{|F|}\), with
one physical source label for every use of the same site and round. Duplicate
cap and raw names into two formal fold roles, retaining their underlying
events. For \(g_C,g_H\) distinct formal names of the two families,
\[
 \Pr(\textup{all requested events})
 \le\gamma_{\rm src}^{g_C/4}p^{g_H/4}
 \le\gamma_{\rm eff}^{g_C+g_H},\qquad
 \gamma_{\rm eff}=\max(\gamma_{\rm src}^{1/4},p^{1/4}).
\]
Gluing complete cap and raw witnesses at their quantum-use
roots and adding the dispatcher stars gives bounded degree, fixed
edge/source size and one Lipschitz image on all admitted short
certificates.
\end{lemma}
\begin{proof}
Each family names at least half as many physical source labels as formal names.
Apply its joint law, then Cauchy--Schwarz between the two possibly
dependent families. This gives the fourth powers displayed above.
Raw clock errors and direct hardware uses are annotations of one
raw source label; raw/statistical/overlay cap types and old/new occurrences
remain annotations of one cap source label.

For a fixed cap coordinate and bin at the quantum line, the first label component \(\omega\), equal to \(x_0-t\),
determines time modulo \(L\), and \(W<L\) permits at most one such
time in the bin. The remaining coordinates fix the quantum root.
The gluing degree is consequently bounded; fixed root-use spurs only
enlarge this bound finitely. Images agree at the common root by the
timing convention of Lemma~\ref{f4:strip}. Raw images use literal
\((\sigma,t)\) followed by \(a_r\). Dispatcher attachment stars have
fixed increments in each formal role, including the unfinished-module
station envelope. Existing source names are retained throughout.
The claimed size, entropy and image properties follow by enlarging
these fixed catalogue constants once. Shared physical flags across
two roles have already been paid by the fourth-root weight.
\end{proof}

\begin{lemma}[Alias-safe source recursion]\label{f4:recursion}
Suppose the scheduled circuit has complete regular boxes of length
\(P_{\rm cal}^\ell\), time height \(T_{\rm cal}^\ell\), fixed child-index halos, at most
\(A_{\rm rec}\) candidate pairs, and two full bad child supports
separated by \(\rho_{\rm gap}P_{\rm cal}^\ell\) whenever a parent is bad.
Root counts are polynomial in \(P_{\rm cal}^\ell\). Choose \(s_{\rm top}\)
larger than the complete halo and \(s_{\rm top}P_{\rm cal}^k\le L/2\).
Let \(z_\ell\) bound the per-box sum of weights of the fully embedded,
recursively pruned certificates at level \(\ell\), as constructed in
the proof below. Let \(B_{\rm tail},d_0,a_{\rm tail}\) be the constants
of the long-witness term in that proof. At a source bound small enough
that \(z_0\le1/(4\bar A)\) and
\(a_{\rm tail}\ge\max(0,\log(4\bar AB_{\rm tail}))
+d_0\log P_{\rm cal}/P_{\rm cal}+2\log2\),
for every level-\(\ell\) box \(b\), \(0\le\ell\le k\), the event
\(N^{\rm sh}_\ell(b)\) that \(b\) is bad in the geometric predicate built
from short-certified base roots satisfies
\[
 \Pr(N^{\rm sh}_\ell(b))\le D\,2^{-2^\ell},\qquad
 D=1/(2\bar A),\quad \bar A=\max(1,A_{\rm rec}).
\]
The exceptional events of Lemmas~\ref{f4:raw} and
\ref{f4:cap-exception} are paid separately.
\end{lemma}
\par\noindent\emph{Proof idea.} After excluding long certificates, use the short complete certificates in the block recursion. Separated children have disjoint formal source-name sets because each name has one image. The joint law therefore bounds the joint probability by the product of their weights.

\begin{proof}
A base root is short-certified if it has one of the admitted complete
certificates. Define the geometric bad-box predicate from these roots
under the original law. Outside the two exceptions this includes all
actual bad base roots. Work with sums of source weights of fully
embedded, recursively pruned certificates, as in Lemma~\ref{d:recursion}.
At a parent of level-\(\ell\) children, if a primitive witness has full
graph length
at least \(f_{\rm sh}P_{\rm cal}^\ell\), discard its entire repeated
subtree and pay that witness once. Bounded anchored entropy and the
joint source law bound this contribution by
\(B_{\rm tail}P_{\rm cal}^{d_0\ell}e^{-a_{\rm tail}P_{\rm cal}^\ell}\).
Choose \(f_{\rm sh}\) so a shorter witness has image reach
below one third of the separated-support gap. The selected two child
certificates then have disjoint formal source-name sets, because every
name has one image. Their joint weight is the product of their
weights. Within a child apply the same pruning recursively. Full halos
are local cycle arcs by the choice of \(s_{\rm top}\); time stays
unquotiented. The resulting sum satisfies
\[
 z_{\ell+1}\le A_{\rm rec}z_\ell^2+
       B_{\rm tail}P_{\rm cal}^{d_0\ell}e^{-a_{\rm tail}P_{\rm cal}^\ell}.
\]
This multiplies weights of disjoint names, not marginal bad-box
probabilities. A witness longer than the scale threshold is still
counted at its full length. If that threshold exceeds the cap cutoff,
there are simply no admitted cap witnesses in that case.

Choose the fixed cap so that \(z_0\le1/(4\bar A)\) and
\[
 a_{\rm tail}\ge\max(0,\log(4\bar A B_{\rm tail}))
                 +d_0\log P_{\rm cal}/P_{\rm cal}+2\log2.
\]
Exactly the comparison in Lemma~\ref{d:recursion} makes the remainder
at most \((D/2)2^{-2^{\ell+1}}\); the squared term has the same bound
since \(A_{\rm rec}D^2\le D/2\). Induction proves the claimed form,
with exponent \(\log2\), uniformly in the window. None of these laws
is conditioned on absence of the exceptional events.
\end{proof}

\section{A reserved exponential window}\label{sec:coexistence}

This appendix uses the closed-code processor estimate of Appendix~\ref{app:code-estimates} (Lemma~\ref{x:exponential}), also used by Theorem~\ref{thm:exponential}, and shows that a decidable exponential-storage window can coexist in four dimensions with linear-storage windows whose storing-label set is undecidable.
Appendices~\ref{sec:fold-controls} and~\ref{sec:folded} supply the part governed by the halting computation.
The proof of Proposition~\ref{p:coexistence} uses the same internal-factor layer convention as
Section~\ref{s:model}.

\begin{proposition}[Coexistence of storage scales]\label{p:coexistence}
There is one computably specified translation-invariant, finite-range,
time-periodic circuit on \(\Z^4\), with a fixed site space of qubits
and elementary inventory
\(\{X,H,\mathrm{CNOT},\mathrm{Toffoli},R_0,D_Z,I\}\), under independent
single-qubit depolarization after every layer, with computable disjoint
rational windows \(I_0,I_1,\ldots\), such that:
\begin{enumerate}
\item On \(I_0\), one total computable rate-independent encoder and
intended-code decoders achieve entanglement fidelity at least \(15/16\)
for every integer \(0\le t\le2^{\lfloor c_{\exp}L\rfloor}\), for all
\(L\) above a computable threshold and one fixed rational \(c_{\exp}>0\).
\item On \(I_{n+1}\), if \(U(n)\uparrow\), Proposition~\ref{q:linear}
holds for label \(n+1\) of this circuit with its threshold
\(L_0(n+1)\); if \(U(n)\downarrow\), the all-encoding fidelity bound of
Theorem~\ref{f:forgetting} holds for the entire circuit, uniformly in
size and rate within the window.
\item There are computable constants \(A,a>0\), \(b\in\N\), an
integer \(C_0\ge2\), and \(L_{\rm EB}\ge C_0/2\) for this circuit,
all independent of \(j\). For every \(j\ge1\), every \(\eps\in I_j\)
and every integer \(L\ge L_{\rm EB}\), the entire channel has a
decomposition of the form \eqref{eq:u-eb} at round \(201L+C_0\).
The labels \(j\) for which
the optimal lifetime at fidelity \(7/8\) is unbounded in \(L\) at
every rate in \(I_j\) (Corollary~\ref{cor:window-undecidable}),
and the labels \(j\) for which the folded encoder and decoder of
Proposition~\ref{q:linear} have fidelity at least \(7/8\) at every
rational rate in \(I_j\), every integer \(L\ge L_0(j)\) and every
integer \(0\le t\le\lfloor cL\rfloor\), both equal
\(\{0\}\cup\{n+1:U(n)\uparrow\}\), a \(\Pi^0_1\)-complete set.
\end{enumerate}
Time counts rounds (Definition~\ref{def:round}). The first window is
reserved for a syntactically looping input.
Whole-site noise is not covered by this proposition.
\end{proposition}

The finite-torus memory estimate used by the auxiliary part is Lemma~\ref{x:exponential} in Appendix~\ref{app:code-estimates}. Local toric-code memory and sweep decoding have precedents in
Ahn \cite[Section~7.2 and Appendices~A\nobreakdash--B]{Ahn2004}, Kubica--Preskill
\cite{KubicaPreskill2019}, and the recent constructions of
Drechsler--Oreg--Aharonov and Ponarovsky \cite{DOA2026,Ponarovsky2026}.
That lemma is proved for the reflected-sweep circuit and the
fault support actually needed here.

\par\medskip\noindent\emph{Proof idea for Proposition~\ref{p:coexistence}.}
Reserve label zero for an explicit infinite loop. Add a separate toric-code memory whose cut lets the data through only when independent noise probes detect the top window. On the other windows the folded circuit supplies linear storage when appropriate, and the forgetting proof covers the whole register.

\begin{proof}[Proof of Proposition~\ref{p:coexistence}]
Rebuild the folded quantum construction of Appendices~\ref{sec:fold-controls}
and~\ref{sec:folded} with the fixed positive-input wrapper
\(V(1)=\text{an explicit infinite loop}\) and \(V(m+2)=U(m)\).
At rank \(a=j+n_0+1\), the scan calculation in
Section~\ref{sec:controller} calls \(V(j+1)\). Hence label zero
is the loop and label \(n+1\) simulates \(U(n)\) exactly.
Its decreasing nominal windows \(\mathcal J_j\) have rational gap
\[
 \delta_{\rm gap}=\inf \mathcal J_0-\sup \mathcal J_1>0,\qquad
 \vartheta=(\inf \mathcal J_0+\sup \mathcal J_1)/2.
\]

Add a separate closed four-dimensional toric-code quantum track
with the processor of Lemma~\ref{x:exponential}. There are no quantum
gates between its core and the folded core. All spatial quantum
work precedes the final cuts. Every retained auxiliary quantum
factor, including work and data-derived records, receives a local
acceptance-controlled cut. Clean rejection replaces it; clean
acceptance is identity. Disposable work resets before reuse.
The fixed influence catalogue and predecessor count \(\kappa\)
are chosen before detector depth or probe counts.

At each site and round use \(R+1\) disjoint banks of \(M\) reset
qubit probes, exposed to \(K_{\rm samp}\) sampling layers. Their
endpoint ORs \(E,D_1,\ldots,D_R\) are independent Bernoulli variables
with the refresh rate \eqref{eq:q-refresh}, independently across
site-rounds. Bank zero supplies the common semantic refresh.
The other banks define an exogenous detector
\[
 \mathsf D_{\rm test}=\one\{\textstyle\sum_{r=1}^R D_r\ge
                    \lfloor R\vartheta\rfloor+1\}.
\]
Choose \(R\) so that \(e^{-R\delta_{\rm gap}^2/2}\le\delta_D\).
Multiplying the centred Bernoulli moment bound from Hoeffding's lemma \cite[Lemma~1]{Hoeffding1963},
\(\mathbb E e^{\lambda(X-\mathbb EX)}\le e^{\lambda^2/8}\),
over the independent banks and optimizing gives
\[
 \sup_{\mathcal J_0}\Pr(\mathsf D_{\rm test}=0)\le\delta_D,\qquad
 \sup_{j\ge1,\,\mathcal J_j}\Pr(\mathsf D_{\rm test}=1)\le\delta_D.
\]
The integer threshold implements the strict comparison with
\(\vartheta\) exactly.

Let \(A\) contain all nonprobe sampling flags and all nonsampling
flags on every wire. It is independent of the complete endpoint
vector. The actual detector \(\mathsf D_{\rm act}\) equals \(\mathsf D_{\rm test}\) on
\(A=0\); on \(A=1\) its output may be arbitrary. Compute banks
serially with two OR bits and a binary test counter. Let \(S_{\rm np}\)
count all nonprobe semantic, mixed ancilla, counter and work wires;
it depends on \(R\), but not on \(M\). After choosing \(M\),
count the physical nonsampling depth \(J\), including both cuts, every
idle probe and every reset. Then
\[
 \nu\le[S_{\rm np}K_{\rm samp}+((R+1)M+S_{\rm np})J]\eps,
 \qquad
 \frac\nu\mu\le\frac{8S_{\rm np}}M+
       8\left(R+1+\frac{S_{\rm np}}M\right)\frac J{K_{\rm samp}}.
\]
Choose \(M\ge\max(S_{\rm np},32S_{\rm np}/\delta_{\rm ov})\), then
\(K_{\rm samp}\ge64(R+2)J/\delta_{\rm ov}\). The ratio is at most
\(3\delta_{\rm ov}/8\). At \(\eps=1/(2MK_{\rm samp})\),
\(\mu\ge1/16\), so all nominal windows below \(1/100\) have
preimages in the proven regime. The independence of the sampling banks uses independent single-qubit depolarization.

Choose the constants in this order. Fix both quantum catalogues,
\(p_{\rm aux}\) and \(\kappa\); choose the folded selector's
\(C_h,B\) and all its positive, negative and wrapped bounds. Put
\[
 \mu_{\rm cap}\le\min\{\text{folded cap},p_{\rm aux}/16,
                  1/[64(\kappa+1)],1/100\}.
\]
Choose \(n_0\) to place all nominal windows below that cap, fixing
the rational gap. Choose
\(\delta_D\le\min\{p_{\rm aux}/16,1/[64(\kappa+1)],1/64\}\)
and \(\delta_{\rm ov}\le1\) below the folded overlay margin. Choose \(R\),
then build the finite palette and all nonprobe hardware, counting
\(S_{\rm np}\). Next choose \(M\), form and count the actual
\(J\), and choose \(K_{\rm samp}\). Detector work is on site and changes depth without changing
the earlier influence catalogue. Finally enumerate rational inner
physical windows in \(\mu^{-1}(\mathcal J_j)\), exactly as in
Section~\ref{s:parameters}. Choose the coefficient \(c\) of
Proposition~\ref{q:linear} and its finite
chart thresholds after this entire circuit is fixed.

The main refresh swaps every semantic factor with its own mixed ancilla and disposes of old contents. Thus the folded part has the
same \(A,E,H=A\lor E\) source interface after recompilation, with
\(p\le2\mu\). It retains Proposition~\ref{q:linear}, including on
the reserved loop. On \(\mathcal J_0\), the auxiliary exogenous source is
\[
 Z=A\lor E\lor(1-\mathsf D_{\rm test}),\qquad
 \Pr(Z=1)\le\nu+\mu+\delta_D
                  \le2\mu_{\rm cap}+\delta_D< p_{\rm aux}.
\]
It is iid over physical site-rounds and independent of the quantum
input. On \(Z=0\) the auxiliary processor and accepting cut are
correct; all departures have the preselected bounded influence.
Lemma~\ref{x:exponential} proves part (i). Initialize the folded
part in any fixed allowed state and decode only the auxiliary part.

On every tail window use instead
\[
 B_{\rm open}=A\lor \mathsf D_{\rm test},\qquad
 \beta_{\rm open}\le\nu+\delta_D,
 \qquad\kappa\beta_{\rm open}<1/4.
\]
A zero of this iid field guarantees a clean rejecting cut of all
auxiliary quantum factors. A surviving path of \(s\) rounds crosses
\(s\) distinct site-round sources. Its probability, summed over the
at most \(\kappa^s\) paths, is at most \(4^{-s}\). Polynomial
unions give a small initial-site ancestor tail.

For a halting label combine this with the folded part's ancestor
tail, using one physical initial-site mask that retains a site whenever
either quantum part has an ancestor there. The two masks may be
correlated through \(A\). Under one joint history law,
the union mask has a small marginal and radius-\(O(s)\) determining
cones. Apply Lemma~\ref{f:domination} to this union, couple its two
disjoint copies, and sample two joint histories conditional
on those masks, as in Proposition~\ref{f:extension}. Route all
quantum factors at a retained initial site together. This gives an
extension of the entire circuit even for initial entanglement across
its two parts, hardware and reference. Iteration and the cloning bound of Lemma~\ref{f:bell} prove part (ii); finitely many smaller tori use simultaneous full
replacement with probability at least \((\inf I_j)^{NL^4}>0\).

On any tail window, the folded exceptional probability at
\(201L+C_0\) is polynomial times exponential. The auxiliary
\(L\)-round open-path exception, unioned over outputs, is at most
\(C'L^4\cdot4^{-L}\). Their union has the same exponential form.
On its complement all initial quantum dependence is gone and initial
autonomous metadata was measured. The entire good channel is therefore
measure-and-prepare. Its exogenous mixture proves part (iii)'s EB
bound and the sharp linear tail scale. This bound is not asserted
on the reserved window.

Finally choose the folded part's explicit linear-time encoder on
every label and ignore the auxiliary output in its decoder. It succeeds
on the loop and nonhalting tail; the whole-circuit halting bound gives
a strict finite counterexample on every halting tail label. The exact
rational evaluation and continuity argument of
Section~\ref{a:certification} applies to these seven-gate circuits.
Thus the certification and memory labels are precisely the stated
\(\Pi^0_1\)-complete set. Outside the selected windows the circuit's behaviour remains unclassified.
\end{proof}

\section{Classical control for the folded code}\label{sec:fold-controls}
This appendix supplies the part governed by the halting computation of the coexistence circuit
in Proposition~\ref{p:coexistence}. It uses the selector of
Section~\ref{sec:controller} in four orientations, sharing physical noise.

\paragraph{Four orientations and a protected classical profile.}
The physical coordinates are \(r=(r_1,r_2,r_3,r_4)\in\Z^4\).
For sheet \(\varsigma=(\varsigma_2,\varsigma_4)\in\{0,1\}^2\), let
\(s_j=(-1)^{\varsigma_j}\). The reference coordinates are
\[
 q_1=r_1,\qquad q_3=r_3,\qquad q_2=s_2r_2+o_{2,\varsigma},\qquad q_4=s_4r_4+o_{4,\varsigma}.
\]
Set
\begin{equation}\label{eq:c-orientations}
 i_\varsigma=q_4-q_2,\qquad P_{1,\varsigma}=q_1,\qquad
 P_{2,\varsigma}=q_2,\qquad Q_\varsigma=q_3-q_1.
\end{equation}
The constants \(o_{2,\varsigma},o_{4,\varsigma}\), which depend on \(\varsigma\)
and \(L\) (Appendix~\ref{sec:folded}), name encoder references only. The update
stencils use the four fixed signed linear parts. The inverse is integer:
\[
 r_1=P_{1,\varsigma},\quad r_2=s_2(P_{2,\varsigma}-o_{2,\varsigma}),\quad
 r_3=Q_\varsigma+P_{1,\varsigma},\quad
 r_4=s_4(i_\varsigma+P_{2,\varsigma}-o_{4,\varsigma}).
\]
Thus each map is a torus automorphism at every
\(L\). Each selector reads only its own chart and ignores the profile,
quantum data and other selectors. Four different initial phases and
finite hidden lifts are allowed.

The fold profile is a pair of finite labels describing the two folded
axes. Its intended value is constant along \(r_1,r_3\). Protect it by
quorum 512 in the three 256-site boxes
\[
 [0,15]^2,\qquad320e_1+[0,15]^2,\qquad320e_3+[0,15]^2
\]
in those axes. A clean provisional output differing from the intended
label needs at least 257 wrong replicas: either a wrong winner has 512
votes, or at most 511 replicas retain the intended label. Such replicas
occupy two boxes. Profile ports use the same one-way transport convention
as the selectors. The provisional profile programs the quantum unitaries;
a later restored retained profile does not excuse an earlier wrong use.
Profile and program masks never read quantum data or feed a received
selector word into a subsequent selector update.

Every retained quantum factor of sheet \(\varsigma\), including factors
outside the intended folded code, is final-gated by its own selector.
The profile cannot turn that gate off. Persistent profile aliases and
program fields are dephased, and other hardware resets before use and
output. These conditions are needed for arbitrary-input forgetting.

\begin{proposition}[Shared-noise oriented source law]\label{c:oriented-source}
The four selectors share one physical exogenous field \(H\). Their
marginal source laws are those of Proposition~\ref{c:source-law}.
For \(g\) distinct sheet-typed labels
\[
 \lambda_\varsigma=(\varsigma,i_\varsigma-t,
       P_{1,\varsigma}+58t,P_{2,\varsigma}+58t,Q_\varsigma,\lfloor t/W\rfloor),
\]
the joint exception probability is at most
\(\gamma_{\rm fold}^g\), where \(\gamma_{\rm fold}=\gamma_{\rm src}^{1/4}\).
Profile and quantum raw events use their sheet's same label and require
no new statistical event family.
\end{proposition}
\begin{proof}
Each fixed orientation is injective, so the original iid innovation law
and private lane ownership hold within that marginal. Different
orientations can reuse the same physical coins. If \(g_\varsigma\) labels
belong to sheet \(\varsigma\), Hölder applied to the four marginal
intersections gives
\[
 \Pr\Bigl(\bigcap_\varsigma A_\varsigma\Bigr)
 \le\prod_\varsigma\Pr(A_\varsigma)^{1/4}
 \le\gamma_{\rm src}^{\sum_\varsigma g_\varsigma/4}.
\]
A profile or quantum raw exception is already included in the \(H\)
source for that sheet and bin. Geometric roles annotate a label without
splitting it into separately charged sources. At a closing fold use one
deck translation per sheet and axis for the whole certificate and all
its classical ancestors; within that sheet the same affine physical
inverse continues outside the active region. Distinct typed labels at
distant images of one physical coin are handled by the Hölder loss,
not by an independence assumption. Table~\ref{q:links} verifies the common characteristic geometry, including
compressed sampling-register gather/consult links and predictor-entry amortization.
\end{proof}

\section{The folded four-dimensional code}\label{sec:folded}

The quantum processor is a closed four-dimensional toric code folded
into four internal sheets of the physical four-dimensional lattice.
The fold shares physical faults between distant virtual points. We
keep their sheet labels and prove a single recovery corrects the
whole coherent span of the permitted errors (Lemmas~\ref{q:universal}
and~\ref{q:span}). Topological-code
storage and local correction provide the context
\cite{BDL2026,DKLP2002,Harrington2004,KubicaPreskill2019}; the
source-dependent estimate used here is proved below.

\subsection{Closed folded geometry}\label{q:geometry}

For sufficiently large \(L\), set
\[
 m=\lfloor L/64\rfloor\ge2,\qquad a_*=\lfloor L/4\rfloor.
\]
Activate physical sites with
\(r_2,r_4\in[a_*,a_*+m-1]\). There are four internal sheet labels
\(\varsigma=(\varsigma_2,\varsigma_4)\in\{0,1\}^2\), each carrying six
plaquette qubits. Put \(x_j=r_j-a_*\) for \(j=2,4\) and identify
\begin{equation}\label{q:fold-map}
 q_1=r_1,\quad q_3=r_3,\qquad
 q_j=\begin{cases}x_j,&\varsigma_j=0,\\
                   2m-1-x_j,&\varsigma_j=1\end{cases}
 \quad(j=2,4).
\end{equation}
This is a bijection from the active registers onto the cubical torus
\[
 \mathbb T_L\times\mathbb T_{2m}\times
 \mathbb T_L\times\mathbb T_{2m}.
\]
In a folded direction its positive edge moves one physical step
forward on sheet zero, turns on site at \(x=m-1\), moves one physical
step backward on sheet one, and turns on site at \(x=0\). The two
fold turns commute. Thus these are closed virtual cycles, and every
virtual incidence has a fixed unit-step or on-site physical template.
The values \(a_*,m\) occur in the intended initial profile and code
identification. The circuit uses only the four fixed signed linear
stencils and finite profile labels.

Put a qubit on each virtual plaquette. In one CSS sector an error is
a \(2\)-chain \(e\) over \(\mathbb F_2\), its syndrome is the
\(1\)-cycle \(\partial e\), and a \(3\)-boundary is a stabilizer.
The dual cellulation supplies the other sector. Reducing each circle
to one vertex and one edge makes the product differential zero, so
\(\dim H_2(\mathbb T^4;\mathbb F_2)=\binom42=6\).
Coordinate and complementary dual \(2\)-tori give six logical Pauli
pairs. Finite binary symplectic elimination encodes one unknown
qubit, with the other five logical qubits fixed. Prepare the outside
data slots in a fixed product state. The final decoder uses this
intended virtual complex and one fixed component-filling convention.

Every possible incidence, source sheet, target sheet, and physical
displacement has its own translated factor-disjoint gate layers.
Different candidate templates run in different layers, even when
correct profile values would make them mutually exclusive. Private
classical mask copies avoid shared-control overlaps. Consequently a
malformed profile still defines a valid circuit. Preparations,
resets, record measurements, and discards are unconditional; only
unitaries are programmed. For the correct profile all active data
incidences realize the closed code, while outside data incidences
and corrections are disabled. Its reference circuit therefore
factorizes into the active code processor and an outside channel.
Active check ancillas and records belong to the active processor.

For a fixed sheet extend \eqref{q:fold-map} affinely beyond the active
box and use
\begin{equation}\label{q:orientation}
 i_\varsigma=q_4-q_2,\quad P_{1,\varsigma}=q_1,\quad
 P_{2,\varsigma}=q_2,\quad Q_\varsigma=q_3-q_1.
\end{equation}
Its integral linear part has determinant \(\pm1\). It defines that
sheet's selector orientation. The four selectors share the same
physical innovations; each retains the original single-selector
marginal law.

\subsection{Common potentials and every link type}

Use the potentials \eqref{q:walls} and the backward increments of Table~\ref{q:links}.

For the selector, a parent moves in virtual coordinates by
\((u_1,u_2,u_1,u_2+\delta_i)\). A negative charge chooses outside its
corresponding nonzero replica box; the sum charge chooses a nonzero
box. For the profile only \(q_1,q_3\) move, and the desired profile
is constant on those replica planes. A wrong clean provisional
profile output needs at least \(257\) wrong replicas, hence two
boxes. Its \(320\)-spacing gives the profile row. Syndrome gains
follow from Lemma~\ref{q:sweep}.

Intrinsic overlay events and sampling-register driver errors are named on the
base characteristic. The gather and consult are compressed together
before making a terminal. Thus no uncancelled \(B\)-sized sampling-register link enters a constant that must be fixed before \(B\).

A clean recentred deep query has formal time \(t+T_{\rm hist}-1\) and replica
coordinate \mbox{\(P-(T_{\rm hist}-1)v\)}. The first three co-moving \(q\)-coordinates
agree with the root, whereas its fourth is smaller by
\(T_{\rm hist}-1\). This gives the predictor-entry row. Such a clean entry
traverses all \(T_{\rm hist}\) noiseless formal levels before reaching actual
old data; an internally corrupted lookup goes directly to its local
source instead. Allocate the entry's \(F_5\) debit over those next
\(T_{\rm hist}\) edges. Each loses less than one unit, so their good gain is
at least \(87\). Successive entered blocks on an arm are disjoint.
An arm beginning inside a block owes no earlier entry debit; a
fork-first bad edge has its loss ceiling enlarged by one. This is
a reassignment of actual increments, leaving telescoping unchanged.
The common good margin is still \(42\).

A wrong programmed gate has bounded physical support. Its support
in each sheet lies within a fixed virtual radius of that sheet's
affine image of its physical origin, even when different sheet
images are far apart. Trace a syndrome arm to the last relevant
wrong gate in its fixed ideal-hardware cone and attach its per-sheet
link. A wrong clean provisional profile use enters its profile
error derivation; a dirty mask use terminates at its local source.
A selector-induced erasure enters that selector. After entering a
classical derivation an arm never returns to syndrome types. Hence
these quantum-to-classical and terminal links occur at most once
per arm. Repeated classical predictor entries are paid by the
separate block argument above.

Apply Lemma~\ref{q:frontier} with five charges after this amortization.
The source labels are
\begin{equation}\label{q:labels}
 \lambda_\varsigma=
 (\varsigma,i_\varsigma-t,P_{1,\varsigma}+58t,
      P_{2,\varsigma}+58t,Q_\varsigma,\lfloor t/W\rfloor).
\end{equation}
They recover all four co-moving virtual coordinates; in particular
\(q_4+57t=(i_\varsigma-t)+(P_{2,\varsigma}+58t)\).

\subsection{Local charts and spatial certificates}

At a closing fold translate one sheet's whole coordinate chart by
\(\pm2m\), once per folded axis for the entire certificate and all
its classical ancestors. At a middle fold the default coordinates
already give adjacent virtual values. The two translations commute.
Fix the ordinary period\nobreakdash-\(L\) lifts in directions one and three in
the same way. A local virtual chart narrower than
\(\min(L,2m)/100\) meets at most one fold per axis. Within one sheet
it has one affine physical inverse. Ancestors leaving the active
box continue this same affine extension. Source names and reference
phase use the same translation, so a typed label has one image.

A joint-field root has the witness from Lemma~\ref{q:frontier};
zero sampling-register corridors collapse to fixed role/bin links. Its projected
graph has bounded degree fixed before the sampling-register length, alphabet,
and on-site circuit depth are selected. The following proof uses
the joint typed-label law with base \(\gamma_{\rm fold}=\gamma_{\rm src}^{1/4}\).
\begin{lemma}[Spatial diameter tail]\label{q:spatial}
For sufficiently small constants chosen as in the controller
construction, the probability that the set of joint-field roots (the
vertices that carry a grouped witness of Lemma~\ref{q:frontier}, the
tracked roots of the proof of Lemma~\ref{q:universal}) has a
prescribed-root spatial path spanning distance \(R\), at one time or
in two adjacent times, is at most \(A_{\rm sp}e^{-a_{\rm sp}R}\). The constants
are computable and uniform over the labels \(j\) with \(V(j+1)\uparrow\), within the
finite legal chart on which the source law applies.
\end{lemma}
\par\noindent\emph{Proof idea.} Cover the path by balls containing complete witnesses and retain a disjoint family. Connecting their enlarged balls gives a spanning certificate whose length is paid for by distinct marked sources.

\begin{proof}
For every joint-field root on a finite path choose the connected
witness of Lemma~\ref{q:frontier}. Its transverse characteristic
projection, including all its terminal labels, lies in a ball centred
at the root with radius at most \(C_1g\). Select these balls greedily
in decreasing radius, keeping a ball if it is disjoint from all
previously kept balls. Every discarded ball is contained in the
threefold enlargement of an intersecting retained ball of at least
its radius. Use \(4+2R_{\rm blk}+1\)-fold enlargements, with
\(R_{\rm blk}\) from Lemma~\ref{q:round-shadow}, so the same covering
also includes the intermediate-stage shadow links. These enlargements cover the polygonal
path. A simple chain of these enlarged balls joins its two endpoints.

The terminal-label sets of the selected witnesses are disjoint:
a common label has one transverse characteristic position and would
belong to two disjoint original balls. This implication concerns
labels; the underlying statistical tests may still overlap.
Connect successive selected centres by coordinate paths, and attach
the two endpoint anchors. Consecutive centres are separated by at
most \(4+2R_{\rm blk}+1\) times the sum of their radii. Since the chain is simple,
each radius is charged a bounded number of times. Thus the union is
a connected certificate containing the prescribed path root, with
\(e\le C_2G\) edges and \(G\) distinct true labels. Its span gives
\(e\ge R/C_3\), where \(C_3\) bounds the graph's spatial edge range.
For the two-time version move the roots to a common characteristic
time by bounded links first.

A doubled-edge traversal of a connected \(e\)-edge graph has length
\(2e\). Its bounded degree, terminal subset, and finite role labels
bound the number of rooted annotated certificates by \(K_{\rm an}^e\) for a
computable \(K_{\rm an}\). A fixed certificate has probability at most
\(\gamma_{\rm fold}^G\le\gamma_{\rm fold}^{e/C_2}\). Therefore the required probability is
at most
\begin{equation}\label{q:animal}
 \sum_{e\ge R/C_3}
       \bigl(K_{\rm an}\gamma_{\rm fold}^{1/C_2}\bigr)^e.
\end{equation}
The construction first fixes these geometric constants and then
chooses its source parameters so that the base is at most \(1/4\).
This gives the claimed exponential tail.
\end{proof}

\subsection{Linear horizon on physical tori}

For the explicit maps write
\[
 F_{n,L,t}(\eps)=F(\mathcal E_{n,L},\mathcal D_{n,L,t};L,t,\eps).
\]

\begin{proposition}[Linear storage]\label{q:linear}
For the independent single-qubit realization of the folded circuit of
Appendices~\ref{sec:fold-controls} and~\ref{sec:folded}, built with a
machine wrapper \(V\) whose label \(j\) simulates \(V(j+1)\) and with
the windows \(I_j\) of the proof of Proposition~\ref{p:coexistence},
there are a computable rational \(c>0\), computable positive \(A_{\rm lin},a_{\rm lin}\),
an integer \(b_{\rm lin}\) and a total computable threshold \(L_0(j)\) such that,
for each label \(j\) with \(V(j+1)\uparrow\) and all \(L\ge L_0(j)\), the explicit
rate-independent encoding and intended-fold recovery have fidelity
\[
 F_{j,L,t}(\eps)\ge1-A_{\rm lin}L^{b_{\rm lin}}e^{-a_{\rm lin}L}
 \ge15/16\qquad(0\le t\le\lfloor cL\rfloor,\ \eps\in I_j).
\]
The constants \(c,A_{\rm lin},a_{\rm lin},b_{\rm lin}\) are independent of \(j\).
The same initial encoder works for all these times and rates.
\end{proposition}
\par\noindent\emph{Proof idea.} Choose the horizon so every relevant cone fits into a small lifted chart. Pay for repeated physical faults across the sheets, and recover using a single decoder whenever there is no dangerous certificate.

\begin{proof}
After fixing the circuit, compute a conservative per-round
velocity and source-support bound \(V_*\), including physical sampling lanes, characteristic drift, every fixed hardware role, and both
old/new times. Fix an additive halo \(A_*\). Choose a rational \(c\)
small enough that all cones through \(cL\), clipped containing balls,
and \(4+2R_{\rm blk}+1\)-fold enlargements have width less than \(m/100\).
For example a sufficiently reduced rational of order
\(1/[10^{10}(V_*+1)]\) works after an effective enlargement of
\(L_0\), since \(m\ge L/128\) for all large \(L\).

Each sheet's active logical \(i_\varsigma\)-interval has width at most
\(2m\). Above the size cutoff its chart seam is placed a distance
at least \(L/4\) away, and Lemma~\ref{c:initialization} covers the whole
required initial history, future and collar cone by a finite cover search. Each selector
may use its own phase and hidden lift. Across the fold, only the live selector output enables storage. The low/high initializer supplies positive-time
source supports, even when \(W\) exceeds \(cL\).

Stop a first dangerous virtual path at its first exit from a small
box at scale \mbox{\(r_{\rm d}=\Theta(m)=\Theta(L)\)}. Its two-time roots,
source supports and enlarged witness balls lie in one
small virtual chart. Use the one-deck-translation-per-sheet
convention above. Within a sheet its affine physical inverse is
injective on these supports. The fourth root in H\"older's inequality accounts for coins repeated across sheets. For each source event use its actual
restricted positive-time atoms; extend missing exterior comparison
atoms independently.

The animal count \eqref{q:animal}, with
\(\gamma_{\rm fold}=\gamma_{\rm src}^{1/4}\), bounds existence of any dangerous
certificate by an exponential in its virtual span. Polynomial
unions over physical roots, \(O(L)\) times, and fixed hardware roles
give
\(\Pr(\mathcal G^c)\le A_{\rm lin}L^{b_{\rm lin}}e^{-a_{\rm lin}L}\).
By Lemma~\ref{q:universal}, on \(\mathcal G\) one intended-fold
recovery corrects every permitted Pauli history. Conditional on the
exogenous classical history, actual wrong programmed unitaries have
exactly the per-sheet operator supports in its catalogue. Physical
noise has its Pauli/replacement unravelling.
Lemma~\ref{q:span} therefore gives exact recovery for the actual
coherent circuit, including any outside correlations it created.
Bad histories have nonnegative entanglement fidelity, giving the bound.

The source parameters and all geometric constants are computable,
so exponential decay supplies a total computable threshold with
failure probability at most \(1/16\) for all larger sizes.
Finite CSS encoding, component filling, and finite hidden-cover
search with fixed fallback are total, independent of the actual
rate and halting answer. 
\end{proof}

\needspace{6\baselineskip}
\section{Reproducibility}\label{app:reproducibility}
Archive and Lean sources:
\href{https://doi.org/10.5281/zenodo.24254845}{doi:10.5281/zenodo.24254845}.

Fix the universal machine and compiler of Sections~\ref{intro:question} and~\ref{s:model}
and the classical imports in Appendix~\ref{app:imports}. For Theorem~\ref{thm:main},
start with the controller of Section~\ref{sec:controller}, use the
compiler and constant choices of Section~\ref{sec:compilation}, and
construct the quantum schedule of Section~\ref{sec:quantum}. For
Theorem~\ref{thm:exponential}, use the affine layout and static chart of
Section~\ref{sec:exponential}, then the parameter order of
Section~\ref{v:parameters}. For Theorem~\ref{thm:four-dimensional}, use the folded geometry of
Section~\ref{f4:geometry-section} and the parameter order of
Section~\ref{f4:constants}. For Theorem~\ref{q1:main} and Corollary~\ref{q1:memory}, use the two classical lanes of
Section~\ref{q1:model}, the quantum schedule of Section~\ref{q1:compiler}, the selector of
Section~\ref{q1:selector} and the order of constants of Lemma~\ref{q1:l16}. The recipe determines the circuits once
the choices allowed by the stated construction bounds are fixed.
Every realization satisfying those bounds obeys the corresponding theorem, with its own constants and windows.

From a circuit choice and label \(n\), compute the physical rate window
by the rational inner-endpoint prescriptions of
Sections~\ref{s:parameters} and~\ref{v:parameters}, Lemma~\ref{f4:encoder}
or Section~\ref{q1:parameters}, as appropriate. From \((n,L)\), the finite initialization searches of
Lemmas~\ref{c:initialization}, \ref{v:encoder} and~\ref{f4:encoder}
give the encoder, using the exact code procedures in Lemma~\ref{d:code}
for the concatenated code; on rings, Section~\ref{q1:storage} and
Lemma~\ref{q1:l15} give the encoder. For the coexistence circuit, start with
the wrapped machine, folded track and closed-code track in the proof
of Proposition~\ref{p:coexistence}. That proof specifies the rational
inner physical windows, the closed-code initialization on \(I_0\)
and the folded encoder used for the tail and for linear certification.

For a readout at \((n,L,t)\), Proposition~\ref{d:storage} and the
proof of Theorem~\ref{f4:detailed}(i) prescribe noiseless completion
with the actual in-flight work followed by the fixed bottom-up recovery;
on rings, Section~\ref{q1:storage} and Lemma~\ref{q1:l15} prescribe completion
at a fixed nominal rate followed by the fixed recovery.
Proposition~\ref{v:certification}(ii) fixes the decoder that uses the
syndrome of the intended complex, including lexicographic box, lift and filling choices and
a fixed failure correction. To query the prescribed fidelity at a
rational rate \(q\), use the finite polynomial calculation of
Proposition~\ref{a:rational}, the six-dimensional adaptation in
Proposition~\ref{v:certification}(ii) and, on rings, the calculation in
Section~\ref{q1:effective}.

These reconstructions are algorithms specified in the paper. The archive
provides no programs other than the Lean developments of
Appendix~\ref{app:lean}.

\section{Imported mathematical results}\label{app:imports}
Table~\ref{tab:imports} specifies the Douglas--Mghirbi preprint inputs of the higher-dimensional constructions and their versions: version~1.4.0 of the classical noise diagram \cite{DMNoise2026}, \cite{DMPhase2026} and \cite{DMFixed2026}. Section~\ref{app:imports-1d} lists the inputs used on rings, from version~2.0.0 of the same noise-diagram preprint \cite{DMNoise1D2026}. These preprints do not use any result proved here. Each archive contains the corresponding paper's PDF. The other imported results are listed below with their source editions.

\begin{table}[!htbp]
\caption{Source statements and proofs used in the construction.}\label{tab:imports}
\centering\small
\begin{tabular}{@{}>{\raggedright\arraybackslash}p{.29\linewidth}>{\raggedright\arraybackslash}p{.39\linewidth}>{\raggedright\arraybackslash}p{.24\linewidth}@{}}
\toprule Construction or input & Source statements and proofs & Interface and use here\\\midrule
Rate-selected causal computation & \cite[Proposition~4.3; Lemmas~4.1--4.2, 6.1--6.3, 10.4, 11.2--11.4, 11.6, Section~11.4 and item~3 of the halting proof in Section~11.7]{DMNoise2026}, version~1.4.0 (including the construction in the proof of Lemma~11.2) & Proposition~\ref{c:selector-import}; Section~\ref{sec:controller}\\\addlinespace[3pt]
Finite scaffold encoding and parent rules & \cite[Definitions~3.1, 3.2 and~3.8, Lemma~3.9, and Lemmas~A.1--A.3]{DMPhase2026}, version~1.2.0 & Sections~\ref{sec:controller} and~\ref{sec:ceiling}, Proposition~\ref{u:scaffold-import}\\\addlinespace[3pt]
Static computation and cutoff & \cite[Lemmas~3.2--3.3, 4.1--4.2, 5.1, 10.3, 10.5 and~11.2, Proposition~4.3]{DMNoise2026}, version~1.4.0; the proofs of Lemmas~3.2 and~11.2 and Proposition~4.3 are adapted to the fixed phase & Proposition~\ref{v:static-import} and Lemmas~\ref{v:encoder}, \ref{v:ancestor} and~\ref{f4:encoder}\\\addlinespace[3pt]
Replica-majority construction and contour method & \cite[Theorem~2.1; Lemmas~12.3--12.4]{DMFixed2026}, version~3.0.0 & Proposition~\ref{c:selector-import}; Lemma~\ref{v:density} gives the two-appearance proof here\\\bottomrule
\end{tabular}
\end{table}

The finite chart cover in Section~\ref{c:chart} uses the northeast-deterministic presentation of the limit set of a \(2\times2\) substitution from \cite[Theorem~3]{LGO2012} (Theorem~4 gives the four-way presentation), with the theorem locator in \href{https://hal.science/hal-00661887v1}{HAL:hal-00661887v1}.

Lemma~\ref{f4:raw} applies the contour existence theorem of \cite[Theorem~22]{SST2026} to the three paired-box types. The source's Definitions~15\nobreakdash--17 and~20\nobreakdash--21 specify the graphs and their embeddings, (2.14) defines the associated monotone automaton, and Lemma~13 identifies the unique trajectory as the maximal one. The proof uses the finite explanation graph from Proposition~31 and Theorem~33, the zero-sum identity of Lemma~41 and the edge counts from the proofs of Lemmas~38 and~40. The locators use the journal numbering and also apply to \href{https://arxiv.org/abs/2202.10999v3}{arXiv:2202.10999v3}. For the marked spanning tree it uses \mbox{\cite[Definition~2.4, Lemma~2.5 and Theorem~2 (Tree Separator)]{Gacs2021}}, \href{https://arxiv.org/abs/2105.05968v1}{arXiv:2105.05968v1}, to extract a bounded-size subtree with many original sink marks.

\subsection{One-dimensional classical inputs}\label{app:imports-1d}
The one-dimensional construction uses version~2.0.0 of
\cite{DMNoise1D2026}, with the numbering of that version.
Definition~\ref*{cl:def:sparse} and the paragraph preceding it give sparsity and
the convention that restricted error sets are empty outside the restriction.
Lemma~\ref*{cl:lem:sparse-witness} supplies only the deterministic mesh
implication; Lemmas~\ref{q1:l1} and~\ref{q1:l12} prove the dependent-key
estimates here. R1 uses the one-point variant of the serialization in
Section~\ref*{cl:sec:packets} specified and proved in
Lemma~\ref{q1:l9}: scalar bits occupy the low word bits of separate
live letters instead of scalar-bit tags, with all other word bits zero;
typed low and high word halves and fixed endpoint padding are retained.
The encoders, parsers and represented evaluators use that same layout.
Lemma~\ref{q1:l9} checks every classical use of it, including
Proposition~\ref*{cl:prop:organization},
Definition~\ref*{cl:def:observations} and its lookahead bound,
Lemmas~\ref*{cl:lem:endpoint} and~\ref*{cl:lem:two-period},
the clean identity of Lemma~\ref*{cl:lem:clean}, and the boundary
construction in the proof of Lemma~\ref*{cl:lem:past}.
In the structural comparison the two-period tests have an empty top
exception, positive served inputs in the specified common tribe and
live prescribed outputs. The proof of
Lemma~\ref*{cl:lem:clock-witness} supplies the initialized nested-test
cover for R2. The proof of Lemma~\ref*{cl:lem:past} also supplies R2's
stationary-helper negative-time comparison.
Lemmas~\ref*{cl:lem:stripes} and~\ref*{cl:lem:receiver-mixing} supply
tunable all-start short and long noise-only tests and whole-helper
coalescence at sufficiently small fixed rate.
Lemma~\ref*{cl:lem:joint} supplies the two-scale decomposition geometry
and short-before-long order, while Lemma~\ref{q1:l12} re-proves the
probability estimates. Lemmas~\ref*{cl:lem:locking},
\ref*{cl:lem:cap-cut} and~\ref*{cl:lem:ascent} supply the immutable
literal lock and arbitrary-row finite halting cap and ascent.
Sections~\ref*{cl:sec:native}--\ref*{cl:sec:embedding},
\ref*{cl:sec:receiver}--\ref*{cl:sec:joint} and
Appendix~\ref*{cl:sec:constants} specify the native algorithms,
geometry and parameter order used in these proofs. The finite compiler
anomalies, rate-band envelope, arbitrary moments, finite-ring chart
and full-register channel estimates are proved in
Sections~\ref{q1:clock}--\ref{q1:effective}.
The quantum inputs are this paper's Lemmas~\ref{d:code}--\ref{d:reduction}, used parametrically in Lemma~\ref{q1:l4}, and Lemmas~\ref{f:factorization}, \ref{f:domination}, \ref{f:iteration}, \ref{f:bell} and Proposition~\ref{f:extension}. Remark~\ref{rem:forgetting-interface} identifies the hypotheses of the latter channel argument; Section~\ref{q1:forgetting} supplies and checks them for the circuit on rings. The exact finite-fidelity and \(\Pi^0_1\) argument of Section~\ref{q1:effective} follows Proposition~\ref{a:rational} and Theorem~\ref{a:complete}.

\section{Lean formalization}\label{app:lean}
The Lean 4 library formalizes Lemmas~\ref{f:iteration} and~\ref{f:bell},
conditional forms of the full-replacement bound \eqref{eq:finite-reset}
and the associated lifetime estimates, a conditional abstract form of
Theorem~\ref{f:forgetting}, the overlap step in the proof of
Lemma~\ref{q1:l18}, and the conditional certification of
Theorem~\ref{lean:conditional}. Table~\ref{tab:lean} gives the declarations.
The one-dimensional development formalizes Lemmas~\ref{q1:l1} and
\ref{q1:l1.1}, the arithmetic of Lemma~\ref{q1:l18} with its fidelity
inputs supplied as hypotheses, and selected exponent calculations
from Lemmas~\ref{q1:l11} and~\ref{q1:l12}.
Theorems~\ref{thm:exponential}, \ref{thm:four-dimensional}, \ref{thm:main} and~\ref{q1:main}, Proposition~\ref{f:extension}, the constructions, the noise analysis and the compilation are not formalized. Theorem~\ref{f:forgetting} is checked from the extension onward: its Lean form assumes the two-output extension of Proposition~\ref{f:extension} and the full-replacement decomposition on the small tori as explicit hypotheses.
Theorem~\ref{lean:conditional} gives the conditional certification statement and proof.

\begin{table}[!htbp]
\caption{The Lean statements.}\label{tab:lean}
\small\centering
\begin{tabular}{@{}>{\raggedright\arraybackslash}p{0.37\textwidth}>{\raggedright\arraybackslash}p{0.59\textwidth}@{}}
\toprule
Paper statement & Lean declaration\\
\midrule
Lemma~\ref{f:iteration}, for every positive \(j\)
& \texttt{QuantumScope.iteration\_extension}\\\addlinespace[3pt]
Lemmas~\ref{f:iteration} and~\ref{f:bell}, the bound \(1/2+2^{-j-1}\)
& \texttt{QuantumScope.iterated\_cloning\_bound}\\\addlinespace[3pt]
Lemma~\ref{f:bell}, for every \(k>0\), CPTP encoder and common decoder
& \texttt{QuantumScope.bell\_sharing\_bound}\\\addlinespace[3pt]
Theorem~\ref{f:forgetting}, the conditional abstract form
& \texttt{QuantumScope.uniform\_forgetting}\\\addlinespace[3pt]
Conditional form of \eqref{eq:finite-reset}, for round or layer channels
with the stated replacement decomposition and weight at least \(p\)
& \texttt{QuantumScope.replacement\_ceiling\_pow},
  \texttt{QuantumScope.replacement\_ceiling}\\\addlinespace[3pt]
Conditional layer estimate under the same hypothesis with \(p>0\),
implying \(\tau\le\log(6/5)/p<1/p\) and hence
\eqref{f4:lifetime-upper} and the bound in Section~\ref{q1:effective}
& \texttt{QuantumScope.replacement\_lifetime\_layer}\\\addlinespace[3pt]
The overlap step in the proof of Lemma~\ref{q1:l18}
& \texttt{QuantumScope.good\_event\_fidelity\_union}\\\addlinespace[3pt]
Theorem~\ref{lean:conditional}, the conditional abstract form
of Theorem~\ref{a:complete}
& \texttt{CertificationScope.abstract\_window\_certification}\\
\bottomrule
\end{tabular}
\end{table}

The channel definitions use finite complex matrices and complex-linear
maps. Complete positivity is positivity after amplification by every
finite ancillary dimension. Trace preservation holds for every matrix.
Every partial-trace marginal of the extension equals the given channel
on all inputs. Entanglement fidelity uses the normalized Bell projector
onto \((|00\rangle+|11\rangle)/\sqrt2\) and leaves the reference qubit
untouched. The conditional certification theorem assumes the physical storage and forgetting results as explicit hypotheses. The conditional forgetting theorem likewise assumes the extension and the replacement decomposition; a full replacement is a convex decomposition \(\rho\mapsto q\Tr(\rho)\sigma+(1-q)\Phi(\rho)\) of the layer channel with \(\Phi\) CPTP.

\paragraph{Library and build.}
In the archive, the \texttt{lean/} subdirectory contains \texttt{lean-toolchain}, \texttt{lake-manifest.json} and the \texttt{QuantumStorage} library. \texttt{Challenge.lean} specifies the fourteen theorems, and \texttt{Solution.lean} imports the proved declarations.
The toolchain is Lean \texttt{4.35.0}. The Mathlib commit is
\begin{verbatim}
56a005855ea17e0b0958f73ec801ae140b0a1679
\end{verbatim}
\needspace{6\baselineskip}
With the pinned toolchain installed (elan installs it from \texttt{lean-toolchain}), run these commands from \texttt{lean/} in a POSIX shell. The optional cache step fetches prebuilt dependencies. Without it,
\texttt{lake build} compiles Mathlib from source, which can take hours.
\begin{verbatim}
export LEAN_NUM_THREADS=1
lake exe cache get
lake build
lake env lean -j1 -M5000 Solution.lean
\end{verbatim}
Each of the fourteen axiom sets printed by the last command consists exactly of \texttt{propext}, \texttt{Classical.choice} and \texttt{Quot.sound}. The expected output, wrapped to fit, is
{\small
\begin{verbatim}
'CertificationScope.abstract_window_certification' depends on axioms:
  [propext, Classical.choice, Quot.sound]
'QuantumScope.bell_sharing_bound' depends on axioms:
  [propext, Classical.choice, Quot.sound]
'QuantumScope.equalMarginalExtension_comp' depends on axioms:
  [propext, Classical.choice, Quot.sound]
'QuantumScope.iteration_extension' depends on axioms:
  [propext, Classical.choice, Quot.sound]
'QuantumScope.iterated_cloning_bound' depends on axioms:
  [propext, Classical.choice, Quot.sound]
'QuantumScope.replacement_fidelity' depends on axioms:
  [propext, Classical.choice, Quot.sound]
'QuantumScope.replacement_ceiling' depends on axioms:
  [propext, Classical.choice, Quot.sound]
'QuantumScope.replacement_ceiling_pow' depends on axioms:
  [propext, Classical.choice, Quot.sound]
'QuantumScope.replacement_lifetime' depends on axioms:
  [propext, Classical.choice, Quot.sound]
'QuantumScope.replacement_lifetime_layer' depends on axioms:
  [propext, Classical.choice, Quot.sound]
'QuantumScope.replacement_eventually' depends on axioms:
  [propext, Classical.choice, Quot.sound]
'QuantumScope.uniform_forgetting' depends on axioms:
  [propext, Classical.choice, Quot.sound]
'QuantumScope.good_event_fidelity' depends on axioms:
  [propext, Classical.choice, Quot.sound]
'QuantumScope.good_event_fidelity_union' depends on axioms:
  [propext, Classical.choice, Quot.sound]
\end{verbatim}
}
Any other axiom, including \texttt{sorryAx}, means the check failed.
\texttt{-M5000} sets the Lean memory budget to 5000 MB for that command.

\paragraph{Checking the statements.}
With the same pinned toolchain and Mathlib, append these lines to
separate copies of \texttt{Challenge.lean} and \texttt{Solution.lean}
in the \texttt{lean/} directory.
\begin{verbatim}
#check @QuantumScope.bell_sharing_bound
#check @CertificationScope.abstract_window_certification
#check @QuantumScope.equalMarginalExtension_comp
#check @QuantumScope.iteration_extension
#check @QuantumScope.iterated_cloning_bound
#check @QuantumScope.replacement_fidelity
#check @QuantumScope.replacement_ceiling
#check @QuantumScope.replacement_ceiling_pow
#check @QuantumScope.replacement_lifetime
#check @QuantumScope.replacement_lifetime_layer
#check @QuantumScope.replacement_eventually
#check @QuantumScope.uniform_forgetting
#check @QuantumScope.good_event_fidelity
#check @QuantumScope.good_event_fidelity_union
\end{verbatim}
Run each copy separately with \texttt{lake env lean -j1 -M5000},
followed by the copy's filename. Compare the fourteen printed declaration
types with the statements in \texttt{Challenge.lean} and the paper.
The challenge file has fourteen intentional \texttt{sorry} bodies to expose
the statements. The proof and axiom checks use \texttt{Solution.lean}. The archive's \texttt{lean/README.md} describes a mechanical comparison of the proved statements, with every definition they use, against \texttt{Challenge.lean}.

\paragraph{The one-dimensional sparsity estimate.}
A second development, in the archive's \texttt{lean-1d/} subdirectory,
formalizes Lemma~\ref{q1:l1} and the classical inputs it uses from
\cite{DMNoise1D2026}. It verifies the sparsity definition
\cite[Definition~\ref*{cl:def:sparse}]{DMNoise1D2026} and its basic
properties, the disjoint-child witness construction of
\cite[Lemma~\ref*{cl:lem:sparse-witness}]{DMNoise1D2026} with its finite
tree counts, and the independent-flag probability bounds of that lemma.
For independent elementary keys with possibly unequal rates at most
\(\eps\), finite stencils of at most \(s_R\) keys and chosen integer radius
bounds, it proves the bound \(z_{R,j}(\eps)\) of Lemma~\ref{q1:l1} with the
displayed constants, and \(C_R \eps^{c_R}\le 1/2\) for
\(0\le \eps\le \eps_{R0}\). It also proves Lemma~\ref{q1:l1.1} with integer primitive-stencil
radius bounds: its explicit cutoff \(\eps_R(d,\delta)\) is positive,
is computable as a rational from \(\lceil d\rceil\) and \(\delta\),
and makes the clock moment summable with sum at most \(\delta\).
It proves the recursion and horizon arithmetic of Lemma~\ref{q1:l18}.
Its fidelity calculation assumes the failure probabilities, the union
bound and the overlap inequality of that proof. The overlap inequality
is established in the companion \texttt{lean/} library. The
one-dimensional development also verifies selected identities and
inequalities in the exponent calculations of Lemmas~\ref{q1:l11}
and~\ref{q1:l12}, as listed in \texttt{lean-1d/README.md}.
Choosing integer radius bounds when the final
circuit is enumerated, which flooring the physical coordinates permits
without changing the constants, and identifying the physical noise with
the product-key model are application steps, not formalized results.
The development also verifies the support-independence and
bounded-overlap estimates underlying
\cite[Lemma~\ref*{cl:lem:joint}]{DMNoise1D2026}, the raw-flag
nested-test sum in its first conclusion, and arithmetic parameter
inequalities of \cite{DMNoise1D2026} listed in \texttt{lean-1d/README.md}. It does not formalize
Lemma~\ref{q1:l1.2}, the probability estimates of Lemmas~\ref{q1:l12}
and~\ref{q1:l18}, the other lemmas of
Sections~\ref{q1:model}--\ref{q1:effective}, Theorem~\ref{q1:main}, the
automaton or the physical compilation. Its theorems use only the axioms
\texttt{propext}, \texttt{Classical.choice} and \texttt{Quot.sound}. The toolchain and
Mathlib commit are those above; build and check instructions and a
statement-by-statement comparison with the printed text are in
\texttt{lean-1d/README.md}.

\end{document}